\documentclass{aa}
\usepackage[varg]{txfonts}

\usepackage{graphicx}
\usepackage{natbib}
\bibpunct{(}{)}{;}{a}{}{,} % to follow the A&A style
\usepackage{color}

\newcommand{\rev}[1]{{#1}}

\newcommand{\lya}{\mbox{Ly$\alpha$}}
\newcommand{\sublya}{\mbox{\scriptsize{Ly$\alpha$}}}

\newcommand{\flcgs}{\mbox{erg s$^{-1}$ cm$^{-2}$}}
\newcommand{\sbl}{\mbox{erg s$^{-1}$ cm$^{-2}$ arcsec$^{-2}$}}
\newcommand{\rlyah}{\mbox{$r_{\mathrm{s,h}}$}}

\begin{document}

\title{Extended Lyman $\alpha$ haloes around individual high-redshift galaxies revealed by MUSE}

\author{L. Wisotzki\inst{1} \and
        R. Bacon\inst{2} \and
        J. Blaizot\inst{2} \and
        J. Brinchmann\inst{3,4} \and
        E. C. Herenz\inst{1} \and
        J. Schaye\inst{3} \and
        N. Bouch\'{e}\inst{5} \and
        S. Cantalupo\inst{6} \and
        T. Contini\inst{7,8} \and
        C. M. Carollo\inst{6} \and
        J. Caruana\inst{1} \and
        J.-B. Courbot\inst{9,2} \and
        E. Emsellem\inst{10,2} \and
        S. Kamann\inst{11} \and
        J. Kerutt\inst{1} \and
        F. Leclercq\inst{2} \and
        S. J. Lilly\inst{6} \and
        V. Patr\'{i}cio\inst{2} \and
        C. Sandin\inst{1} \and
        M. Steinmetz\inst{1} \and
        L. A. Straka\inst{3} \and
        T. Urrutia\inst{1} \and
        A. Verhamme\inst{2,12} \and
        P. M. Weilbacher\inst{1} \and
        M. Wendt\inst{13,1}
        }

\titlerunning{Lyman $\alpha$ haloes around high-redshifts galaxies}
\authorrunning{L. Wisotzki et al.}

\date{Received <date>; accepted <date>}

\institute{%
	Leibniz-Institut f\"{u}r Astrophysik Potsdam (AIP), An der Sternwarte 16, 14482 Potsdam, Germany \\
	e-mail: \texttt{lwisotzki@aip.de}
	\and
	CRAL, Observatoire de Lyon, CNRS, Universit\'{e} Lyon 1, 9 avenue Ch. Andr\'{e}, 69561 Saint Genis-Laval Cedex, France
	\and
	Leiden Observatory, Leiden University, PO Box 9513, 2300 RA Leiden, The Netherlands
	\and
	Instituto de Astrof\'{i}sica e Ci{\^e}ncias do Espa\c{c}o, Universidade do Porto, CAUP, Rua das Estrelas, 4150-762 Porto, Portugal
	\and
        Institut de Recherche en Astrophysique et Plan\'{e}tologie (IRAP), CNRS, 9 Avenue Colonel Roche, 31400 Toulouse, France
	\and
	ETH Z\"{u}rich, Institute of Astronomy, Wolfgang-Pauli-Str. 27, 8093 Z\"{u}rich, Switzerland
	\and
        Institut de Recherche en Astrophysique et Plan\'{e}tologie (IRAP), CNRS, 14 avenue \'{E}douard Belin, 31400 Toulouse, France
	\and
	Universit\'{e} de Toulouse, UPS-OMP, 31400 Toulouse, France
	\and
	ICube, Universit\'{e} de Strasbourg, CNRS, 67412 Illkirch, France
	\and
	ESO, European Southern Observatory, Karl-Schwarzschild Str. 2, 85748 Garching bei M\"{u}nchen, Germany
	\and
	 Institut f\"{u}r Astrophysik, Universit\"{a}t G\"{o}ttingen, Friedrich-Hund-Platz 1, 37077 G\"{o}ttingen, Germany
	 \and
	 Geneva Observatory, University of Geneva, 51 Chemin des Maillettes, 1290 Versoix, Switzerland
	 \and
	Institut f\"ur Physik und Astronomie, Universit\"at Potsdam, Haus 28, 	Karl-Liebknecht-Str. 24/25, 14476 Golm (Potsdam), Germany
	}

\abstract{%
We report the detection of extended \lya\ emission around individual star-forming galaxies at redshifts $z = 3$--6 in an ultradeep exposure of the Hubble Deep Field South obtained with MUSE on the ESO-VLT. The data reach a limiting surface brightness ($1\sigma$) of $\sim$$1\times 10^{-19}$\sbl\ in azimuthally averaged radial profiles, an order of magnitude improvement over previous narrowband imaging. Our sample consists of 26 spectroscopically confirmed \lya-emitting, but mostly continuum-faint ($m_\mathrm{AB} \gtrsim 27$) galaxies. In most objects the \lya\ emission is considerably more extended than the UV continuum light. While 5 of the faintest galaxies in the sample show no significantly detected \lya\ haloes, the derived upper limits suggest that this is just due to insufficient S/N. \lya\ haloes therefore appear to be (nearly) ubiquitous even for low-mass ($\sim$$10^8$--$10^9\:M_\odot$) star-forming galaxies at $z>3$. We decompose the \lya\ emission of each object into a compact `continuum-like' and an extended halo component, and infer sizes and luminosities of the haloes. The extended \lya\ emission approximately follows an exponential surface brightness distribution with a scale length of a few kpc. While these haloes are thus quite modest in terms of their absolute sizes, they are larger by a factor of 5--15 than the corresponding rest-frame UV continuum sources as seen by HST. They are also much more extended, by a factor $\sim$5, than \lya\ haloes around low-redshift star-forming galaxies. Between $\sim$40\% and $\gtrsim$90\% of the observed \lya\ flux comes from the extended halo component, with no obvious correlation of this fraction with either the absolute or the relative size of the \lya\ halo. Our observations provide direct insights into the spatial distribution of at least partly neutral gas residing in the circumgalactic medium of low to intermediate mass galaxies at $z > 3$. 
}

\keywords{Galaxies: high-redshift - galaxies: evolution - galaxies: formation - cosmology: observations - intergalactic medium}

\maketitle

%%%%%%%%%%%%%%%%%%%%%%%%%%%%%%%%%%%%%%%%%%%%%%%%%%%%%%%%%%%%%%%%%%%%%%%%%%%%%%%%

%\input{LAE-ext_1-introduction}

\section{Introduction}

A major observational challenge in the investigation of high-redshift galaxies lies in determining the spatial distribution of their gaseous components, especially in relation to the already assembled stellar aggregates. Many of the established tracers of neutral and ionized gas in and around low-$z$ galaxies are unavailable at high redshifts, such as \ion{H}{i} 21~cm emission \citep[e.g.][]{Obreschkow:2011ee}, or hard to come by with ground-based observations, such as H$\alpha$ recombination radiation for $z\ga 3$. While absorption lines in the spectra of background sources are very sensitive to even very low column densities of both \ion{H}{i} and metals in the vicinities of galaxies \citep[e.g.][]{Chen:2001hj,Steidel:2010go,Turner:2014gf}, they cannot provide spatially resolved information for individual objects. Among the observable gas tracers in high-$z$ galaxies, the \ion{H}{i} \lya\ emission line is copiously produced in star-forming galaxies and has the advantage of being accessible to ground-based telescopes over a broad redshift range. However, due to the resonant nature of the transition, \lya\ photons are prone to scattering by hydrogen atoms, and the spatial distribution of observed \lya\ emission does not necessarily reflect the regions of its origin. The propagation of \lya\ photons through the interstellar and circumgalactic medium of a galaxy (and into intergalactic space) is a very complex problem with many uncertainties. Theoretical studies have shown that random walks over several kpc are possible until such photons escape \citep{Laursen:2007kl,Zheng:2011kz,Dijkstra:2012ju,Lake:2015vh}. These large path lengths also greatly enhance the chances of photon destruction due to absorption by dust, although the magnitude of this effect may depend on details of the clumping and the geometry of the gas distribution \citep{Neufeld:1991ir,Laursen:2013bd}. The \lya\ escape fractions estimated from comparing \lya\ and UV continuum or H$\alpha$ luminosities are uncertain, but on average they are probably much below unity at least for relatively luminous star-forming galaxies \citep[e.g.][]{Hayes:2010kt,Blanc:2011fp}. Furthermore, models including 3D radiative transfer of \lya\ through an inhomogeneous medium suggest that for any given galaxy the escape fraction may vary strongly with the viewing direction \citep{Laursen:2007kl,Verhamme:2012kb,Behrens:2014ee,Zheng:2014ix}.

Clues to the physical conditions that govern the escape of \lya\ photons can be found in nearby galaxies by comparing \lya\ images obtained by UV satellites with the distribution of starlight and other emission lines. After the first successful \lya\ maps were obtained \citep{Kunth:2003bq,Hayes:2005ew,Ostlin:2009jp}, recently the LARS collaboration \citep[`\lya\ Reference Sample':][]{Ostlin:2014bs, Hayes:2013jc, Hayes:2014jv} presented observations of several nearby galaxies that \emph{would} be selected as \lya\ emitters (LAEs) if placed at high redshifts. In most of those cases the \lya\ emission is not co-spatial with the star-forming regions or with the H$\alpha$ emission, but surrounds the galaxy in a diffuse halo that is clearly more extended than the stellar UV continuum. These results unambiguously demonstrate the presence of a complex circumgalactic medium around many galaxies, revealed through its \lya\ emission.

Conducting similar studies with high-redshift galaxies, i.e.\ mapping the spatial distribution of \lya\ emission in relation to the starlight, is however extremely difficult, because of limitations in both sensitivity and spatial resolution. 
\rev{Exceptions are the relatively bright and sometimes huge \lya\ nebulae found around high-$z$ radio galaxies and (mostly radio-loud) quasars \citep[e.g.][and references therein]{Heckman:1991ey,VillarMartin:2003fb,Cantalupo:2014ig,Herenz:2015kb}. Since such nebulae are heavily influenced by their central AGN, in terms of an enhanced UV radiation field as well as a possible jet- or outflow-related origin of the circumgalactic material, they must be considered as special cases and not representative for the galaxy population at large.}

\rev{For normal (but star-forming and \lya-emitting) galaxies at $z\ga 2$ it was noted already in some of the first narrowband imaging observations that} the galaxies appeared more extended in \lya\ than in the rest-frame UV continuum \citep{Moller:1998eo,Fynbo:2001io}. \rev{However,} the evidence was marginal, and no attempts to quantify the actual sizes or spatial profiles of the \lya\ emission were made. \citet{Swinbank:2007er} presented a single but convincing case of a gravitationally lensed galaxy at $z=4.9$ with extended \lya\ emission. A significant step forward in terms of statistics was the study by \citet{Rauch:2008jy} who identified 27 faint LAEs between $z\simeq 2.6$ and 3.8 in an ultradeep longslit exposure of 92 hours coadded observing time on the ESO-VLT. They found that the majority of their LAEs had spatial profiles along the slit broader than a reference point source, with `radii' of up to $\sim$4\arcsec\ (30~kpc) at their limiting surface brightness of $1\times 10^{-19}$~\sbl. The uncertainties of their measurements were however considerable, given the low S/N of many of their objects, the inevitable slit losses, and also the lack of deep continuum data. 

In past years, most activities in this field employed imaging with narrowband filters, thus focusing on specific redshifts. \citet{Hayashino:2004ke} discovered LAEs embedded in large-scale extended \lya\ emission and presented a first composite \lya\ image providing clear evidence of excess flux beyond the PSF. \citet{Nilsson:2009ib} compared the widths of a large sample of $z\simeq 2.2$ LAE candidates and noted that they were `generally more extended in the narrowband image than their broad-band counterparts'. \rev{Similar observations with the Hubble Space Telescope (HST) gave mixed results: While \citet{Bond:2010hv} concluded from narrowband imaging of a few bright LAEs at $z\approx 3.1$ that the \lya\ emission regions were compact and largely coincident with the UV continuum sources, \citet{Finkelstein:2011he} reported the detection of spatially resolved \lya\ emission around a $z=4.4$ LAE.}

In order to overcome the surface brightness limitations of the narrowband technique, which rarely reaches below $10^{-18}$~\sbl, \citet{Steidel:2011jk} increased the effective sensitivity by an order of magnitude by stacking the images of 92 relatively bright ($R_\mathrm{AB}\simeq 24.5$) Lyman Break Galaxies (LBGs) at $z\simeq 2.3$--3 in \lya\ narrowband as well as broadband filters. The \lya\ emission in their stacked LBG image extends much beyond the mean size of the UV continuum, out to $\sim$10\arcsec\ (80 proper kpc) at a surface brightness level of $\sim$10$^{-19}$~\sbl. They argued that the observed extended \lya\ emission is mainly powered by star formation inside the galaxy, but then scattered outwards by an extended partly neutral circumgalactic medium (which is also detectable in stacked absorption spectra of galaxy-galaxy pairs; \citealt{Steidel:2010go}).

More recently, other groups adopted the stacking approach \citep{Matsuda:2012fp,Feldmeier:2013fx,Momose:2014fe}. Note that in contrast to \citet{Steidel:2011jk}, the targets in those experiments are mostly classical LAEs, selected by their emission lines and showing much fainter (sometimes undetected) continuum counterparts. \citet{Matsuda:2012fp} resolved extended \lya\ emission around the mean of $\sim$2000 LAEs at $z\simeq 3$ with high significance, but found that the sizes and luminosities of their average \lya\ haloes depend strongly on the richness of the environment; while they could reproduce the Steidel et al.\ results for LBGs residing in putative protoclusters, the \lya\ haloes of field LAEs came out to be much fainter and smaller. \citet{Feldmeier:2013fx} argued that the systematic errors of such stacking experiments were previously underestimated, and in their own data they found only marginal evidence for the \lya\ emission of LAE stacks at $z\simeq 2$--3 to be more extended than the continuum. \citet{Momose:2014fe} followed up on the work by \citet{Matsuda:2012fp} and expanded the dataset to $\sim$4500 LAEs in 5 redshifts slices between 2.2 and 6.6. They concluded that all stacked subsets -- except one at $z=3.7$ -- showed with high significance that LAEs have extended \lya\ haloes, with typical exponential scale lengths of $\sim$5--10~kpc.

In this paper we present new observational data that for the first time reveal the 2-dimensional distribution of spatially extended \lya\ emission around high-redshift galaxies on an individual object-by-object basis. This sensitivity improvement was made possible by the MUSE (Multi-Unit Spectroscopic Explorer) instrument, which we recently commissioned as a new facility instrument at the ESO-VLT \citep{Bacon:2014wp}; here we focus on results from early MUSE observations in the Hubble Deep Field South \citep{Bacon:2015eh}.

The paper is organised as follows: In Sect.~\ref{sec:obs} we describe the observations, the sample construction and the extraction of images used in the study. Section~\ref{sec:prof} presents the examination of the data with respect to the fundamental question of whether or not extended \lya\ emission is detected, including a careful assessment of the error budget. We show that most (if not all) LAEs are surrounded by \lya\ haloes that are considerably more extended than the UV continuum. This is followed in Sect.~\ref{sec:sbmod} by a detailed analysis involving 2-dimensional surface brightness modelling of the resolved \lya\ images. This section is quite technical and may be skipped by readers mainly interested in the results. In Sect.~\ref{sec:sizes} we compare the sizes of our detected \lya\ haloes with other observables and with the results of other studies, whereas in Sect.~\ref{sec:lum} we similarly consider the luminosities of the haloes. We discuss some consequences of our findings in Sect.~\ref{sec:disc} and present our summary and conclusions in Sect.~\ref{sec:concl}. 

All cosmological quantities in this paper are calculated assuming a flat universe with $H_0 = 70$~km s$^{-1}$ Mpc$^{-1}$, $\Omega_\mathrm{m} = 0.3$ and $\Omega_\Lambda = 0.7$. All quoted sizes are expressed as proper transverse distances.

%%%%%%%%%%%%%%%%%%%%%%%%%%%%%%%%%%%%%%%%%%%%%%%%%%%%%%%%%%%%%%%%%%%%%%%%%%%%%%%%

% \input{LAE-ext_2-observations}

\section{Observational data}
\label{sec:obs}

\subsection{Observations and data reduction}
\label{sec:obs-obs}

The data were taken with the MUSE instrument \citep{Bacon:2010jn} at the ESO-VLT between July 25 and August 3, 2014, during the last commissioning run before releasing MUSE to the community. We observed a single $1'\times 1'$ field in the Hubble Deep Field South (HDFS), with a combined on-sky exposure time of 27 hours. The observations and data processing steps are described in greater detail in \citet[][hereafter B2015]{Bacon:2015eh}; here we only very briefly summarise the main aspects. Note that we have released the reduced datacube, the source catalogue as well as an interactive source and spectra browser to the public.%
\footnote{\texttt{http://muse-vlt.eu/science/hdfs-v1-0/}}

The combined MUSE-HDFS dataset consists of 54 individual exposures of 30 minutes each. Between subsequent exposures the instrument adaptor rotation angle was advanced by 90\degr, so that the field was captured at four different position angles, from 0\degr to 270\degr. With this strategy, each position in the mapped field (except for the field centre) falls onto four completely different locations in the instrument, providing a maximum of spatial decorrelation without losses in field of view. Additionally, small dithering offsets (randomly chosen within a dither box of 2\arcsec\ in both $\alpha$ and $\delta$) were applied to each exposure. 

The data reduction followed closely the procedure described in section 3.1 of B2015, but with the following  refinements: (i) The per-slice self-calibration process was improved by weighting the slice flux at each wavelength by the inverse of the corresponding average sky flux. This prevented the additive correction to be overly biased towards longer wavelengths were the sky is much brighter, and thus it made the overall self-calibration more achromatic. (ii) For the sky subtraction we used a revised version of ZAP (Soto et al., in prep.) which incorporates a more sophisticated pre-processing before applying the principal component analysis. More eigenspectra were used to remove the correlated signal, which resulted in lower sky subtraction residuals. (iii) 
The accessible field of view of each exposure was defined in a cleaner way by trimming the field edges.

The individual reduced and registered cubes were coadded into a final datacube of $326 \times 331$ spatial pixels (`spaxels'), each with 3641 spectral pixels ranging from 4750~\AA\ to 9300~\AA. Because of the rotational and translational dithering, spaxels near the field edge received less than the full exposure time; this was recorded in a separate 3-dimensional exposure cube.  Defining the `useful' field of view as the region receiving at least 50\% of the full exposure, an area of exactly $1'\times 1'$ ($300 \times 300$ spaxels at a spatial scale of $0\farcs2 \times 0\farcs2$ per spaxel) was covered. The spectral resolution of the data is $\sim$2.5~\AA\ FWHM, at a sampling of 1.25~\AA\ per spectral pixel. The effective seeing in the combined cube is 0\farcs66 at 7000~\AA\ (FWHM of a Moffat fit to the brightest star in the field), and about 10\% better (worse) at the red (blue) end of the spectral range, respectively. The flux scale established by non-simultaneous observations of spectrophotometric standard stars is consistent with HST photometry of the stars in the HDFS-MUSE field of view to within $\pm 0.05$~mag. While a cube with formally propagated pixel variances was also created in the reduction process, we made no use of this for the current paper, for two reasons: (i) The propagated variances are correlated between adjacent pixels because of the resampling in the cube creation. (ii) Imperfect flat-fielding and in particular sky subtraction produced high-frequency residuals somewhat similar to random noise. Below in Sect.~\ref{sec:prof-err-noise} we describe a self-calibration procedure to determine the `effective noise' in the data including all the relevant effects.

\begin{table*}
\caption[]{Basic sample properties. ID: Running source identifier in the catalogue by B2015. $\alpha_{2000}$, $\delta_{2000}$: coordinates in B2015. $z$: Redshift estimated from peak of \lya\ emission line. $m_{814}$: Continuum AB magnitude in the HST/WFPC2 F814W filter band, taken from the GALFIT models described in Sect.~\ref{sec:prof-cont}. $M_{\mathrm{UV}}$: Absolute UV magnitude. $F_{\mathrm{Ly}\alpha}$: Total \lya\ flux in $10^{-18}$ erg s$^{-1}$ cm$^{-2}$, integrated over an aperture of 3\arcsec\ radius. $\log L$: Decadic logarithm of the \lya\ luminosity in erg s$^{-1}$. EW$_\mathrm{t}$: Total \lya\ rest frame equivalent width in \AA. 
}
\begin{center}
\begin{tabular}{rrrrr@{\hspace{0.2em}$\pm$\hspace{0.2em}}lrr@{\hspace{0.2em}$\pm$\hspace{0.2em}}lrr@{\hspace{0.2em}$\pm$\hspace{0.2em}}l}
\hline\hline\noalign{\smallskip} 
\multicolumn{1}{c}{ID} & \multicolumn{1}{c}{$\alpha_{2000}$} & \multicolumn{1}{c}{$\delta_{2000}$} & \multicolumn{1}{c}{$z$} & \multicolumn{2}{c}{$m_{814}$} & \multicolumn{1}{c}{$-M_{\mathrm{UV}}$} & \multicolumn{2}{c}{$F_{\mathrm{Ly}\alpha}$} & \multicolumn{1}{c}{$\log L_{\mathrm{Ly}\alpha}$} & \multicolumn{2}{c}{EW$_{\mathrm{t}}$}\\ 
\noalign{\smallskip}\hline\noalign{\smallskip} 
 43 & 22 32 52.08 & $-$60 33 42.6 & 3.290 & $24.67$ & $ 0.01$ & $21.02$ & 34.0 &  1.1 & 42.54\hspace{0.4em} &  15.1 &   0.5\\ 
 92 & 22 32 54.72 & $-$60 34 14.1 & 4.580 & $25.76$ & $ 0.02$ & $20.49$ & 22.2 &  1.3 & 42.69\hspace{0.4em} &  35.0 &   2.1\\ 
 95 & 22 32 58.56 & $-$60 34 09.0 & 4.225 & $25.94$ & $ 0.02$ & $20.18$ & 12.7 &  1.4 & 42.37\hspace{0.4em} &  22.1 &   2.4\\ 
 112 & 22 32 57.60 & $-$60 33 48.5 & 3.908 & $26.13$ & $ 0.02$ & $19.86$ & 26.4 &  0.8 & 42.61\hspace{0.4em} &  51.5 &   1.6\\ 
 139 & 22 32 55.44 & $-$60 33 40.2 & 3.349 & $26.58$ & $ 0.02$ & $19.13$ & 19.6 &  0.8 & 42.32\hspace{0.4em} &  51.4 &   2.1\\ 
 181 & 22 32 59.04 & $-$60 33 25.5 & 3.337 & $27.19$ & $ 0.03$ & $18.52$ & 27.1 &  0.7 & 42.45\hspace{0.4em} & 124.6 &   3.4\\ 
 200 & 22 32 56.40 & $-$60 33 22.1 & 3.349 & $27.10$ & $ 0.05$ & $18.61$ &  8.3 &  1.1 & 41.94\hspace{0.4em} &  35.4 &   4.7\\ 
 216 & 22 32 56.64 & $-$60 33 38.5 & 4.017 & $27.25$ & $ 0.04$ & $18.78$ & 12.9 &  1.5 & 42.32\hspace{0.4em} &  71.9 &   8.2\\ 
 232 & 22 32 52.56 & $-$60 33 39.9 & 5.215 & $28.77$ & $ 0.12$ & $17.69$ &  4.3 &  1.3 & 42.12\hspace{0.4em} & 122.7 &  37.9\\ 
 246 & 22 32 56.40 & $-$60 33 30.5 & 5.680 & $27.93$ & $ 0.06$ & $18.68$ & 12.6 &  2.6 & 42.66\hspace{0.4em} & 175.2 &  36.1\\ 
 294 & 22 32 52.80 & $-$60 33 34.9 & 3.992 & $28.82$ & $ 0.13$ & $17.20$ &  7.3 &  0.9 & 42.07\hspace{0.4em} & 172.8 &  20.5\\ 
 308 & 22 32 58.08 & $-$60 33 42.3 & 4.018 & $27.70$ & $ 0.09$ & $18.33$ &  7.6 &  1.5 & 42.09\hspace{0.4em} &  64.8 &  12.4\\ 
 311 & 22 32 57.12 & $-$60 33 51.7 & 3.888 & $28.44$ & $ 0.15$ & $17.53$ &  5.0 &  1.0 & 41.88\hspace{0.4em} &  82.2 &  16.1\\ 
 325 & 22 32 52.32 & $-$60 33 46.7 & 4.701 & $27.91$ & $ 0.15$ & $18.39$ & 11.5 &  0.8 & 42.43\hspace{0.4em} & 133.9 &   8.9\\ 
 393 & 22 32 57.84 & $-$60 33 29.1 & 4.189 & $28.69$ & $ 0.19$ & $17.41$ &  7.1 &  1.4 & 42.11\hspace{0.4em} & 156.2 &  31.6\\ 
 422 & 22 32 52.08 & $-$60 34 10.9 & 3.129 & $28.46$ & $ 0.15$ & $17.13$ &  6.6 &  1.0 & 41.77\hspace{0.4em} &  92.9 &  14.1\\ 
 437 & 22 32 55.92 & $-$60 34 06.6 & 3.120 & $28.46$ & $ 0.10$ & $17.14$ & 10.9 &  0.7 & 41.99\hspace{0.4em} & 152.7 &   9.8\\ 
 489 & 22 32 57.12 & $-$60 33 44.5 & 2.956 & $28.87$ & $ 0.13$ & $16.62$ &  5.1 &  1.4 & 41.60\hspace{0.4em} & 100.9 &  27.3\\ 
 543 & 22 32 57.36 & $-$60 33 48.6 & 3.633 & \multicolumn{2}{l}{$>29.0$} & $<16.9$ &  6.0 &  0.8 & 41.88\hspace{0.4em} &  \multicolumn{2}{l}{$>155$}\\ 
 546 & 22 32 53.76 & $-$60 33 41.1 & 5.710 & \multicolumn{2}{l}{$>29.0$} & $<17.6$ &  8.0 &  2.4 & 42.47\hspace{0.4em} &  \multicolumn{2}{l}{$>301$}\\ 
 547 & 22 32 54.00 & $-$60 34 06.2 & 5.710 & \multicolumn{2}{l}{$>29.0$} & $<17.6$ & 10.7 &  3.3 & 42.60\hspace{0.4em} &  \multicolumn{2}{l}{$>400$}\\ 
 549 & 22 32 55.68 & $-$60 33 40.7 & 4.672 & \multicolumn{2}{l}{$>29.0$} & $<17.3$ &  4.9 &  0.7 & 42.05\hspace{0.4em} &  \multicolumn{2}{l}{$>154$}\\ 
 553 & 22 32 52.56 & $-$60 33 55.9 & 5.079 & \multicolumn{2}{l}{$>29.0$} & $<17.4$ &  9.3 &  0.9 & 42.42\hspace{0.4em} &  \multicolumn{2}{l}{$>315$}\\ 
 558 & 22 32 54.48 & $-$60 34 02.2 & 3.126 & \multicolumn{2}{l}{$>29.0$} & $<16.6$ &  6.1 &  0.9 & 41.74\hspace{0.4em} &  \multicolumn{2}{l}{$>141$}\\ 
 563 & 22 32 52.32 & $-$60 34 01.2 & 3.826 & \multicolumn{2}{l}{$>29.0$} & $<16.9$ &  6.6 &  1.7 & 41.98\hspace{0.4em} &  \multicolumn{2}{l}{$>178$}\\ 
 568 & 22 32 53.76 & $-$60 33 35.6 & 4.664 & \multicolumn{2}{l}{$>29.0$} & $<17.3$ &  4.6 &  0.9 & 42.03\hspace{0.4em} &  \multicolumn{2}{l}{$>145$}\\ 
\noalign{\smallskip}\hline\noalign{\smallskip} 
\end{tabular}
\end{center}
\label{tab:table1}
\end{table*}

\begin{figure}
\includegraphics[width=\hsize]{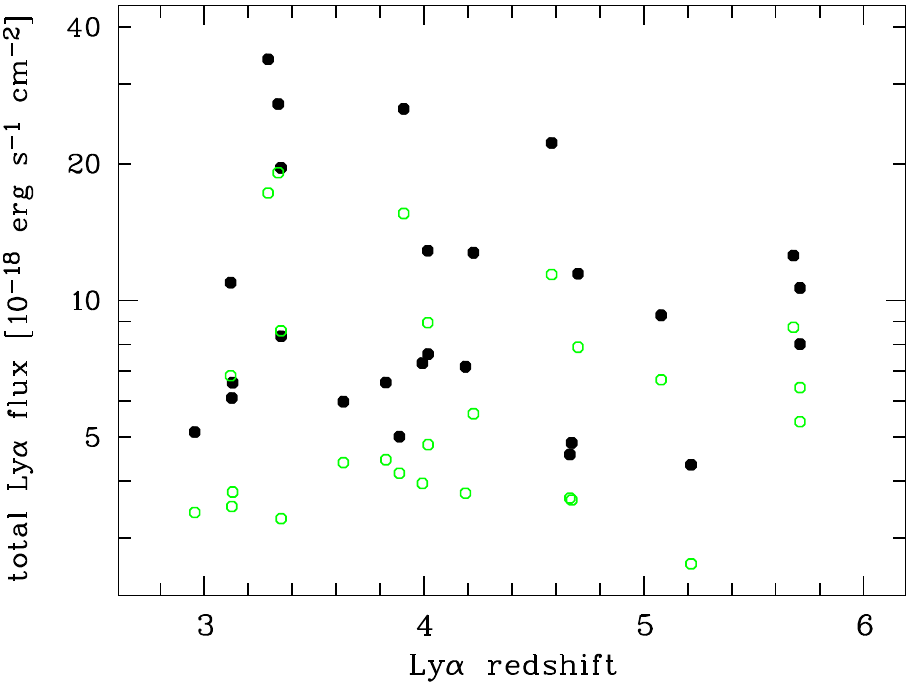}
\caption[]{Integrated \lya\ fluxes and redshifts of the 26 LAEs in our sample. The black filled symbols show the fluxes integrated within apertures of 3\arcsec\ radius, the green open symbols show the fluxes within $r = 1\arcsec$.}
\label{fig:sample}
\end{figure}

\subsection{The sample of Lyman $\alpha$ emitters}
\label{sec:obs-sample}

Together with the MUSE datacube, B2015 presented a catalogue of 189 objects with extracted spectra and redshifts. Most of the entries were taken from  \citet{Casertano:2000fn}, with some additional MUSE-detected pure emission line objects. The 89 sources classified as $z>2.9$ galaxies in the B2015 catalogue form the parent sample for the current study. We restricted the sample further by applying the following additional criteria:

\begin{enumerate}\setlength{\itemsep}{0.5ex}

\item We excluded all sources closer than 4\arcsec\ to the edges of the MUSE field of view, in order to be able to construct radial profiles and growth curves over 360\degr\ of azimuth.

\item We removed all same-redshift object pairs closer to each other than 50~kpc of projected transverse separation; there were 8 such pairs with velocity differences (estimated from \lya) of less than 1000~km/s. While such binary systems are certainly interesting in their own right, we decided to focus the present investigation on isolated galaxies and avoid cases where strong interactions between close neighbours might produce extended emission caused by tidal features.

\item We removed one curious case of a bright LAE spatially coinciding with an [\ion{O}{ii}] emitter (ID\#71%
  \footnote{For brevity, we denote individual objects by their running identifiers in the B2015 source catalogue.}
; cf.\ Fig.~15 in B2015) because of the obvious confusion. 

\item We furthermore deleted object ID\#290, which is listed as an LAE at $z=6.28$ in B2015, but is detected in the HST/WFPC2 F450W band, so it is either a misclassification or again a superposition of objects at different redshifts.

\item We also removed one likely AGN (ID\#144 at $z=4.017$) -- in fact the strongest \lya\ emitter in the field. 

\item We finally imposed a \lya\ flux cut, requiring a minimal signal-to-noise ratio (S/N) of 3 in the large aperture (radius of 3\arcsec) defined in Sect.~\ref{sec:prof-lya} for measuring the total fluxes. Note that the spectra used by B2015 for source classification have much higher S/N values than this, as they were extracted from the datacube using an aperture radius of 0\farcs7 for isolated sources. However, because of the spatially extended \lya\ emission the line fluxes in the released catalogue are significantly biased low. We quantify this point in Sect.~\ref{sec:prof-lya}.

\end{enumerate}

We did thus not consider the physical galaxy pair ID\#40 and ID\#56 at $z=3.01$, the brightest and the 3rd-brightest of the $z>3$ galaxies in the HDFS-MUSE field ($I_{814} = 24.5$ and 25.0, respectively), separated by just 2\arcsec\ and already for that reason not part of the sample. ID\#40 shows only a feeble \lya\ emission line, while ID\#56 has a pure absorption spectrum. These galaxies are therefore not LAEs, but would probably qualify as LBGs. Upon further examination of the MUSE data around this complex, a low surface brightness \lya\ nebula emerges that subtends over several arcsec. We plan a separate publication dedicated to this remarkable group of objects (Cantalupo et al., in prep.).

After applying the above selection criteria we were left with a sample of 26 isolated non-AGN galaxies, with total \lya\ line fluxes ranging from $\sim 4.5\times 10^{-18}$~erg~s$^{-1}$~cm$^{-2}$ up to $\sim 3\times 10^{-17}$~erg~s$^{-1}$~cm$^{-2}$ and covering a redshift range from 2.96 to 5.71. Their spectra can be inspected via the HDFS data release web interface (Sect.~\ref{sec:obs-obs}). All objects except one have total rest-frame \lya\ equivalent widths $> 20$~\AA\ and thus are LAEs in the sense of the typical selection criterion used in narrowband searches. The sample and some basic object properties are listed in Table~\ref{tab:table1}. Notice that the HST/WFPC2 F814W magnitudes are not identical to those from B2015 (taken in turn from \citealt{Casertano:2000fn}), but were redetermined by us using GALFIT modelling as described in Sect.~\ref{sec:prof-cont}. The differences between these two magnitude estimates are however small ($< 0.1$~mag) except for some of the faintest objects. Absolute magnitudes and equivalent widths were derived assuming a UV continuum slope $\beta = -2$ ($f_\lambda \propto \lambda^\beta$). 

Figure~\ref{fig:sample} shows the distribution of redshifts and \lya\ fluxes, with the latter being provided for two different extraction apertures with radii of 1\arcsec\ and 3\arcsec, respectively. On average the wide aperture measurements have 60\% higher fluxes, while for a perfect point source this aperture difference would range between 15\% and 23\% over the MUSE wavelength.
The differences are evident and a first strong hint at the extended nature of the \lya\ emission.  Notice also that despite this sample containing only the brightest LAEs in the HDFS, very few of these objects are \lya-luminous enough to be detectable in a conventional narrowband imaging survey.

\subsection{Ly$\alpha$ images}
\label{sec:obs-extr}

\begin{figure*}
\includegraphics[width=\hsize]{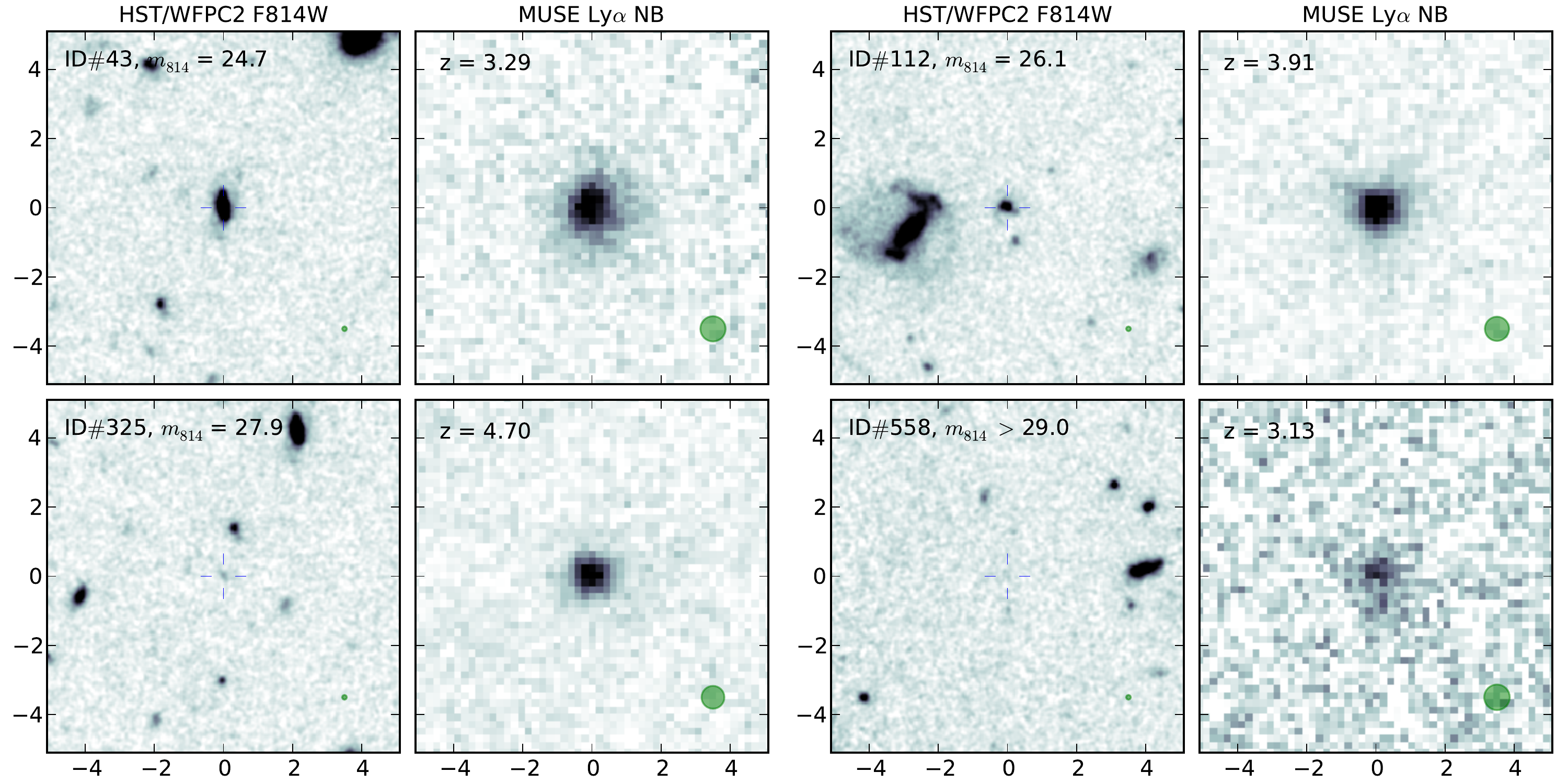}
\caption[]{Example images of four of our Ly$\alpha$ emitters, showing broadband (HST/WFPC2 F814W) and MUSE continuum-subtracted pseudo-narrowband data (see text). Each panel displays an area of $10\arcsec\times 10\arcsec$ centred on the LAE. The labels provide identifiers, broadband magnitudes, and redshifts. The location of each object in the HST data is marked by a blue crosswire;  note that object ID\#558 is not significantly detected by HST. The green circle in the lower right corner of each panel indicates the spatial resolution (FWHM of the respective point spread function).
}
\label{fig:images}
\end{figure*}

Using the MUSE datacube we constructed `pseudo-narrowband' (NB) images of our objects, each centred on the position and wavelength of the corresponding \lya\ line. We used the B2015 spectra to inspect the \lya\ emission lines and define the bandwidths of the NB images. This was done interactively, chosing each band limit such that about 90--95\% of the total line flux was included. The bandwidths came out to be mostly between 5 and 10 spectral pixels; the median was 7 pixels or 8.75\,\AA. 

Before extracting the NB images we performed another preprocessing step. At the low flux levels of interest in this study, source crowding becomes a serious issue for many objects. Nearly all objects in our sample have projected close neighbours within a few arcsec in the HST data. However, since these neighbours are typically at other redshifts than the LAEs, they contaminate the NB signal only with their continuum emission. A traditional way to remove the continuum would be by subtracting a suitably scaled off-band image. We adopted a related method that takes better advantage of having a datacube: We first median-filtered the datacube in the spectral direction with a very wide filter window of $\pm$150 spectral pixels; this produced a continuum-only cube with all line emission removed and with the continuum spectra of real objects being heavily smoothed. We then subtracted this filtered cube from the original data and thus obtained an essentially pure emission line cube which was (to first order) free from any continuum signal. By visual inspection we found 4 instances where a bright foreground source showed a significant spectral feature at the same wavelength as the \lya\ line of the LAE, leading to a hump or a dip in the cleaned cube at the location of the foreground object. In those cases we manually masked the affected region. In all other objects we saw no significant residuals of continuum objects remaining after the cleaning. 

From the continuum-subtracted cube we extracted, for each object in turn, small NB images of typically $51\times 51$ spatial pixels ($10\arcsec\times 10\arcsec$) centred on the \lya\ emission of each LAE, summing over the spectral bandwidths defined above. For some objects these images extended slightly outside the MUSE field of view; the affected pixels were then masked. In addition to the NB images we produced also cutouts from the HST image of the HDFS in the F814W band. Examples of HST broadband and MUSE \lya\ NB images of four of our LAEs are shown in Fig.~\ref{fig:images}. These images demonstrate that our continuum removal procedure generally performed very well, and that the various foreground galaxies disappear without any detectable trace. Figure~\ref{fig:images} also indicates the spatial resolution by depicting a schematic PSF. The \lya\ NB and HST broadband images of all objects in the sample are presented in a condensed form in Fig.~\ref{fig:ima+prof}, while Fig.~\ref{fig:lya-grey+red} provides a more detailed view at the detected spatially extended \lya\ emission.

%%%%%%%%%%%%%%%%%%%%%%%%%%%%%%%%%%%%%%%%%%%%%%%%%%%%%%%%%%%%%%%%%%%%%%%%%%%%%%%%

%\input{LAE-ext_3-profiles}

\begin{figure}
\includegraphics[width=\hsize]{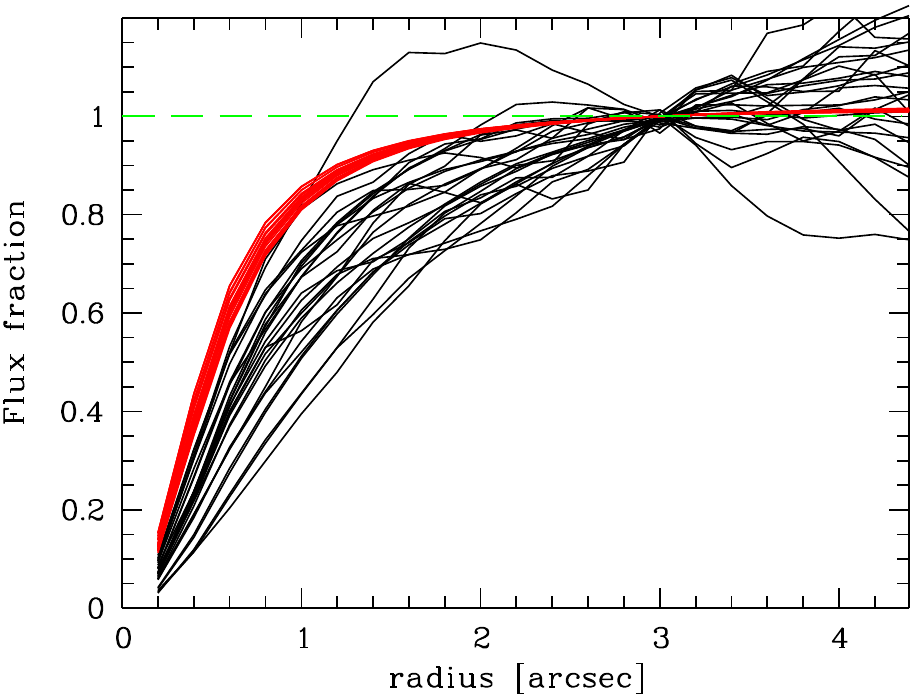}
\caption[]{\lya\ growth curves of all LAEs in the sample, each normalised to the integrated flux within 3\arcsec. The corresponding growth curves for the individually estimated monochromatic PSFs are shown in red.}
\label{fig:gc}
\end{figure}

\section{Analysis of radial surface brightness profiles}
\label{sec:prof}

The \lya\ surface brightness distribution in and around a star-forming galaxy depends on the production mechanism and escape paths of the \lya\ photons. For recombination radiation from a fully photoionized medium that is optically thin to \lya, the line emission should directly follow the ionizing continuum flux, or approximately the observable FUV continuum. If resonant scattering by neutral hydrogen, or additional \lya\ production channels such as UV fluorescence from neutral gas or cooling radiation are relevant, the galaxy may appear more extended in \lya\ than in the continuum, which should show up as a difference between the azimuthally averaged surface brightness profiles of \lya\ emission and UV continuum. This is the most commonly employed diagnostic method to search for extended \lya, and we follow here the same approach. We first demonstrate that the radial profiles of many of our LAEs indeed show strong evidence for being considerably more extended than the UV continuum profiles. We then go through several technical details that are relevant to assert the robustness of this result. After deriving realistic error bars we quantify the statistical significance for or against extended Ly$\alpha$ profiles on an individual object-by-object basis, for all galaxies in the sample.

\begin{figure*}
\setlength{\unitlength}{1mm}
\begin{picture}(170,0190)
\put(0,145){\includegraphics[width=4.2cm]{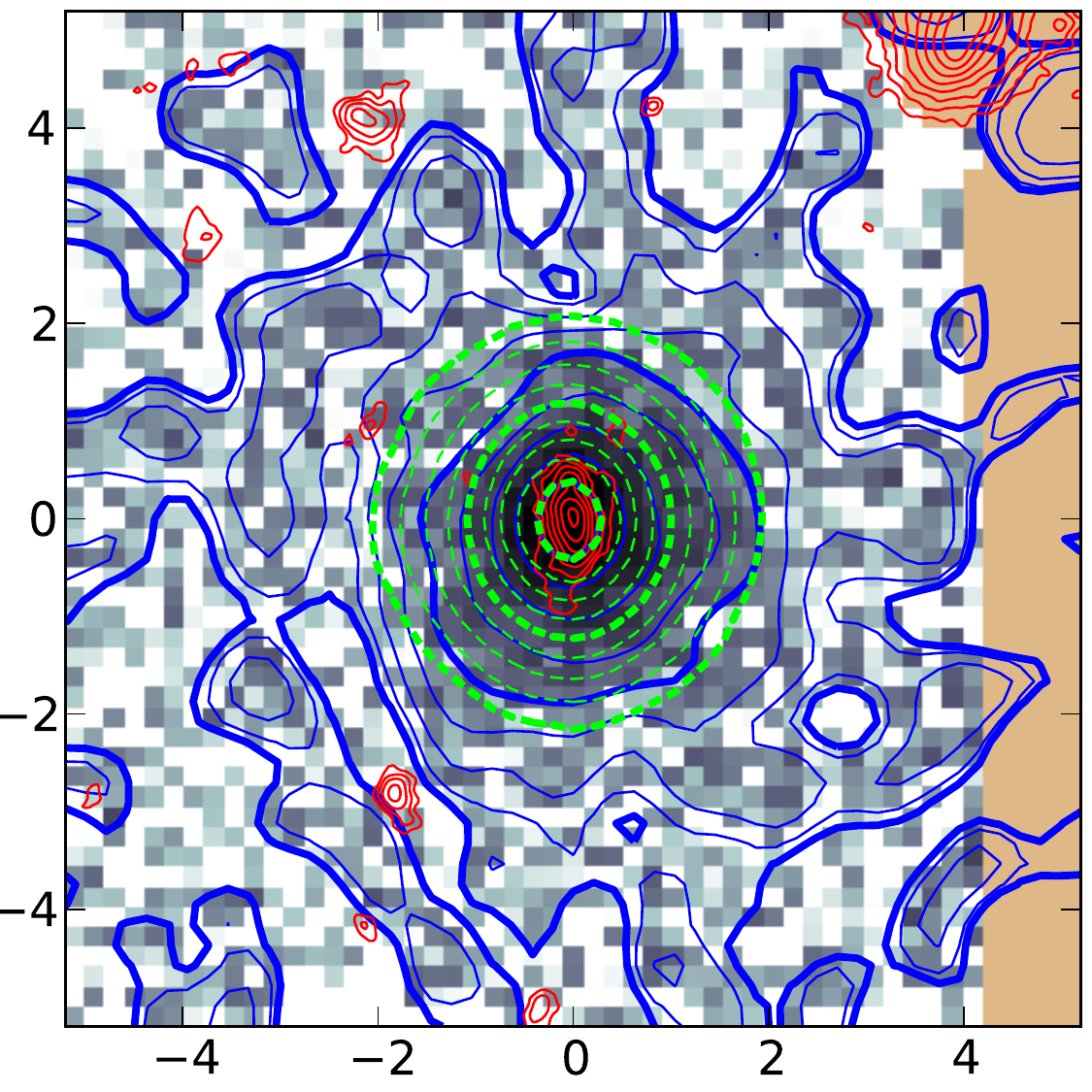}}
\put(44,144){\includegraphics[width=4.5cm]{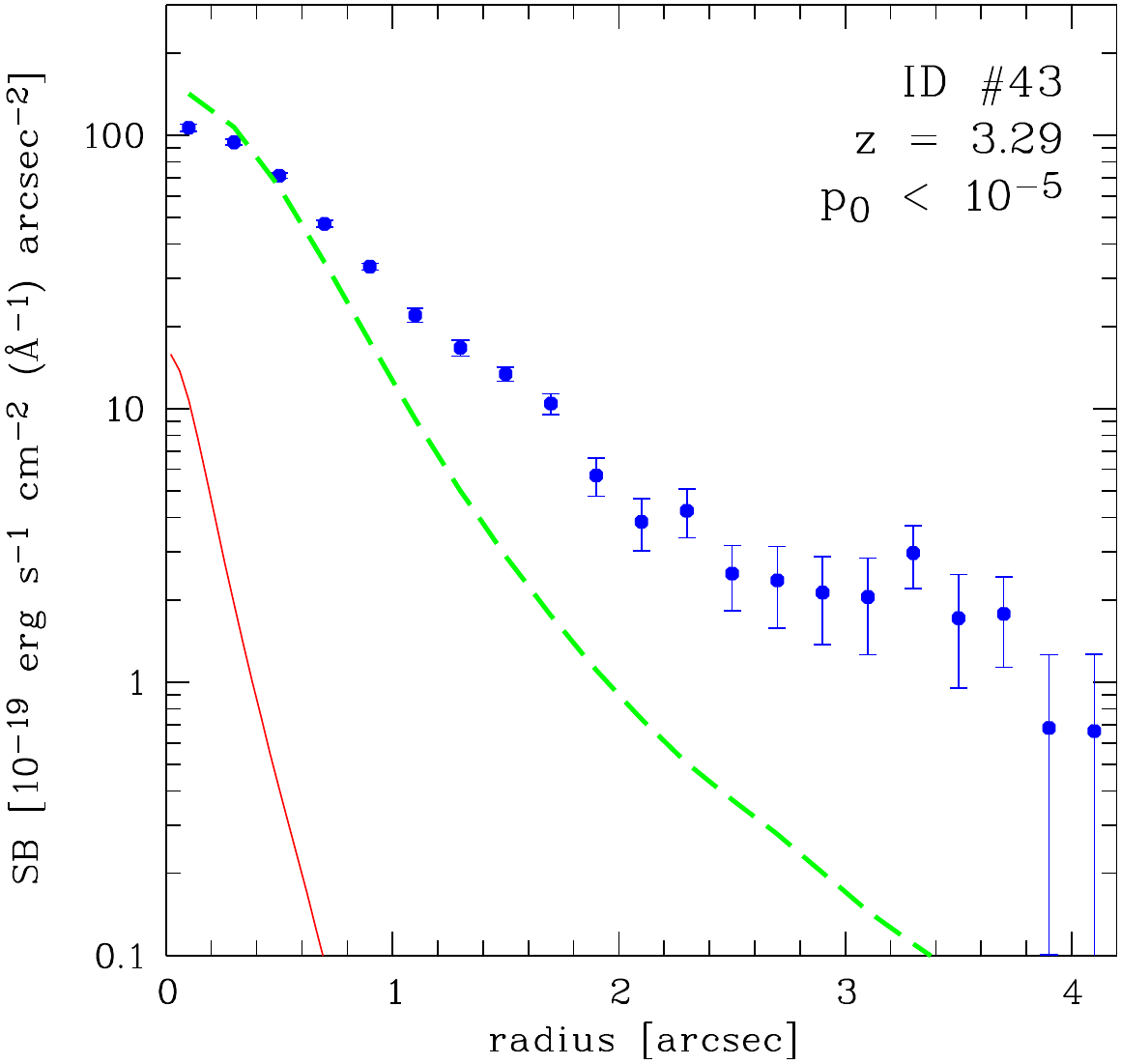}}
\put(93,145){\includegraphics[width=4.2cm]{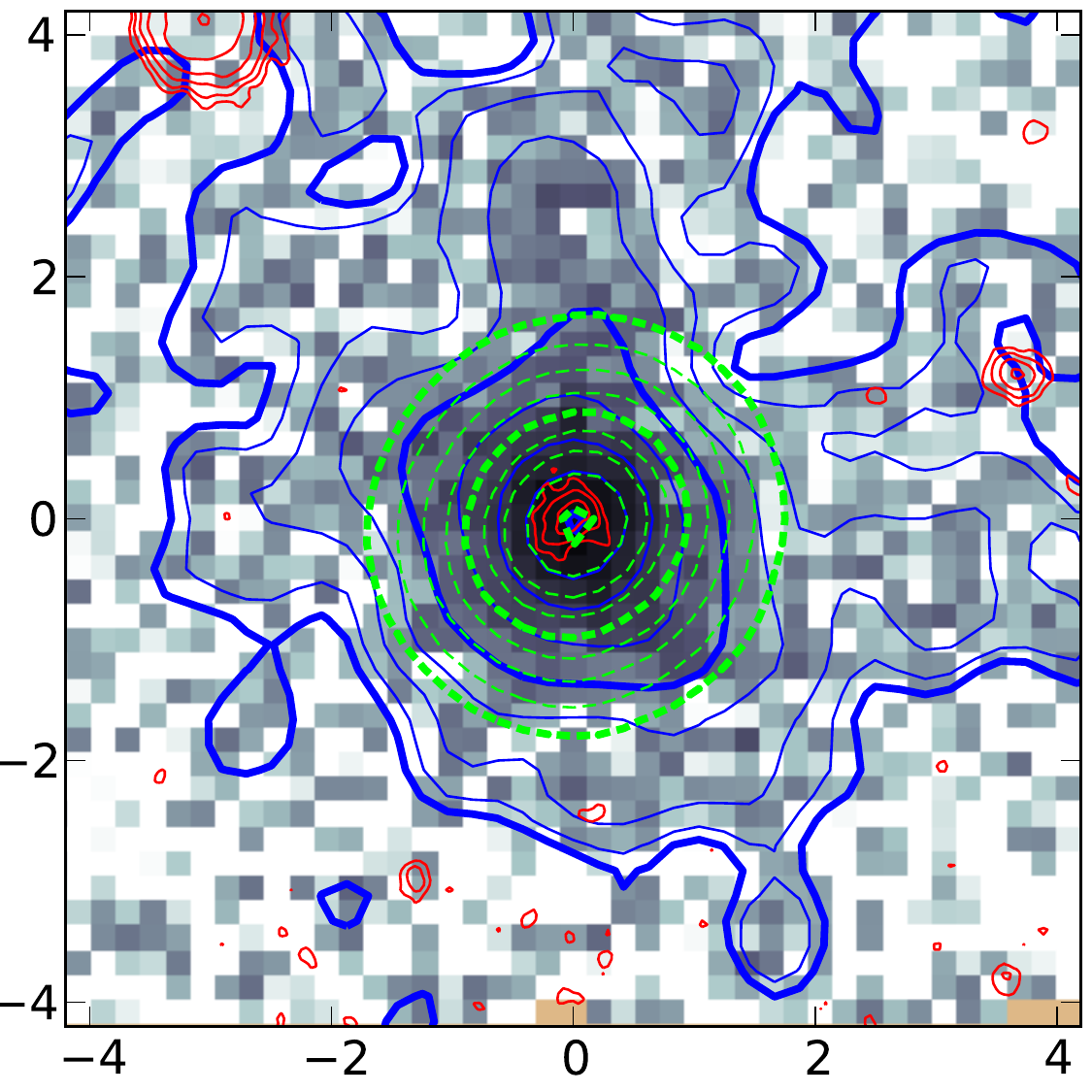}}
\put(137,144){\includegraphics[width=4.5cm]{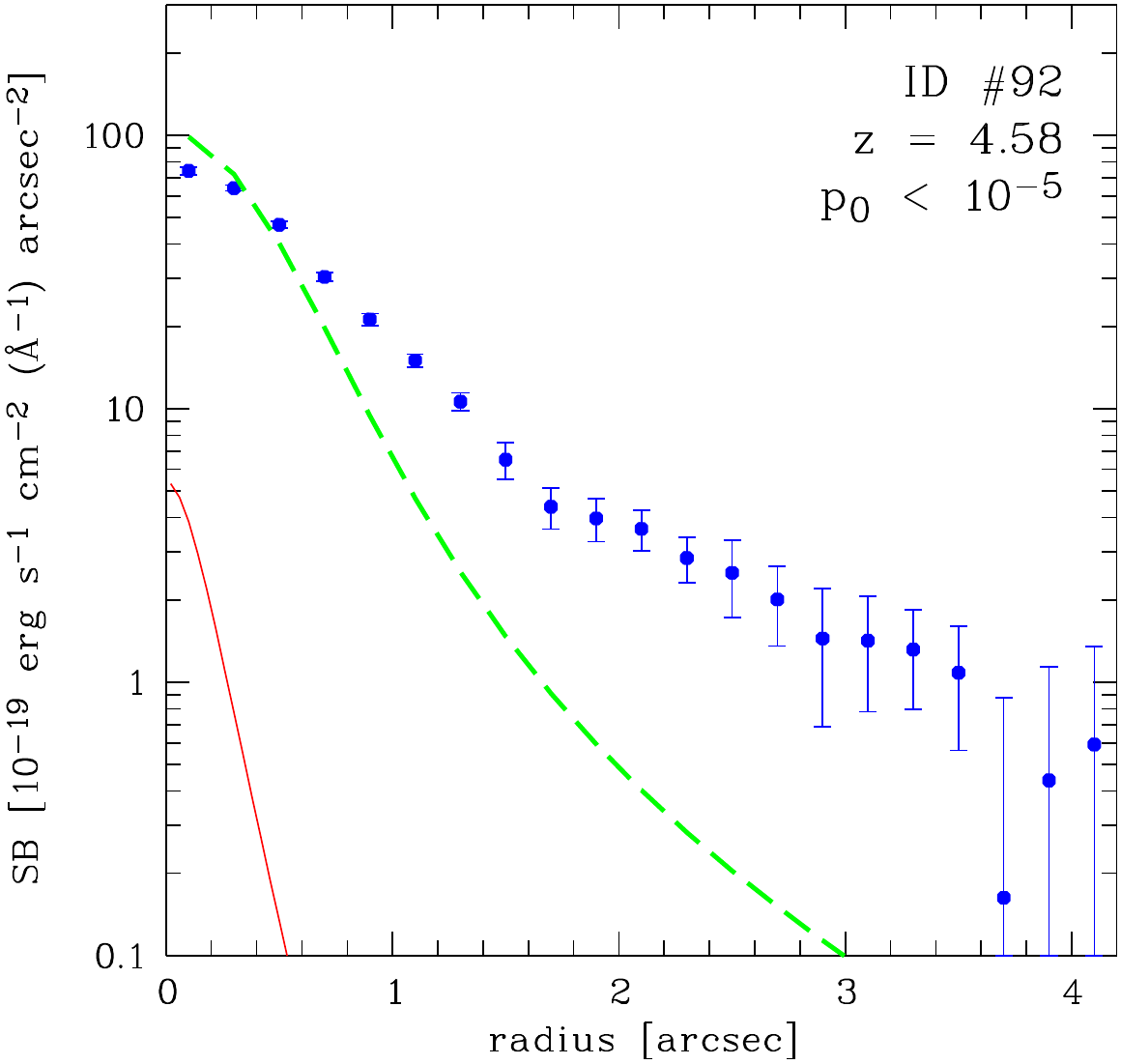}}
\put(0,97){\includegraphics[width=4.2cm]{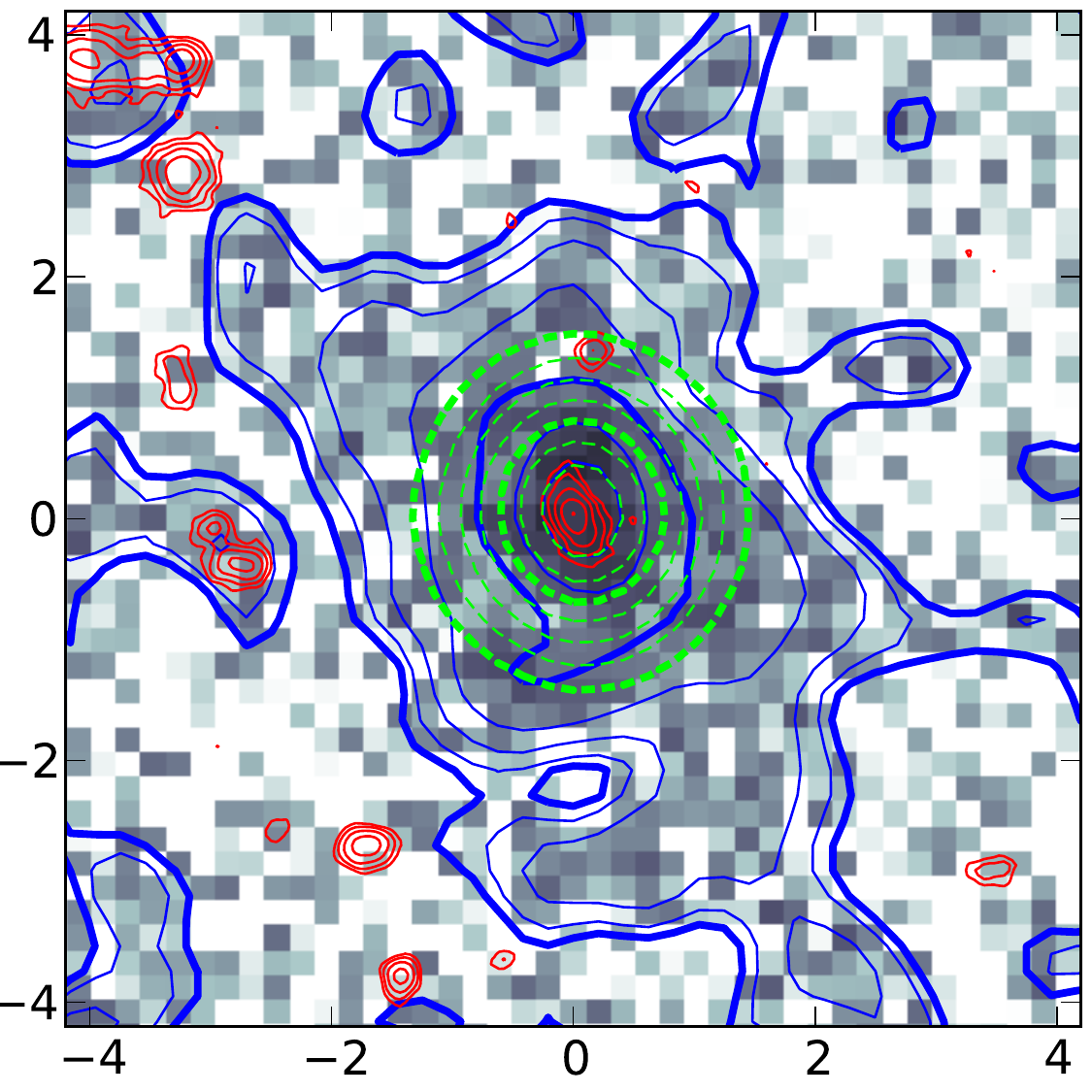}}
\put(44,96){\includegraphics[width=4.5cm]{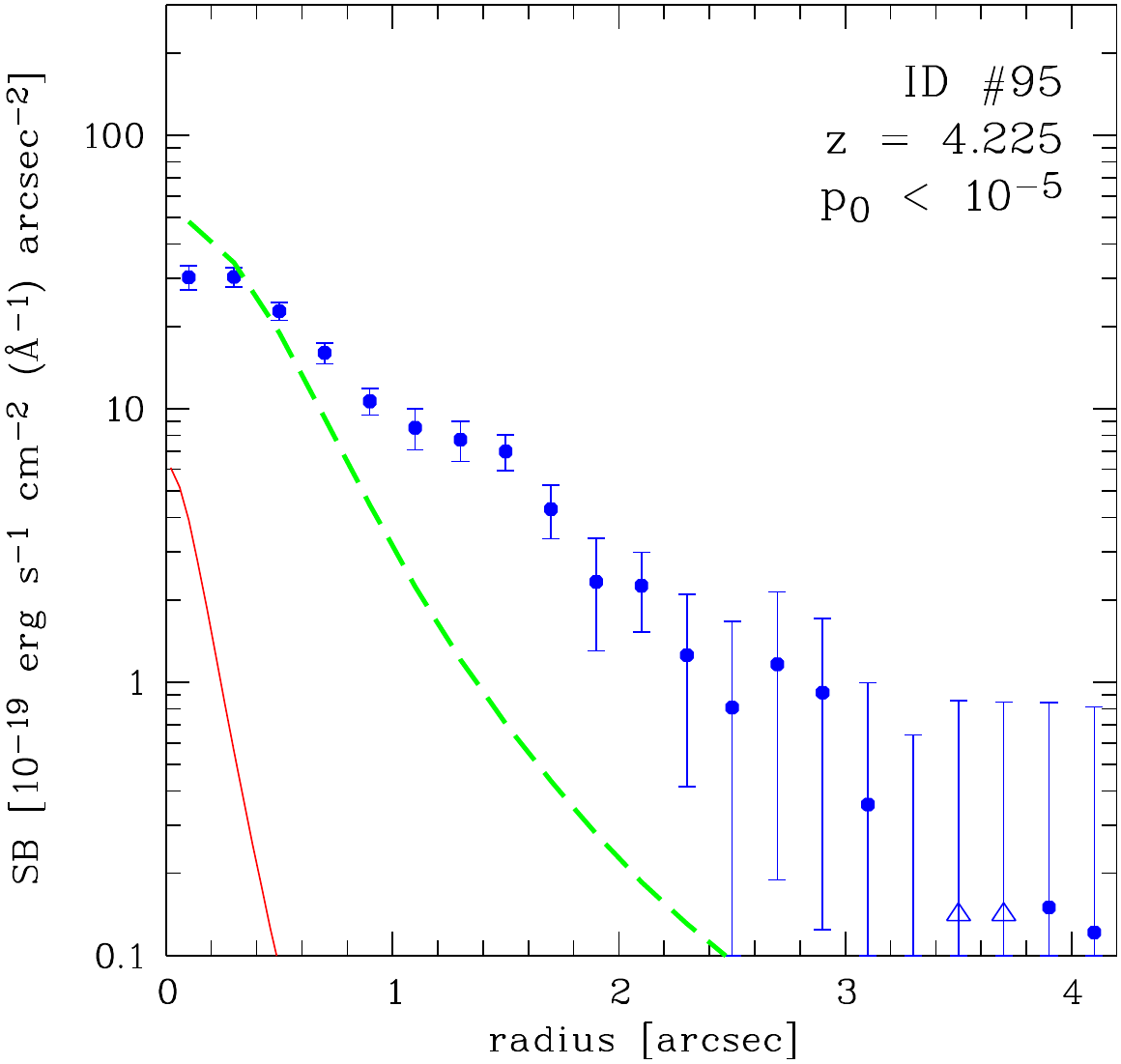}}
\put(93,97){\includegraphics[width=4.2cm]{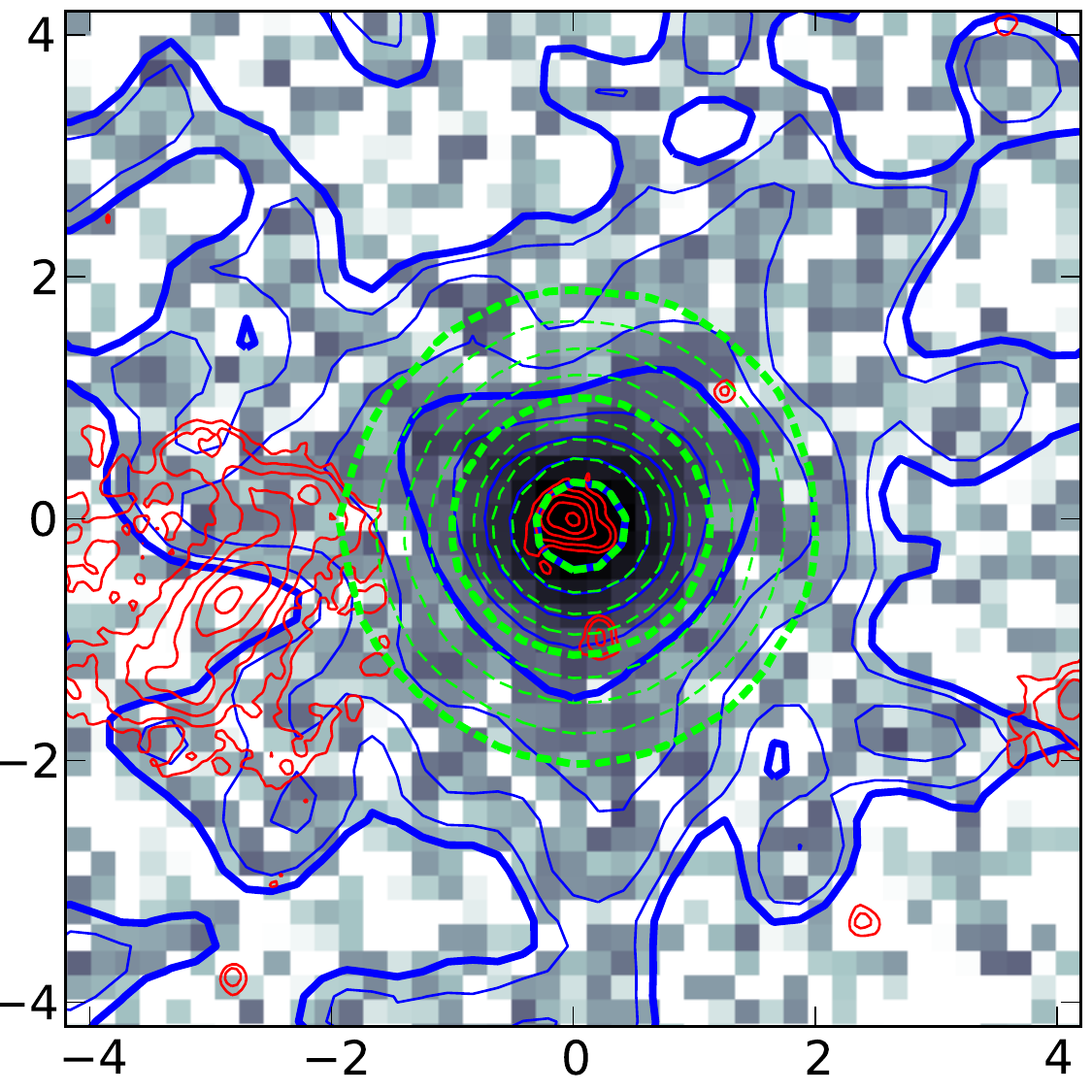}}
\put(137,96){\includegraphics[width=4.5cm]{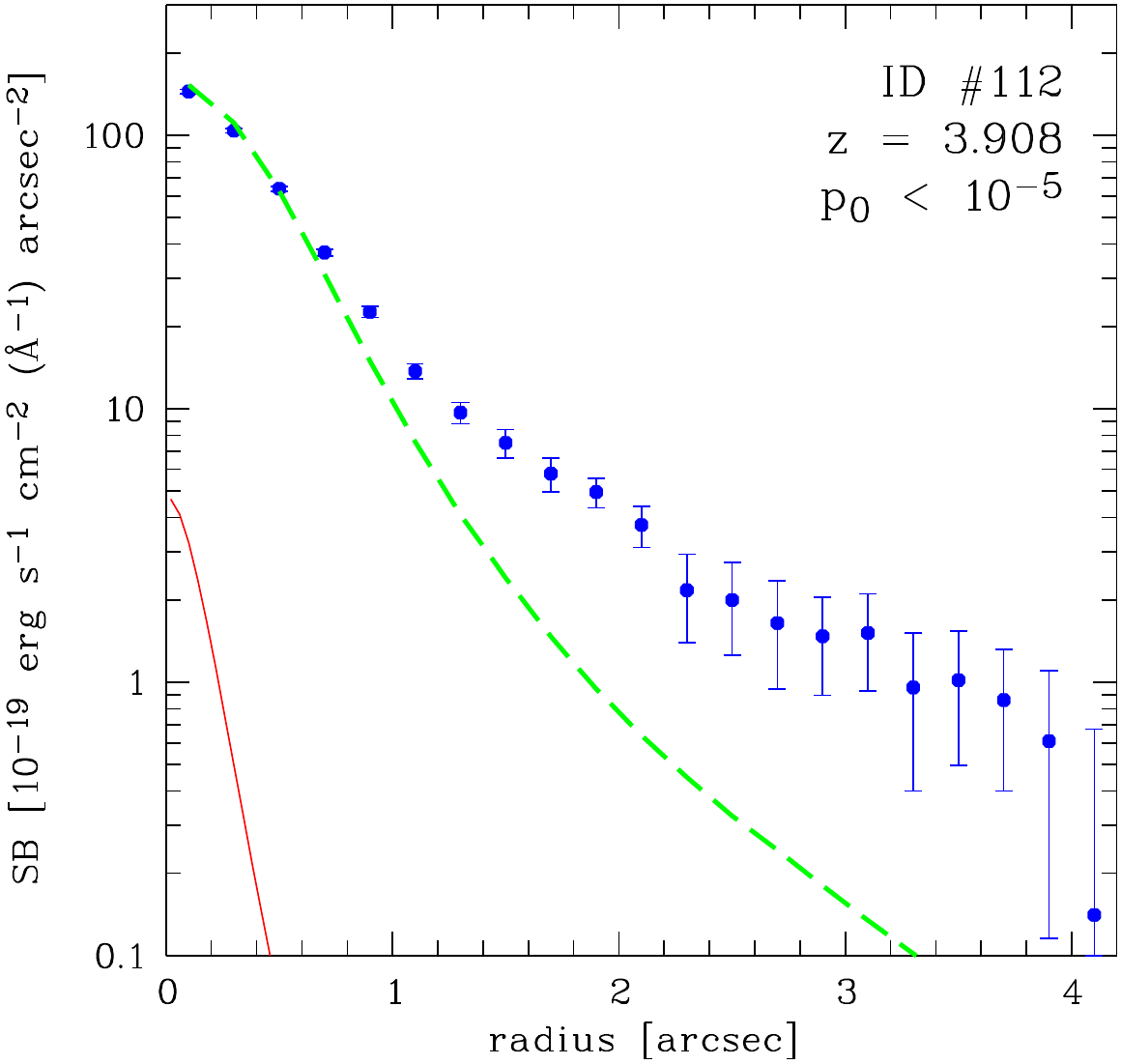}}
\put(0,49){\includegraphics[width=4.2cm]{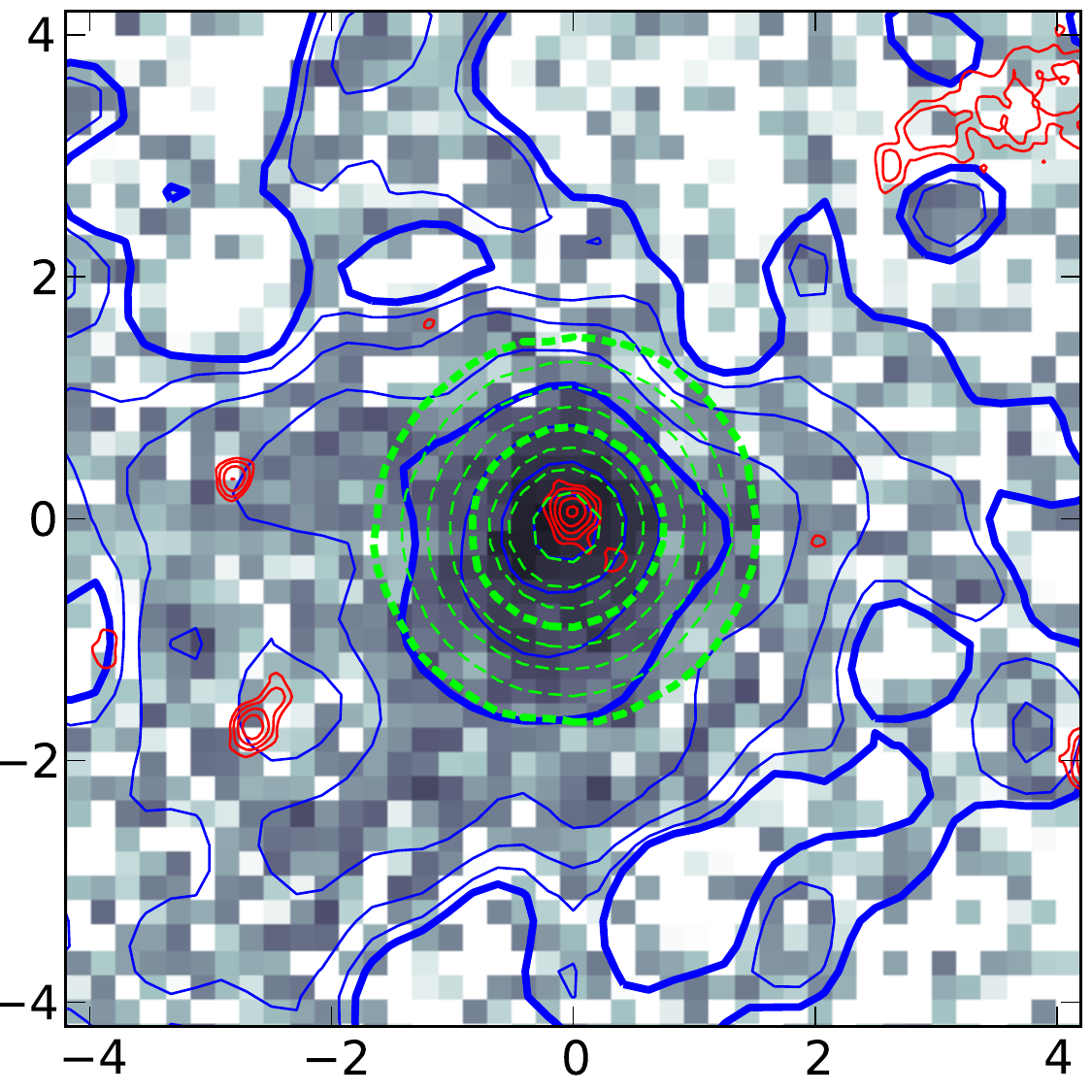}}
\put(44,48){\includegraphics[width=4.5cm]{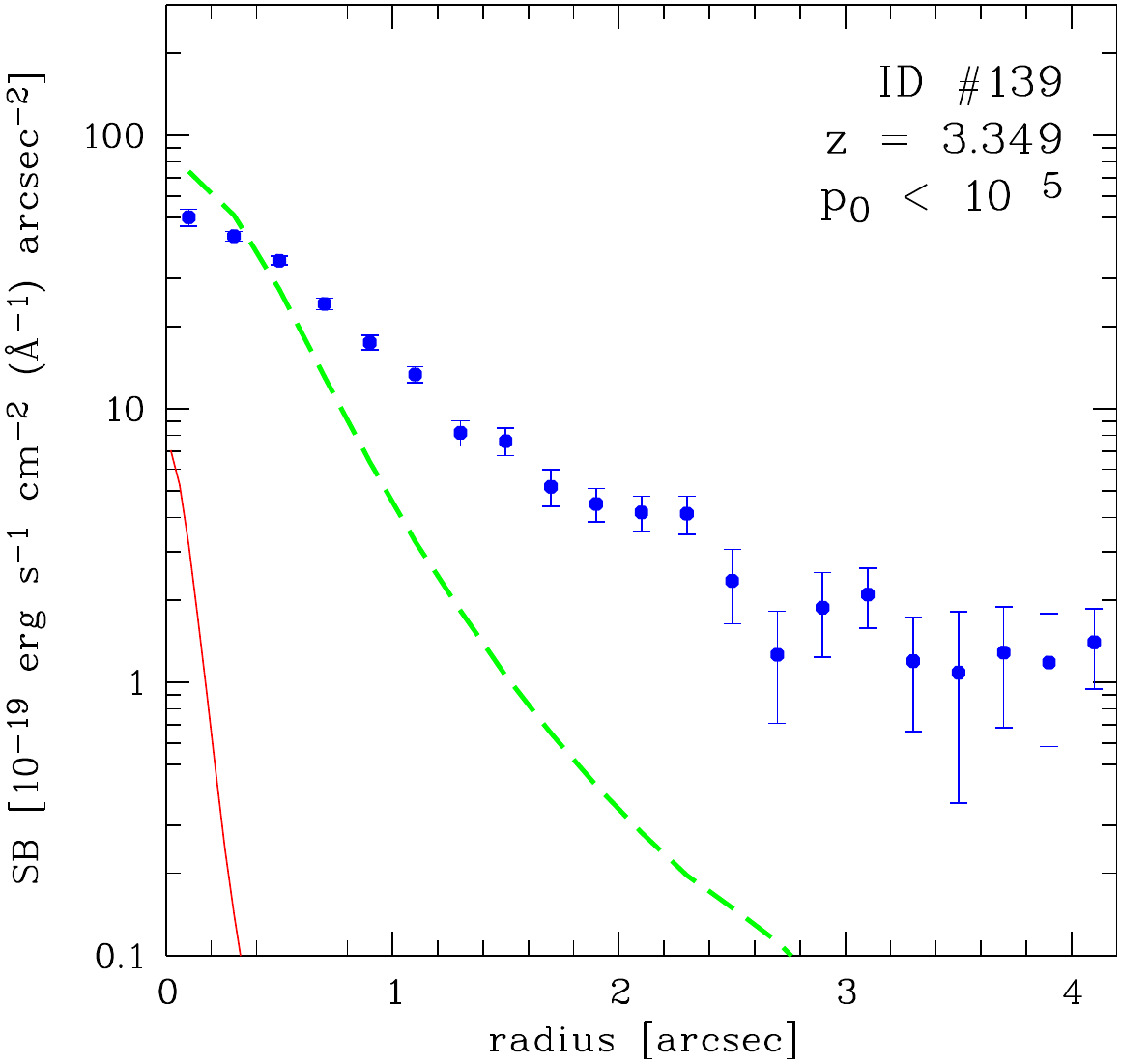}}
\put(93,49){\includegraphics[width=4.2cm]{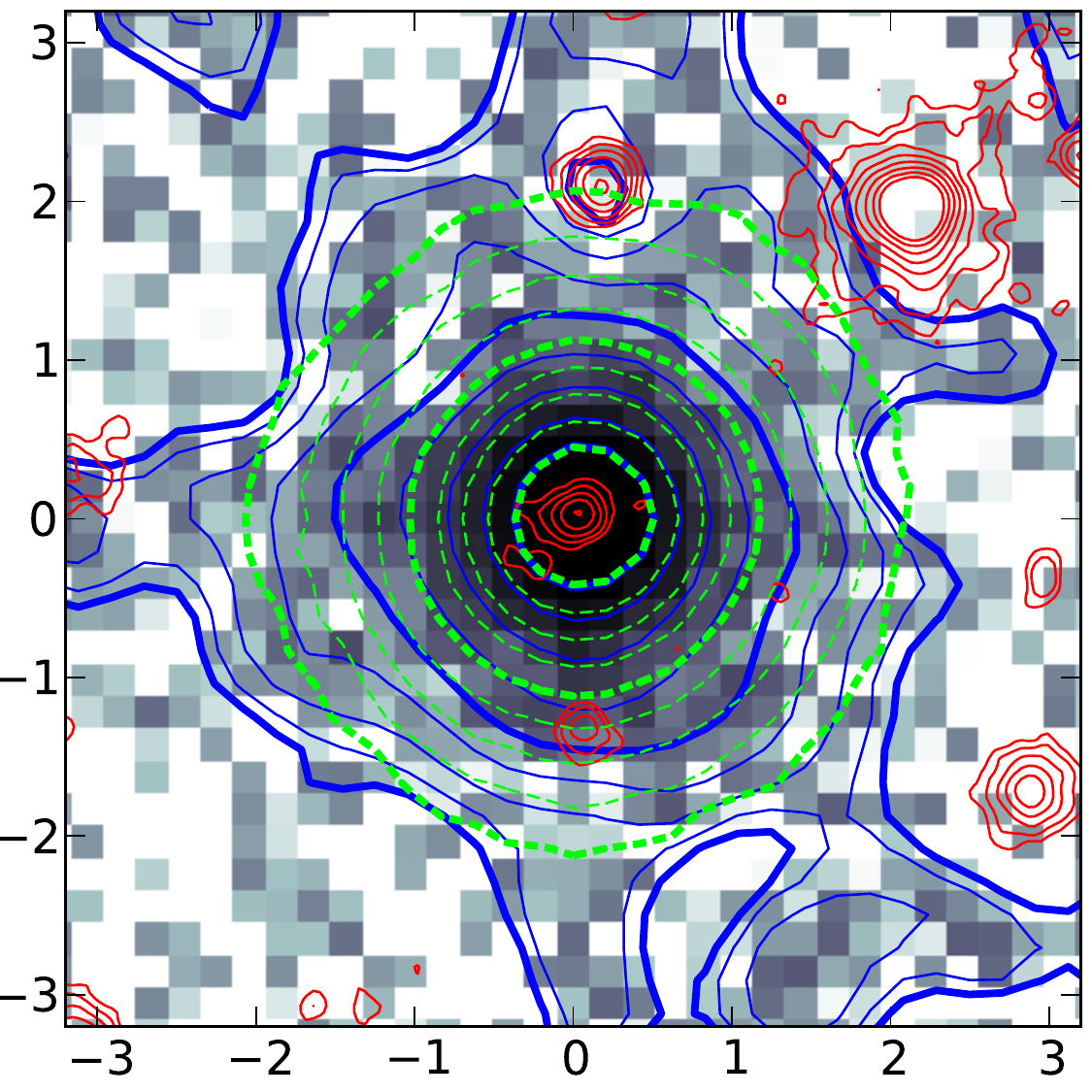}}
\put(137,48){\includegraphics[width=4.5cm]{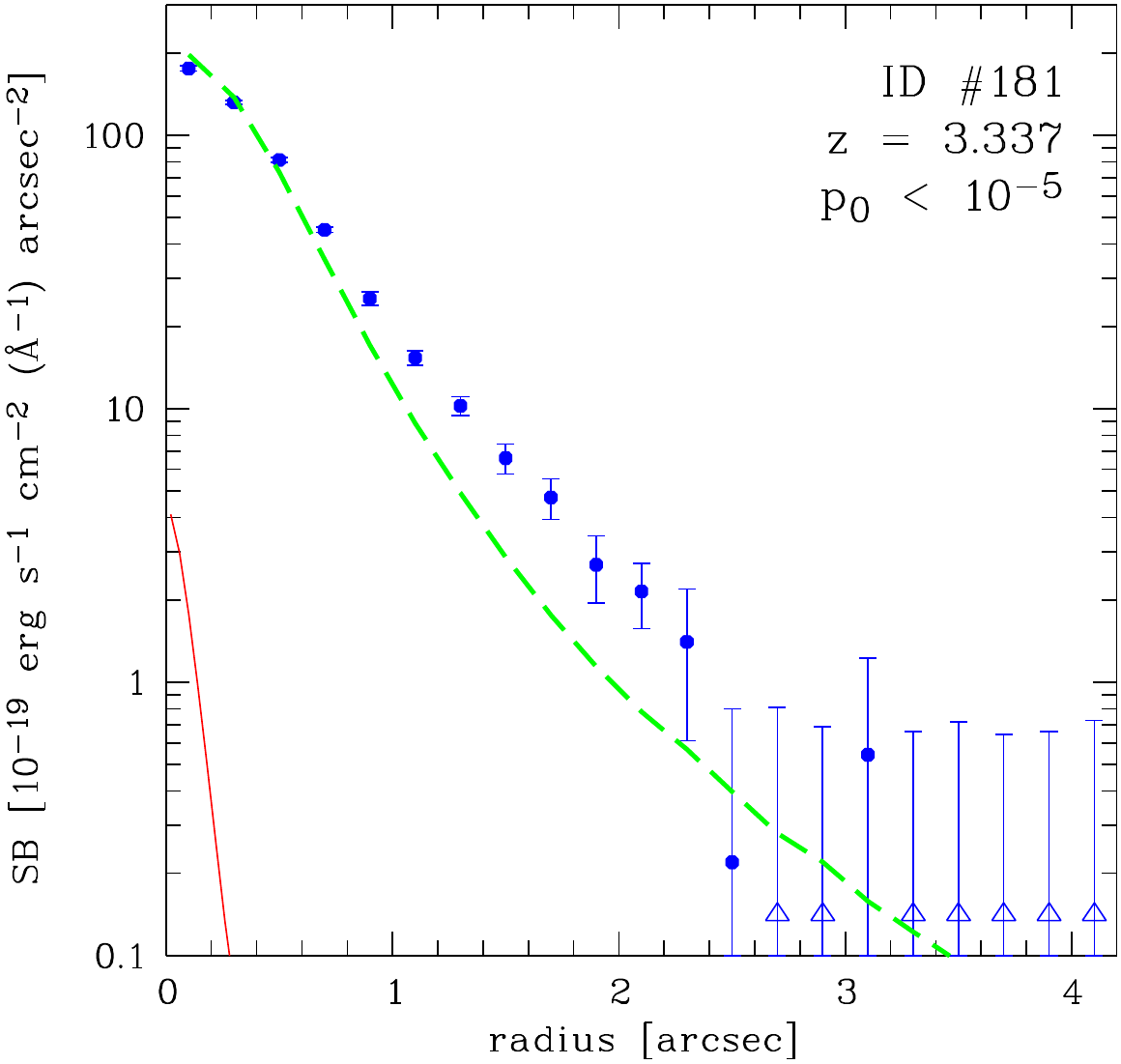}}
\put(0,1){\includegraphics[width=4.2cm]{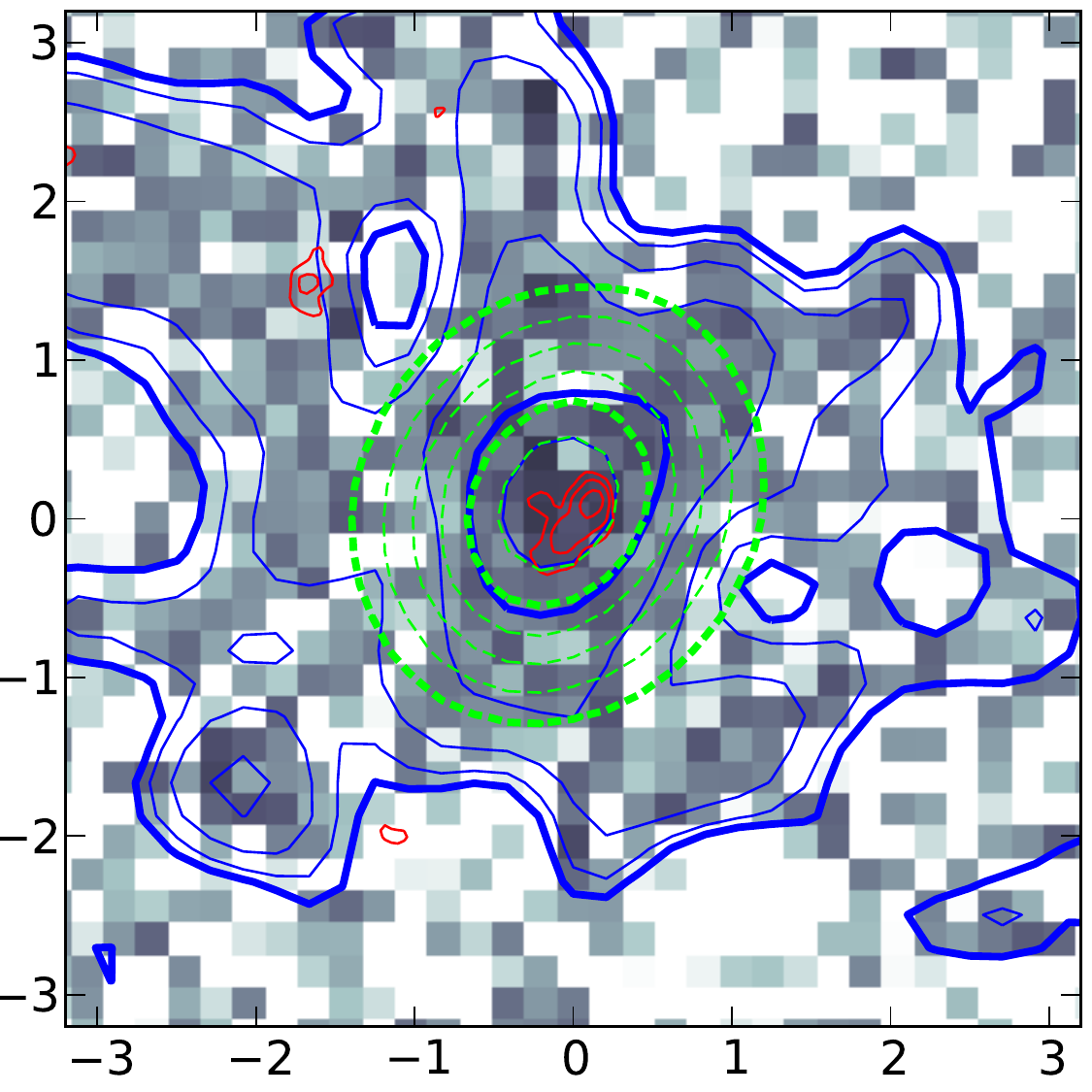}}
\put(44,0){\includegraphics[width=4.5cm]{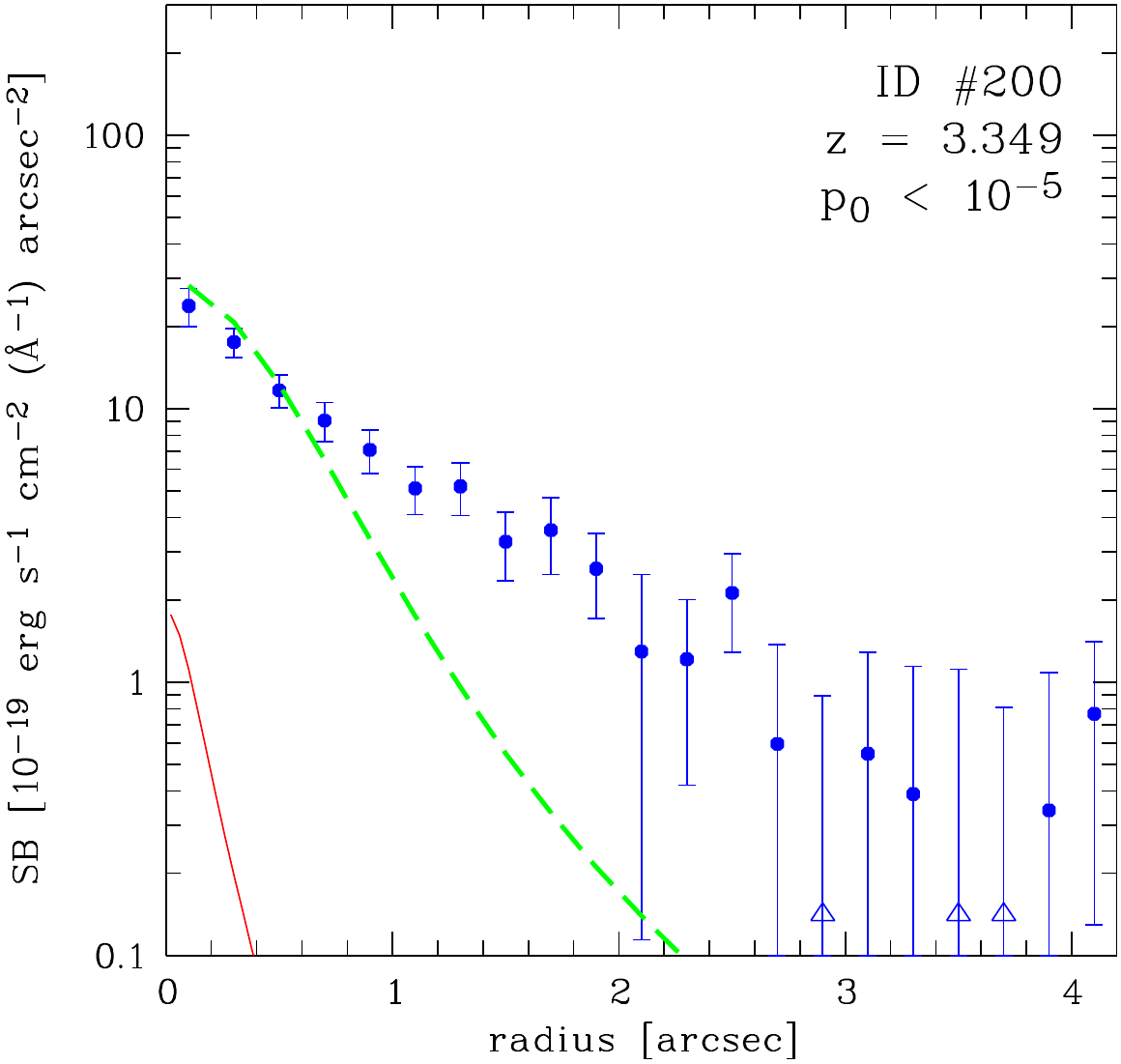}}
\put(93,1){\includegraphics[width=4.2cm]{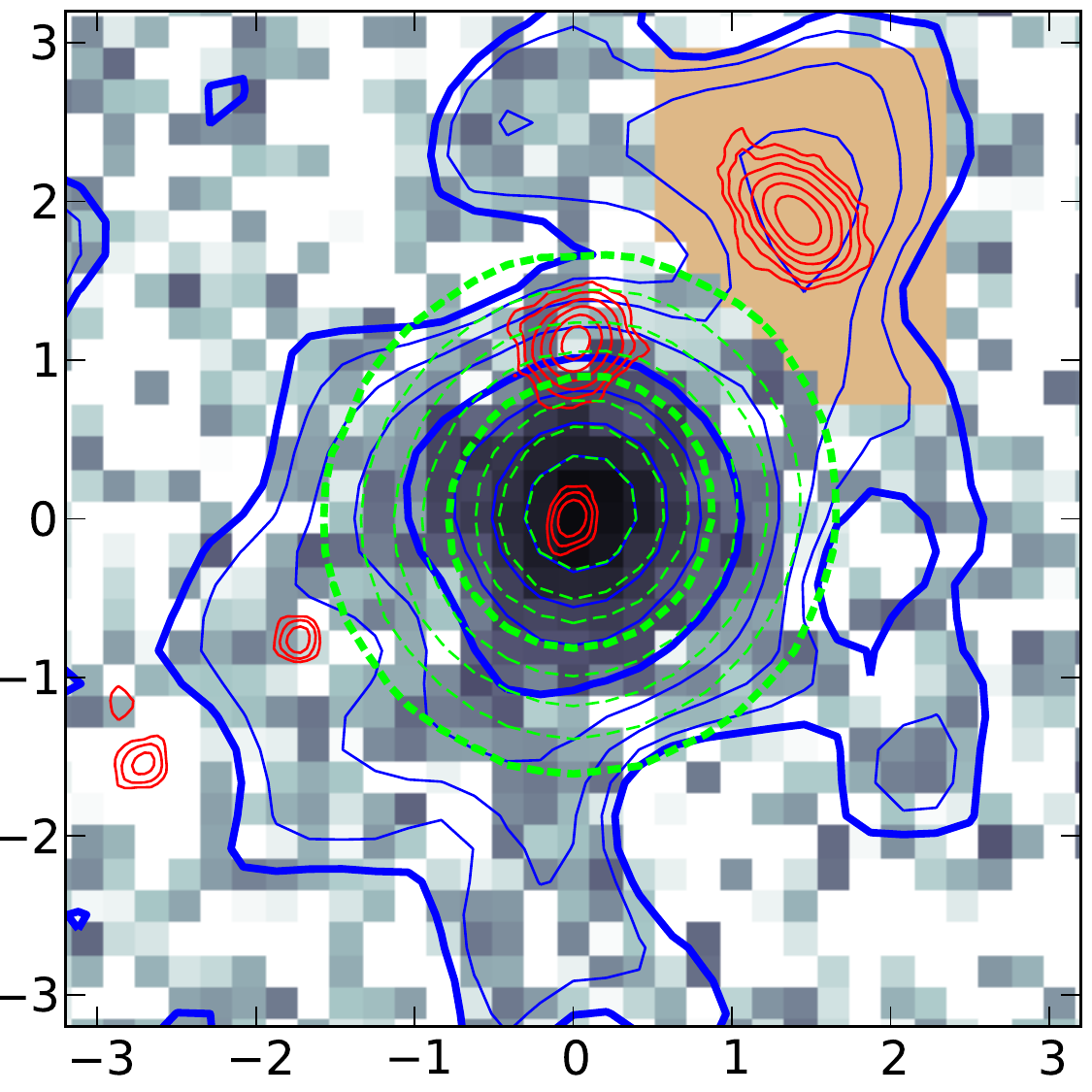}}
\put(137,0){\includegraphics[width=4.5cm]{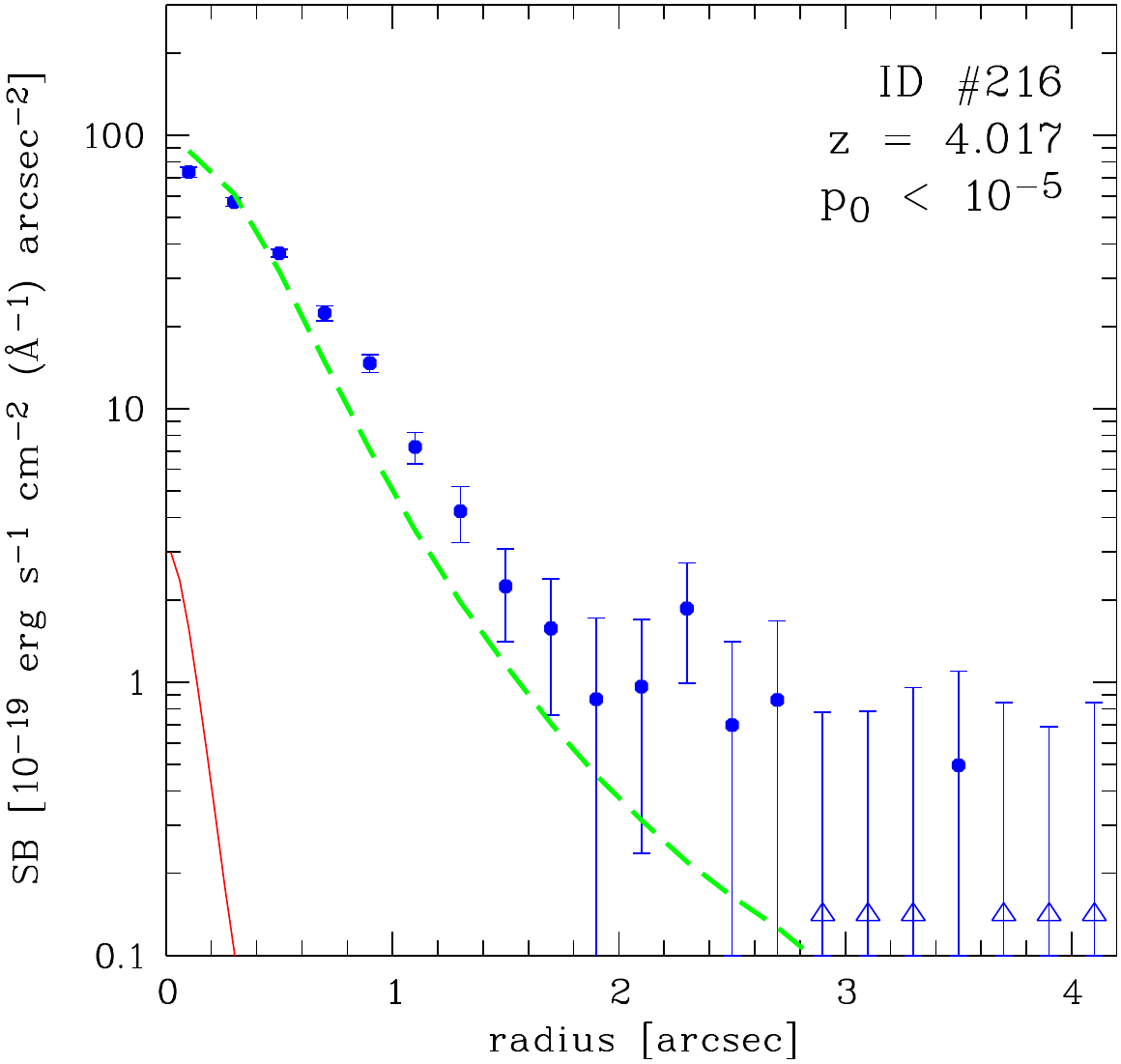}}
\end{picture}
\caption[]{\lya\ pseudo-narrowband images and radial surface brightness profiles of the LAEs in the sample, ordered by their MUSE-HDFS identifiers. Left-hand panels in each column: The greyscale pixel data show \lya\ surface brightness in asinh stretch, with equal cut levels for all objects. The spatial scale in arcsec is given by the axes labels; notice that the scale varies between different objects. The blue contours also show the \lya\ emission, but after smoothing to $\approx 1\arcsec$ resolution to emphasize the overall distribution. The contours are spaced logarithmically by 0.25~dex, with the lowest contour level always at $1\times 10^{-19}$~erg~s$^{-1}$ cm$^{-2}$ arcsec$^{-2}$, given by the outermost thick line.  Overlayed in red contours are the WFPC2 F814W images at HST resolution. The green dashed contours represent seeing-convolved UV continuum models of the central galaxies, scaled to match the \lya\ emission under the null hypothesis (Sect.~\ref{sec:prof-comp1}). The surface brightness levels are the same as for the \lya\ contours. Light brown areas indicate regions that were masked out as explained in Sect.~\ref{sec:obs-extr}. Right-hand panels in each column: The blue points show the azimuthally averaged \lya\ surface brightnesses measured in concentric circular annuli (triangles indicate negative values), with  $1\sigma$ error bars derived as described in the text. The overplotted red lines represent the circularised UV continuum profiles measured in the HST data, in monochromatic flux density units of $10^{-19}$~erg~s$^{-1}$ cm$^{-2}$ \AA$^{-1}$ arcsec$^{-2}$ (note the difference in units!). A vertical red line indicates that the object is unresolved by HST and was modelled as a point source; this line is short-dashed when no counterpart to the LAE was detected in the HST image. The green dashed curves correspond to the green dashed contours in the image panels and show the modelled continuum profiles after convolution with the MUSE PSF and rescaling to match the \lya\ profile under the null hypothesis. The inset labels provide object identifiers, redshifts, and the probabilities $p_0$ of the null hypothesis that the \lya\ emission follows the shape of the UV continuum, as explained in Sect.~\ref{sec:prof-stat}.
}
\label{fig:ima+prof}
\end{figure*}

\addtocounter{figure}{-1}
\begin{figure*}
\setlength{\unitlength}{1mm}
\begin{picture}(170,0238)
\put(0,193){\includegraphics[width=4.2cm]{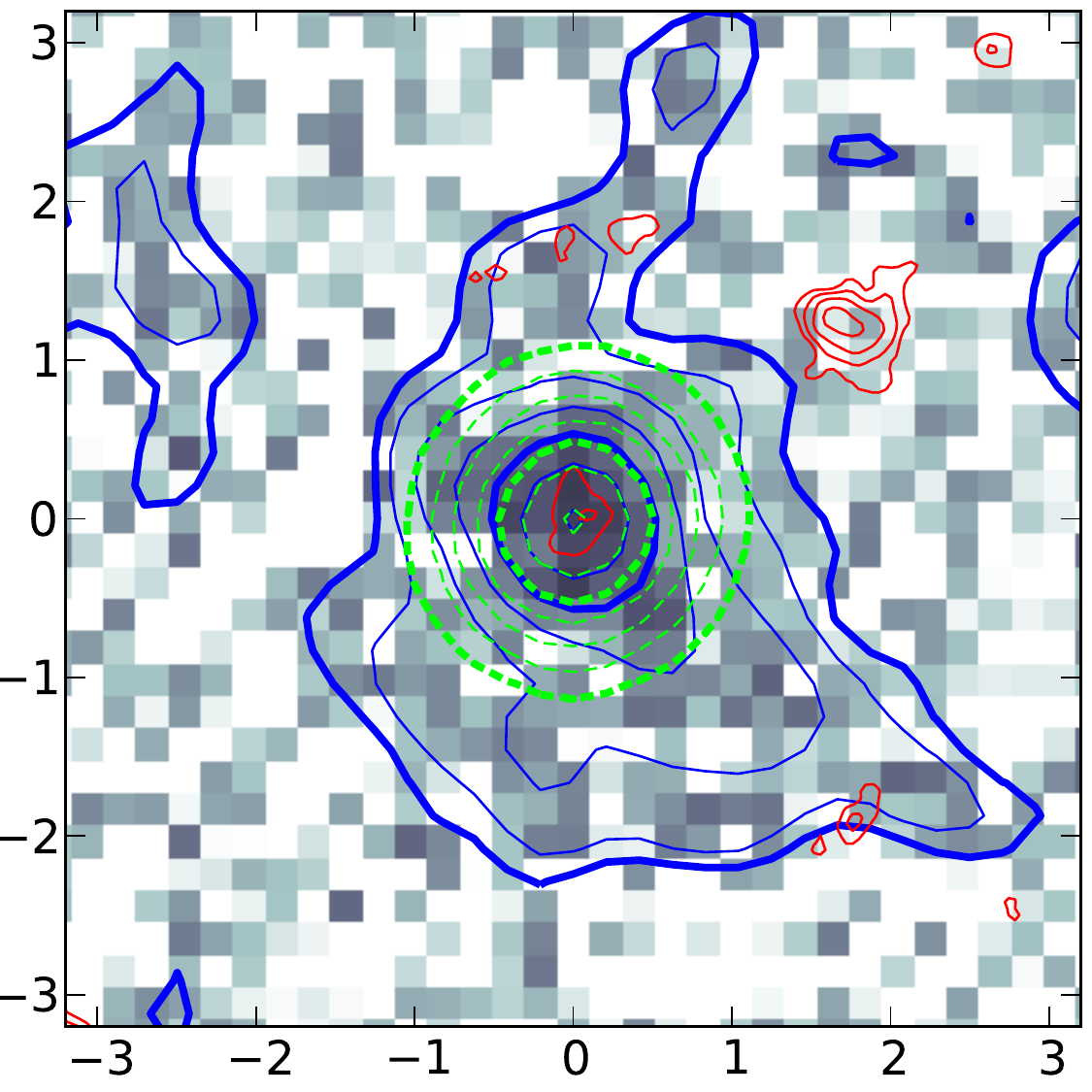}}
\put(44,192){\includegraphics[width=4.5cm]{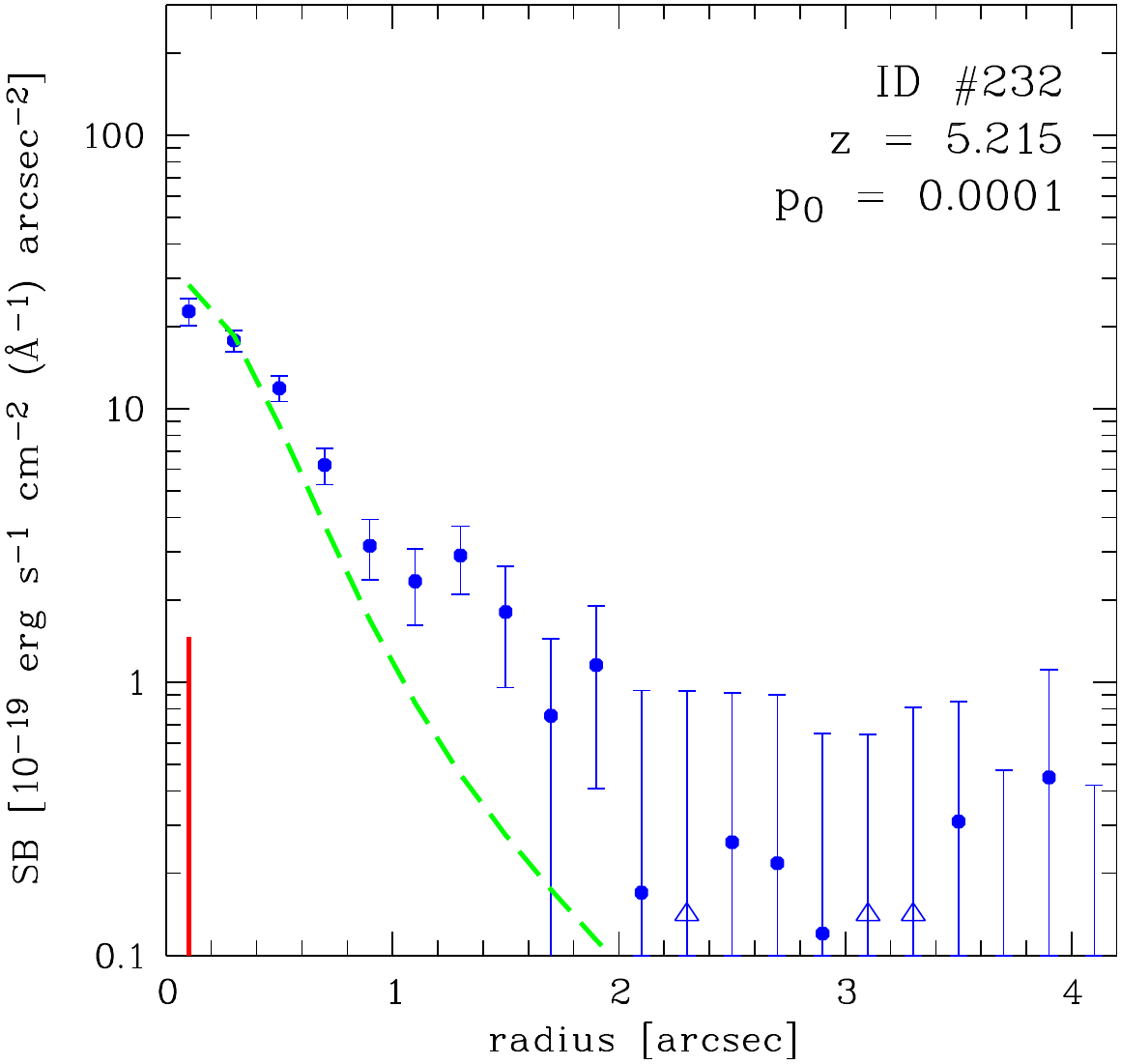}}
\put(93,193){\includegraphics[width=4.2cm]{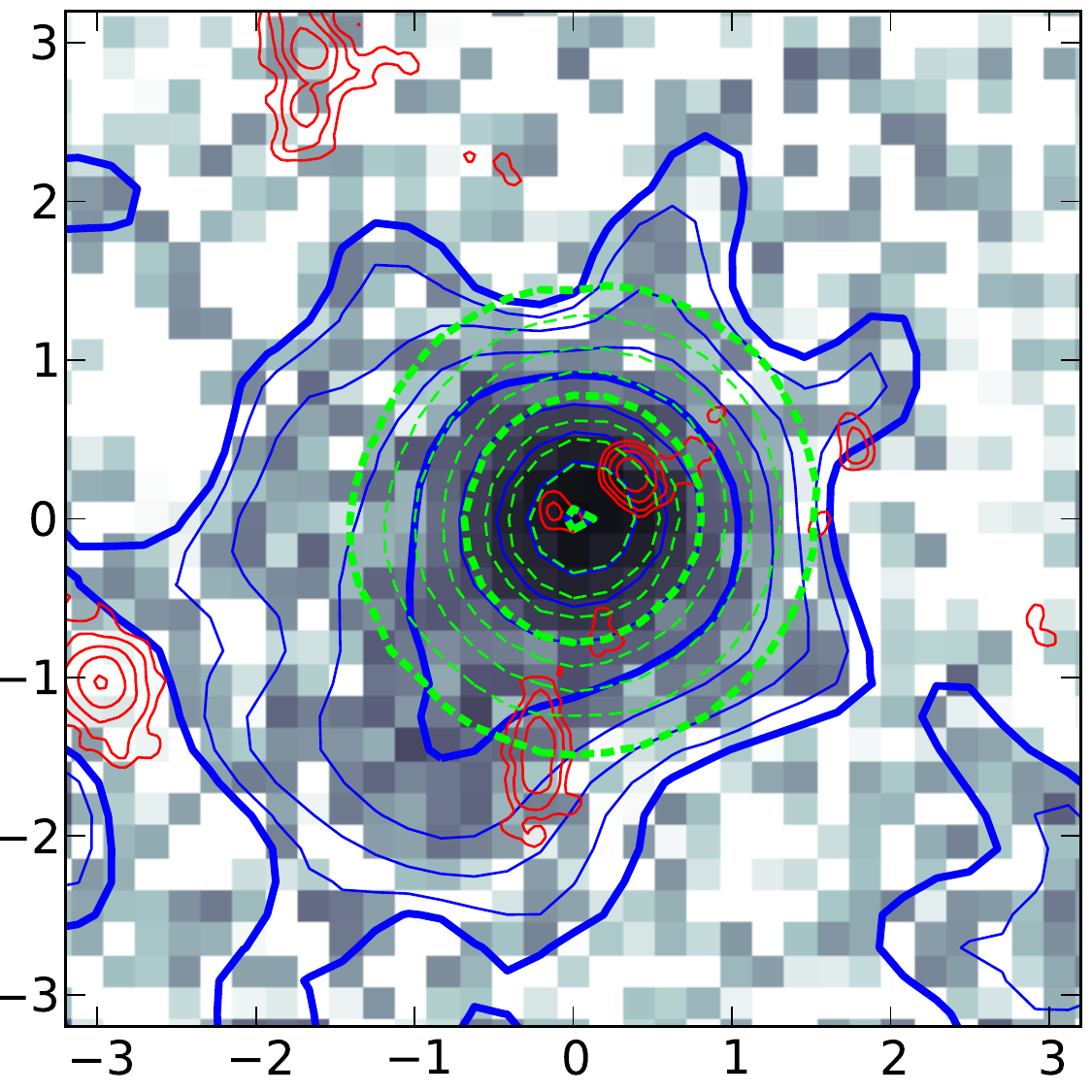}}
\put(137,192){\includegraphics[width=4.5cm]{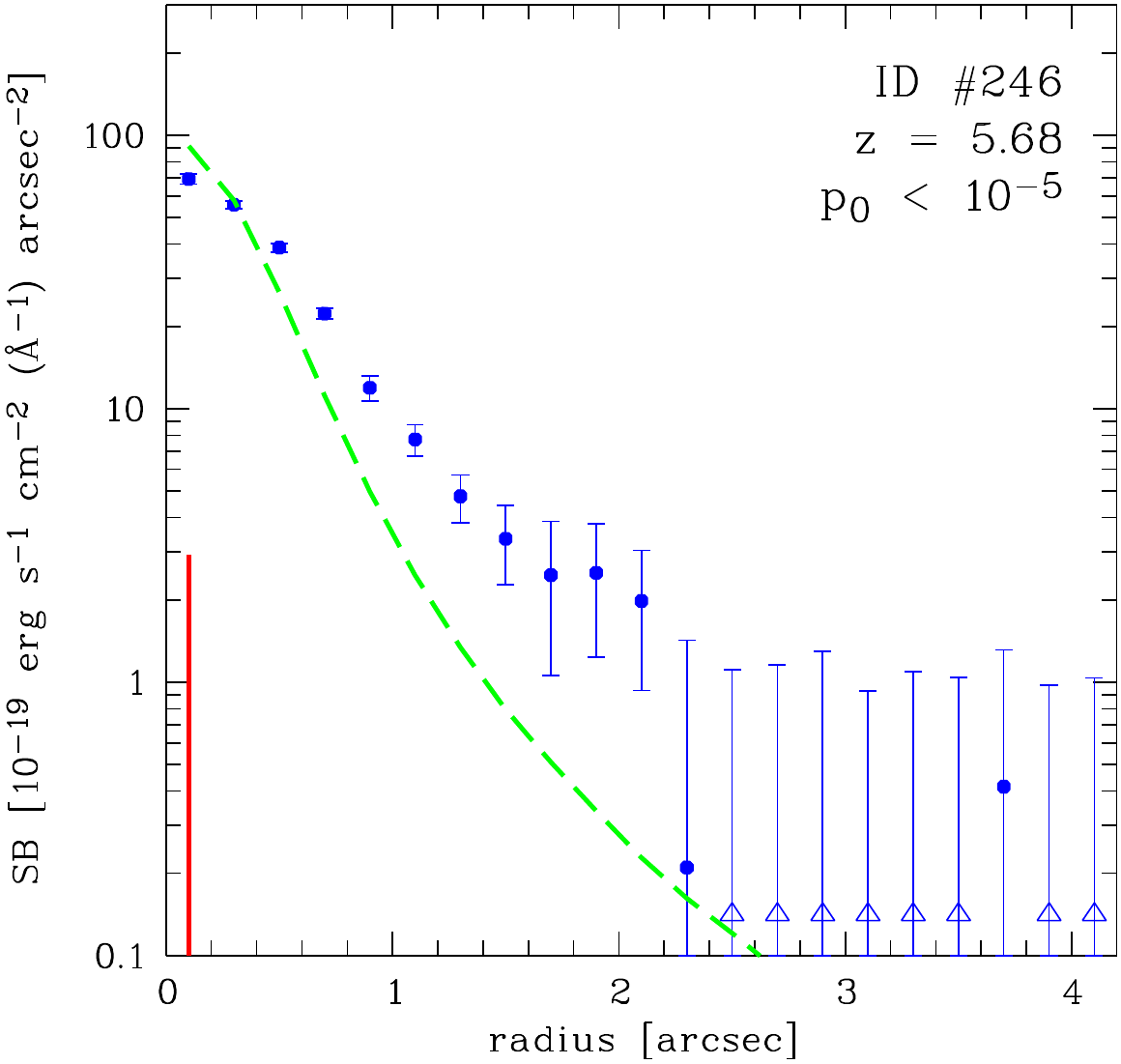}}
\put(0,145){\includegraphics[width=4.2cm]{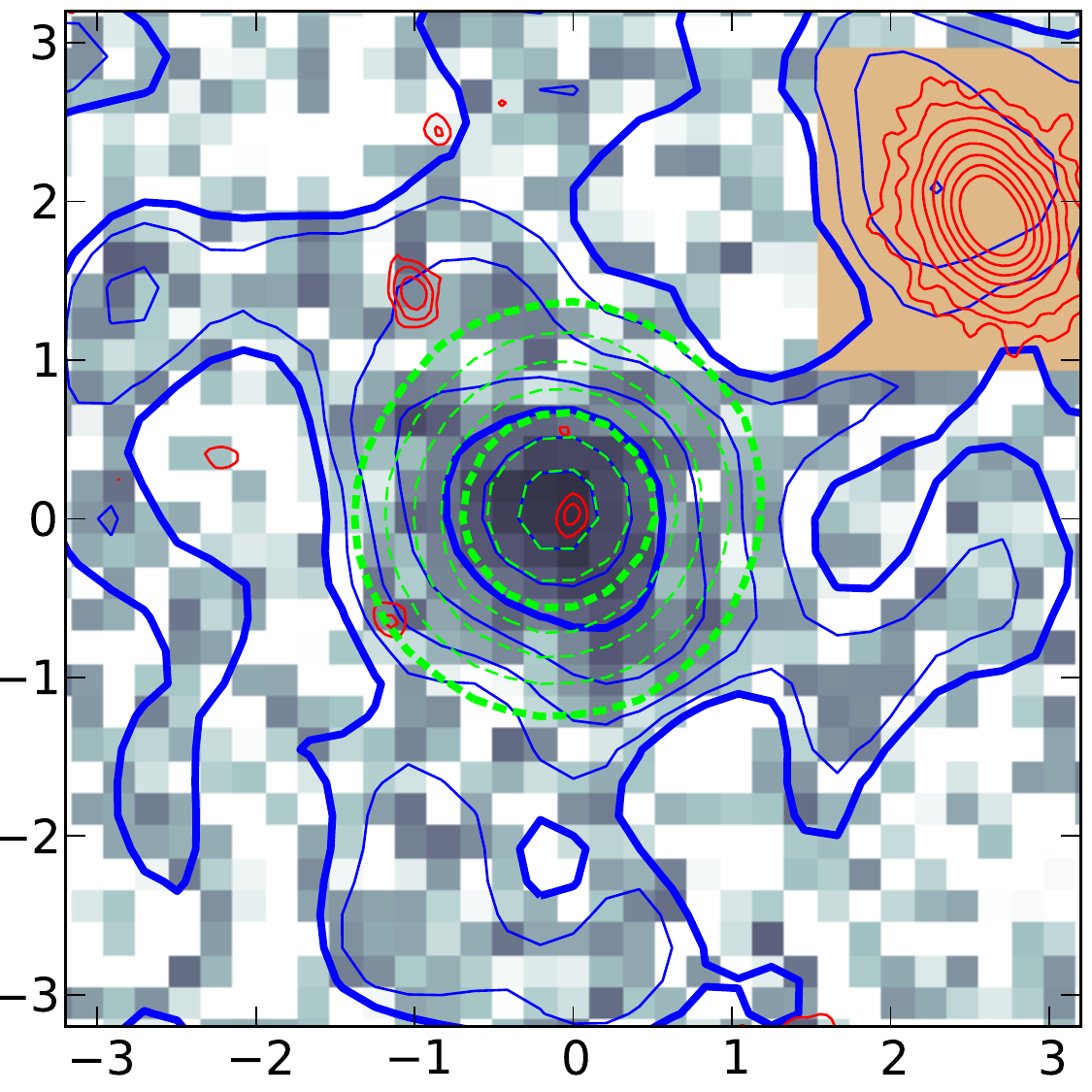}}
\put(44,144){\includegraphics[width=4.5cm]{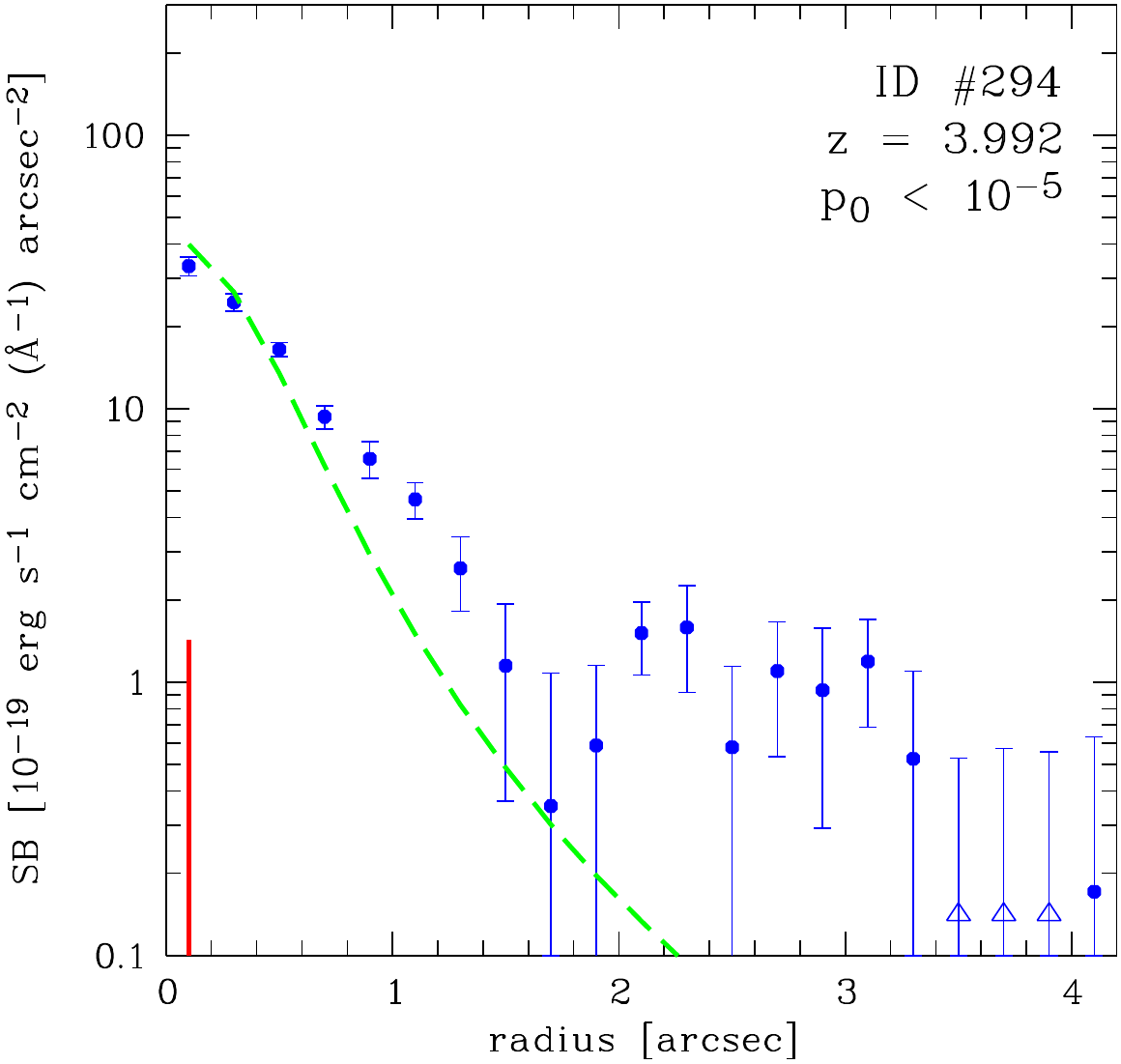}}
\put(93,145){\includegraphics[width=4.2cm]{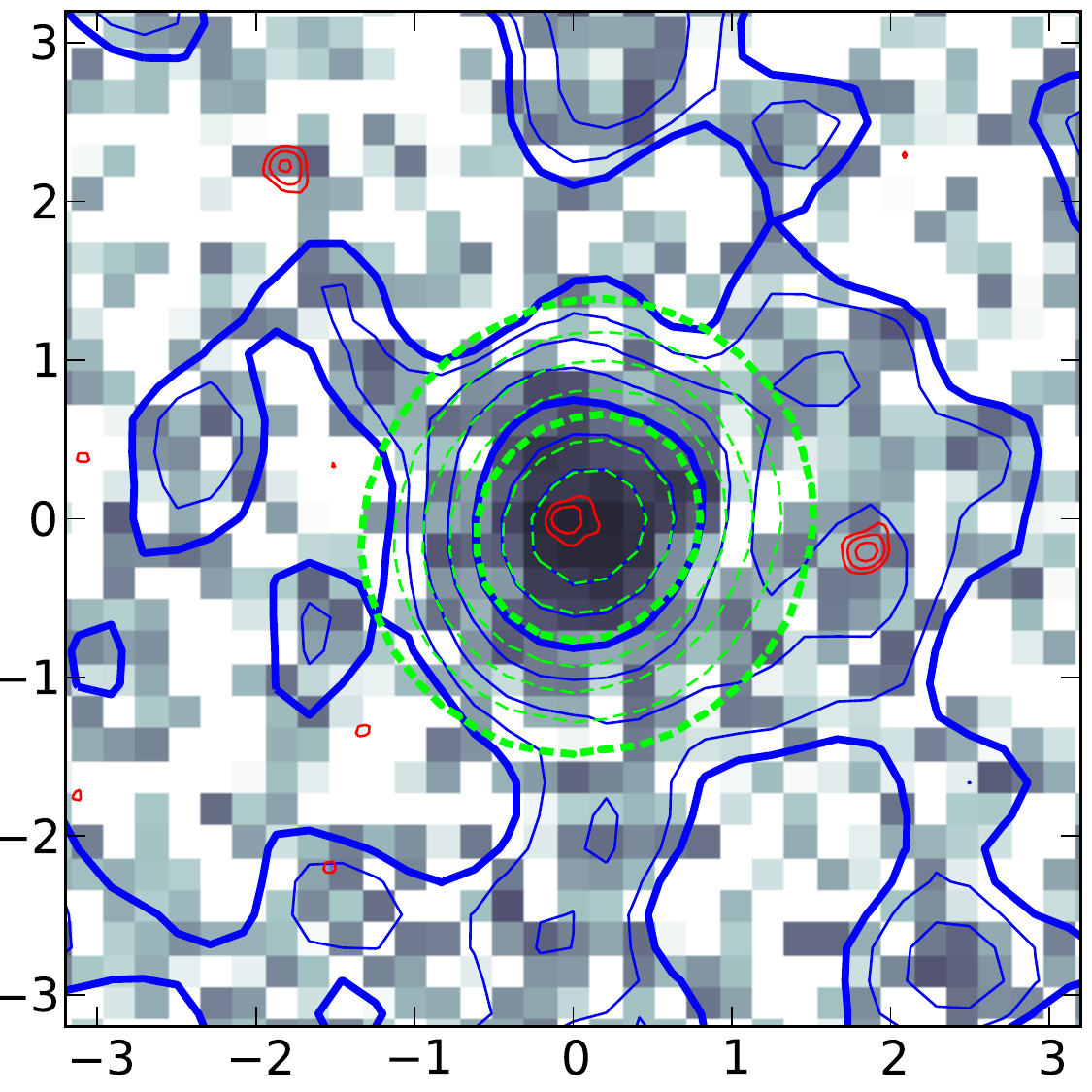}}
\put(137,144){\includegraphics[width=4.5cm]{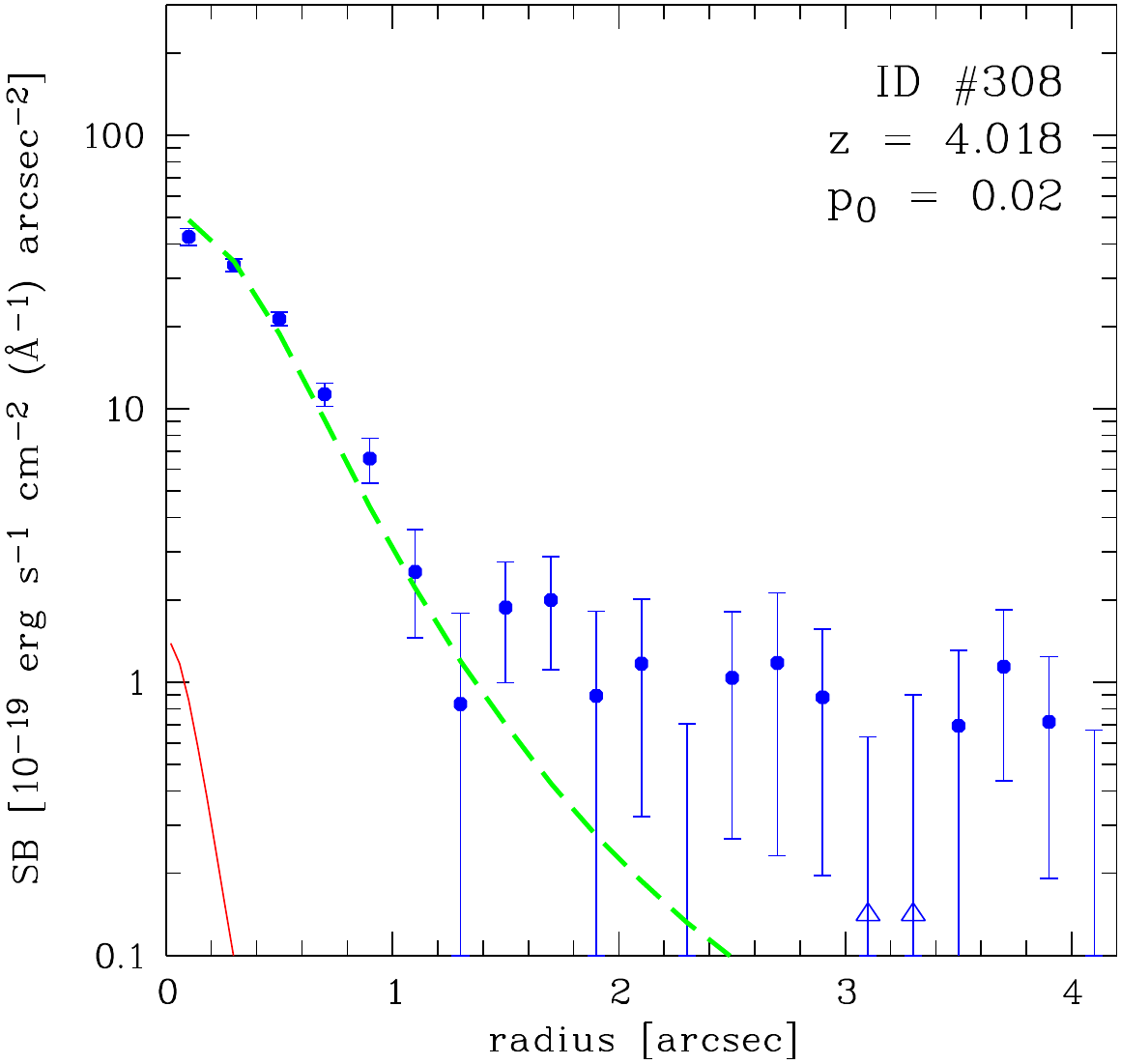}}
\put(0,97){\includegraphics[width=4.2cm]{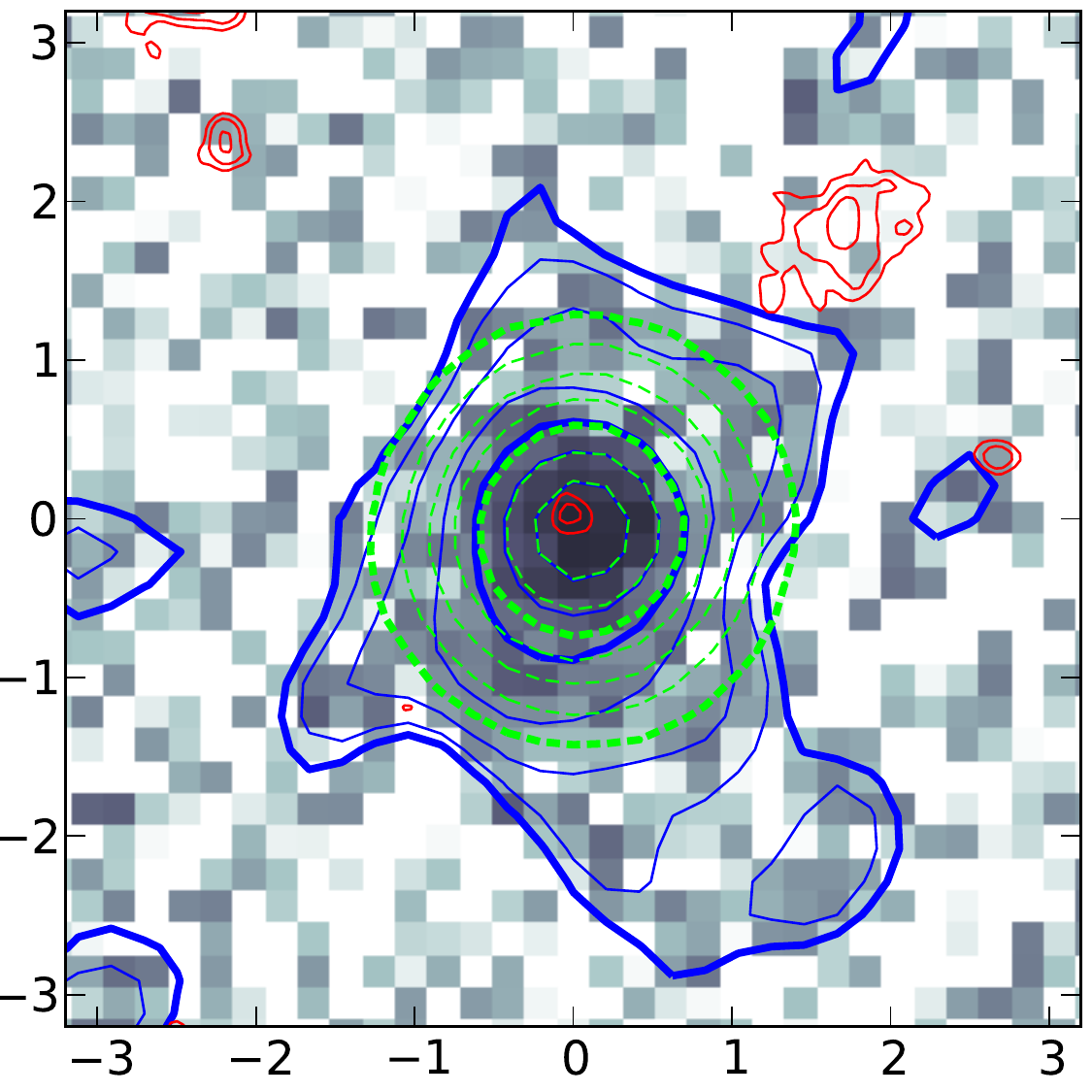}}
\put(44,96){\includegraphics[width=4.5cm]{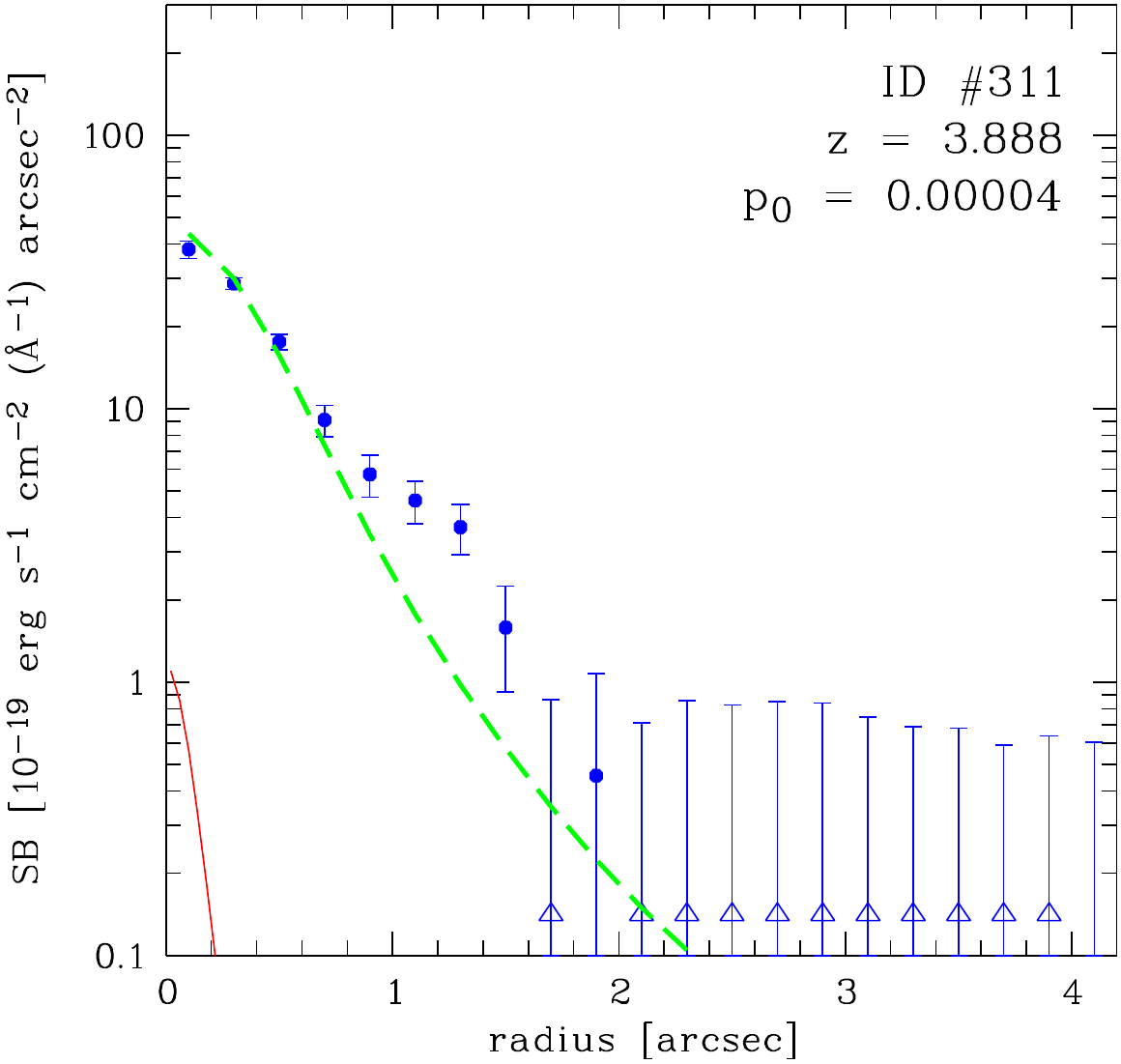}}
\put(93,97){\includegraphics[width=4.2cm]{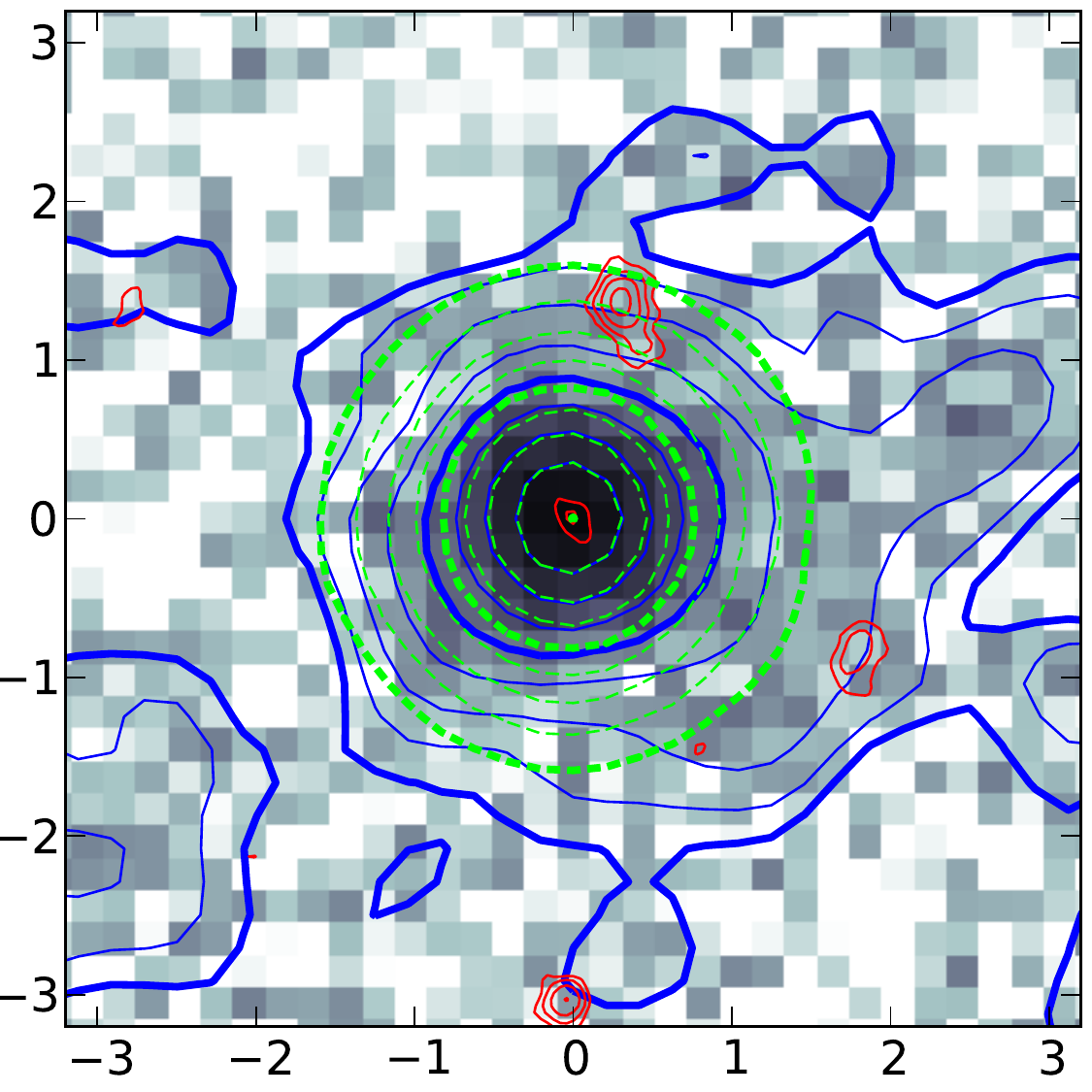}}
\put(137,96){\includegraphics[width=4.5cm]{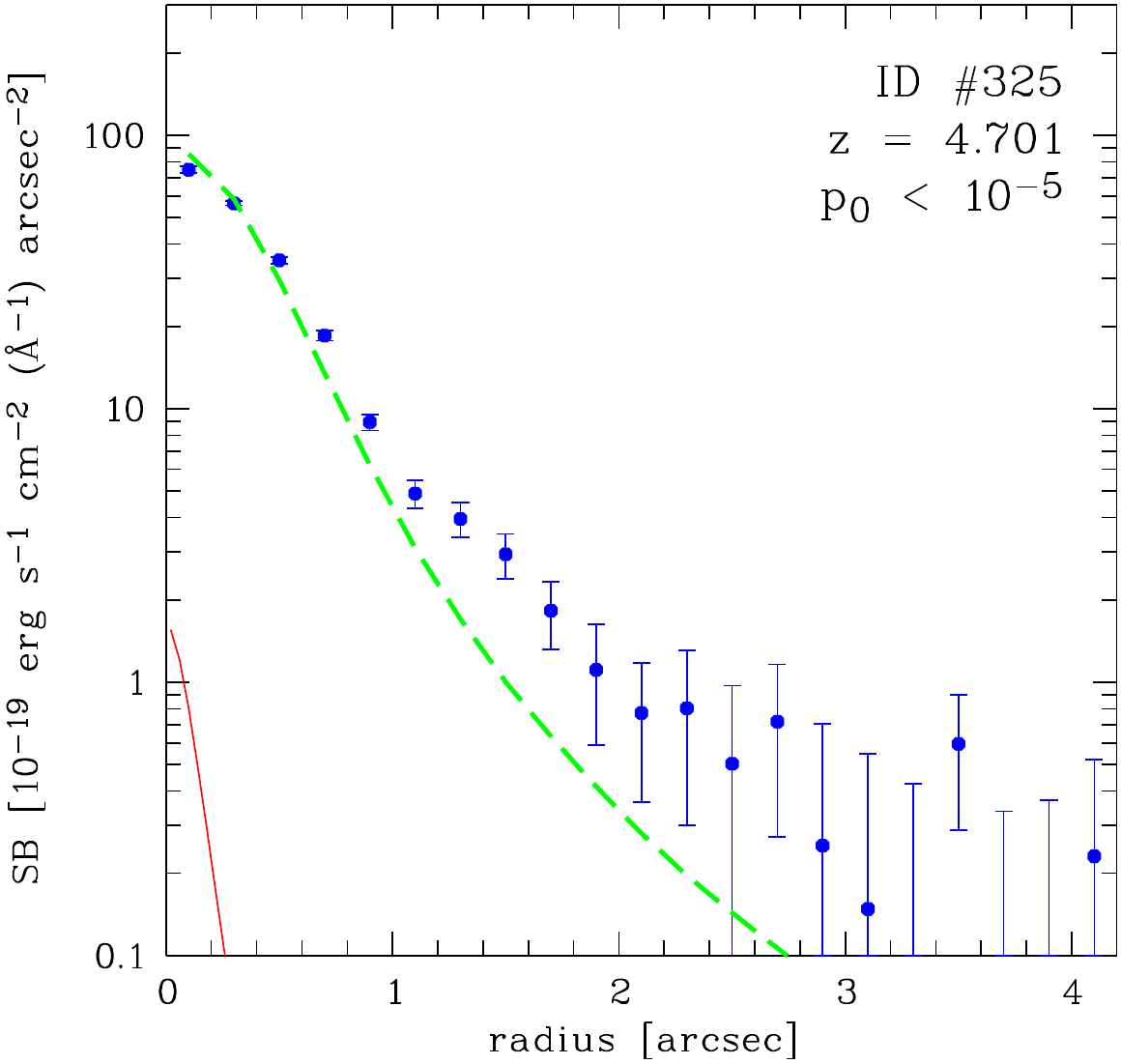}}
\put(0,49){\includegraphics[width=4.2cm]{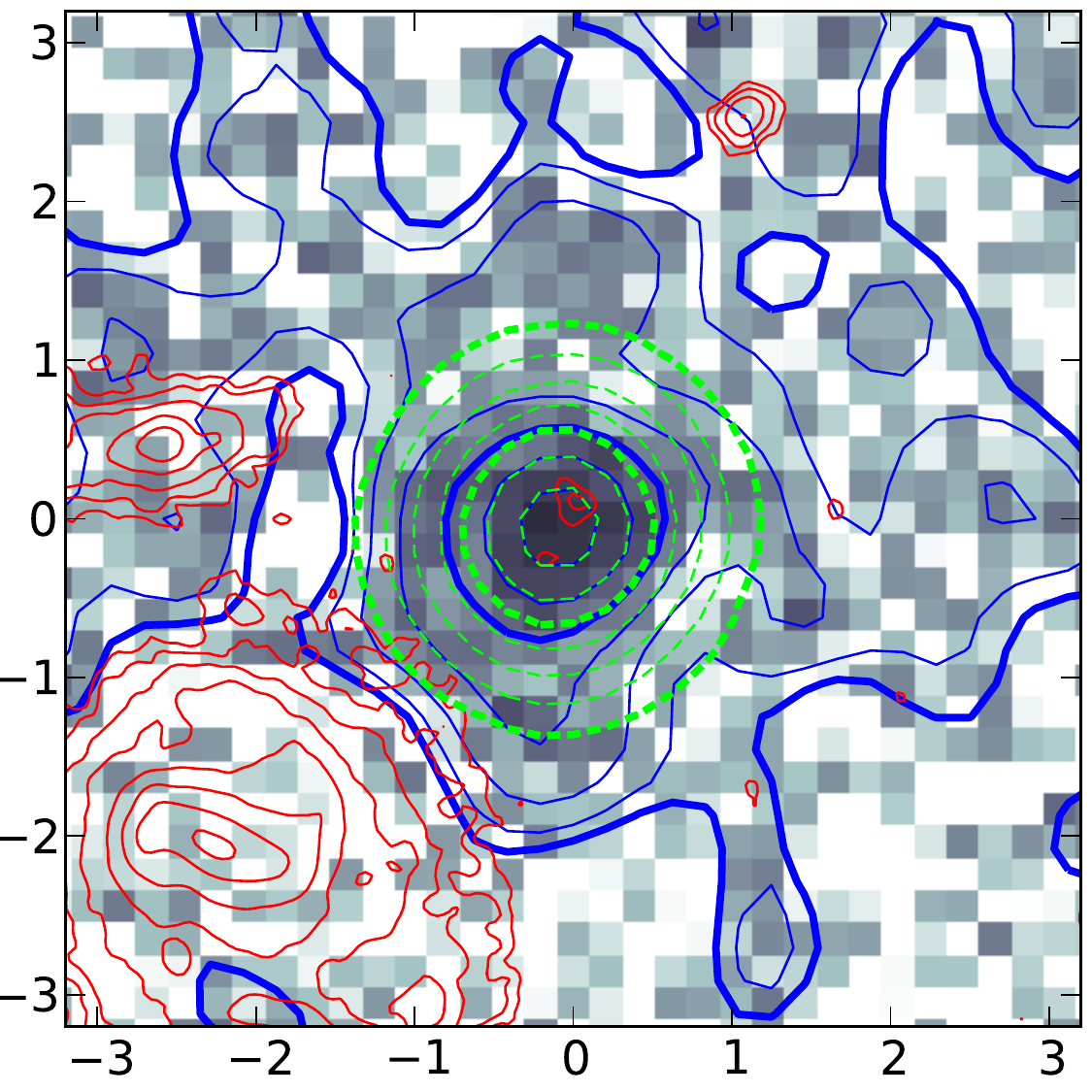}}
\put(44,48){\includegraphics[width=4.5cm]{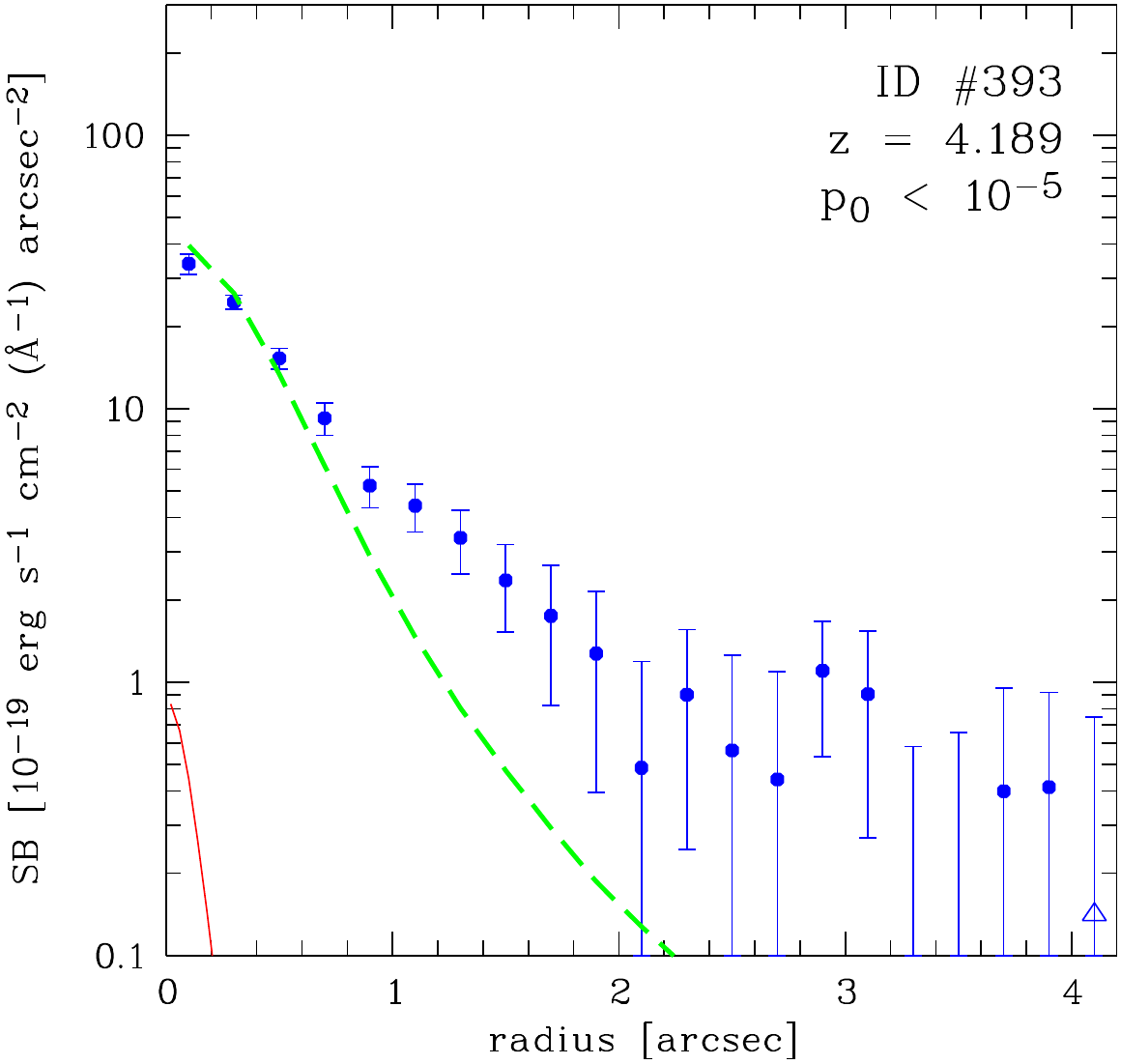}}
\put(93,49){\includegraphics[width=4.2cm]{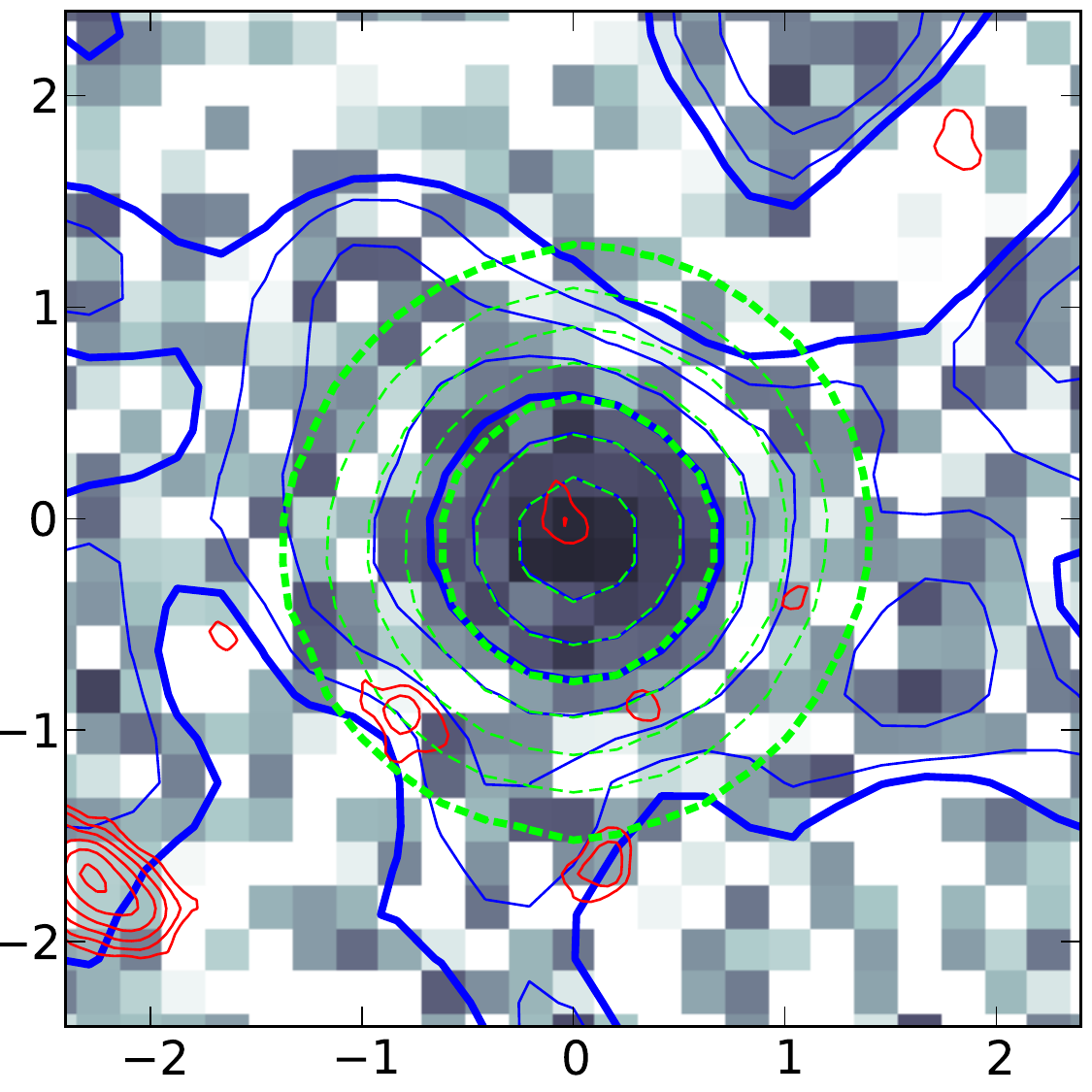}}
\put(137,48){\includegraphics[width=4.5cm]{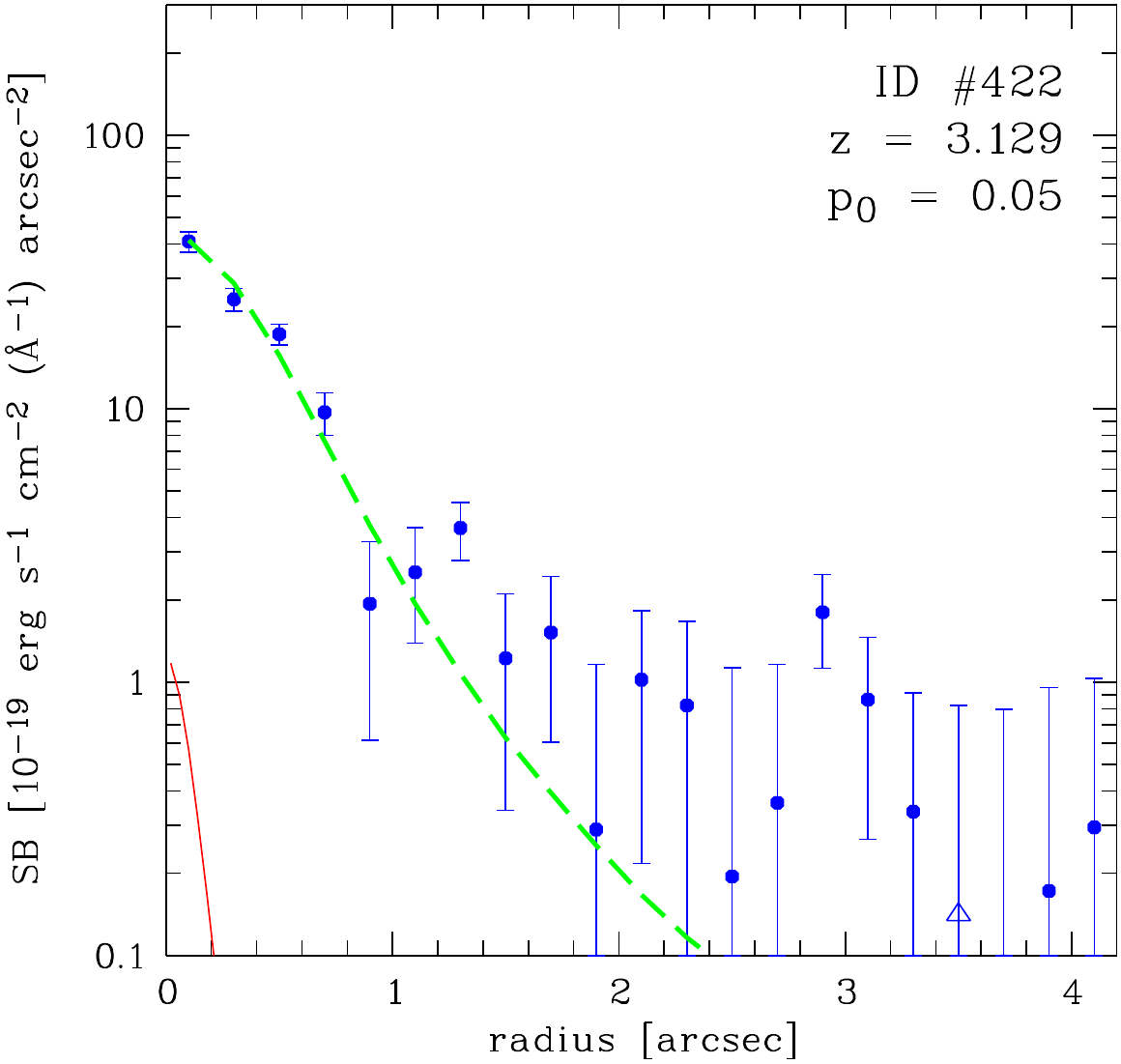}}
\put(0,1){\includegraphics[width=4.2cm]{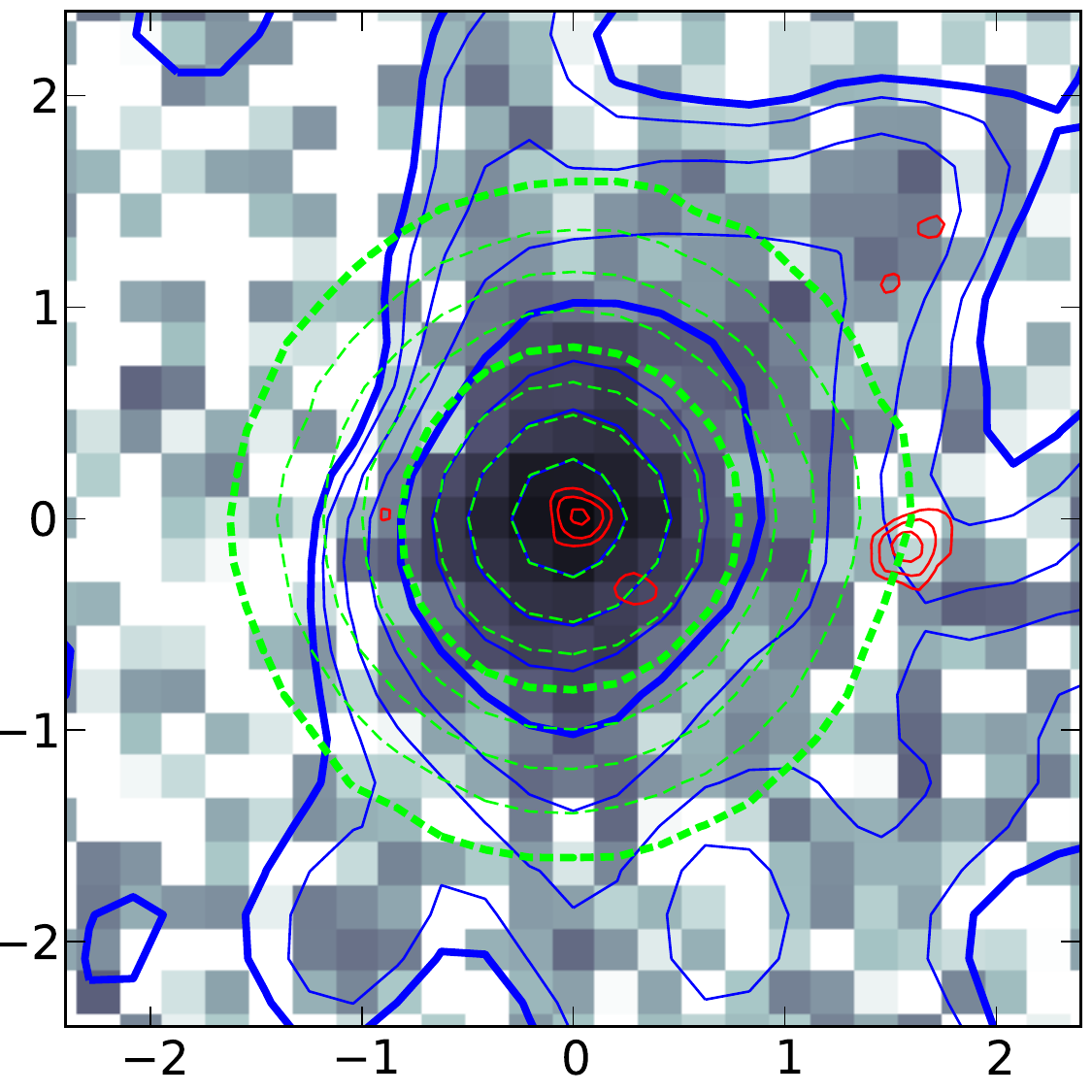}}
\put(44,0){\includegraphics[width=4.5cm]{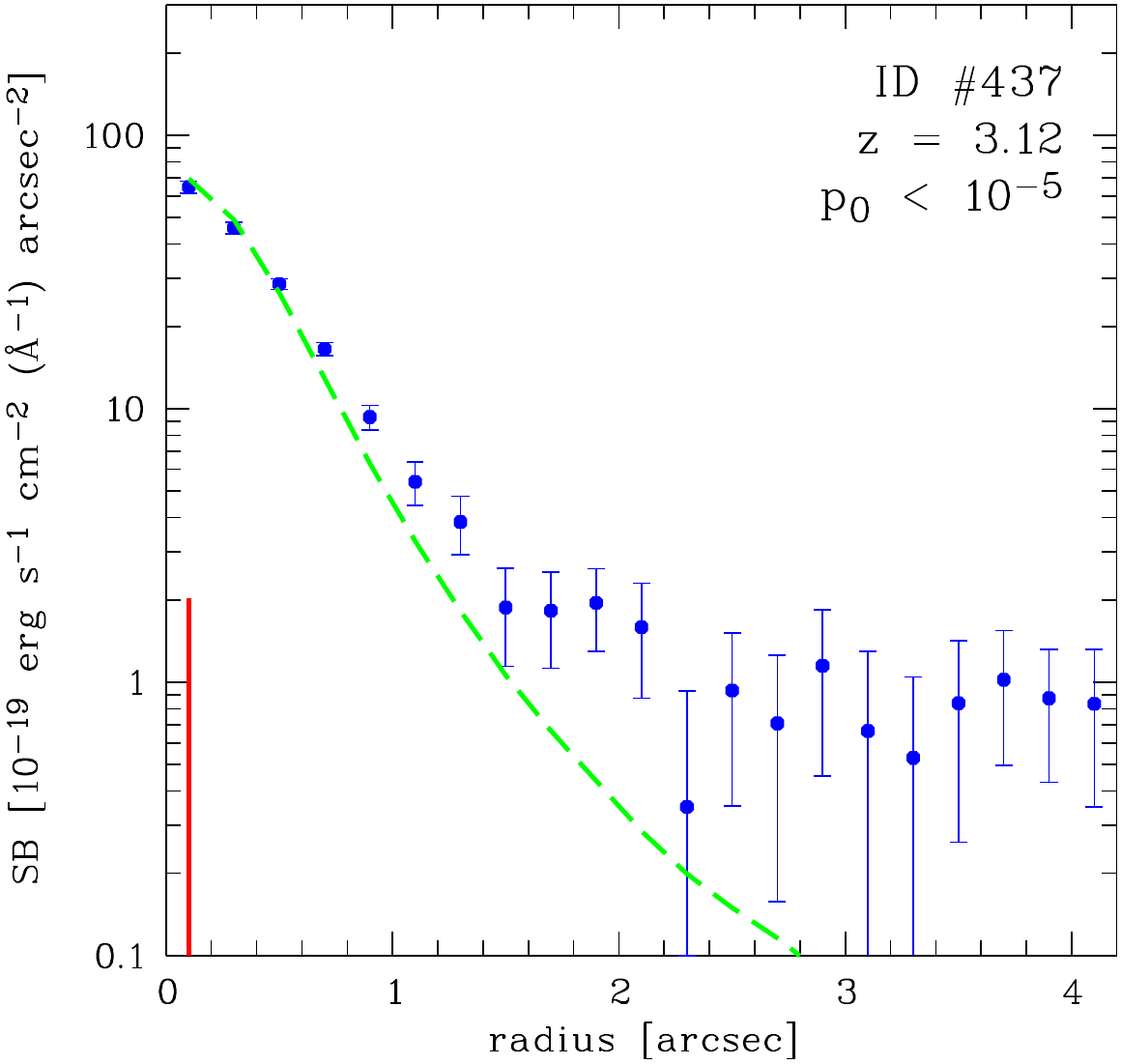}}
\put(93,1){\includegraphics[width=4.2cm]{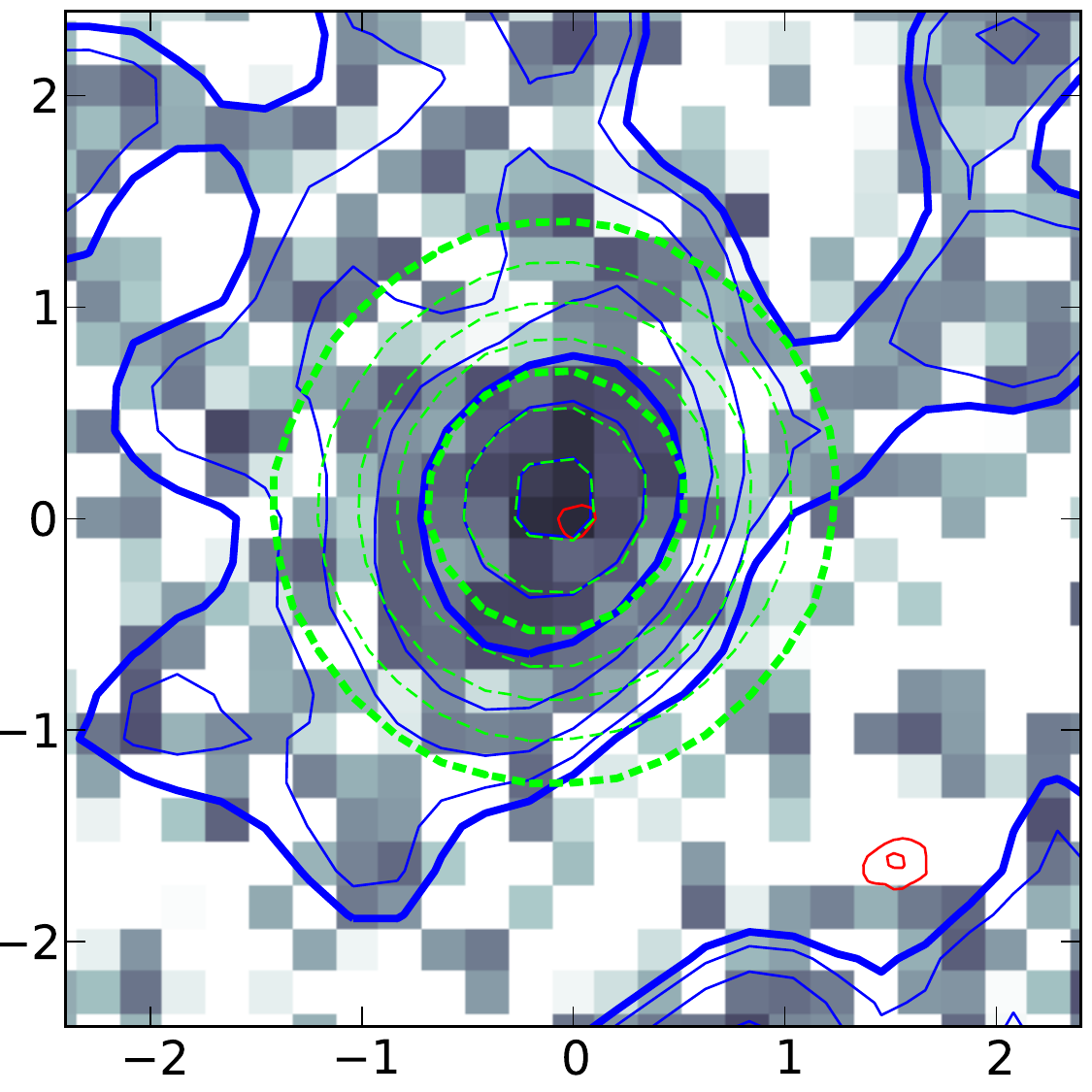}}
\put(137,0){\includegraphics[width=4.5cm]{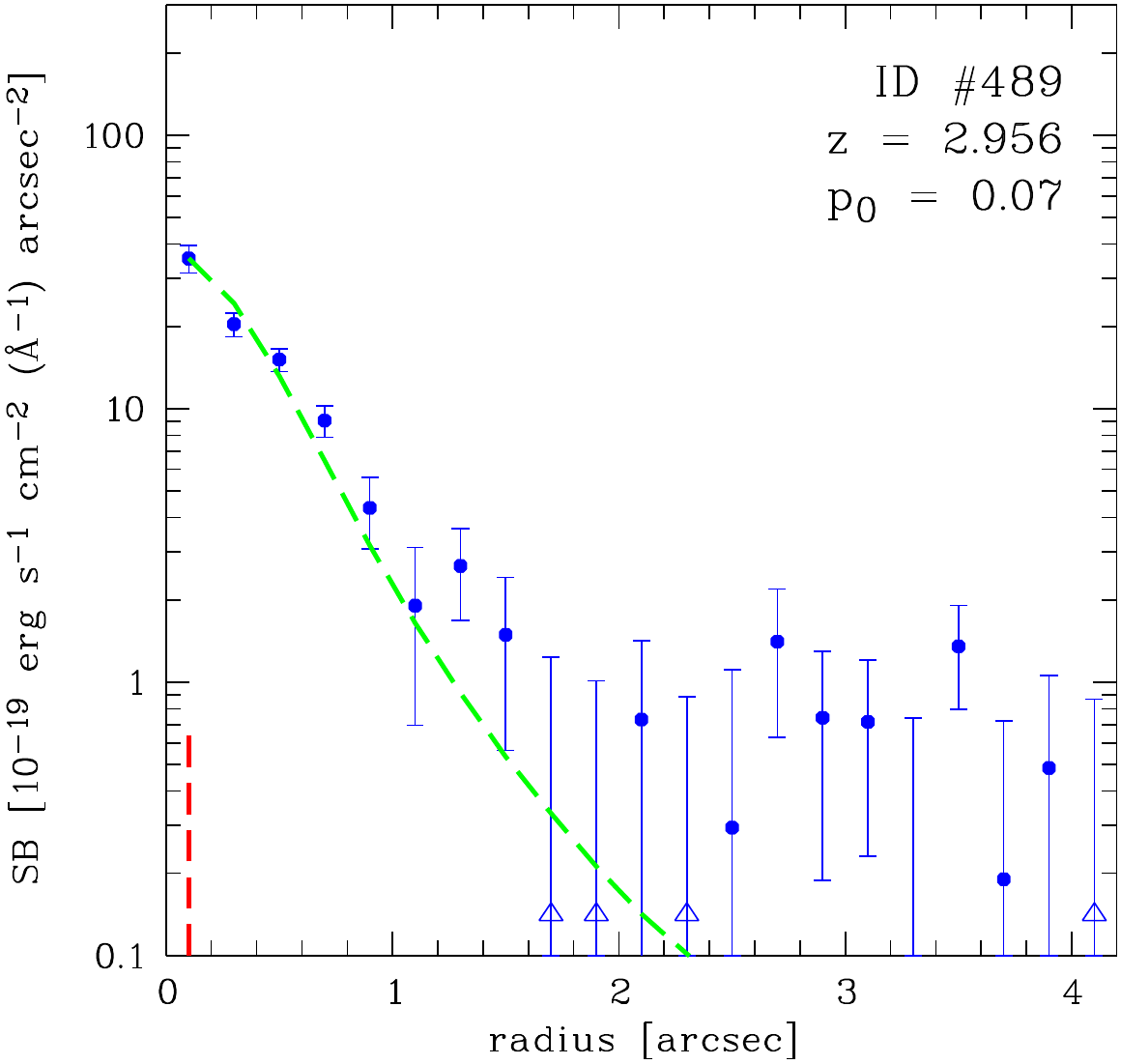}}
\end{picture}
\caption[]{(continued)}
\end{figure*}

\addtocounter{figure}{-1}
\begin{figure*}
\setlength{\unitlength}{1mm}
\begin{picture}(170,0190)
\put(0,145){\includegraphics[width=4.2cm]{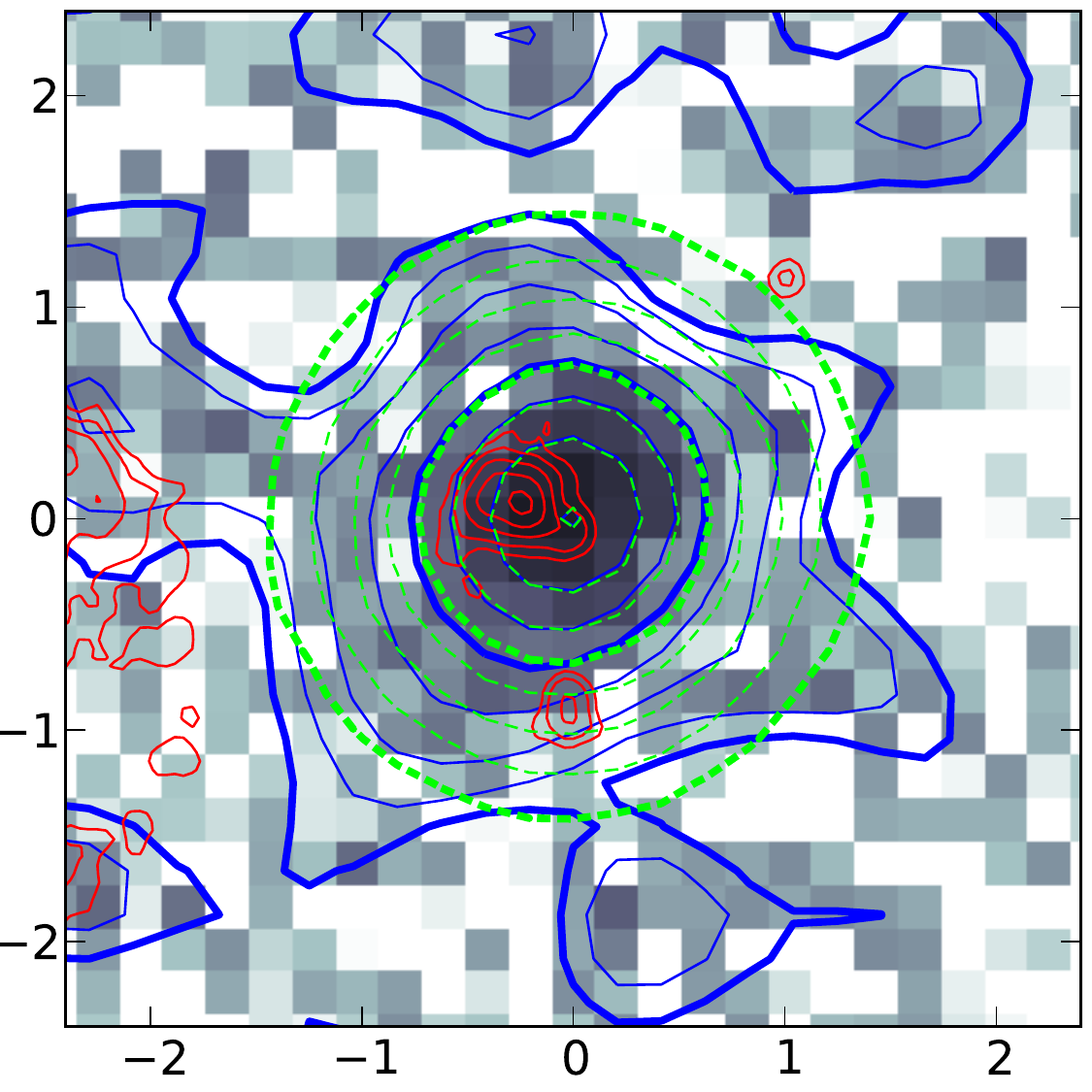}}
\put(44,144){\includegraphics[width=4.5cm]{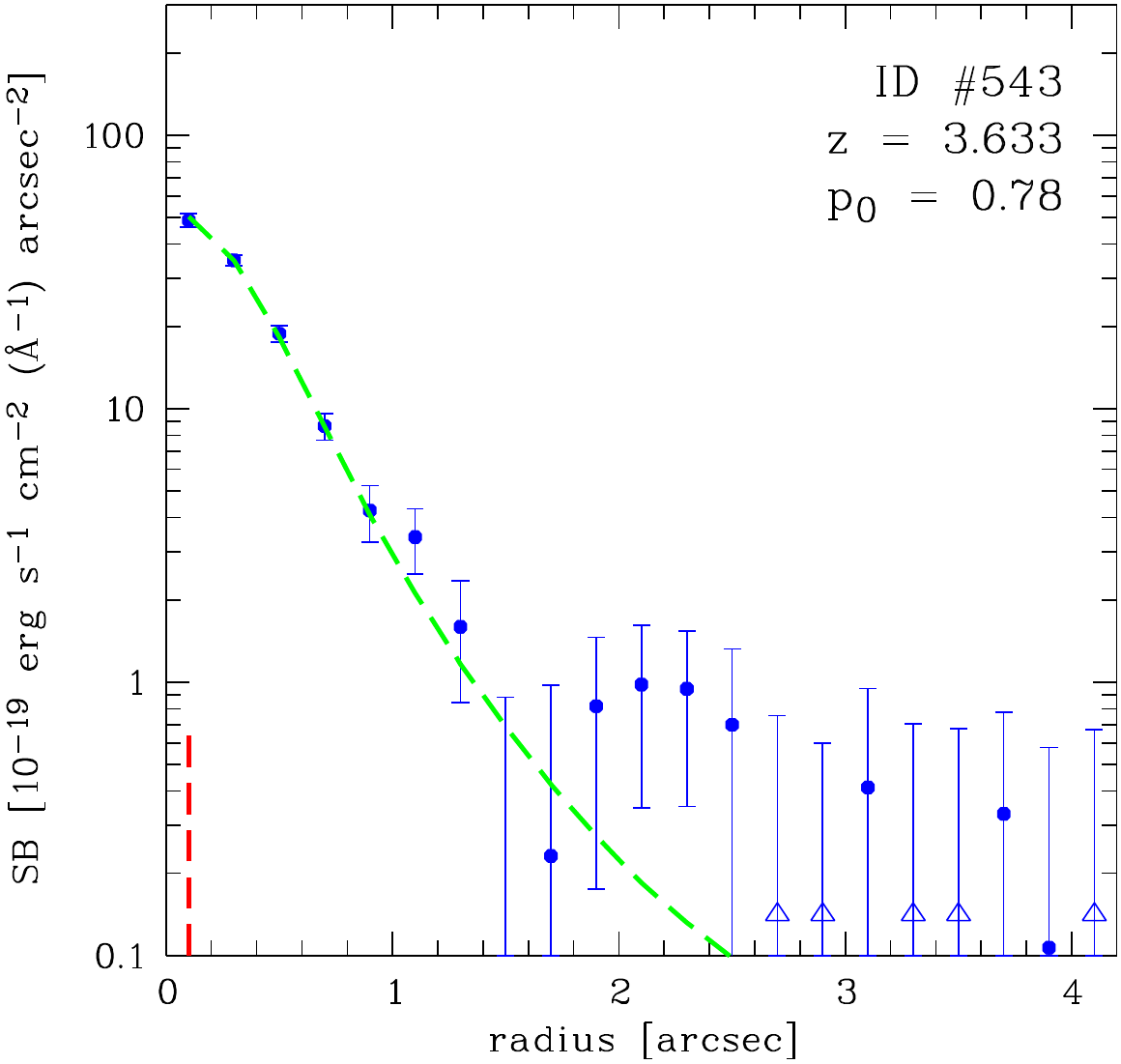}}
\put(93,145){\includegraphics[width=4.2cm]{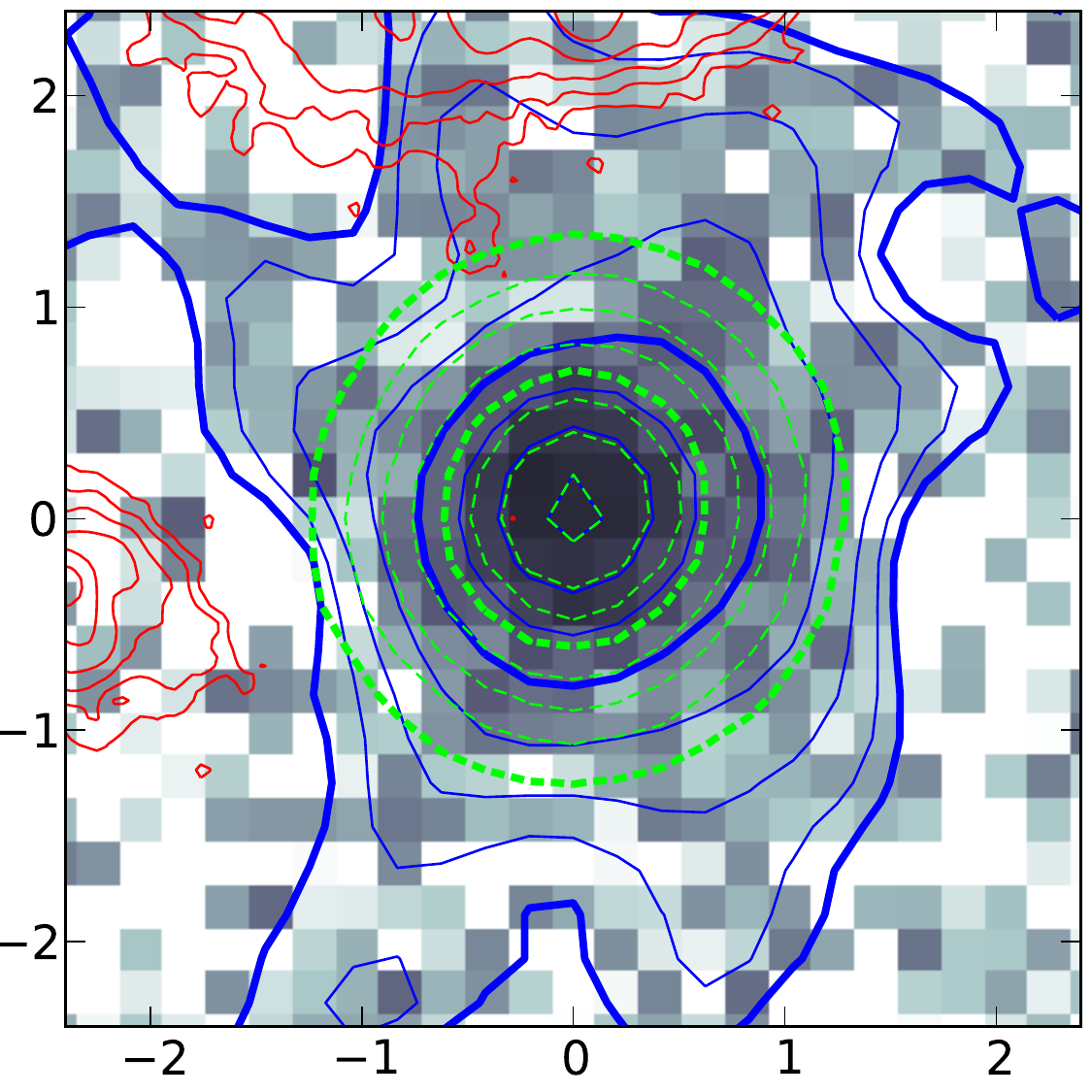}}
\put(137,144){\includegraphics[width=4.5cm]{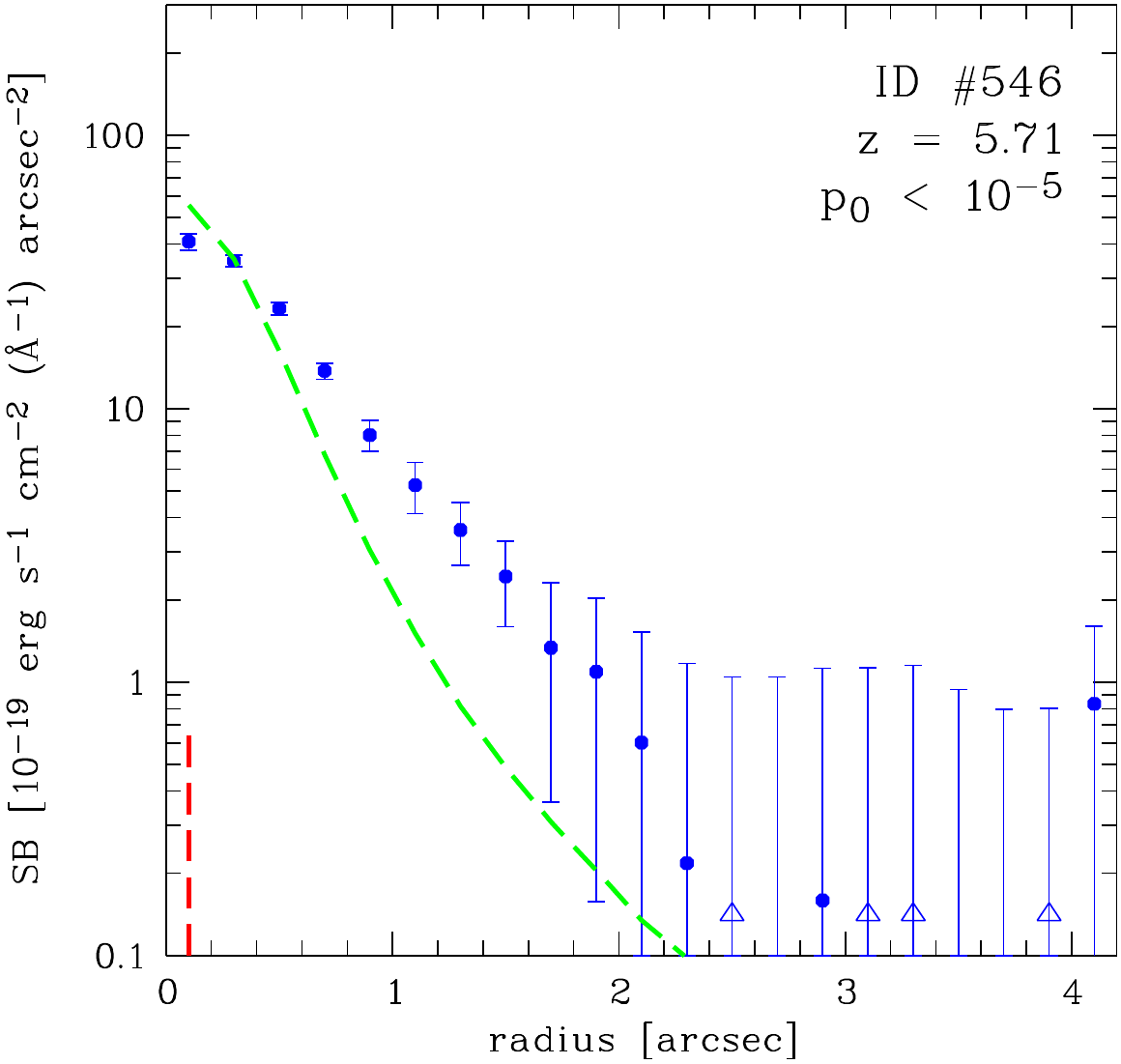}}
\put(0,97){\includegraphics[width=4.2cm]{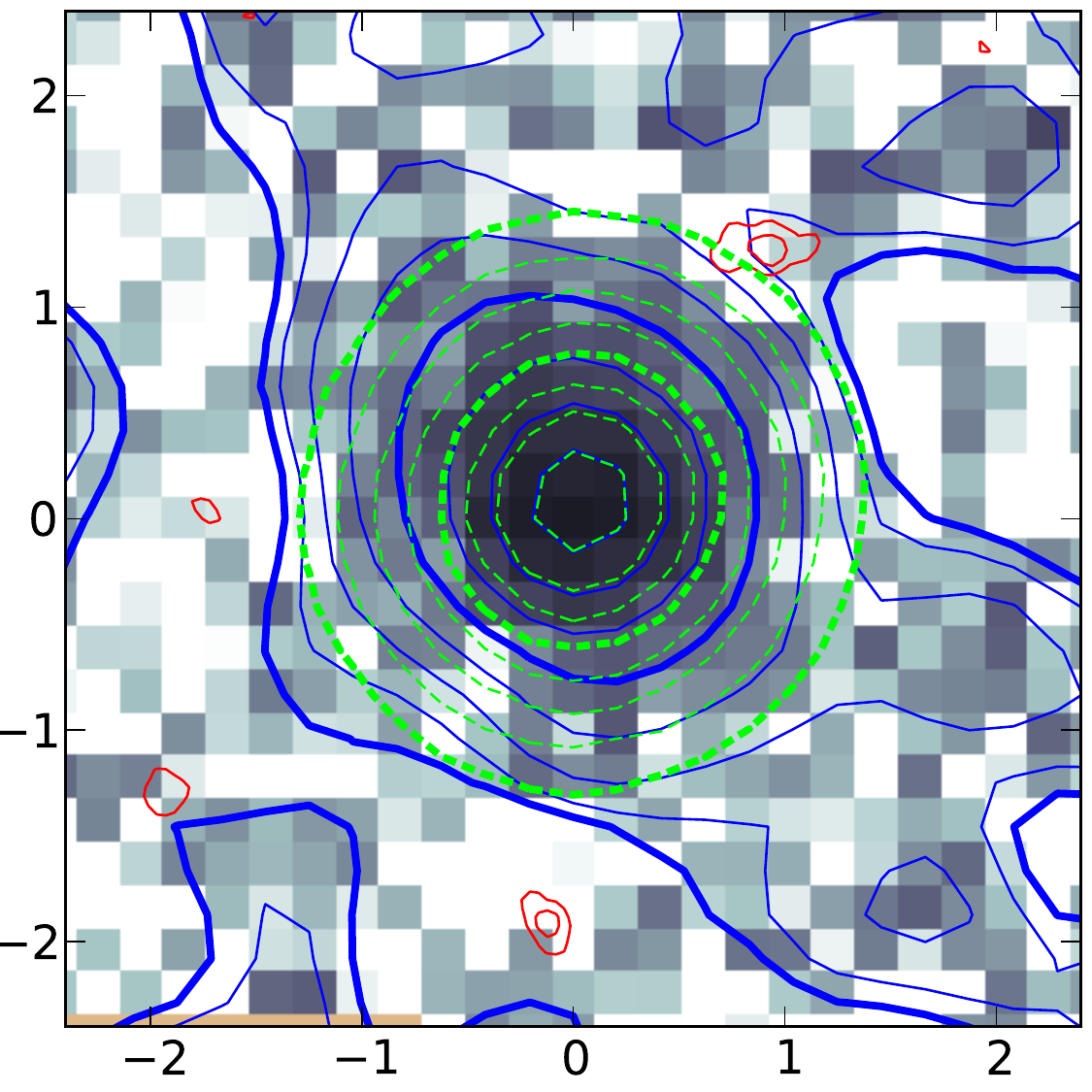}}
\put(44,96){\includegraphics[width=4.5cm]{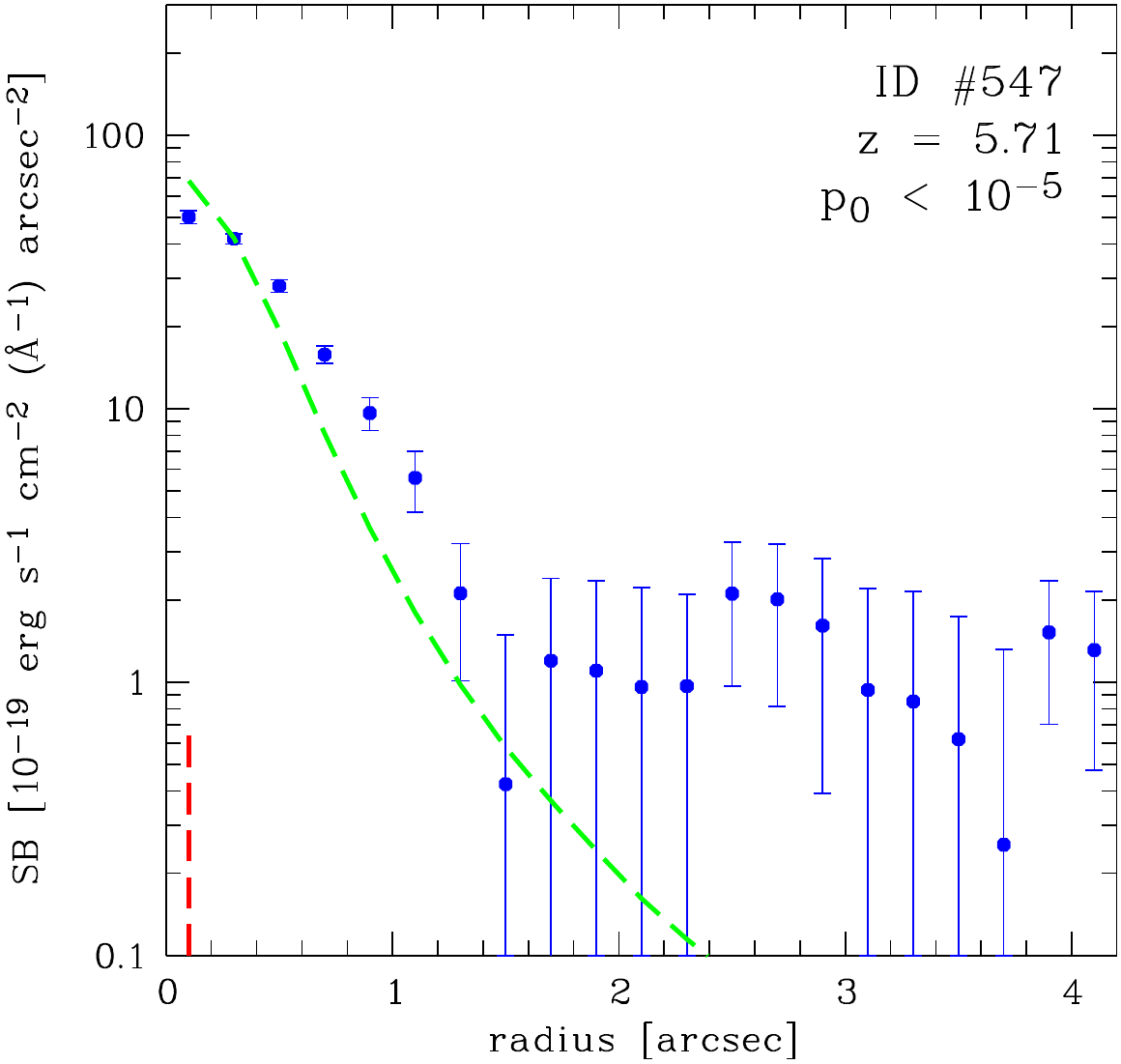}}
\put(93,97){\includegraphics[width=4.2cm]{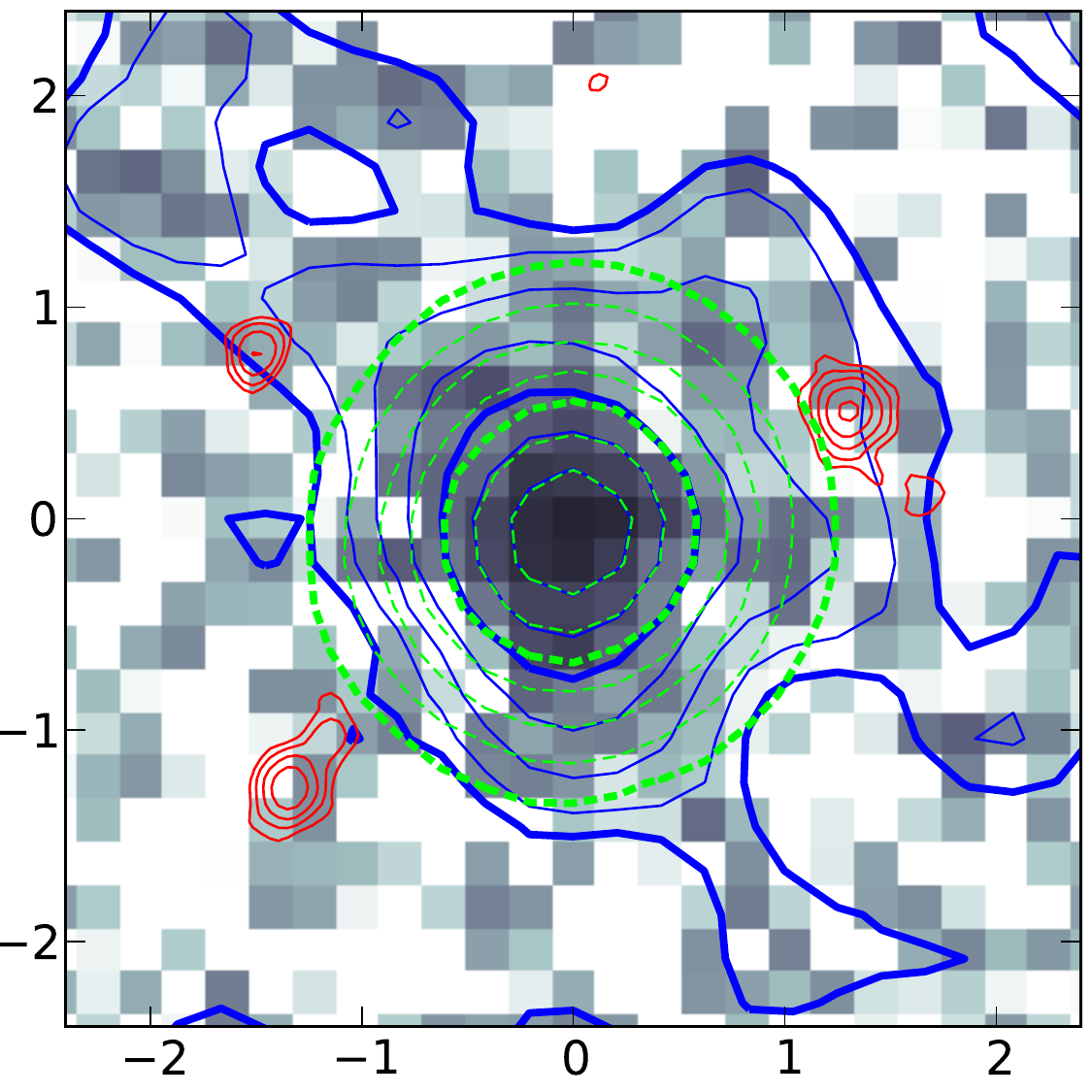}}
\put(137,96){\includegraphics[width=4.5cm]{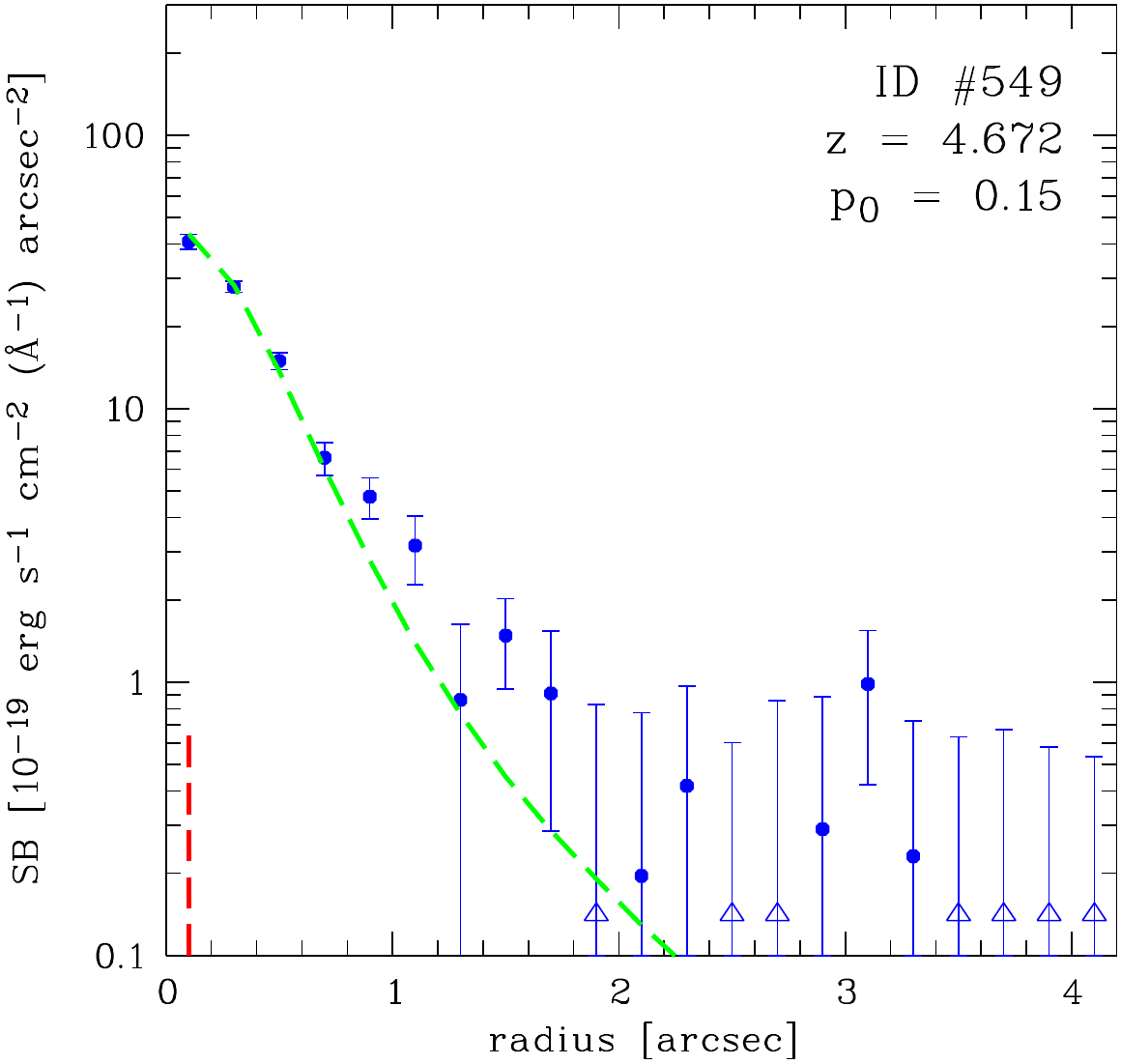}}
\put(0,49){\includegraphics[width=4.2cm]{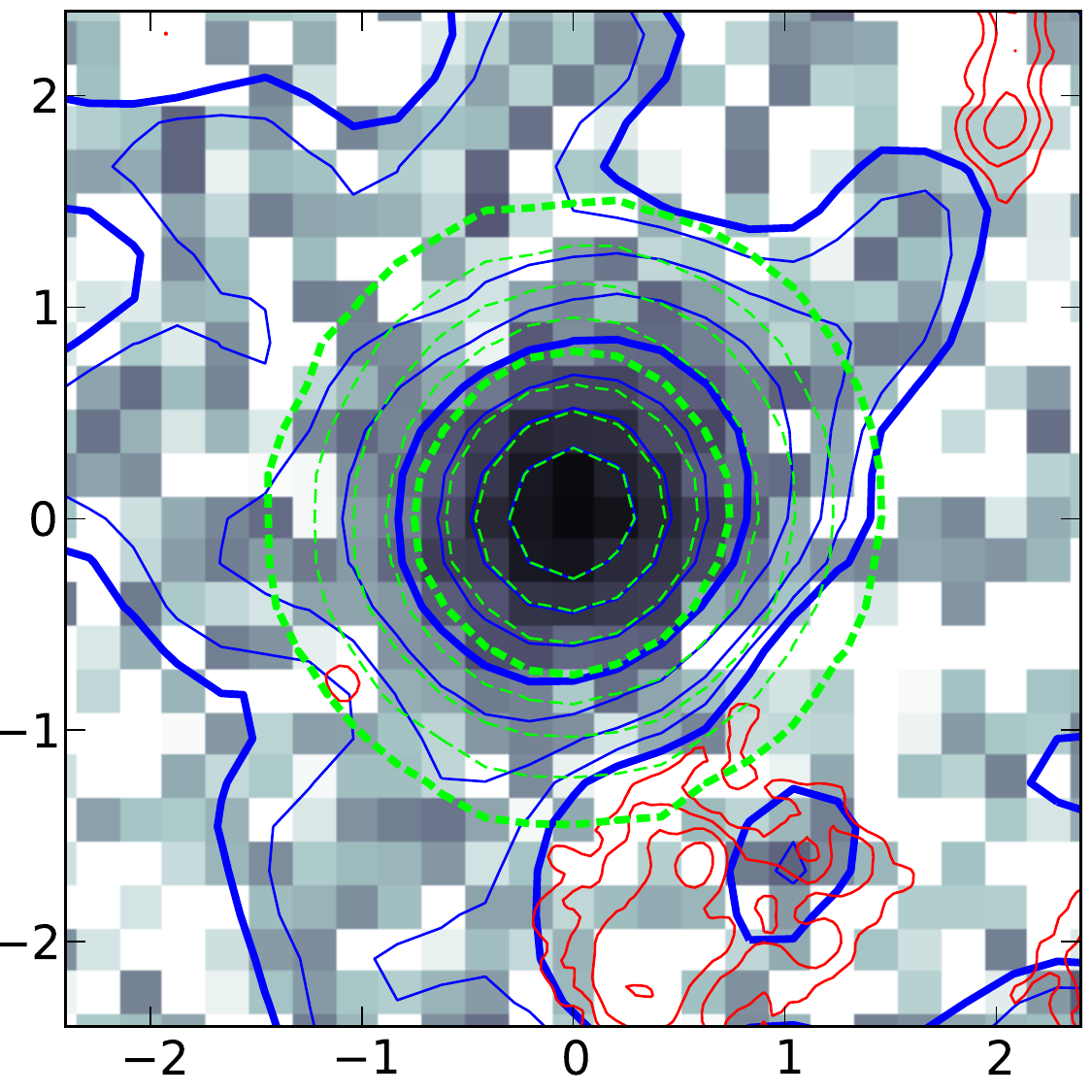}}
\put(44,48){\includegraphics[width=4.5cm]{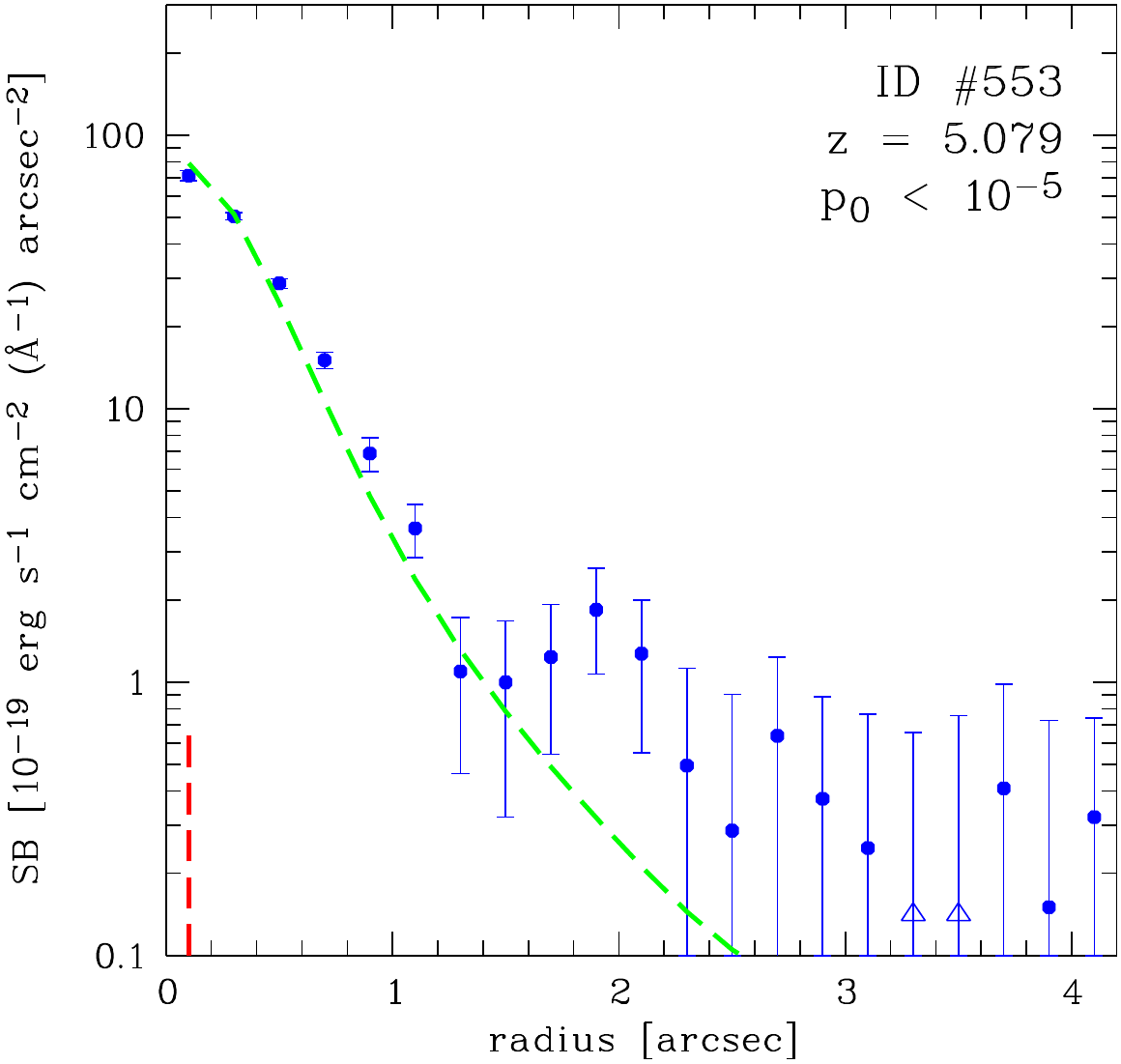}}
\put(93,49){\includegraphics[width=4.2cm]{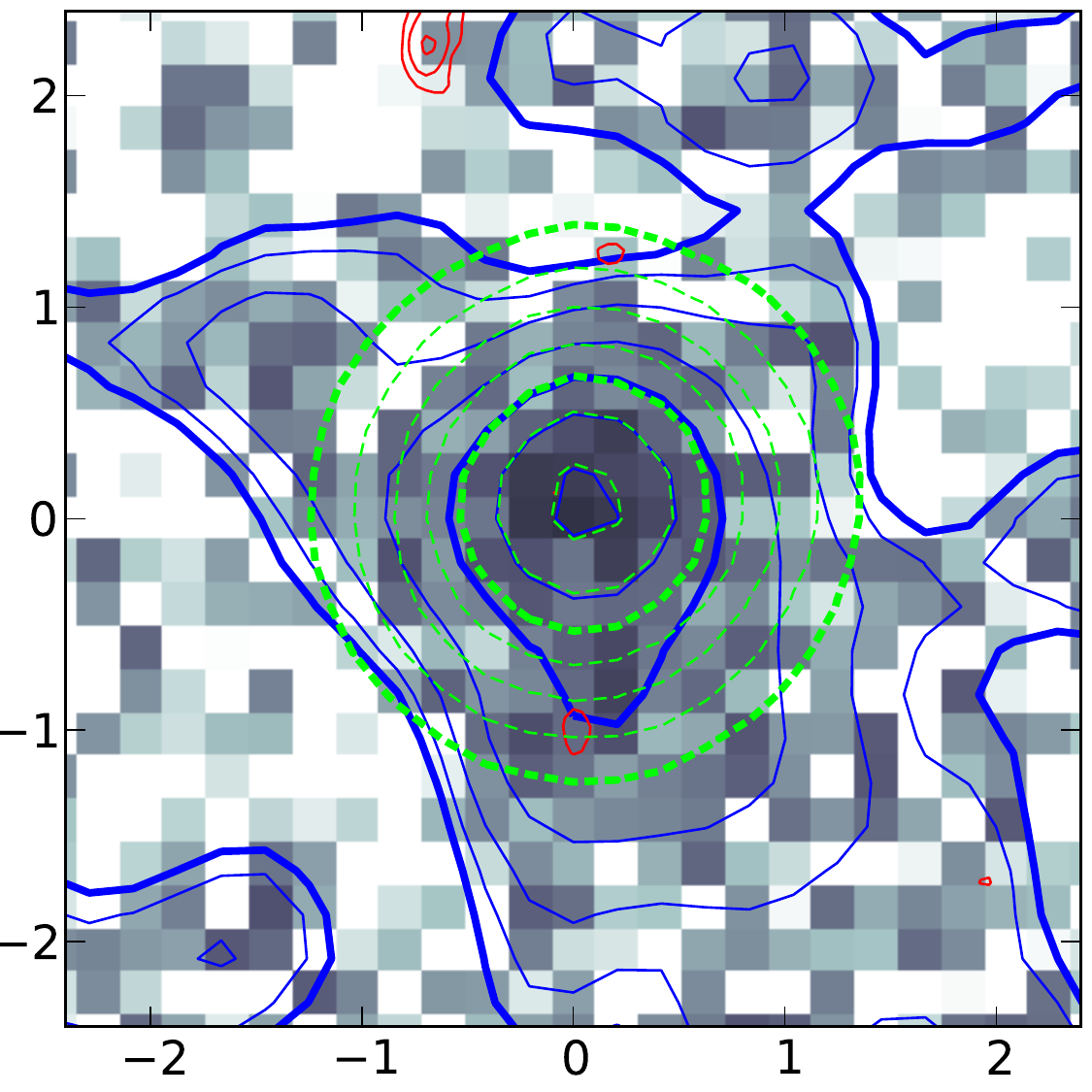}}
\put(137,48){\includegraphics[width=4.5cm]{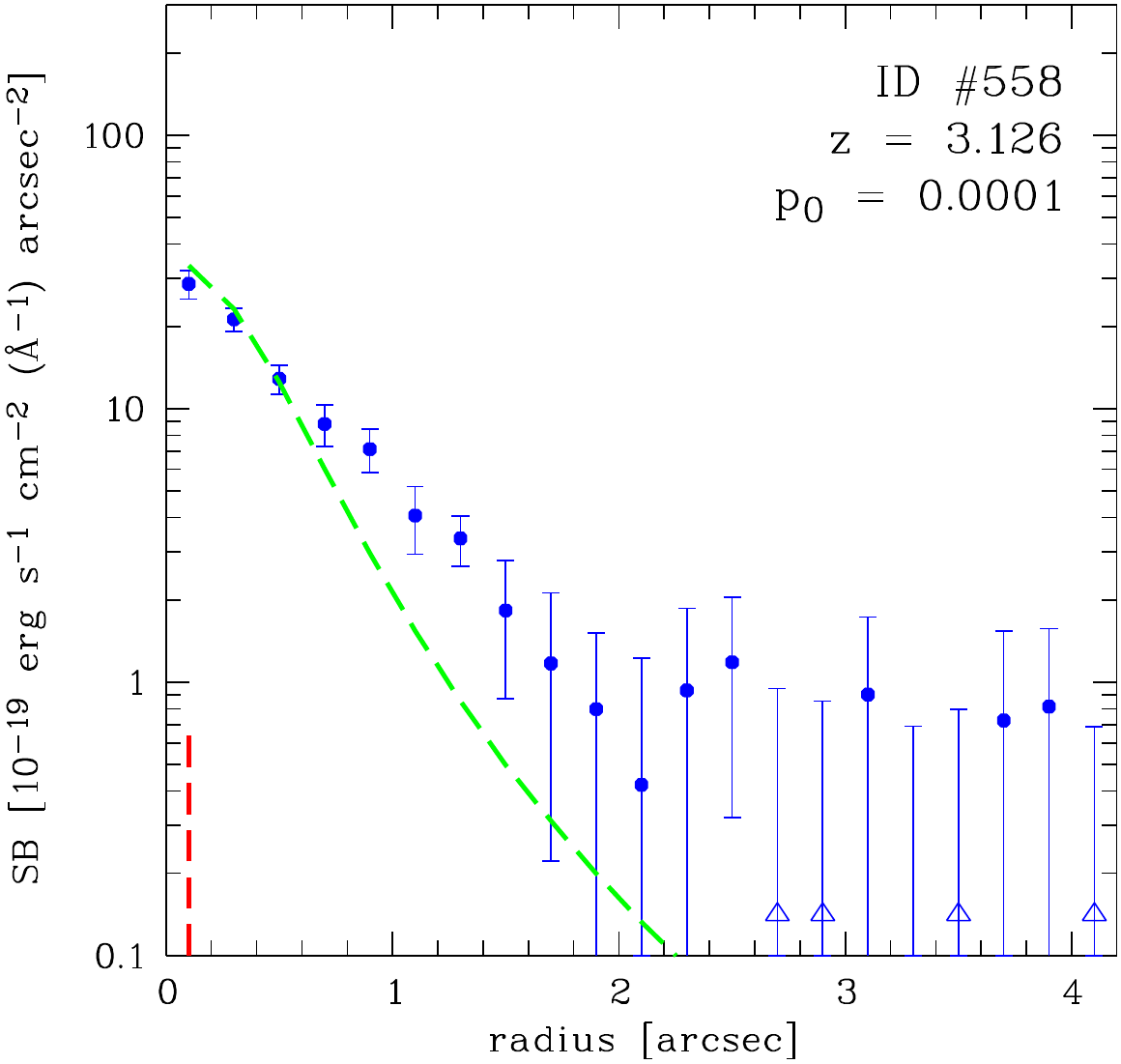}}
\put(0,1){\includegraphics[width=4.2cm]{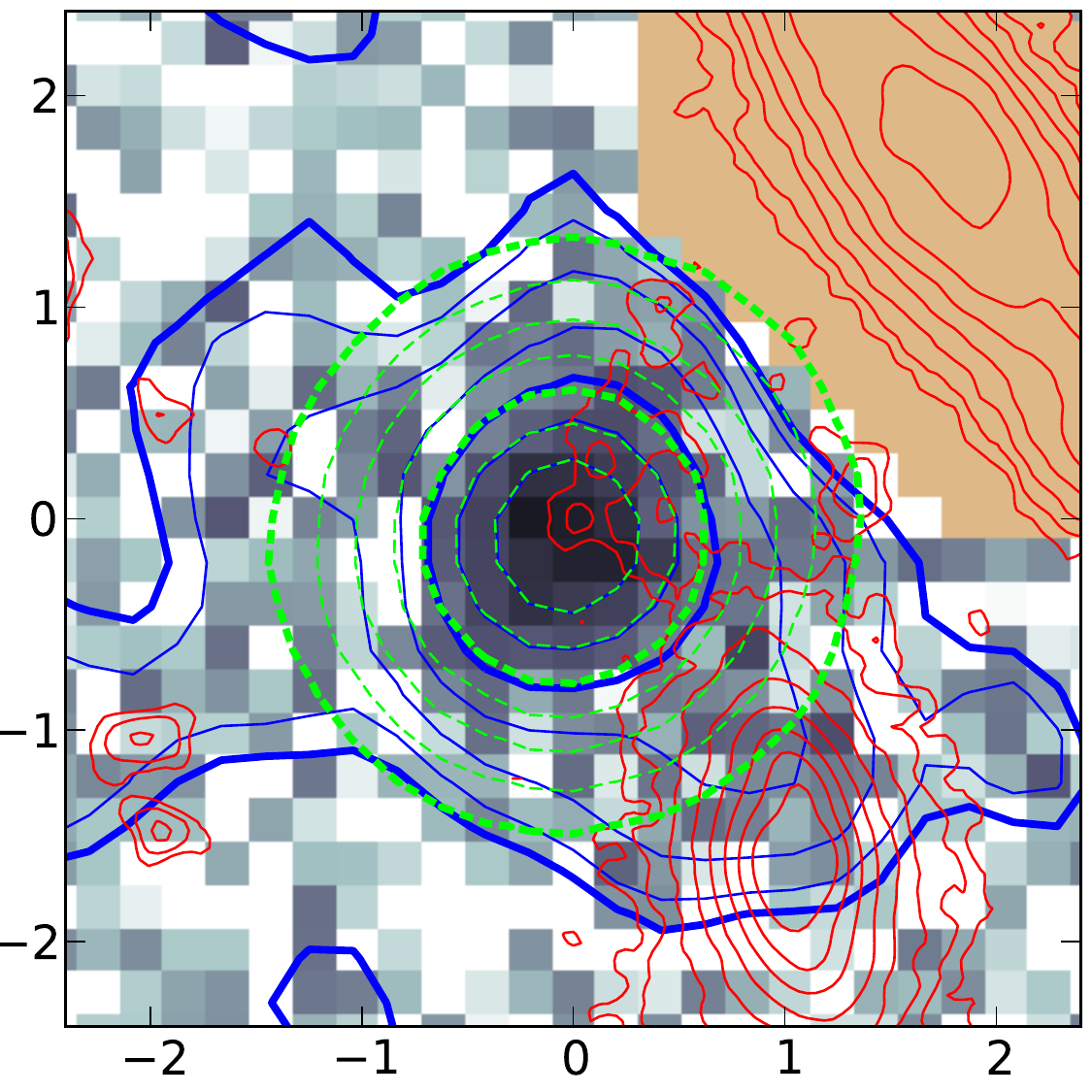}}
\put(44,0){\includegraphics[width=4.5cm]{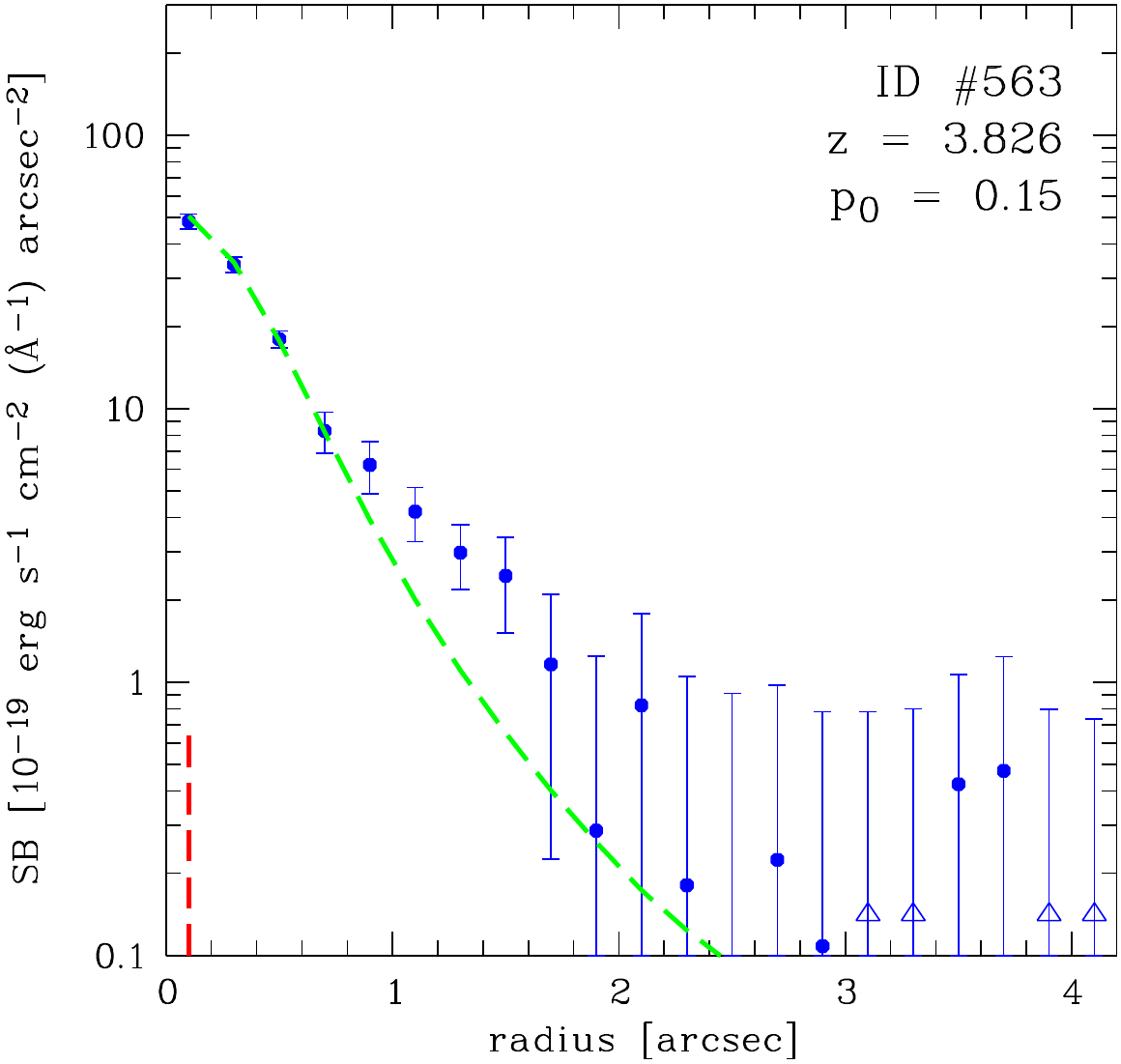}}
\put(93,1){\includegraphics[width=4.2cm]{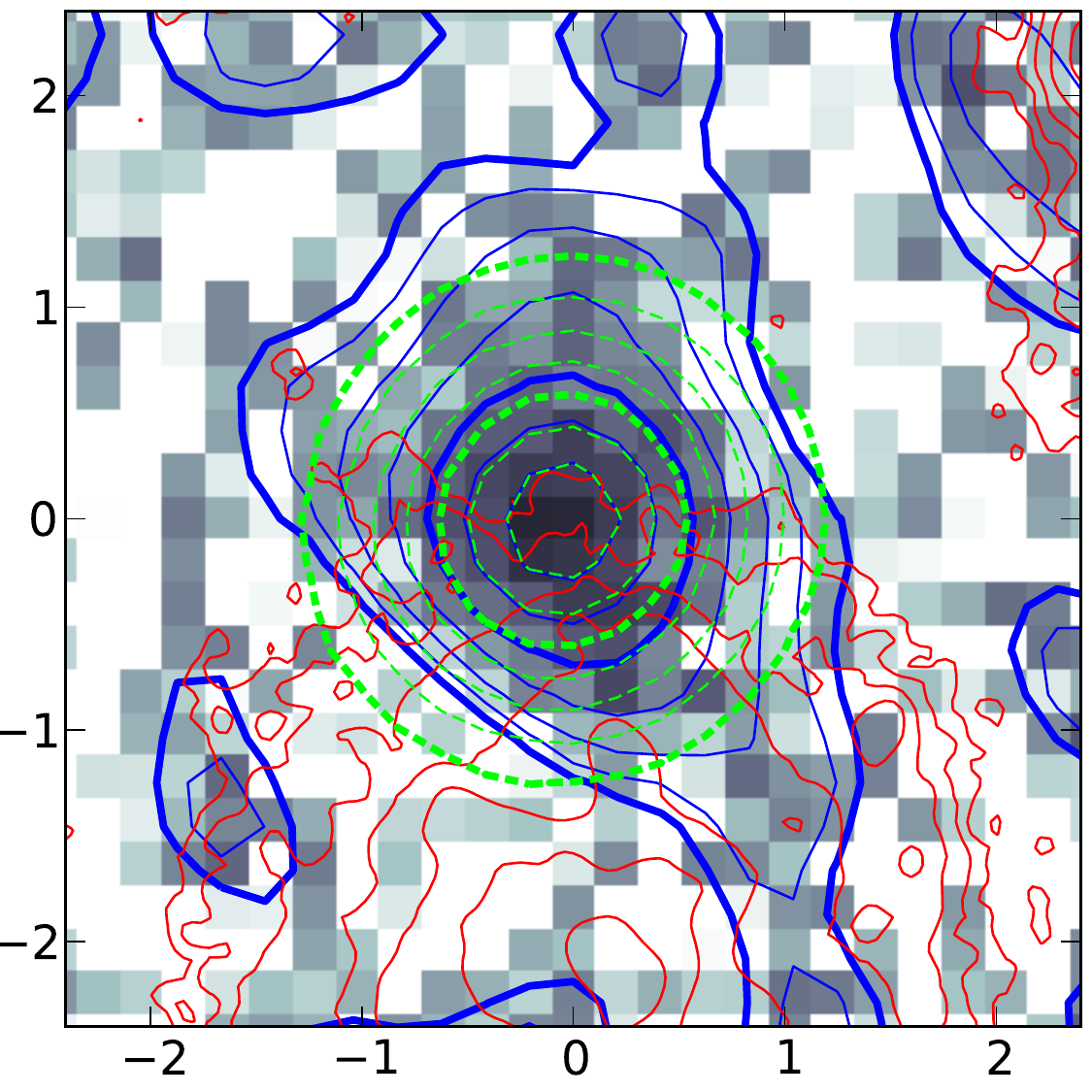}}
\put(137,0){\includegraphics[width=4.5cm]{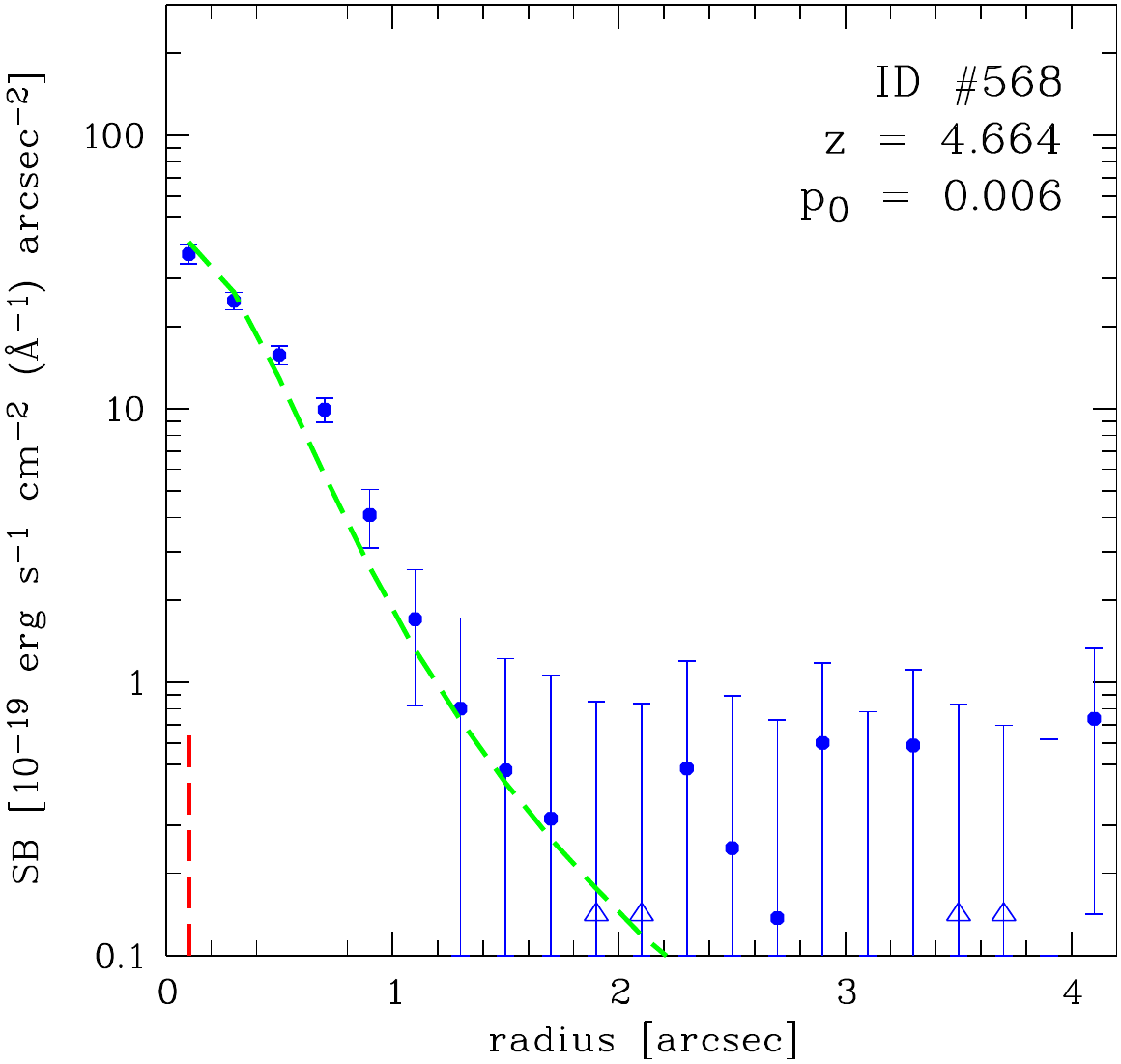}}
\end{picture}
\caption[]{(continued)}
\end{figure*}

\subsection{Profile construction}
\label{sec:prof-prof}

\subsubsection{Ly$\alpha$ profiles}
\label{sec:prof-lya}

We first determined the centroids of each galaxy in the \lya\ NB images. This was necessary since the astrometric registration from the MUSE-HDFS datacube to the HST images is currently not accurate enough to start directly from the HST centroids; furthermore, several of our LAEs have no detected HST counterparts anyway. We explain the details of our centroiding procedure and the resulting errors in Sect.~\ref{sec:prof-err-cen} below. 

We then extracted radial surface brightness (SB) profiles from the NB images by computing the average pixel values in concentric circular annuli (leaving out masked pixels). By integrating the SB profiles outwards we also constructed the corresponding growth curves. A visual inspection of these curves, and comparing them with the growth curves of perfect point sources, showed immediately that nearly all of our LAEs are spatially resolved, and that the growth curves converge towards a constant  integrated flux only for radii $\gtrsim 3\arcsec$. As a compromise between the need to capture most of the flux and the wish not to degrade the S/N by too much, we adopted
a circular aperture of 3\arcsec\ radius for estimating the total fluxes. 

Figure \ref{fig:gc} shows the 26 \lya\ growth curves after normalisation by the total \lya\ fluxes, together with the growth curves of the corresponding point spread functions constructed as described in Sect.~\ref{sec:prof-psf}. All growth curves cross unity level at $r = 3\arcsec$ by design, but both below and above that radius there is substantial dispersion between the objects. A few growth curves actually turn over around 3\arcsec\ (one object -- ID\#311 -- already at 2\arcsec), due to slightly negative SB values in the averaged profiles at these radii. However, after several tests we decided \emph{not} to apply any local background adjustment; the growth curves and SB profiles were always taken \emph{as measured} in the (continuum-subtracted) NB images. 

The adopted 3\arcsec\ flux integration aperture is significantly wider than what has been used in most previous LAE surveys. For several objects it is nevertheless still conservative: As can be seen in Fig.~\ref{fig:gc}, the growth curves for the majority of our LAEs continue to increase beyond that radius, although the errors become very large for $r > 3\arcsec$. In Sect.~\ref{sec:disc} we briefly discuss some implications of these significant aperture effects for LAE demographic studies. 

The resulting \lya\ SB profiles of our 26 galaxies are presented in Fig.~\ref{fig:ima+prof}, together with the images. The error bars for the profiles were estimated as described in Sect.~\ref{sec:prof-err-noise}. 

The measured \lya\ profiles detect emission at $\ga 1\sigma$ significance at least down to $\mathrm{SB} \sim 1\times 10^{-19}$~erg~s$^{-1}$ cm$^{-2}$ arcsec$^{-2}$ for most objects. This is an order of magnitude increase in sensitivity over most narrowband studies of LAEs. In other words, our MUSE observations reach surface brightness levels \emph{in individual galaxies} that in previous NB imaging could be achieved only by the stacking of $\sim$100 or more objects \citep{Steidel:2011jk,Momose:2014fe}.

\subsubsection{UV continuum profiles}
\label{sec:prof-cont}

Determining the spatial distribution of rest-frame UV continuum radiation from faint galaxies at $z>3$ is not straightforward. In past studies of this topic it was often assumed that the galaxies are point sources in the continuum \citep{Rauch:2008jy,Feldmeier:2013fx}, which is probably a good approximation especially for ground-based observations under moderate seeing conditions. Alternatively, the stacking of broadband images can be used to produce sample averages of the light distributions in the continuum \citep{Steidel:2011jk,Momose:2014fe}, albeit blurred by the seeing. 

Since in this paper we study individual sources, we need also individual continuum profile estimates whenever possible. Of the two available HST/WFPC2 filters overlapping with the MUSE spectral range, F814W is clearly the better choice, as the F606W band is affected by Lyman forest attenuation, and for objects at $z<5$ it may also be contaminated by \lya\ line emission. The F814W images show identifiable counterparts for 18 of the 26 objects in the sample. 3 of these are extremely faint and consistent with being point sources. We also assumed all HST-undetected continuum counterparts as well as all $z>5$ objects (see Sect.~\ref{sec:prof-err-cont}) to be point-like.

For the remaining 13 objects with counterparts resolved by HST we fitted simple parametric models to the 2-dimensional light distributions in the HST images. Such models have the advantage over directly extracting the HST pixel data that at least some physically motivated outward extrapolation beyond the pixel-by-pixel detection limits of HST is possible. Furthermore, the modelling made it easier to deblend the light distributions from close projected neighbours. Details of the procedure and a discussion of the reliability of the reconstructed continuum profiles are given below in Sect.~\ref{sec:prof-err-cont}. The integrated magnitudes from these fits are reported in Table~\ref{tab:table1}, the estimated continuum shape parameters (scale lengths and axis ratios) are listed in Table~\ref{tab:table2}.

The continuum-emitting regions of our LAE galaxies are all very compact, with exponential scale lengths between 1~kpc and $\la$200~pc (i.e., unresolved by HST). Given the faint absolute magnitudes, the obtained scale lengths are however consistent with other studies of high-$z$ galaxies in the literature \citep[e.g.][]{Morishita:2014ih}. It is important to realise that at the seeing-limited resolution of MUSE, all our objects are very nearly point sources in the UV continuum.

We then constructed azimuthally averaged radial profiles from the modelled 2-dimensional SB distributions in the F814W band. For those galaxies resolved by HST, we show the reconstructed radial continuum light distribution at HST resolution as the red curves in the profile plots of Fig.~\ref{fig:ima+prof}, whereas HST-unresolved and -undetected objects are represented by vertical bars. We subsequently convolved these source models with the point spread function of MUSE to predict their profiles at MUSE resolution and to compare them with the \lya\ brightness distributions.

\subsubsection{Do the \lya\ and continuum profiles differ?}
\label{sec:prof-comp1}

If \lya\ and UV continuum were to trace each other perfectly, the continuum profiles would be consistent with the \lya\ profiles apart from a single scaling factor per object, which is proportional to the equivalent width (EW) of the emission line. Admittedly this is a rather simplistic assumption, and there may be astrophysical reasons for an EW that varies with radius even without scattering of the \lya\ photons, but for now we restrict the discussion to the simple question whether or not the two profiles are different. Our \emph{null hypothesis} is that they are \emph{not} different except for noise. In that case the above scaling factor can be obtained by globally matching the PSF-convolved continuum to the \lya\ data in a minimum-$\chi^2$ sense. The resulting scaled continuum light distributions are shown in Fig.~\ref{fig:ima+prof} by the green/dashed lines in the profile plots and by the green/dashed contours in the image panels. We reiterate that the physical dimensions of these galaxies -- indicated by the red profiles -- are very small, and that the extended wings of the green profiles are exclusively due to the convolution with the MUSE PSF (cf.\ Sect.~\ref{sec:prof-psf}).

It is clear at first glance that in most objects, the \lya\ data\-points are nearly always above the scaled-up continuum. Only in the very central regions the scaled continuum agrees with, and sometimes even slightly overshoots the \lya\ data. (Note that this overshooting does \emph{not} indicate any central absorption, but is a simple consequence of the different shapes of the \lya\ and continuum profiles, together with the $\chi^2$ minimization criterion used to determine the scaling factor.)

Browsing visually through the sample, the \lya\ emission appears clearly more extended than the continuum profile in the first 11 objects (in order of appearance in Fig.~\ref{fig:ima+prof}), and in several more. Only for very few of our LAEs are the \lya\ profiles consistent -- within the error bars -- with the seeing-convolved continuum profiles. This is so far a subjective impression; it will be replaced by a proper statistical assessment in Sect.~\ref{sec:prof-stat} below.

\subsection{Error budget}
\label{sec:prof-err}

\subsubsection{Point spread function: Construction and accuracy}
\label{sec:prof-psf}

It is well known although not always fully appreciated that the point spread function (PSF) in astronomical images is usually poorly described by a Gaussian, and that is has very extended wings \citep[for a recent compilation see][]{Sandin:2014fg}. In challenges such as ours, where the spatial extent of marginally resolved objects is under investigation, it is thus imperative to obtain a good knowledge of the PSF over the full radial range of interest.

The MUSE pointing in the HDFS contains one bright and isolated star (ID\#0, $V = 18.4$) which is excellently suited as a PSF calibrator for the MUSE cube. Since the PSF changes with wavelength across the MUSE spectral range (see Fig.~2 in B2015), we determined a separate monochromatic PSF for each LAE in the sample by extracting narrowband images centred on this star, at the same wavelengths and with the same bandwidths as for the \lya\ NB images (see Sect.~\ref{sec:obs-extr}), but of course from the original, i.e.\ not continuum-subtracted cube. Our LAE-specific PSFs are thus not parametrised model fits, but were extracted pixel by pixel directly from the MUSE cube. Even so, the signal-to-noise ratio in the PSF is so much higher than in any of our LAE datapoints that we can safely neglect PSF uncertainties in the error budget. For all objects assumed to be point sources in the UV continuum, the green/dashed lines in Fig.~\ref{fig:ima+prof} directly represent the azimuthally averaged PSF profiles. Note that the PSF is well defined out to a radial distance of 5\arcsec\ and beyond.

Another relevant question is whether the PSF varies across the field. In our MUSE commissioning observations of globular clusters we did not see any evidence for significant spatial variations \citep{Husser:2015}. It is not possible to test this issue with comparable accuracy in a field as empty as the HDFS. The B2015 catalogue lists only 7 additional stars in the entire MUSE pointing, all of which are at least 4~mag fainter than star ID\#0 and do not constrain the PSF to similarly low SB levels. From an inspection of their radial profiles we can at least confirm that there is no evidence for spatial variations of the width of the PSF \emph{core} across the field of view. 

We note in passing that the modelling of the HST data (Sect.~\ref{sec:prof-err-cont}) also required a PSF. Since star ID\#0 is saturated in the HST images, we selected one of the fainter stars (ID\#21), which we found to be adequate for the purpose of this paper.

\subsubsection{Robustness of the continuum profile estimation}
\label{sec:prof-err-cont}

The HST/WFPC2 F814W band provides the best constraints on the continuum morphology in the rest-frame UV longwards of \lya\ for the 21 objects at $z < 4.8$. While our sample also contains 5 LAEs with $z > 4.8$, three of them are undetected in F814W and were thus anyway \emph{assumed} to be point sources in the continuum. The remaining two $z>5$ objects might show some \lya\ contribution to the flux in the HST band, and although we estimate this contamination to be below 20\% for both objects (judged from the measured \lya\ luminosities), we decided to treat them as point sources in the modelling.  

We used GALFIT \citep{Peng:2002di,Peng:2010eh} to fit a 2-dimensional light distribution to each detected galaxy in the HST data. The baseline model was an elongated exponential disc involving 4 free parameters (total magnitude $m_{814,\mathrm{c}}$, scale length $r_{\mathrm{s,c}}$, axis ratio $q_{\mathrm{c}}$, and position angle $\phi_{\mathrm{c}}$), to be convolved with the HST point spread function. We obtained meaningful models for 13 objects, including 3 cases where to obtain a converged fit we had to enforce the axis ratios to be 1. For the remaining objects we measured PSF-matched magnitudes, or adopted a $3\sigma$ upper limit of $m_{814}>29$ \citep{Casertano:2000fn} for those undetected by HST. The fitted continuum magnitudes are given in Table~\ref{tab:table1}; scale lengths and axis ratios are listed in Table~\ref{tab:table2}. The error estimates were taken as provided by GALFIT, based on the curvature of the $\chi^2$ hypersurface.

One might argue that instead of using HST, we could have obtained the continuum profiles directly from the MUSE data. That approach is however unfeasible for most of our objects, for two reasons: (i) Only very few of our LAEs actually show any significantly detected continuum signal in the spectroscopic data\-cube, even after massive spectral binning (this is actually expected from the high \lya\ equivalent widths of most objects; see Table~\ref{tab:table1}). (ii) Moreover, several objects have close projected neighbours, distinctly visible in HST but overlapping at MUSE resolution, making it virtually impossible to isolate the spatial continuum profiles of the LAEs without resorting to model assumptions. Even in the few cases where a reasonably clean continuum profile can be extracted from the MUSE data, the blurring by the seeing erases most of the source-specific details, and one obtains essentially a very noisy and slightly broadened version of the PSF. We did perform this experiment for the 6 continuum-brightest galaxies in our sample and confirm that the profiles reconstructed from HST and convolved to MUSE resolution are consistent with, but more robust and of higher quality than the direct estimates from the MUSE cubes.

A possible concern about our modelling approach could be that we \emph{assumed} an exponential law for the intrinsic radial SB distribution. A Sersic (\citeyear{Sersic:1968ta}) profile with shape index $n > 1$ has more extended wings which might then affect the relation between continuum and \lya\ profiles. However, constraining the shapes beyond estimating scale lengths is barely possible at the low S/N of our galaxies in the HST data and restricted to the few brightest objects. With those we investigated Sersic model fits and found that even when allowing $n$ to be a free parameter, we obtained $n \la 1$ in most and $n<2$ in all cases. This is in agreement with systematic studies of galaxy sizes and shapes at high redshifts \citep[e.g.][]{Morishita:2014ih,Shibuya:2015bj} which find Sersic indices around unity especially for low-mass star-forming galaxies. At any rate, the differences between Sersic models with $n=1$ and $n=2$ for the small intrinsic sizes of our galaxies become essentially invisible after convolution with the MUSE PSF.

\subsubsection{Centroiding uncertainties}
\label{sec:prof-err-cen}

Since inaccurate centroiding of a faint object may broaden its radial profile, we were particularly careful about the centroiding process and the propagation of centroiding errors into the error bars of the profiles. The determination of the \lya\ centroid in a NB image was done as follows: Starting with a visual guess for the centroid, we extracted a radial profile in concentric circular annuli. We then reconstructed a grid of circularised 2d images interpolated from this profile, but subpixel-shifted against the original NB image. We measured the $\chi^2$ of the residuals after subtracting the circularised profile images from the NB data and took the grid point with the minimum $\chi^2$ as the new centroid, which was in turn used to produce a new radial profile. This was iterated a few times with refined grids of subpixel shifts, until a well-defined `confidence valley floor' around the minimum in the $\chi^2$ distribution could be identified. Depending on the brightness but also on the `peakiness' of an LAE, between 3 and 5 iterations were required. As centroiding uncertainty we adopted the size of the rectangle encompassing the $\Delta\chi^2 = 2.3$ contour.

We then quantified the effects of inaccurate centroiding on the extracted \lya\ profiles. Each profile was extracted 100 times in an identical manner, but with randomly varied central positions assuming normally distributed centroiding errors. The rms variations of the profile points at fixed radii were taken as the relevant profile error, later to be combined in quadrature with other error terms. 

While potentially important, centroiding uncertainties turned out to be a very minor contribution to our error budget. Even the faintest objects in the sample could be centred with errors in $x$ and $y$ of less than one pixel. The errors were largest for LAEs that are both faint and fuzzy, but exactly for those fuzzy objects the propagation of centroiding errors had only a very small impact on the radial profiles.

A caveat to our centroiding procedure is the fact that we did not allow for the possibility of spatial offsets between \lya\ and UV continuum emission. While the MUSE-HDFS datacube is registered to the HST astrometry, the registration is not perfect, with small local deviations that are currently not well understood. This issue is under investigation within the MUSE consortium using star cluster observations. The question how well \lya\ emission and continuum line up is certainly an interesting one, but cannot be addressed from this dataset at present. We note that \citet{Shibuya:2014dn} found spatial offsets of typically below 0\farcs2 and mostly consistent with zero for a large sample of $z=2.2$ LAEs. In the following we simply \emph{assume} that the centroids of \lya\ and UV continuum are perfectly aligned; the uncertainty of this assumption is not included in the error budget.

\subsubsection{Imperfect flatfielding and sky subtraction}
\label{sec:prof-err-sys}

A relevant contribution to the error budget originates from residual instrumental signatures in the MUSE cube. These are dominated by two effects: (1) The splitting into 24 individual spectrograph channels and further into 1152 mini-slits by the image slicers leaves low-level traces that currently cannot be removed entirely by the flatfielding process. These flatfielding residuals are most visible in a wavelength-collapsed `white-light' image (see Fig.~3 in B2015). (2) Sky subtraction residuals also follow the instrumental slice and channel structures, but mainly occur in the wings of strong night sky emission lines. Both these effects are substantially smeared out and thus reduced in their relative amplitudes by the rotational and translational dithering, but they are not removed this way. Because of the 4 rotation angles used for the observations, all linear features such as slice or channel traces are converted into wavelength-dependent criss-cross patterns in the combined dataset that cover a broad range of spatial frequencies. In some ways these residuals act similarly to random noise, but of course without actually being random in origin. 

Low-level flatfielding and sky-subtraction systematics could, in principle, produce artefacts that are spatially coherent over a few arcseconds and disturb the extracted \lya\ radial profiles -- possibly in extreme cases to a degree that a profile might appear extended, even though it is not. The question is how the flatfielding residuals compare with the usual photon and readout noise. Visual inspection of the NB \lya\ images (see Fig.~\ref{fig:images}) suggests that systematics do at least not dominate over `normal' noise. Nevertheless it is clear that they must be taken into account. We now outline our procedure to quantify these hidden systematics, not just globally, but on an object-by-object basis.

\subsubsection{Estimation of the `effective noise'}
\label{sec:prof-err-noise}

\begin{figure}[tb]
\includegraphics[width=\hsize]{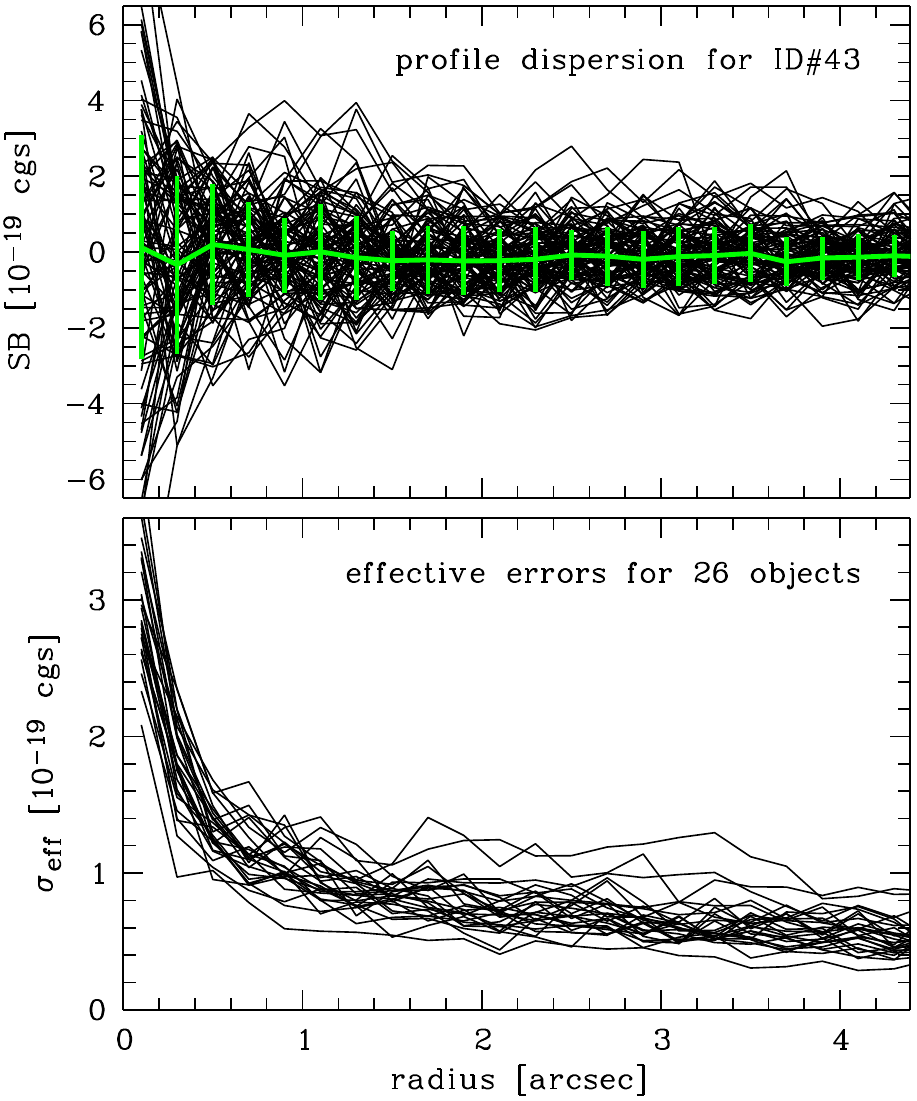}
\caption[]{Determination of object-specific `effective errors' in the \lya\ radial profile. Top panel:  The black lines show the extracted profiles from 100 empty locations for the same wavelength bands as the \lya\ NB image of the indicated object. The overplotted green line delineates the median at each radius, and the green error bars show the estimated `effective noise' $\sigma_\mathrm{eff}$ (see text). This procedure was repeated for each object. Bottom panel: 1$\sigma$ surface brightness sensitivities for the \lya\ NB profiles of all objects in our sample, within circular concentric annuli of width 0\farcs2 (= size of 1 MUSE spaxel) as a function of radius.}
\label{fig:efferr}
\end{figure}

Since the MUSE field of view is much larger than the sizes of individual LAEs, we can use empty regions as `noise calibrators'. We interactively defined 100 locations in the HST image where no source was even marginally detectable within a few arcseconds. These locations are more or less uniformly distributed over the MUSE field, except that they avoid the proximities of star ID\#0 and of bright galaxies. The same empty locations were previously employed for the analysis of global noise properties in B2015. For all objects in our sample we extracted NB images at these locations using the same bandwidths and central wavelengths as the corresponding \lya\ NB images. We then constructed radial profiles relative to an arbitrary central coordinate in each empty location, thus producing 100 noise calibrator profiles for each LAE.

The upper panel of Fig.~\ref{fig:efferr} shows the superposition of all these noise calibrator profiles for one particular object in our sample. The variance between the profiles clearly decreases as the area of each annulus increases. A closer examination reveals that the distribution of surface brightness values at given radius is typically close to, but not quite Gaussian, with somewhat broader wings (but with very few catastrophic outliers). We obtained a robust estimate of the width of each of these distributions by taking the difference between the 1st and the 3rd quartile. We then divided this difference by 1.35 (the quartile distance for a normal distribution with a $\sigma$ of unity) and adopted that as our estimates of the `effective noise' $\sigma_\mathrm{eff}$, for each radial bin and each LAE separately. These estimated $\sigma_\mathrm{eff}$ values are shown as error bars in the upper panel of Fig.~\ref{fig:efferr}, plotted around the median in each radial bin. 

At the flux level of the LAEs of interest here, the contribution of photon shot noise from the objects themselves is negligible compared to shot noise from the sky and the readout noise. We can therefore directly interpret $\sigma_\mathrm{eff}$ as a measure of the dispersion that would be measured in the profiles of real LAEs implanted at these empty locations. 

It is interesting to compare the decrease of $\sigma_\mathrm{eff}$ with increasing radius to the $1/\sqrt{r}$ dependence expected for ideal random noise when measured in concentric circular annuli of constant width. We found that $\sigma_\mathrm{eff}(r)$ follows $1/\sqrt{r}$ very closely over the entire range, for all objects. The statistical averaging over many pixels reduces the `effective noise' in nearly the same way as for ideal noise.

We also compared the $\sigma_\mathrm{eff}$ with the formally expected errors from only photon and readout noise, provided by the variance cubes. The ratios between `measured' and `formally expected' noise in the radial profiles are between 1.2 and $\sim$2.5, with a pronounced maximum around 1.5. Broadly speaking, the contributions of shot/readout noise and of systematics appear to be roughly equal, assuming that they can be approximatly added up in quadrature.%
  \footnote{The MUSE variance cubes are known to somewhat underestimate the true noise for aperture-integrated quantities because of correlations between neighbouring pixels, as a consequence of the rebinning in the cube construction. However, this effect is greatly reduced when averaging in annuli of 1 pixel width, as each pixel typically has only 2 relevant neighbours, not 8.} 
High ratios between effective and the formal propagated noise are always due to \lya\ being located at wavelengths close to bright sky lines. 

Finally, we added the errors resulting from centroiding uncertainties (Sect.~\ref{sec:prof-err-cen}) quadratically to the effective errors determined above (which caused only minute corrections, see Sect.~\ref{sec:prof-err-cen}), and assigned the results as errors to the extracted \lya\ profiles. Our error bars thus include all error sources that we are aware of, accounting for the usual random noise but also incorporating the effects of residual high-frequency systematics. 

The lower panel of Fig.~\ref{fig:efferr} shows the 1$\sigma$ surface brightness sensitivities of our observations, for each object as a function of radius. There are several reasons for variations in sensitivity between different objects: (i) The MUSE throughput varies significantly with wavelength. (ii) Objects with \lya\ overlapping at least partly with sky lines will be noisier already due to photon shot noise. (iii) The impact of systematics varies from object to object, in particular when the wavelength range of the \lya\ NB image is affected by sky lines. (iv) The bandwidths of the Ly$\alpha$ NB images differ by up to a factor of 2. These relative sensitivities are compressed in a single number per object when one considers the errors of the total \lya\ fluxes listed in Table~\ref{tab:table1}, derived from integrating over a fixed 3\arcsec\ aperture. Since we neglect object shot noise in the error budget, the differences between the flux errors reflect directly the different effective noise levels in the NB images.

The median sensitivity of our observations, expressed as $1\sigma$ error bar of the SB profile at a radius of $r=1\arcsec$, is $1\times 10^{-19}$~\sbl. At $r=3\arcsec$ a median sensitivity of $6\times 10^{-20}$~\sbl\ is reached.

\subsection{Statistical significance of extended Ly$\alpha$ emission}
\label{sec:prof-stat}

\begin{table*}
\caption[]{Summary of measured \lya\ halo and continuum quantities. All scale lengths and sizes are in proper kpc. ID: running source identifier. $r_{\mathrm{s,c}}$: Exponential scale length of the UV continuum; upper limits are given if the object is a point source in HST (italics for objects not detected by HST). $q_\mathrm{c}$: Fitted intrinsic axis ratio of the UV continuum source model ($q_\mathrm{c}\equiv 1.00$ implies that this value was enforced in the fit). $p_0$: Probability of the null hypothesis that the \lya\ radial profile follows the spatial shape of the UV continuum. $r_{\mathrm{s,h}}$: Seeing-corrected exponential scale length of the \lya\ halo from the decomposition into two components. $F_{\mathrm{cl}}$: Integrated flux of the `continuum-like' \lya\ component, in $10^{-18}$~erg s$^{-1}$ cm$^{-2}$. $F_{\mathrm{h}}$: Integrated flux of the \lya\ halo, in $10^{-18}$~erg s$^{-1}$ cm$^{-2}$. $r_{-19}$: Isophotal radius at SB = $1\times 10^{-19}$~erg~s$^{-1}$ cm$^{-2}$ arcsec$^{-2}$. $r_\mathrm{P20,c}$: Petrosian radius of UV continuum. $r_\mathrm{P20,Ly}\alpha$: Petrosian radius of total \lya\ emission. EW$_\mathrm{cl}$: Rest frame equivalent width of continuum-like component in \AA.
}
\begin{center}
\begin{tabular}{rr@{\hspace{0.2em}$\pm$\hspace{0.2em}}lr@{\hspace{0.2em}$\pm$\hspace{0.2em}}l@{\hspace{0.2em}}rr@{\hspace{0.2em}$\pm$\hspace{0.2em}}lr@{\hspace{0.2em}$\pm$\hspace{0.2em}}lr@{\hspace{0.2em}$\pm$\hspace{0.2em}}lrrr@{\hspace{0.5em}}r@{\hspace{0.2em}$\pm$\hspace{0.2em}}l}
\hline\hline\noalign{\smallskip} 
\multicolumn{1}{c}{ID} & \multicolumn{2}{c}{$r_{\mathrm{s,c}}$} & \multicolumn{2}{c}{$q_{\mathrm{c}}$} & \multicolumn{1}{c}{$p_{0}$} & \multicolumn{2}{c}{$r_{\mathrm{s,h}}$} & \multicolumn{2}{c}{$F_{\mathrm{cl}}$} & \multicolumn{2}{c}{$F_{\mathrm{h}}$} & \multicolumn{1}{c}{$r_{-19}$} & \multicolumn{1}{c}{$r_{\mathrm{P20,c}}$} & \multicolumn{1}{c}{$r_{\mathrm{P20,Ly}\alpha}$} & \multicolumn{2}{c}{EW$_{\mathrm{cl}}$}\\ 
\noalign{\smallskip}\hline\noalign{\smallskip} 
 43 & $1.06$ & $0.01$ & $0.25$ & $0.01$ & $<10^{-5}$ & 4.50 & 0.38 &  8.1 &  0.8 & 27.2 &  0.9 & 23.6 & 3.44 & 17.5\hspace{0.6em} & $  3.6$ & $ 0.3$\\ 
 92 & $0.71$ & $0.03$ & $0.67$ & $0.04$ & $<10^{-5}$ & 3.44 & 0.43 &  4.3 &  0.8 & 17.9 &  1.0 & 18.0 & 2.90 & 14.5\hspace{0.6em} & $  6.8$ & $ 1.1$\\ 
 95 & $0.73$ & $0.02$ & $0.34$ & $0.02$ & $<10^{-5}$ & 4.55 & 0.98 &  1.7 &  0.6 & 12.4 &  2.3 & 18.6 & 2.68 & 18.6\hspace{0.6em} & $  2.9$ & $ 1.0$\\ 
 112 & $0.62$ & $0.03$ & $0.70$ & $0.04$ & $<10^{-5}$ & 6.10 & 0.95 & 14.7 &  0.6 & 14.3 &  1.4 & 22.5 & 2.73 & 16.3\hspace{0.6em} & $ 28.6$ & $ 0.6$\\ 
 139 & $0.27$ & $0.02$ & $0.20$ & $0.15$ & $<10^{-5}$ & 4.94 & 0.50 &  2.6 &  0.4 & 18.2 &  1.1 & 22.5 & 1.86 & 20.3\hspace{0.6em} & $  6.8$ & $ 1.0$\\ 
 181 & $0.26$ & $0.03$ & $0.11$ & $0.28$ & $<10^{-5}$ & 2.52 & 0.29 & 14.8 &  1.0 & 12.9 &  1.1 & 17.7 & 1.84 & 11.3\hspace{0.6em} & $ 68.1$ & $ 3.9$\\ 
 200 & $1.12$ & $0.12$ & $0.12$ & $0.06$ & $<10^{-5}$ & 6.39 & 2.00 &  1.6 &  0.7 &  8.3 &  2.2 & 19.5 & 3.52 & 24.3\hspace{0.6em} & $  6.6$ & $ 1.8$\\ 
 216 & $0.41$ & $0.04$ & $0.24$ & $0.13$ & $<10^{-5}$ & 1.84 & 0.40 &  3.3 &  1.5 &  9.4 &  1.2 & 13.4 & 2.00 & 10.4\hspace{0.6em} & $ 18.2$ & $ 7.2$\\ 
 232 & \multicolumn{2}{l}{\hspace{0em}$<0.28$} & \multicolumn{2}{c}{--} & \hspace{0.7em} 0.0001 & 2.86 & 1.61 &  1.4 &  0.4 &  3.1 &  1.4 & 10.2 & 1.33 & 11.7\hspace{0.6em} & $ 39.6$ & $ 12$\\ 
 246 & \multicolumn{2}{l}{\hspace{0em}$<0.26$} & \multicolumn{2}{c}{--} & $<10^{-5}$ & 1.51 & 0.32 &  1.0 &  1.2 & 11.4 &  0.9 & 11.1 & 1.28 &  8.7\hspace{0.6em} & $ 14.4$ & $ 15$\\ 
 294 & \multicolumn{2}{l}{\hspace{0em}$<0.31$} & \multicolumn{2}{c}{--} & $<10^{-5}$ & 3.39 & 1.28 &  2.2 &  0.5 &  4.5 &  0.8 & 13.2 & 1.51 & 13.7\hspace{0.6em} & $ 52.7$ & $ 9.1$\\ 
 308 & $0.62$ & $0.11$ & $0.31$ & $0.13$ & \hspace{0.7em} 0.02 & 6.90 & 4.39 &  4.8 &  0.4 &  3.6 &  2.7 & 14.0 & 2.45 & 14.9\hspace{0.6em} & $ 40.5$ & $ 2.4$\\ 
 311 & $0.23$ & $0.09$ & \multicolumn{2}{l}{$1.00$!} & \hspace{0.7em} 0.00004 & 2.40 & 1.11 &  2.8 &  0.9 &  3.3 &  1.0 & 11.7 & 1.85 & 10.9\hspace{0.6em} & $ 45.1$ & $ 13$\\ 
 325 & $0.44$ & $0.08$ & $0.10$ & $0.41$ & $<10^{-5}$ & 1.40 & 0.26 &  3.6 &  1.1 &  6.8 &  1.0 & 11.3 & 1.95 &  8.6\hspace{0.6em} & $ 42.5$ & $ 14$\\ 
 393 & $0.23$ & $0.11$ & \multicolumn{2}{l}{$1.00$!} & $<10^{-5}$ & 5.02 & 1.60 &  2.7 &  0.5 &  4.9 &  1.3 & 14.6 & 1.83 & 17.1\hspace{0.6em} & $ 58.3$ & $ 5.1$\\ 
 422 & $0.21$ & $0.10$ & \multicolumn{2}{l}{$1.00$!} & \hspace{0.7em} 0.05 & \multicolumn{2}{c}{--} & \multicolumn{2}{c}{--} & \multicolumn{2}{c}{--} & 10.1 & -- & --\hspace{0.6em} & \multicolumn{2}{c}{--}\\ 
 437 & \multicolumn{2}{l}{\hspace{0em}$<0.34$} & \multicolumn{2}{c}{--} & $<10^{-5}$ & 4.94 & 1.85 &  6.5 &  0.5 &  4.8 &  1.0 & 16.6 & 1.66 & 14.4\hspace{0.6em} & $ 90.2$ & $ 5.3$\\ 
 489 & \multicolumn{2}{l}{\hspace{0em}$<0.35$} & \multicolumn{2}{c}{--} & \hspace{0.7em} 0.07 & \multicolumn{2}{c}{--} & \multicolumn{2}{c}{--} & \multicolumn{2}{c}{--} &  9.9 & -- & --\hspace{0.6em} & \multicolumn{2}{c}{--}\\ 
 543 & \multicolumn{2}{l}{\hspace{0em}$<\mathit{0.32}$} & \multicolumn{2}{c}{--} & \hspace{0.7em} 0.78 & \multicolumn{2}{c}{--} & \multicolumn{2}{c}{--} & \multicolumn{2}{c}{--} &  9.9 & -- & --\hspace{0.6em} & \multicolumn{2}{c}{--}\\ 
 546 & \multicolumn{2}{l}{\hspace{0em}$<\mathit{0.26}$} & \multicolumn{2}{c}{--} & $<10^{-5}$ & \textit{1.69} & \textit{0.53} & \textit{ 0.8} & \textit{1.05} & \textit{ 7.1} & \!\textit{ 1.2} & 10.4 & -- &  9.1\hspace{0.6em} & \multicolumn{2}{l}{$\mathit{> 31}$}\\ 
 547 & \multicolumn{2}{l}{\hspace{0em}$<\mathit{0.26}$} & \multicolumn{2}{c}{--} & $<10^{-5}$ & \textit{1.26} & \textit{0.79} & \textit{ 0.0} & \textit{0.87} & \textit{ 8.7} & \!\textit{ 1.3} & 10.4 & -- &  9.1\hspace{0.6em} & \multicolumn{2}{l}{$\mathit{\ge  0}$}\\ 
 549 & \multicolumn{2}{l}{\hspace{0em}$<\mathit{0.29}$} & \multicolumn{2}{c}{--} & \hspace{0.7em} 0.15 & \multicolumn{2}{c}{--} & \multicolumn{2}{c}{--} & \multicolumn{2}{c}{--} &  8.0 & -- & --\hspace{0.6em} & \multicolumn{2}{c}{--}\\ 
 553 & \multicolumn{2}{l}{\hspace{0em}$<\mathit{0.28}$} & \multicolumn{2}{c}{--} & $<10^{-5}$ & \textit{0.92} & \textit{0.57} & \textit{ 2.6} & \textit{1.90} & \textit{ 5.8} & \!\textit{ 1.5} &  9.6 & -- &  7.4\hspace{0.6em} & \multicolumn{2}{l}{$\mathit{> 89}$}\\ 
 558 & \multicolumn{2}{l}{\hspace{0em}$<\mathit{0.34}$} & \multicolumn{2}{c}{--} & \hspace{0.7em} 0.0001 & \textit{3.75} & \textit{1.32} & \textit{ 1.9} & \textit{0.69} & \textit{ 4.4} & \!\textit{ 1.0} & 14.5 & -- & 15.6\hspace{0.6em} & \multicolumn{2}{l}{$\mathit{> 44}$}\\ 
 563 & \multicolumn{2}{l}{\hspace{0em}$<\mathit{0.32}$} & \multicolumn{2}{c}{--} & \hspace{0.7em} 0.15 & \multicolumn{2}{c}{--} & \multicolumn{2}{c}{--} & \multicolumn{2}{c}{--} &  9.5 & -- & --\hspace{0.6em} & \multicolumn{2}{c}{--}\\ 
 568 & \multicolumn{2}{l}{\hspace{0em}$<\mathit{0.29}$} & \multicolumn{2}{c}{--} & \hspace{0.7em} 0.006 & \textit{1.09} & \textit{1.55} & \textit{ 0.9} & \textit{1.49} & \textit{ 3.7} & \!\textit{ 1.0} &  9.0 & -- &  8.2\hspace{0.6em} & \multicolumn{2}{l}{$\mathit{> 29}$}\\ 
\noalign{\smallskip}\hline\noalign{\smallskip} 
\end{tabular}
\end{center}
\label{tab:table2}
\end{table*}

\begin{figure}[tb]
\includegraphics[width=\hsize]{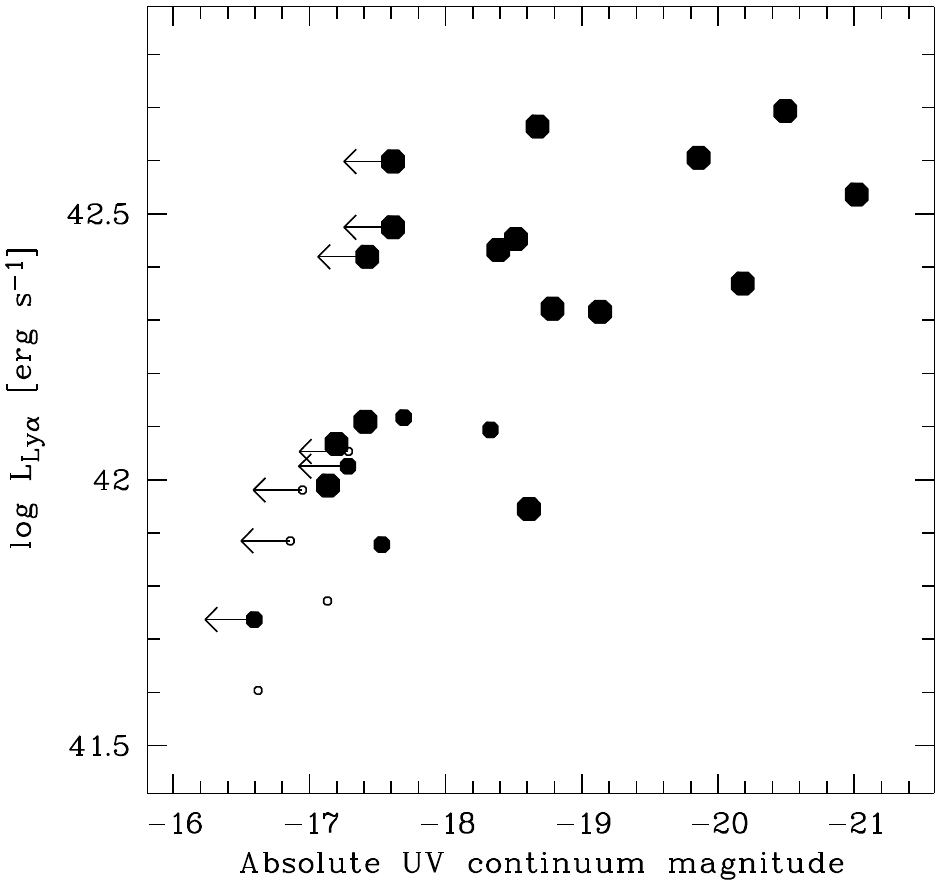}
\caption[]{Statistical evidence for the presence of extended \lya\ emission in relation to UV continuum and \lya\ line luminosities, expressed as probability $p_0$ of the null hypothesis that a given \lya\ profile is consistent with the continuum distribution. Large filled circles: Objects with $p_0 < 10^{-5}$ (i.e.\ very strong evidence for extended \lya); medium-sized filled circles: Objects with $10^{-5} < p_0 < 0.05$ (weaker evidence); small open circles: Objects with $p_0 > 0.05$ (no significant extended \lya\ emission detected). 
}
\label{fig:sample-ext}
\end{figure}

We now proceed to a quantitative assessment of the detection rate of extended \lya\ emission in the MUSE data. We tested, for each object in the sample, the null hypothesis that the \lya\ radial profile is statistically indistinguishable from the scaled UV continuum profile plus random noise. A straighforward $\chi^2$ test comparing the two profiles yielded a probability $p_0$ of the null hypothesis being true. In order to reject the null hypothesis for an object we demand $p_0$ to be small.

The results of this test are documented in Table~\ref{tab:table2}; the $p_0$ values are also displayed in Fig.~\ref{fig:ima+prof}. 21 (out of 26) objects have $p_0 < 0.05$, which we adopt as the nominal threshold for a detection. For 16 objects, $p_0$ is even below $10^{-5}$, rejecting the null hypothesis with very high significance. 

Figure~\ref{fig:sample-ext} shows that the brightest objects in the sample all display highly significant extended \lya\ emission, irrespective of whether we rank by \lya\ or by UV continuum luminosity. Among these are several galaxies with relatively low \lya\ equivalent widths, but also three luminous LAEs without any HST-detected continuum counterparts and rest-frame EW $\gtrsim$ 200~\AA\ (all of them at $z > 5$).

All 5 objects with large $p_0$ values (i.e., no significant extended emission) are located in the bottom left of Fig.~\ref{fig:sample-ext} and thus relatively faint in both \lya\ and UV continuum. This \emph{could} indicate that low-luminosity galaxies are less prone to show extended \lya. However, it is also possible that the S/N and/or the spatial resolution are simply insufficient to discriminate any existing extended \lya\ from the wings of the continuum radial profile caused by the PSF blurring. In fact there is tentative evidence that more objects might be judged as extended if just the error bars were lower: For objects ID\#489 and ID\#549, all points in the profile within $r<1\farcs5$ -- except for the innermost two -- are located $\sim$1--2$\sigma$ \emph{above} the scaled continuum profiles. Object ID\#563 also may show some low-significance extended \lya\ emission, but the object has a higher than average noise level and is furthermore disturbed by two large foreground galaxies. The only two objects without \emph{any} visible trace of an extended \lya\ profile are ID\#422 and ID\#553. We checked their extracted MUSE spectra again to see whether they might have been misclassified, but both objects display the characteristically asymmetric \lya\ line profiles and are confidently classified as LAEs.

In the following we focus only on those objects where we have statistically firm evidence for extended \lya\ emission. We stress however that the present non-detections should be taken as absence of statistical significance, not as significant evidence for an absence of extended \lya. We return to the non-detections in Sect~\ref{sec:sbmod-ul} to estimate upper limits for any putative extended \lya\ component.

When comparing these formal test results with the visual impression from Fig.~\ref{fig:ima+prof} it is important to realise that a high significance for extended emission does \emph{not} depend much on the very extended, very faint, and therefore quite noisy features delineated by the outermost contours (which the human eye -- trained to define an object by its edges -- tends to pick up). We emphasize that $\chi^2$ and $p_0$ are mainly driven by those datapoints that deviate most strongly \emph{in terms of S/N} from the scaled continuum profiles; thus typically by points with small error bars. The main evidence for extended \lya\ emission in our sample comes therefore from relatively high SB levels around $10^{-18}$ \sbl, at radii around $\sim$1\arcsec\ from the centre of each object.

Before entering the detailed analysis in the next sections, it is useful to recap briefly which objects in the sample are available for what type of analysis. Our sample has 26 objects, and in 21 of them we detected a \lya\ halo%
\footnote{For brevity we adopt from now on the term `halo' for the \lya\ emission of a galaxy that extends beyond its continuum-emitting region.}.
Of the 18 galaxies in the sample with a UV continuum counterpart detected by HST, a \lya\ halo was detected in 16 cases, but scale lengths of the continuum sources could be measured only for 12 of them. This restricted availability of supplementary information is of course a limitation for the statistical representativity of the current analysis. We augment the measured quantities with estimated upper or lower limits whenever possible.

%%%%%%%%%%%%%%%%%%%%%%%%%%%%%%%%%%%%%%%%%%%%%%%%%%%%%%%%%%%%%%%%%%%%%%%%%%%%%%%%

%\input{LAE-ext_4-sbmodelling}

\section{Two-dimensional surface brightness modelling}
\label{sec:sbmod}

In the previous section we established the reality of extended \lya\ emission around most of the galaxies in our sample. In the following we characterise the properties of the detected \lya\ haloes in quantitative terms; in this, we also take the blurring by the atmospheric seeing into account. We employ a forward modelling approach, fitting simple surface brightness distributions to the MUSE \lya\ data. These fits are further constrained by using priors on the continuum light distribution from our HST image analysis in \ref{sec:prof-cont}.

\begin{figure*}
\includegraphics[width=\hsize]{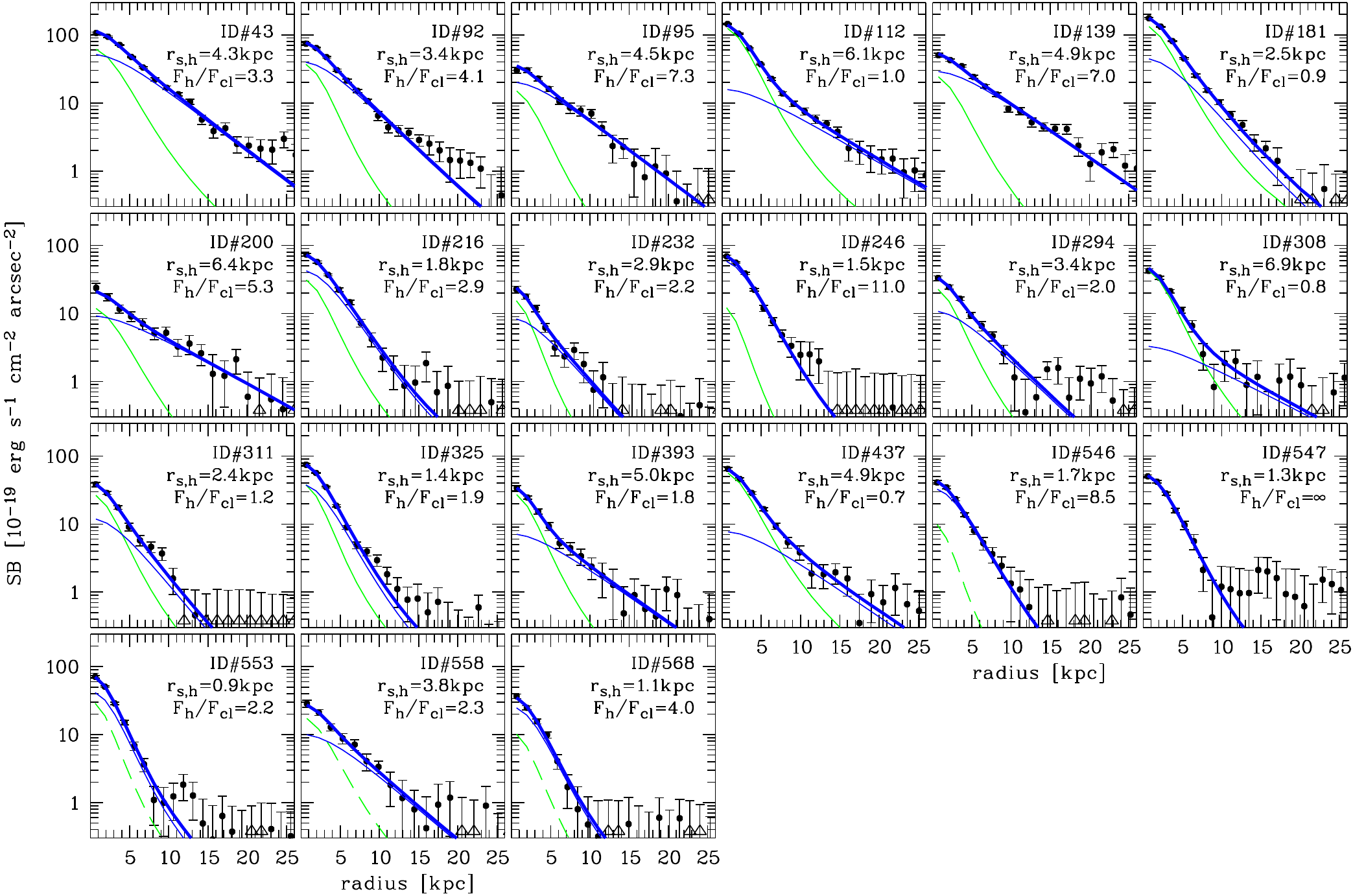}
\caption[]{Decomposition of the observed \lya\ surface brightness distributions into a compact (`continuum-like') and an extended exponential component (`halo'), as described in the text. Only the objects with detected haloes ($p_0 < 0.05$) are shown. The datapoints are the same as in Fig.~\ref{fig:ima+prof}, but now plotted against linear radii. The green curves show the profiles of the `continuum-like' components (dashed for the objects without HST counterparts and thus assumed to be point sources), the thin blue lines represent the extended haloes; both convolved with the MUSE PSF. The thick blue lines show the sum of the two components.}
\label{fig:expfits}
\end{figure*}

\subsection{Model specifications}
\label{sec:sbmod-model}

Following \citet[][hereafter cited as S2011]{Steidel:2011jk} it has become customary to parametrise the \lya\ radial profiles as single exponential functions with just two free parameters, a scale length $r_{\mathrm{s}}$ and a central surface brightness $C$. In S2011 and other studies based on stacked images, these parameters were estimated from simple linear fits to the observed profiles, avoiding the inner regions dominated by the seeing. For objects with sizes of the order of the seeing disc, this approach unfortunately discards exactly those datapoints that have the highest S/N ratios, and it thus misses valuable information. It may also bias the results towards larger scale lengths, especially when the source is only marginally resolved, because of extended PSF wings that may remain significant out to several times the FWHM.

Visual inspection of our observed profiles shows furthermore that describing the radial \lya\ SB distributions by single exponentials is often inadequate, and this is confirmed by quantitative tests. Most of the measured \lya\ profiles in Fig.~\ref{fig:ima+prof} feature a more or less pronounced inwards upturn around a radius of $\sim$1\arcsec, suggesting either a superposition of a compact emission component on top of a more extended one, or alternatively a more complicated radial law. Modelling such curved profiles as single exponentials would yield spurious values for the scale lengths that strongly depended on the chosen radial range for the fitting.

We adopted a two-component model as the functional form fitted to the \lya\ SB distributions of each object, decomposing it into two distinct components. The first was taken to resemble that of the compact UV continuum; we call it the `continuum-like' component and assume that its \emph{shape} parameters are precisely known from our GALFIT modelling of the HST images (i.e., as elongated exponential discs or as point sources), except for a single flux scaling factor related to the \lya\ EW of this component. The \lya\ halo, on the other hand, was again parametrised as a single exponential. Since in this paper we only aim at obtaining robust estimates of the azimuthally averaged halo sizes, we enforced the halo model to be circularly symmetric; we also assumed both components to be concentric (see Sect.~\ref{sec:sbmod-asym} for a brief discussion of the fitting results when these restrictions were relaxed). There were thus 3 free parameters to be optimised in each fit: The integrated fluxes $F_{\mathrm{cl}}$ and $F_{\mathrm{h}}$ of both \lya\ components, and the exponential scale length $r_{\mathrm{s,h}}$ of the halo. We present the results from these fits in the next subsection.

Our two-component model can be seen as a minimalistic extension of the conventional single exponential profile, but it may also represent a physically meaningful hypothesis: If some fraction of the \lya\ luminosity produced within the star forming regions escapes (more or less) directly -- possibly suffering from dust attentuation but without large-scale spatial redistribution -- then this \lya\ emission will roughly follow our `continuum-like' component. The `halo' component, on the other hand, describes the \lya\ emission more extended than the continuum, irrespective of whether it has its origin inside the galaxy and then got scattered outwards, or whether it is produced by mechanisms other than recombination in \ion{H}{ii} regions. 

An alternative approach to model the \lya\ SB distribution would be to replace the separation into two components by a more flexible single SB law for the \lya\ profile such as a Sersic function. Note that a circular Sersic profile has again 3 free parameters to be fitted (total flux, half-ligh radius and Sersic index). We explored such fits as well and briefly report about the outcome below.

\subsection{Fitting results}
\label{sec:sbmod-results}

\subsubsection{Two-component models}
\label{sec:sbmod-results-twocomp}

We again used GALFIT to obtain the model parameters for the observed \lya\ NB images, always incorporating the appropriate monochromatic MUSE PSF (cf.\ Sect.~\ref{sec:prof-psf}). From the fits we constructed separate radial profiles of the individual seeing-convolved components, which are presented in Fig.~\ref{fig:expfits} for the 21 objects with a formally significant detection of a \lya\ halo. The resulting fit parameter values are listed in Table~\ref{tab:table2}. We note that for one object (ID\#547; no HST counterpart) the best-fit is in fact a single-component exponential, and the parameter $F_{\mathrm{cl}}$ consistently converged to zero.

We obtained error estimates for the parameters from a procedure that resembled a Monte-Carlo approach, except that we used the 100 `noise calibrator' empty fields to vary the effective noise. After adding each of these `noise images' to any of the noise-free model galaxies, this image was fitted again using the same setup as for the fitting of the observed NB images. The dispersion of the 100 refitted parameters around the `true' value was taken as the error of that parameter and listed in Table~\ref{tab:table2}. In some objects certain parameter pairs are correlated (such as halo flux and scale length) or anticorrelated (in particular the fluxes of continuum-like and halo components); such correlations were then clearly visible in the distribution of the refitted values. However, these degeneracies are generally not strong, and we neglect them in the error budget.

Remarkably, our restricted two-component model provides in all cases a statistically satisfactory representation of the observed profiles. Expressed as probabilities $p_2$ that the measured $\chi^2$ values in the profiles are consistent with the error bars, all objects have $p_2 > 0.05$ (i.e., are within $2\sigma$), and 16/21 have $p_2 > 0.32$ (within $1\sigma$). Any discernible deviations between the data and the model profiles are restricted to the lowest SB levels where the error bars are large. 

We reiterate the point made already in Sect.~\ref{sec:prof-stat} that it is \emph{not} the low SB tails of the observed profiles that drive the fits, but the high S/N datapoints at small to intermediate radii. The obtained halo scale lengths $r_{\mathrm{s,h}}$ are all below 1\arcsec, many of them even below 0\farcs5. While the \lya\ haloes in most our objects are clearly resolved, they typically have sizes comparable to the MUSE PSF.

\subsubsection{Sersic models}
\label{sec:sbmod-results-sersic}

We applied the same fitting procedure as before to obtain alternatively single-component model fits adopting a circular Sersic profile function. We did this for two reasons: We wanted to avoid committing ourselves prematurely fully to the two-component model (which we preferred in the end, see the discussion below), and we wanted to at least briefly consider the functional form that is most widely used nowadays for the overall SB modelling of galaxies (for example, \citealt{Hayes:2014jv} used Sersic fits to compare the UV continuum with the extended \lya\ emission in their low-redshift LARS galaxies).

In terms of producing statistically acceptable fits of the observed profiles, the Sersic model performed nearly equally well as the two-component model. However, in about half the sample, the Sersic shape parameter $n$ took implausibly high values of $n \gtrsim 10$, and in 8 cases it even reached the maximally permitted value of $n=20$ imposed by GALFIT. Since a large $n$ implies very extended wings of the profile, the total flux obtained by integrating the Sersic formula exceeded the measured flux in these objects by more than a factor of 2. Consequently also the half-light radii were unrealistically high. This happened exactly for those objects where the two-component fit indicated a relatively strong continuum-like component. It is well known from AGN host galaxy studies (and easy to imagine) that a central point source can drive the Sersic index in a single-component fit to very large values; this is most probably what happened in these cases. We interpret this outcome of the Sersic model fits as support for the notion that there really are (at least) two distinct emission components for \lya. For this reason, and because of the unrealistic values of the fit parameters for several objects, we have not considered the Sersic models any further.

\subsection{Undetected haloes and upper limits}
\label{sec:sbmod-ul}

\begin{figure}
\includegraphics[width=\hsize]{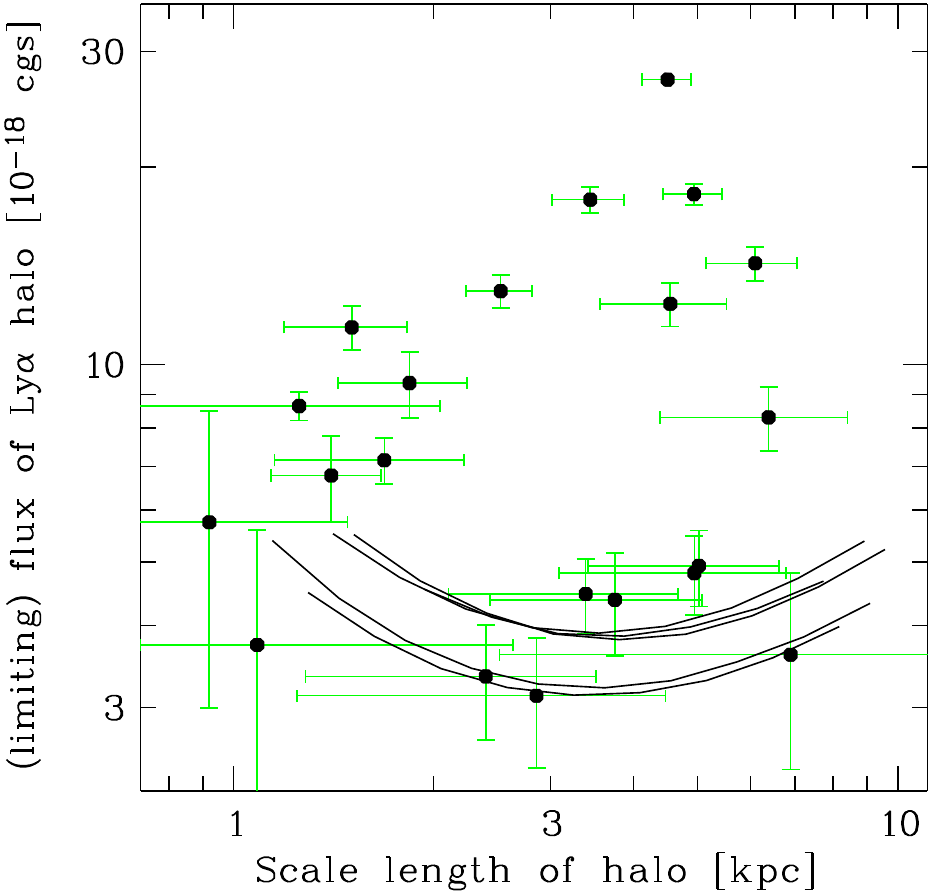}
\caption[]{Estimated upper limits for the five objects with formally undetected \lya\ haloes ($p_0 > 0.05$). Each curve shows the minimal flux that an exponential halo would need to have for a marginal detection on top of the pure continuum prediction for a given object, as a function of its scale length. For comparison we show the measured values of the detected haloes by the black filled symbols and green error bars.
}
\label{fig:upperlim}
\end{figure}

We now estimate upper limits to the halo fluxes of those five objects where no significant halo was detected, using the toolbox developed in this section. Obviously such a limit cannot be expressed as a single flux value, as an undetected \lya\ halo may be too faint to be significant, or too small to be separated from the continuum-like signal, or some combination of the two. In our parametrisation of \lya\ haloes as exponentials superimposed on a compact component, there must be a separate halo flux limit for each assumed scale length.

The starting point for each object was the pure `continuum-like' model profile scaled to the measured \lya\ datapoints as described in Sect.~\ref{sec:prof-cont} (green/dashed profile curves in Fig.~\ref{fig:ima+prof}). We then constructed artificial haloes of different scale lengths and integrated fluxes and added them to the continuum-like component such that the total \lya\ flux inside an aperture of 3\arcsec\ was preserved at the observed value. For each assumed halo we performed the $\chi^2$ consistency test as in Sect.~\ref{sec:prof-stat}. The limiting flux at given scale length was reached when the $\chi^2$ probability (corresponding to $p_0$ in the observed data) dropped below 0.05. The results are presented by the solid lines in Fig.~\ref{fig:upperlim}, which also contains the measured \lya\ halo fluxes of the detected objects for comparison.

As expected, the limiting fluxes increase for smaller scale lengths. They also increase again for large scale lengths; this happens because the total flux is preserved, so the surface brightness and thus detectability goes down. The highest contrast is achieved at assumed halo scale lengths of $\sim$4~kpc, which corresponds approximately to an angle of 0\farcs6, or one seeing disc (FWHM) in our MUSE data. Fortuitously, many of the detected haloes have sizes in that regime.

We argued already in Sect.~\ref{sec:prof-stat} that the objects without detected haloes are all among the faintest in the sample. This is underlined by the fact that the upper limit curves in Fig.~\ref{fig:upperlim} pretty much straddle the lower bound of the observed haloes. A few measurements (all with very large error bars) are below these limits, which is explained by the spread in noise levels between objects (see Sect.~\ref{sec:prof-err-noise}). Also, our procedure to estimate upper limits has been conservative for those objects where some tentative indications for a weak halo are already present (Sect.~\ref{sec:prof-stat}), which was neglected in the present analysis.

The upper limits derived here show that the affected objects are not necessarily special. The limiting halo fluxes correspond to limiting halo/continuum-like flux ratios of $\gtrsim$1--3, fully within the range of the observed values for the brighter objects. Of course it is possible that some of these galaxies really do not have any \lya\ haloes, but this is not constrained by our data.

\subsection{Asymmetries in the \lya\ distribution}
\label{sec:sbmod-asym}

\begin{figure*}
\setlength{\unitlength}{1mm}
\begin{picture}(170,0188)
\put(0,143){\includegraphics[width=91.0mm]{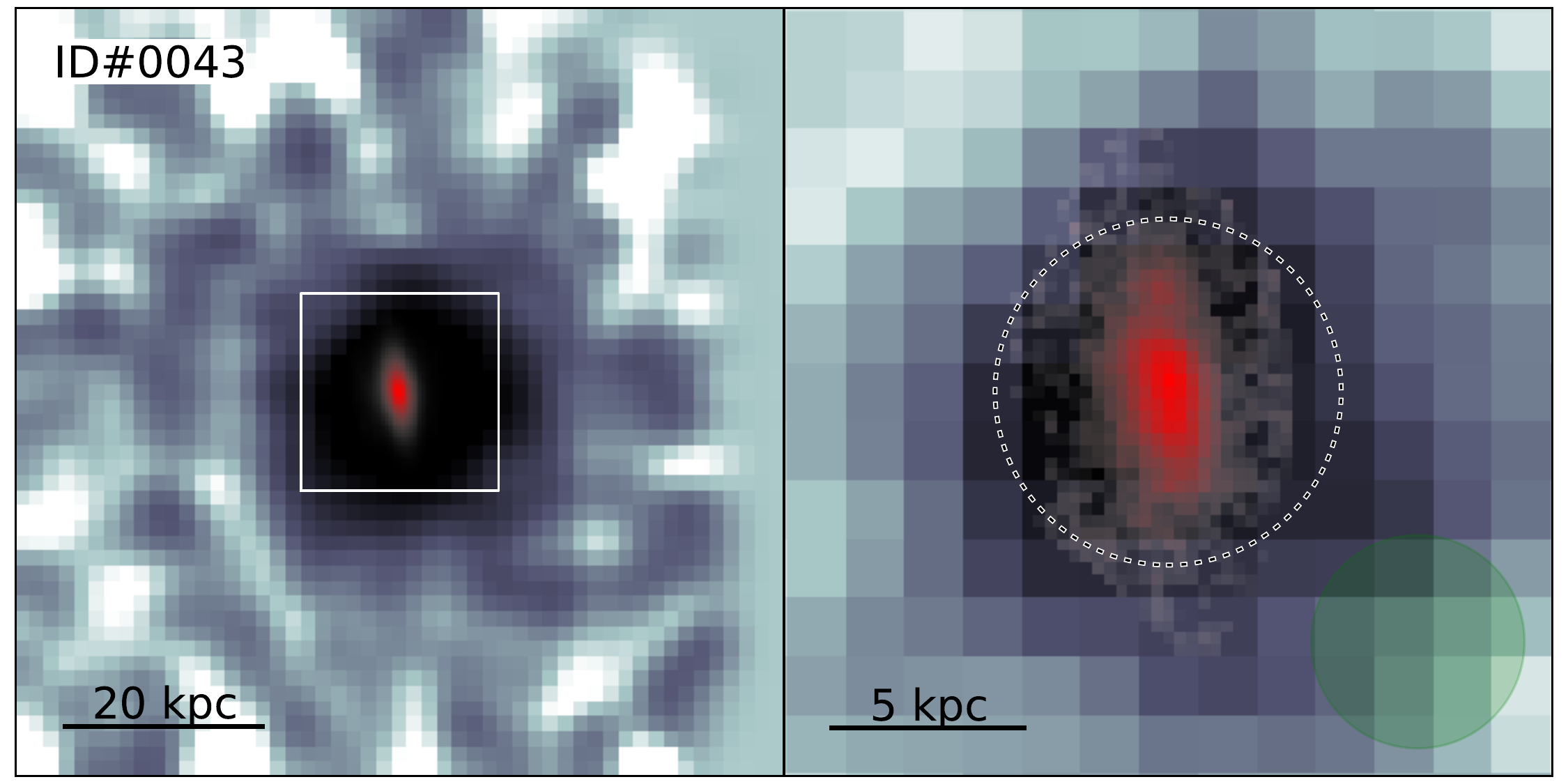}}
\put(93,143){\includegraphics[width=91.0mm]{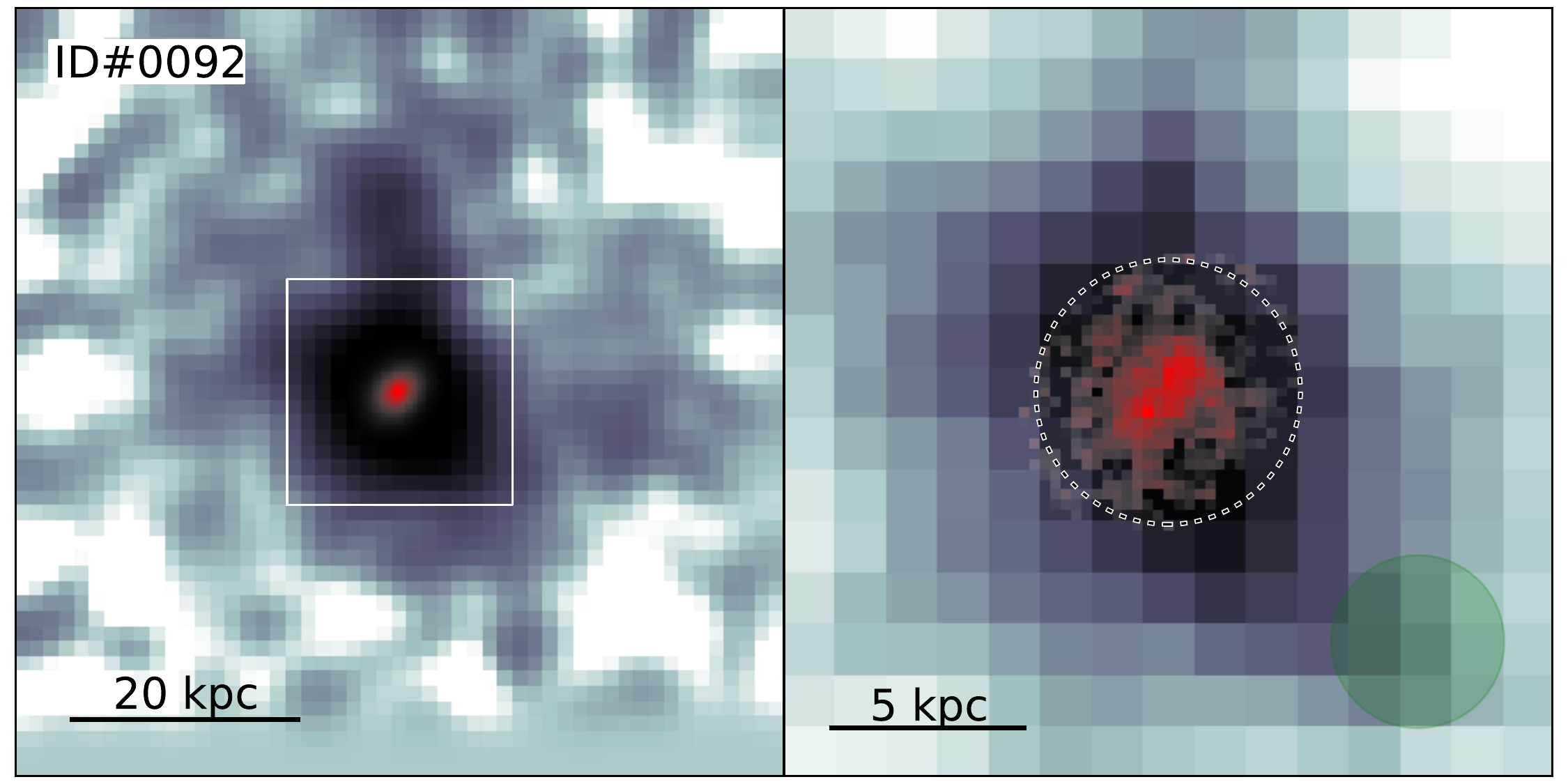}}
\put(0,95){\includegraphics[width=91.0mm]{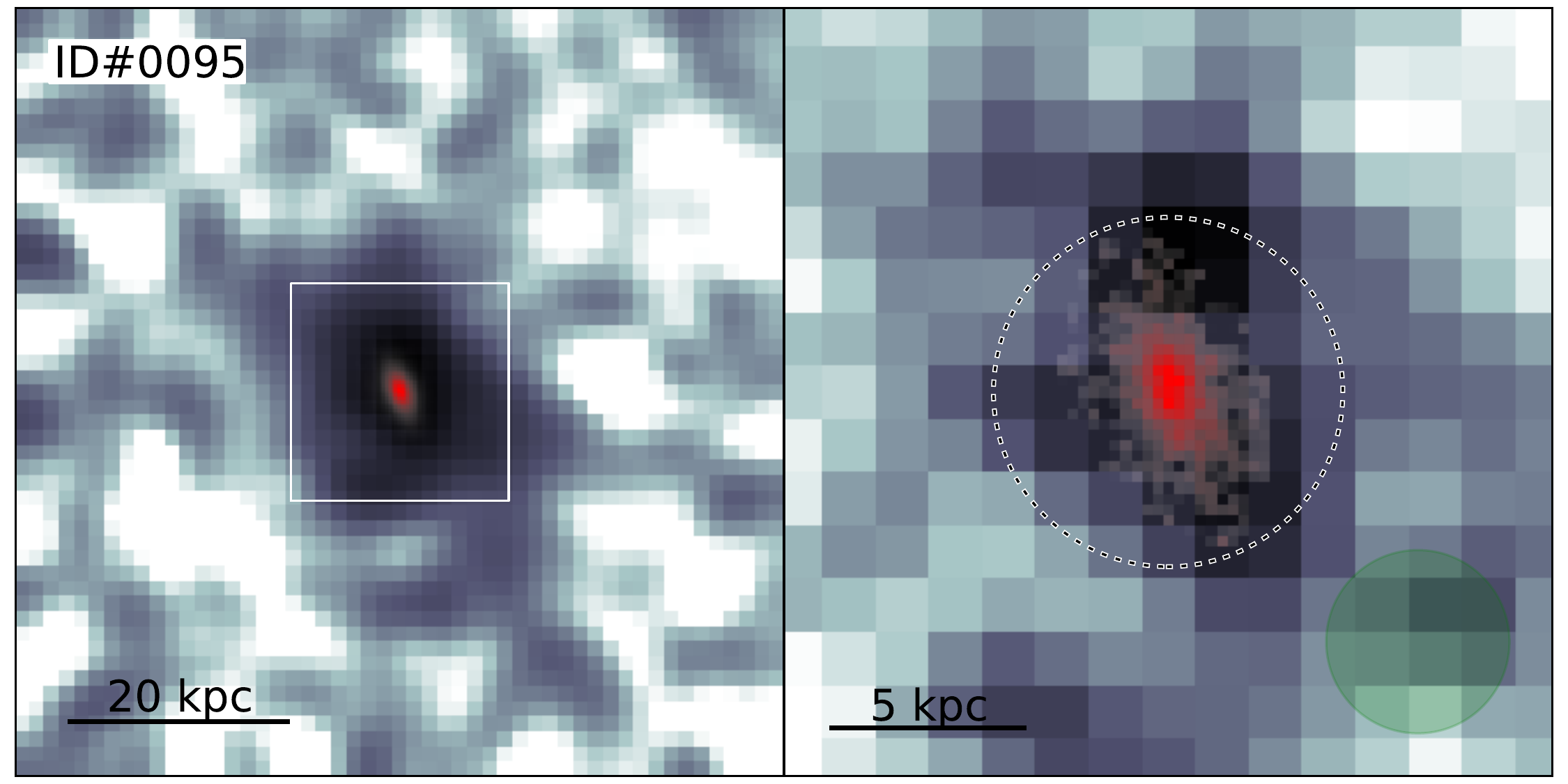}}
\put(93,95){\includegraphics[width=91.0mm]{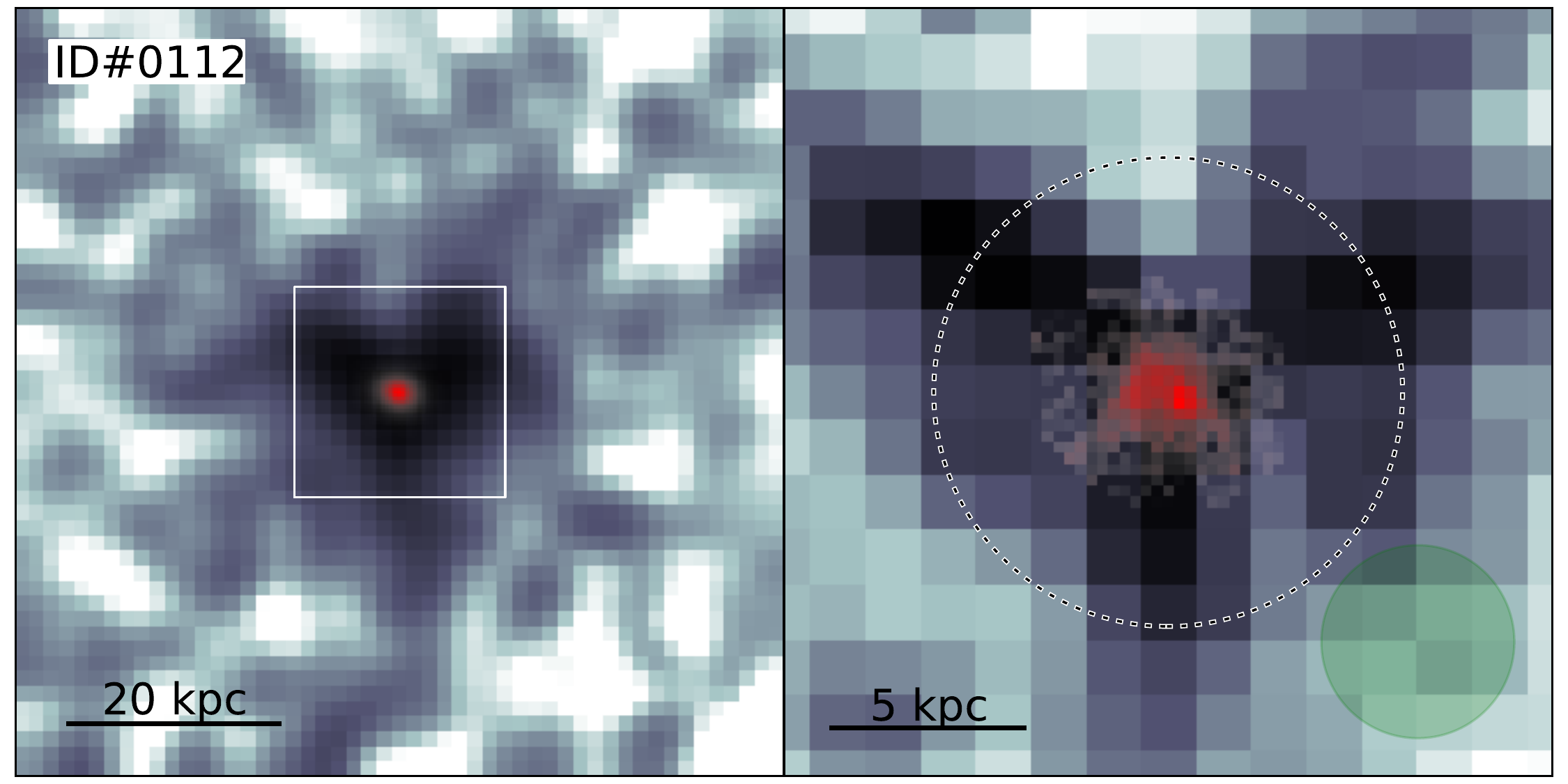}}
\put(0,47){\includegraphics[width=91.0mm]{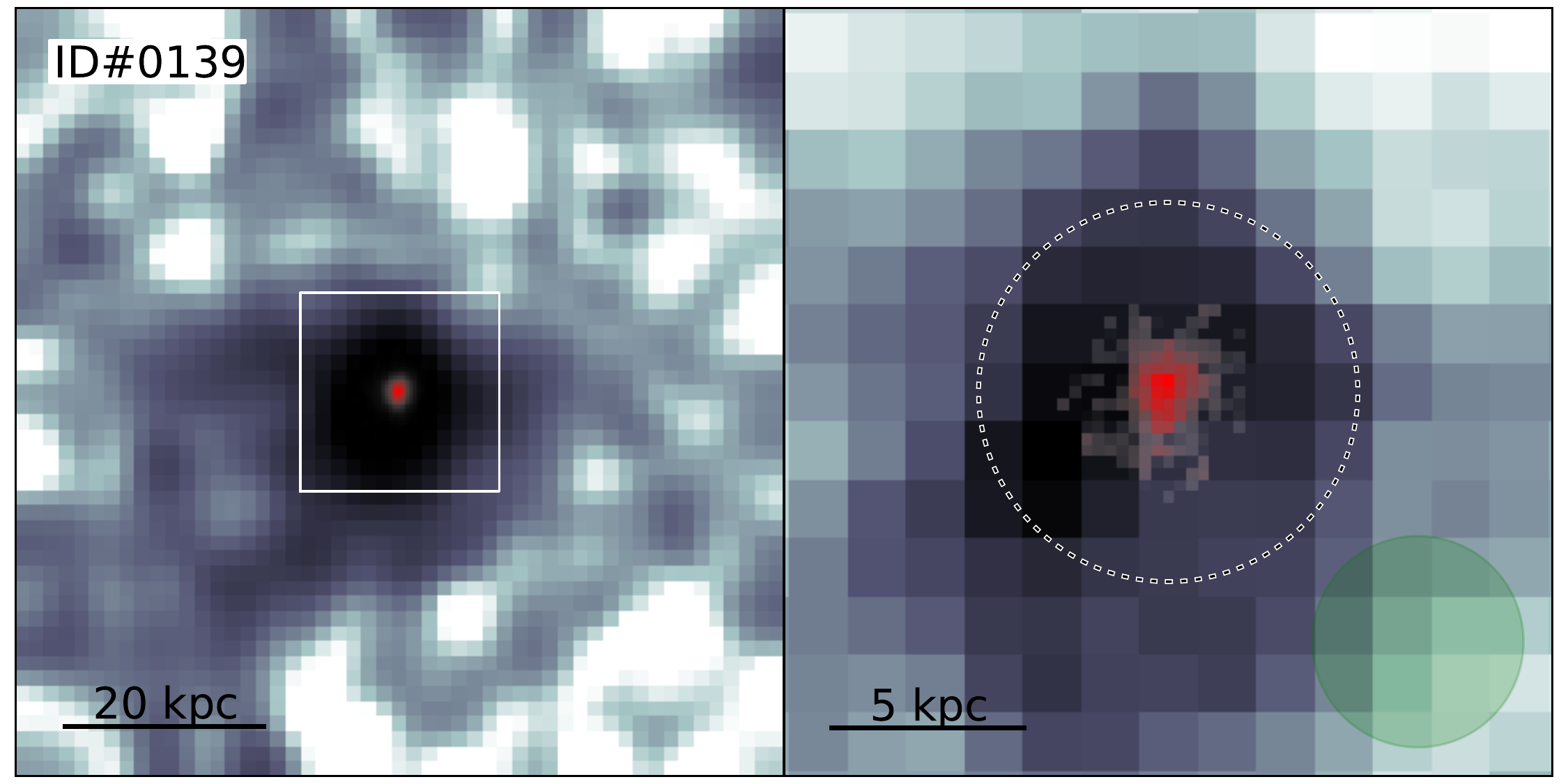}}
\put(93,47){\includegraphics[width=91.0mm]{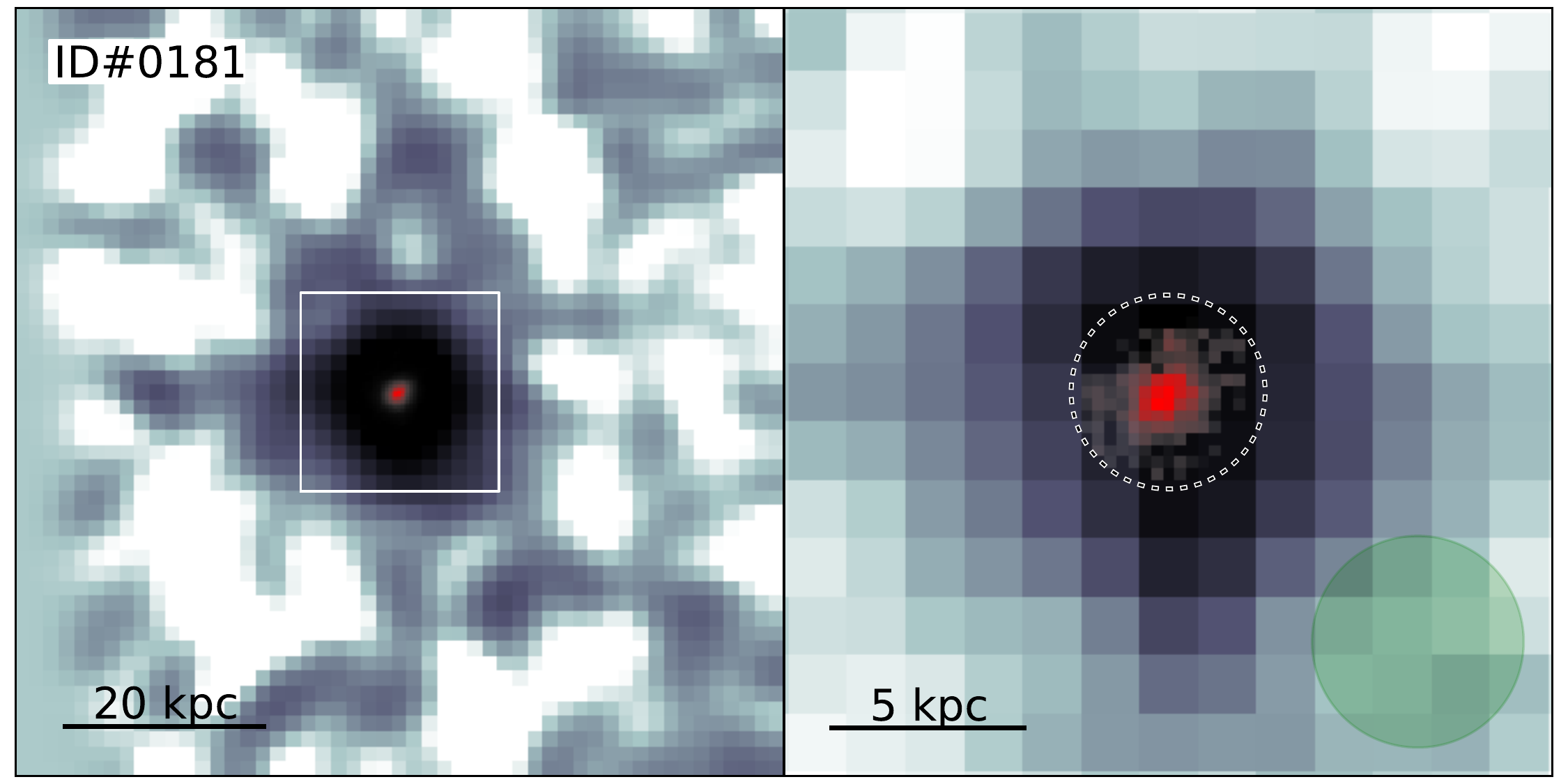}}
\put(0,-1){\includegraphics[width=91.0mm]{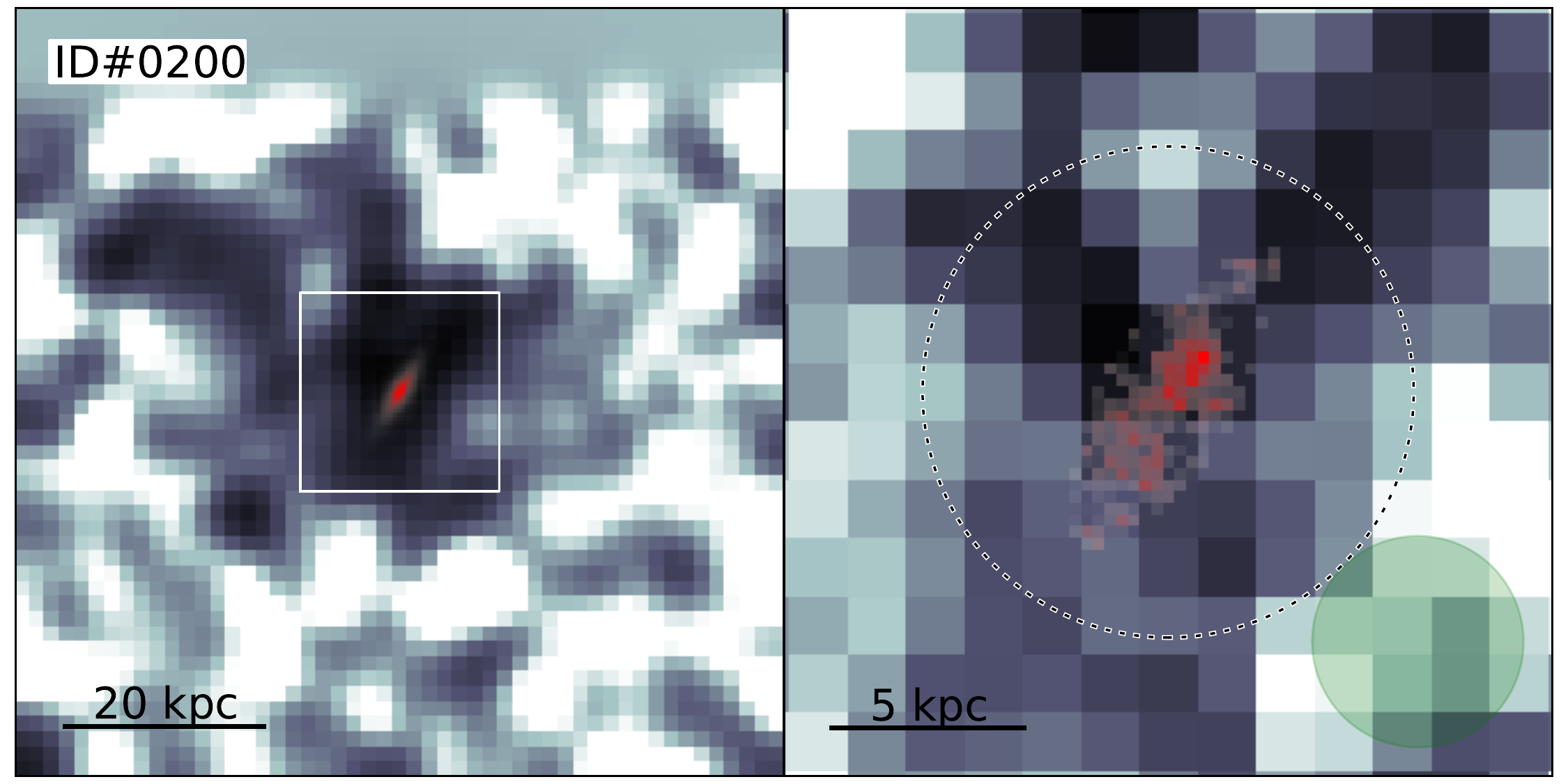}}
\put(93,-1){\includegraphics[width=91.0mm]{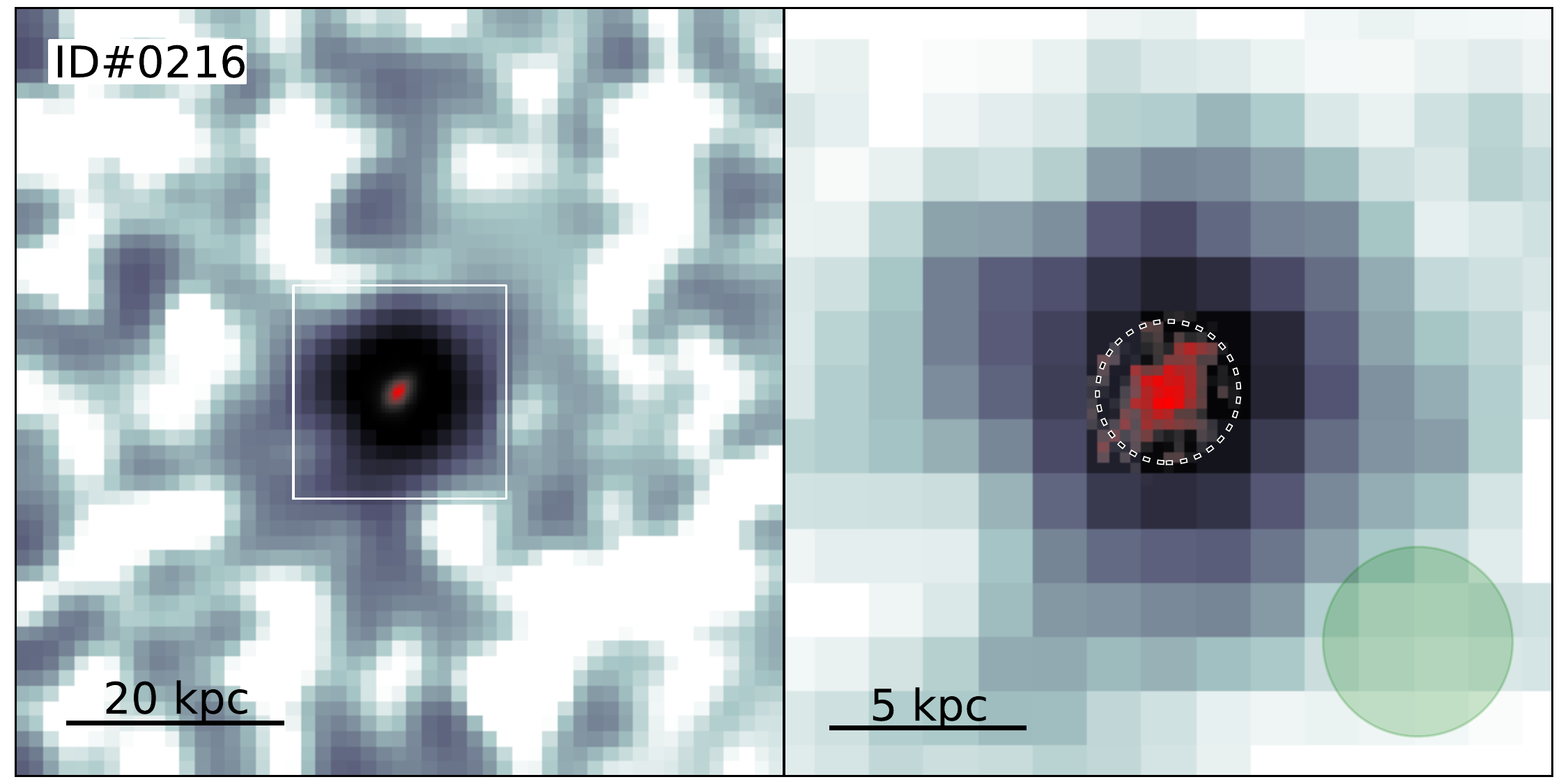}}
\end{picture}
\caption[]{Synopsis of all detected \lya\ haloes, presented in two views per object. Each image shows in greyscale the \lya\ surface brightness of only the extended halo component, with a model of the continuum-like component subtracted. Left panel in each column: Full-size $5\arcsec\times5\arcsec$ image, smoothed as described in the text, and displayed with fixed SB cuts from $-1$ to $+20$ $\times$ $10^{-19}$~\sbl\ to emphasize the outskirts of the halo and the transition into noise. The scale bar in the lower left indicates a transverse distance of 20~kpc. A model image of the HST counterpart is superposed in red when available. The white square specifies the zoom window. Right panel in each column: Zoomed 20~kpc $\times$ 20~kpc view of the same object, very slightly smoothed (see text) and with individually adjusted cut levels from 0 to the maximum pixel value, to show the bright inner region of the halo. The dotted circle centred on the object indicates the exponential scale length $r_\mathrm{s,h}$ of the \lya\ halo. The bar in the lower left shows a distance of 5~kpc. The semitransparent green circle in the lower right represents the FWHM of the MUSE PSF. Superimposed in red are the HST F814W pixel data of the object, with fore- and background sources masked out. 
}
\label{fig:lya-grey+red}
\end{figure*}

\addtocounter{figure}{-1}
\begin{figure*}
\setlength{\unitlength}{1mm}
\begin{picture}(170,0190)
\put(0,145){\includegraphics[width=91.0mm]{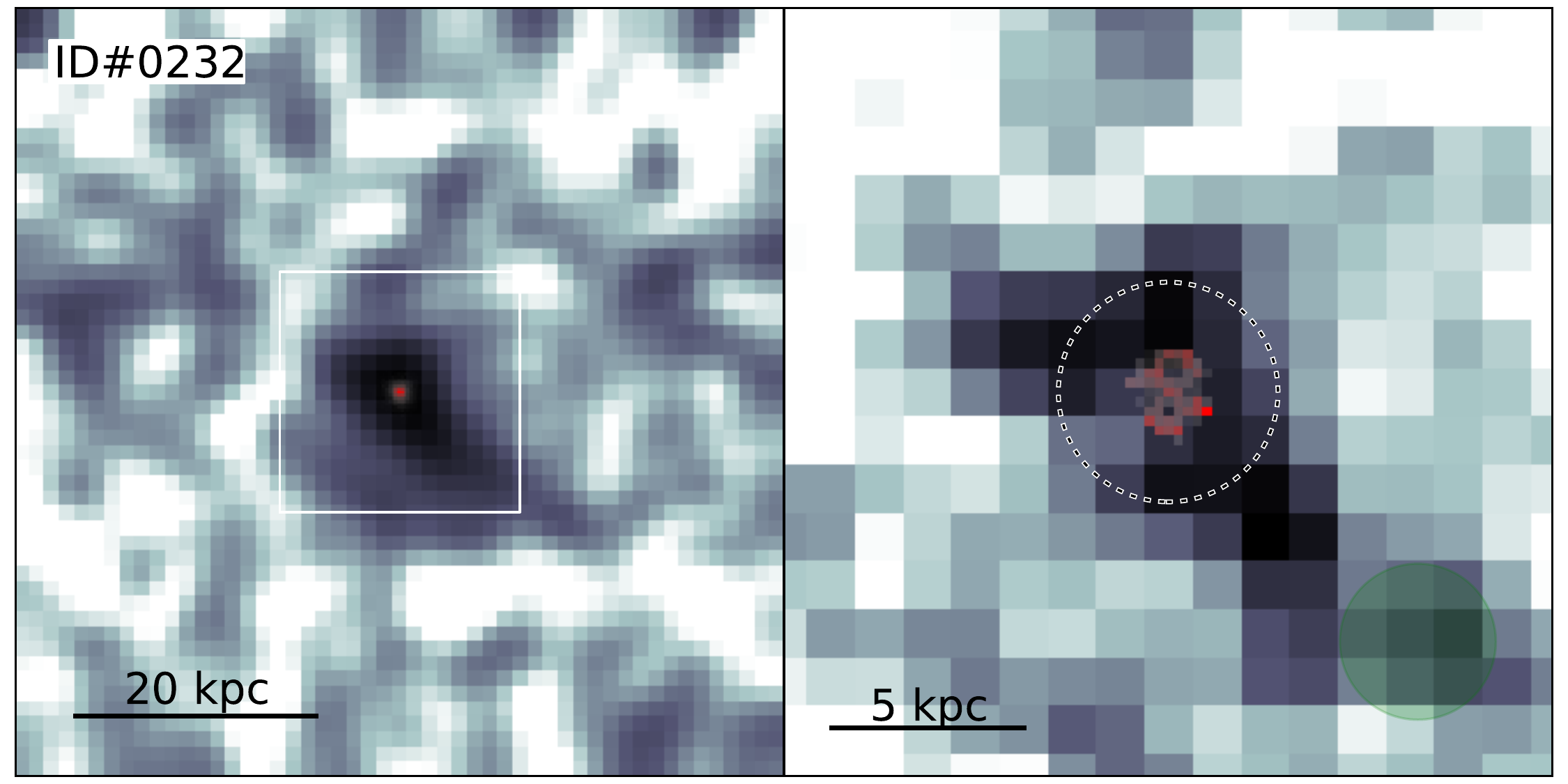}}
\put(93,145){\includegraphics[width=91.0mm]{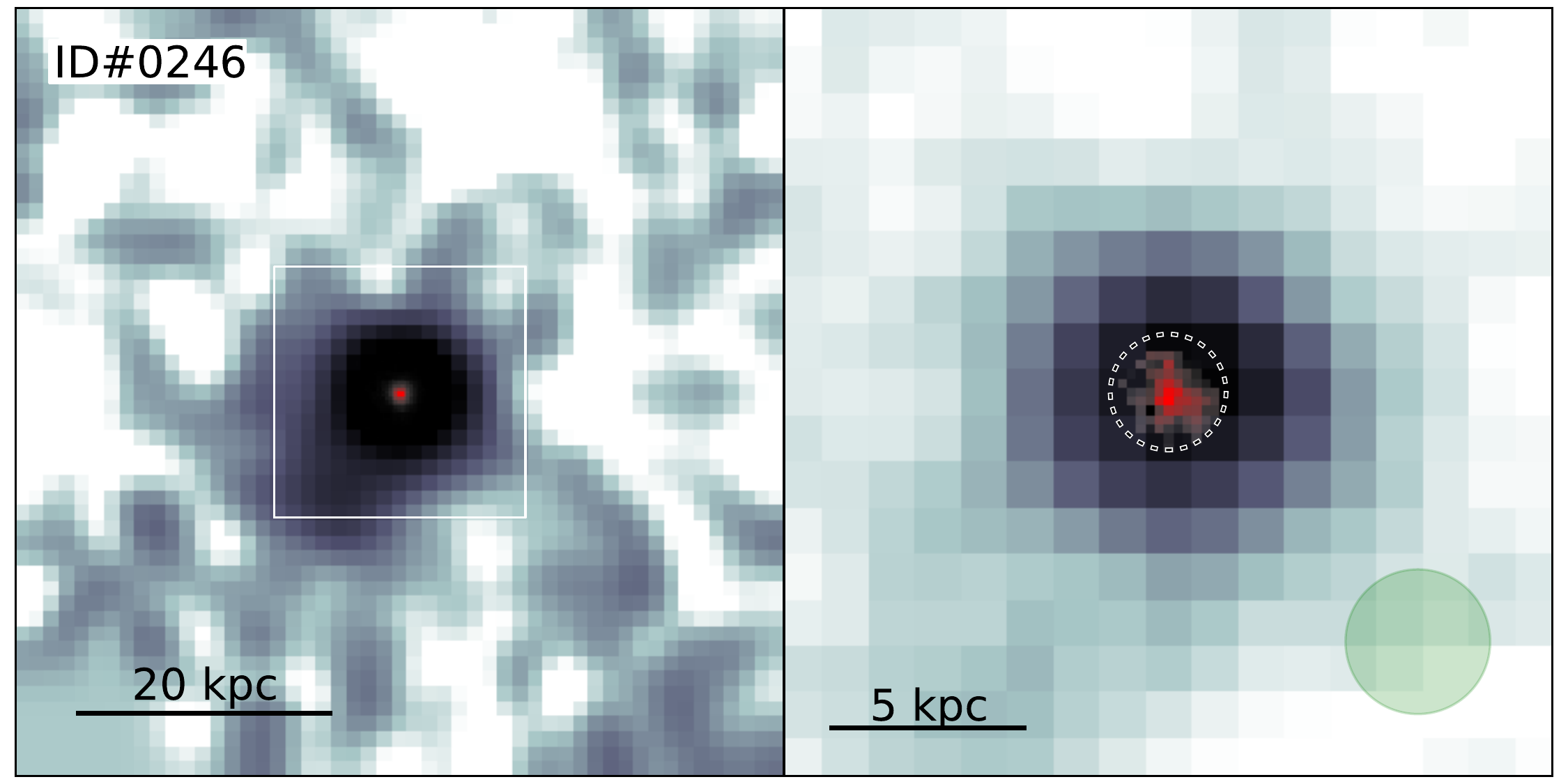}}
\put(0,97){\includegraphics[width=91.0mm]{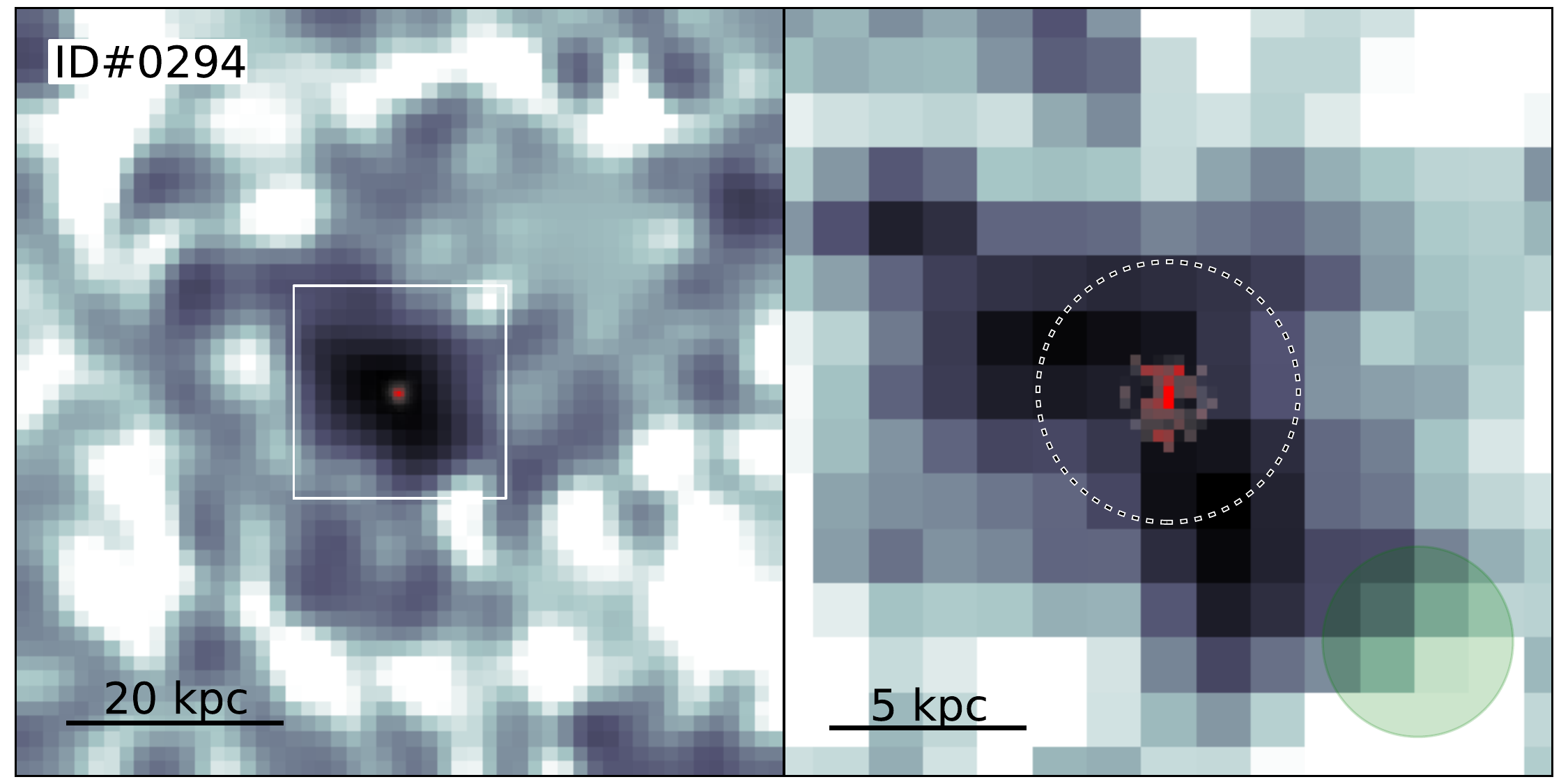}}
\put(93,97){\includegraphics[width=91.0mm]{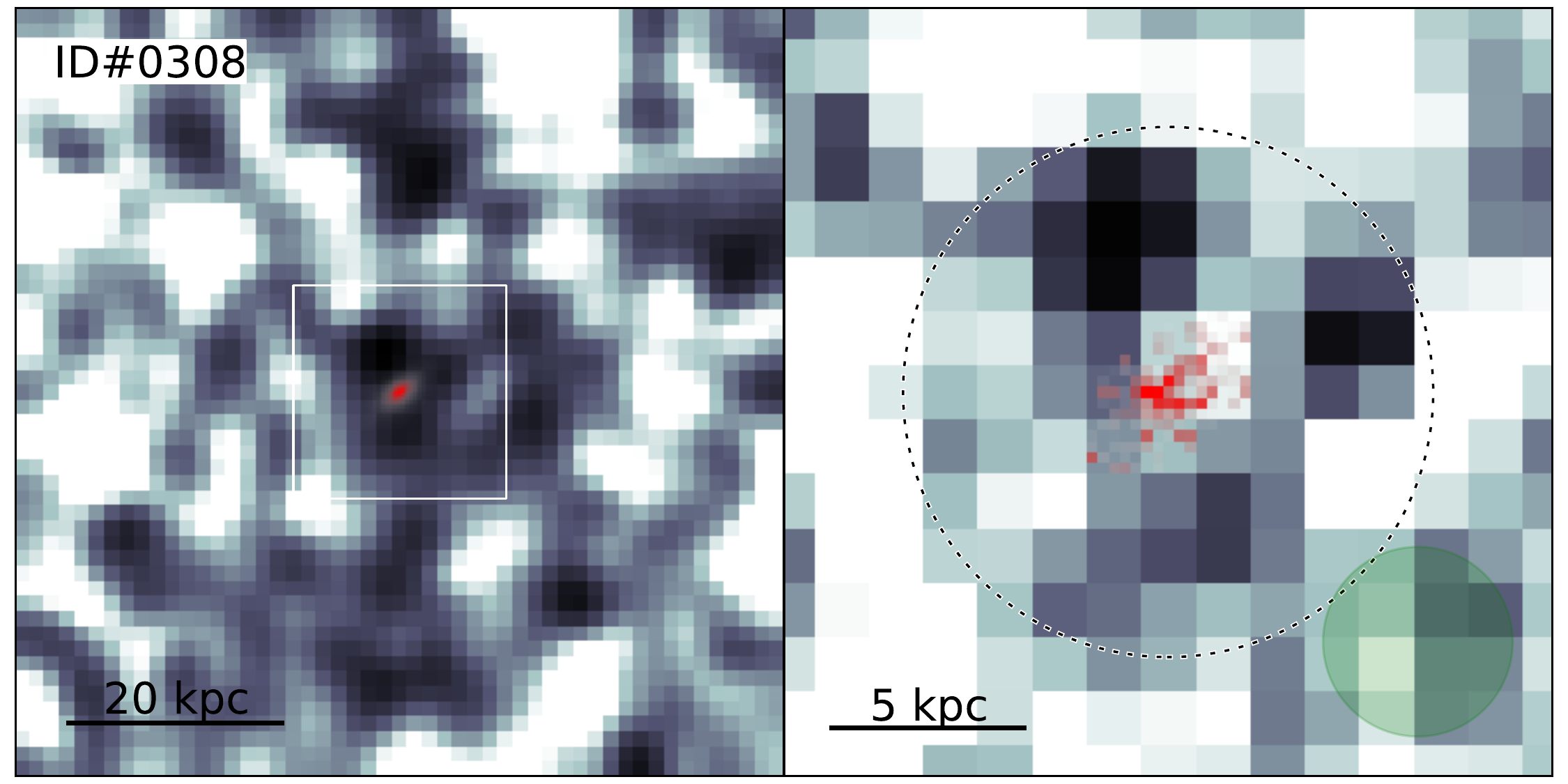}}
\put(0,49){\includegraphics[width=91.0mm]{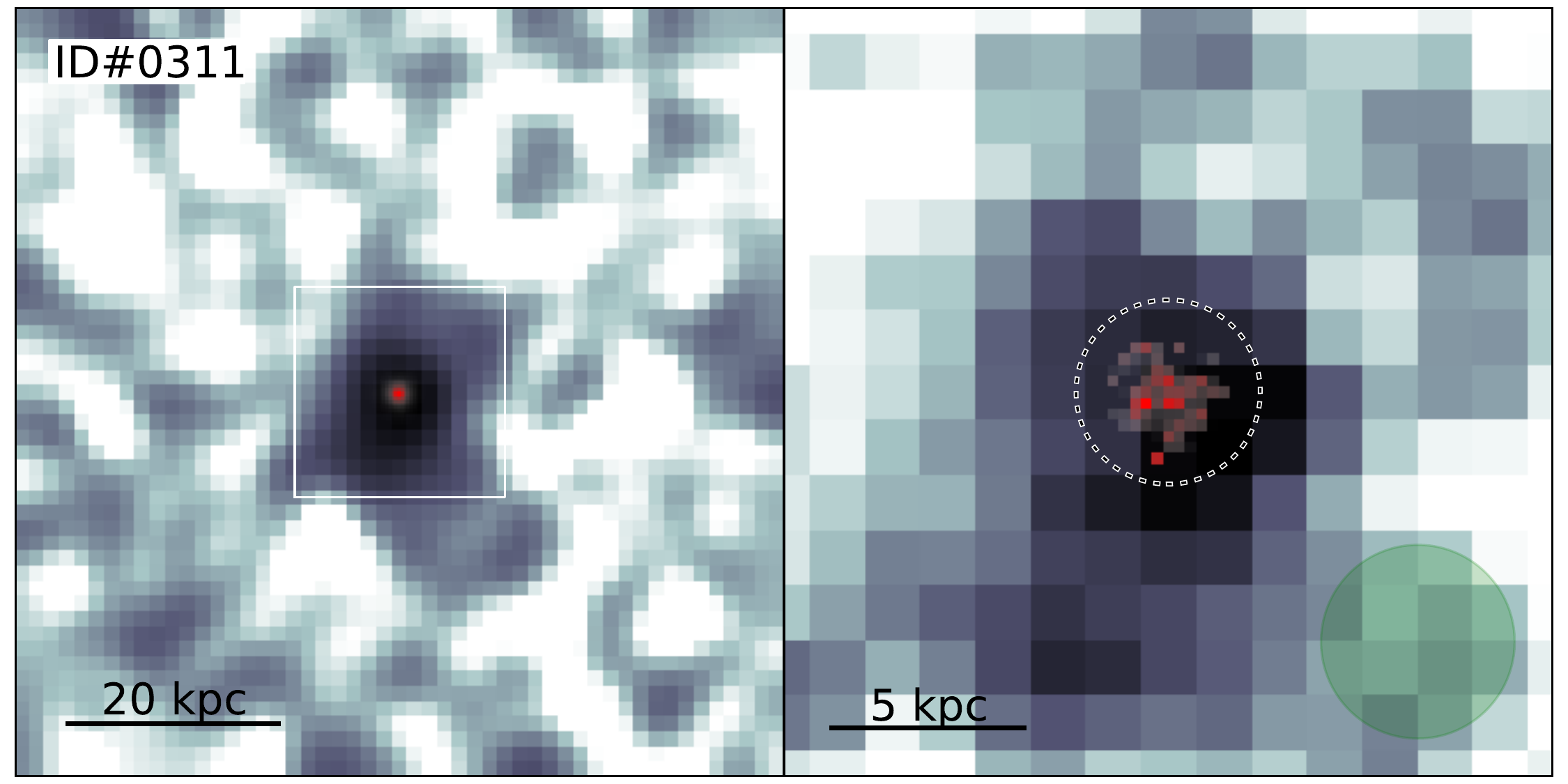}}
\put(93,49){\includegraphics[width=91.0mm]{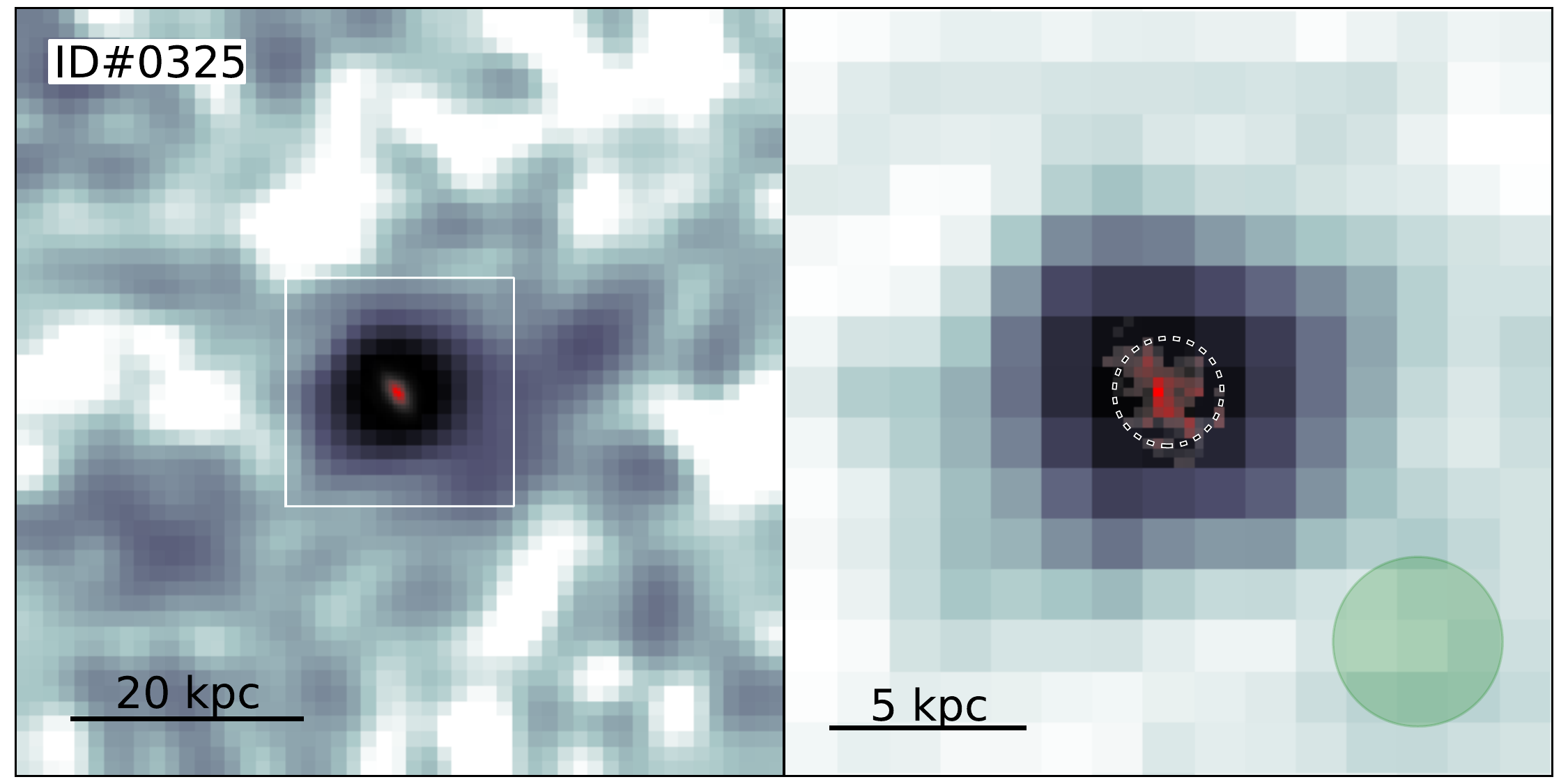}}
\put(0,1){\includegraphics[width=91.0mm]{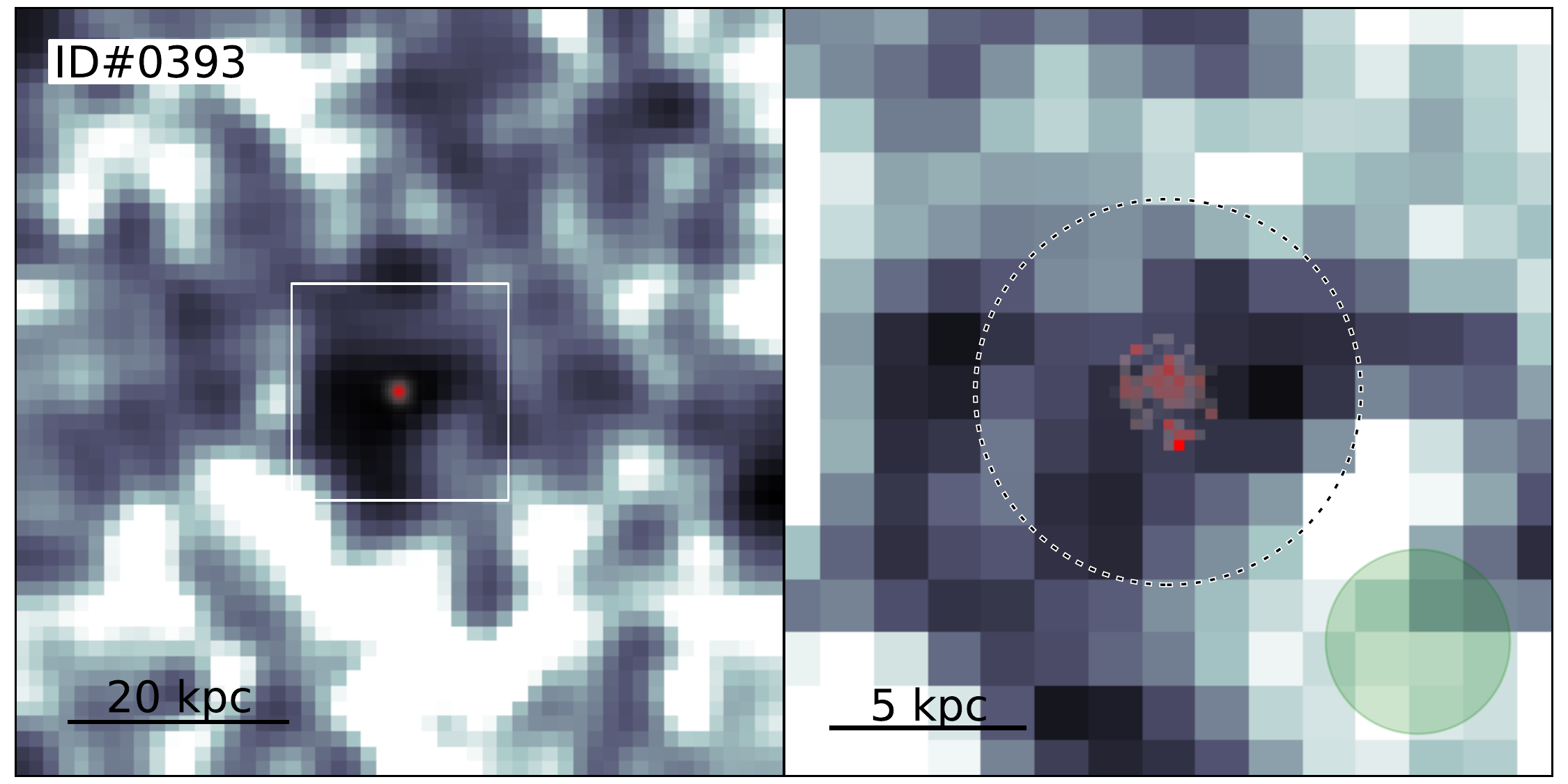}}
\put(93,1){\includegraphics[width=91.0mm]{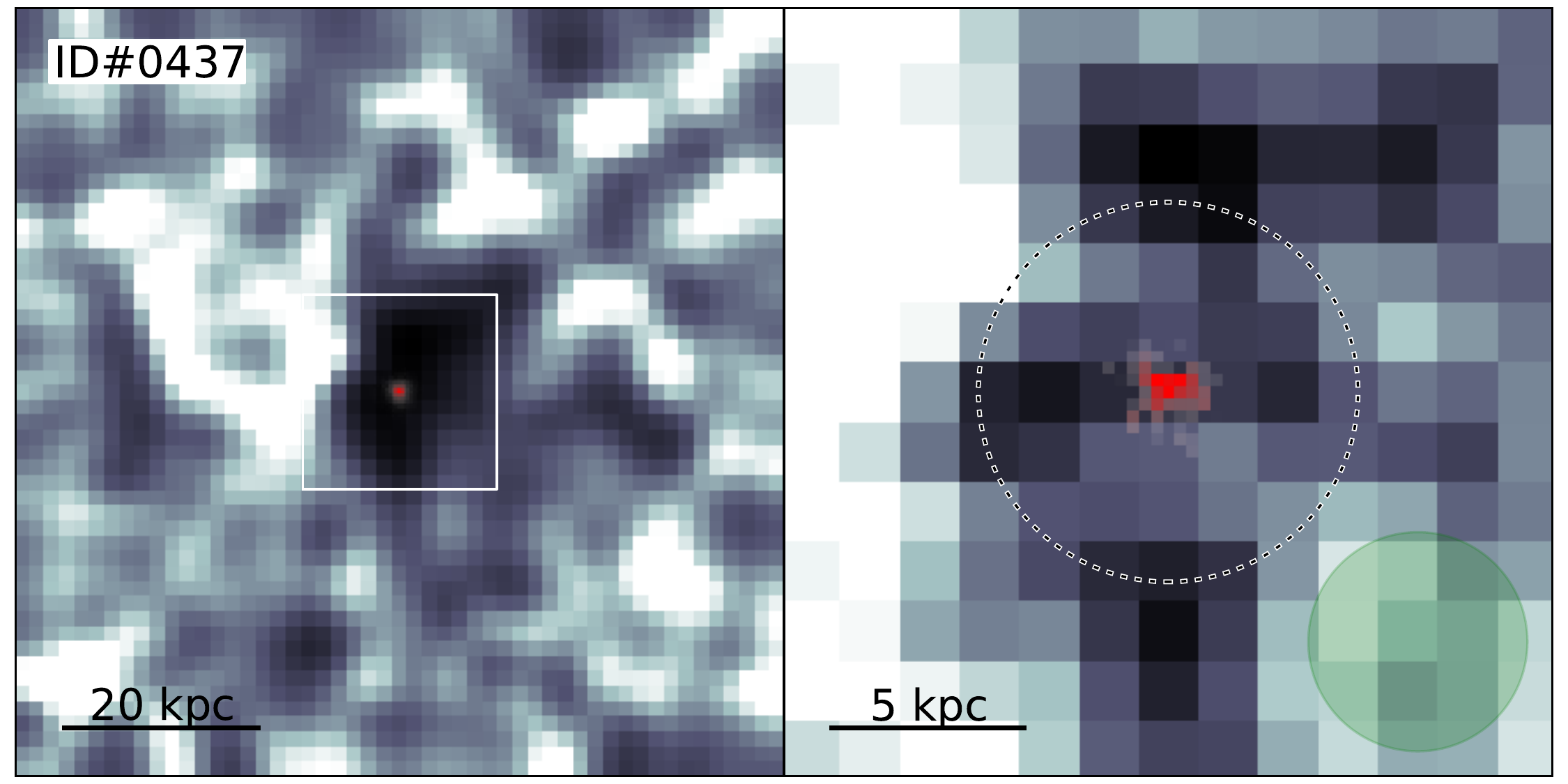}}
\end{picture}
\caption[]{\rev{(continued)}}
\end{figure*}

\addtocounter{figure}{-1}
\begin{figure*}
\setlength{\unitlength}{1mm}
\begin{picture}(170,0142)
\put(0,97){\includegraphics[width=91.0mm]{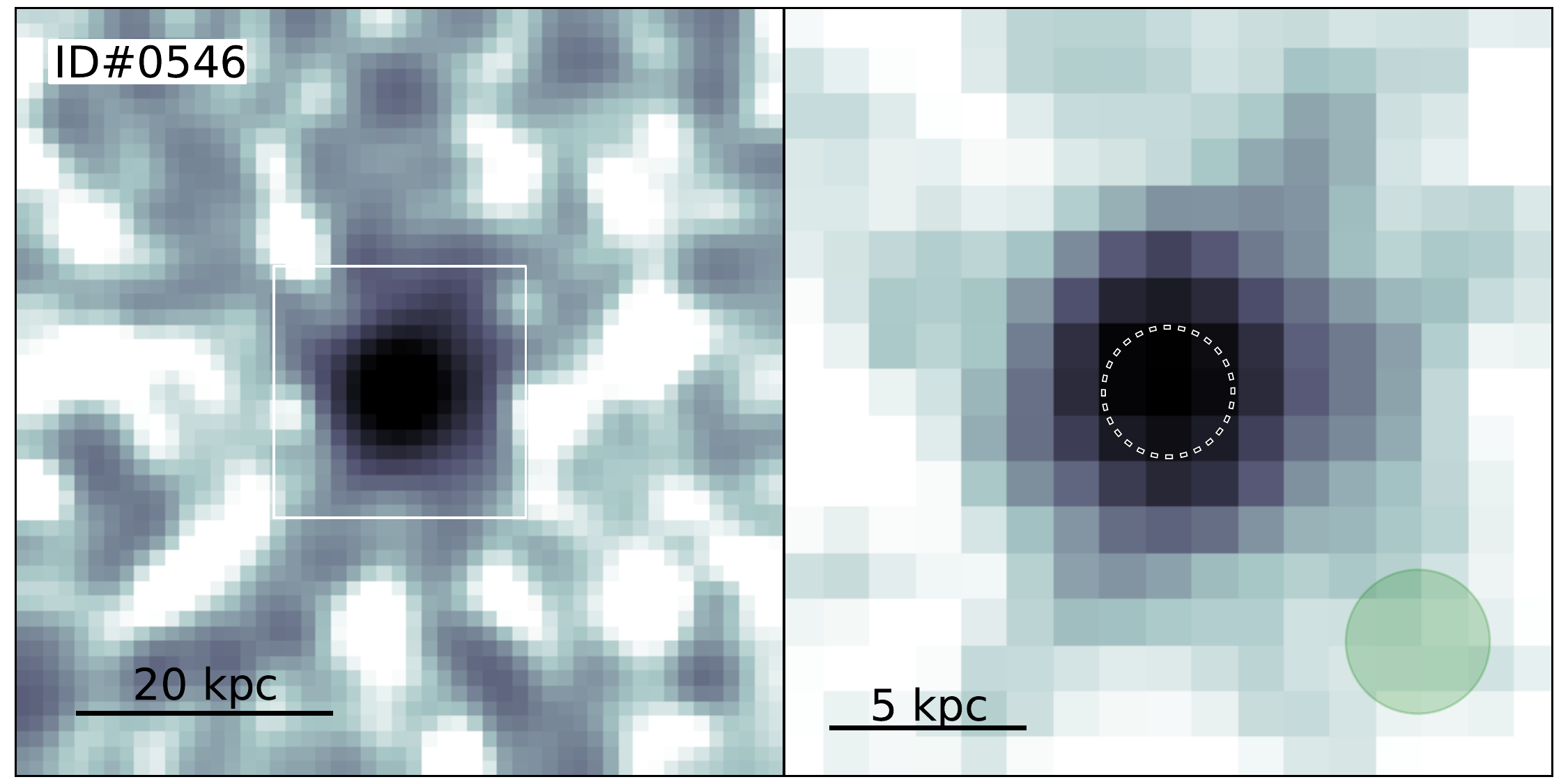}}
\put(93,98){\includegraphics[width=91.0mm]{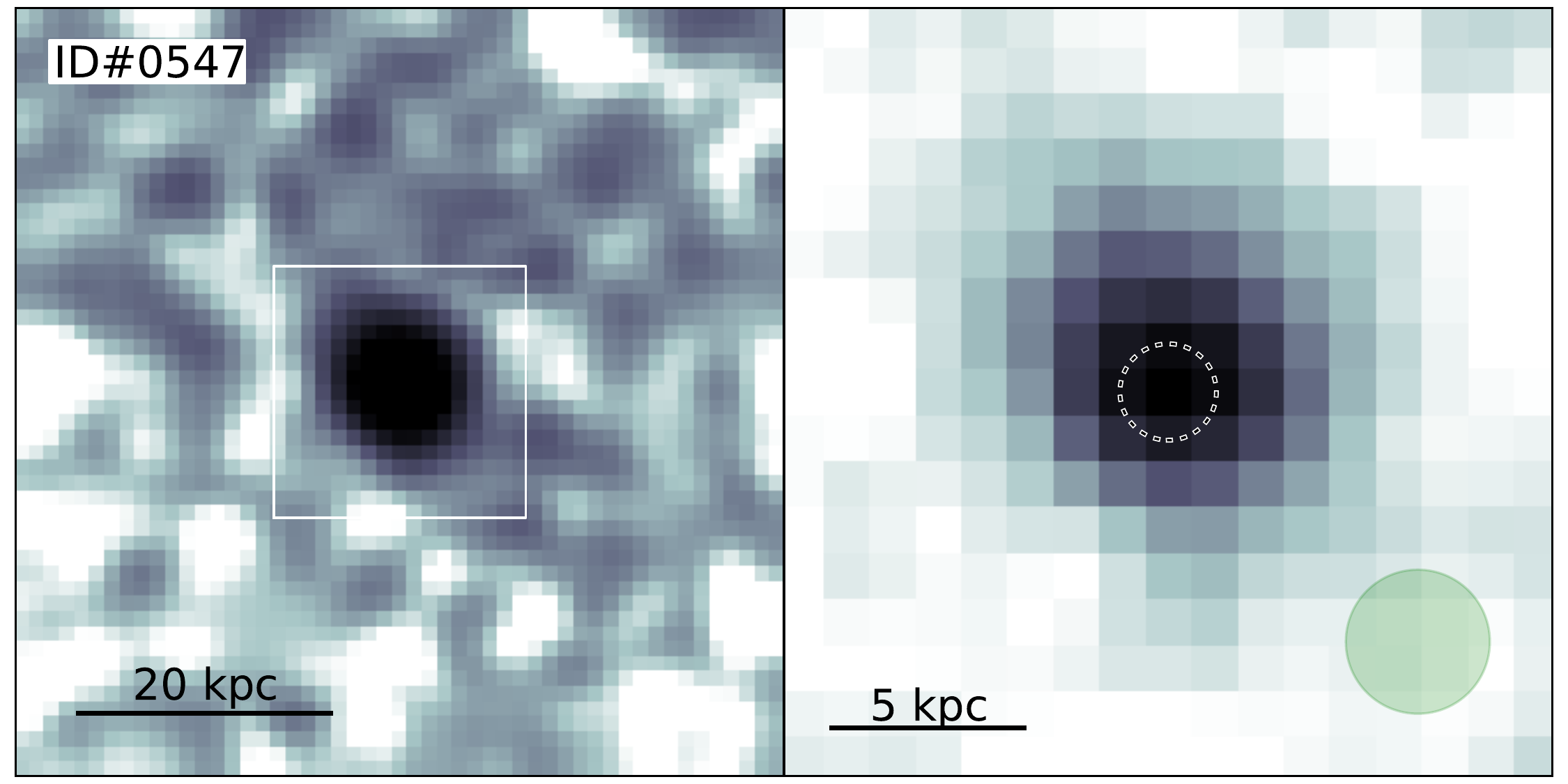}}
\put(0,49){\includegraphics[width=91.0mm]{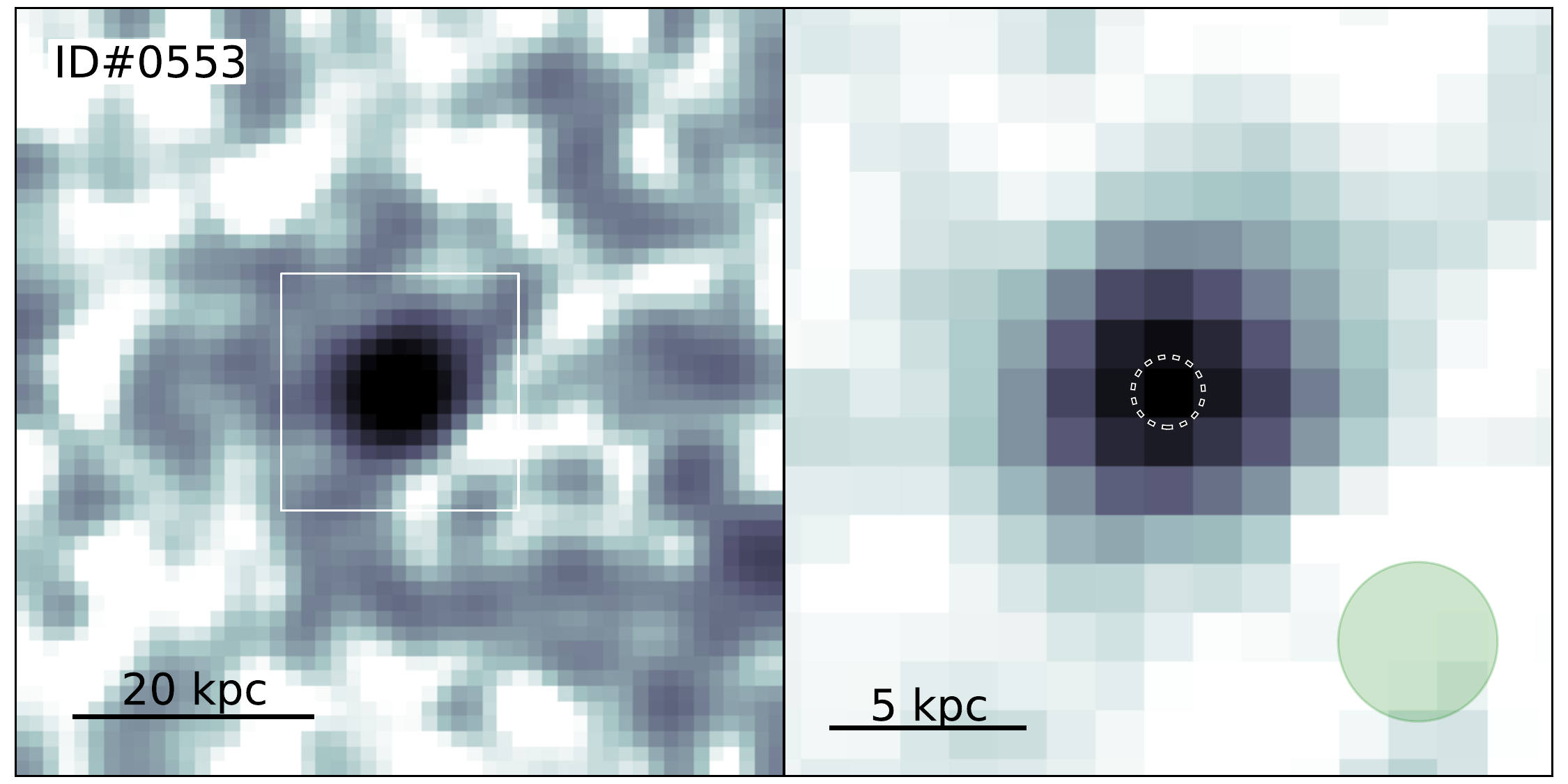}}
\put(93,50){\includegraphics[width=91.0mm]{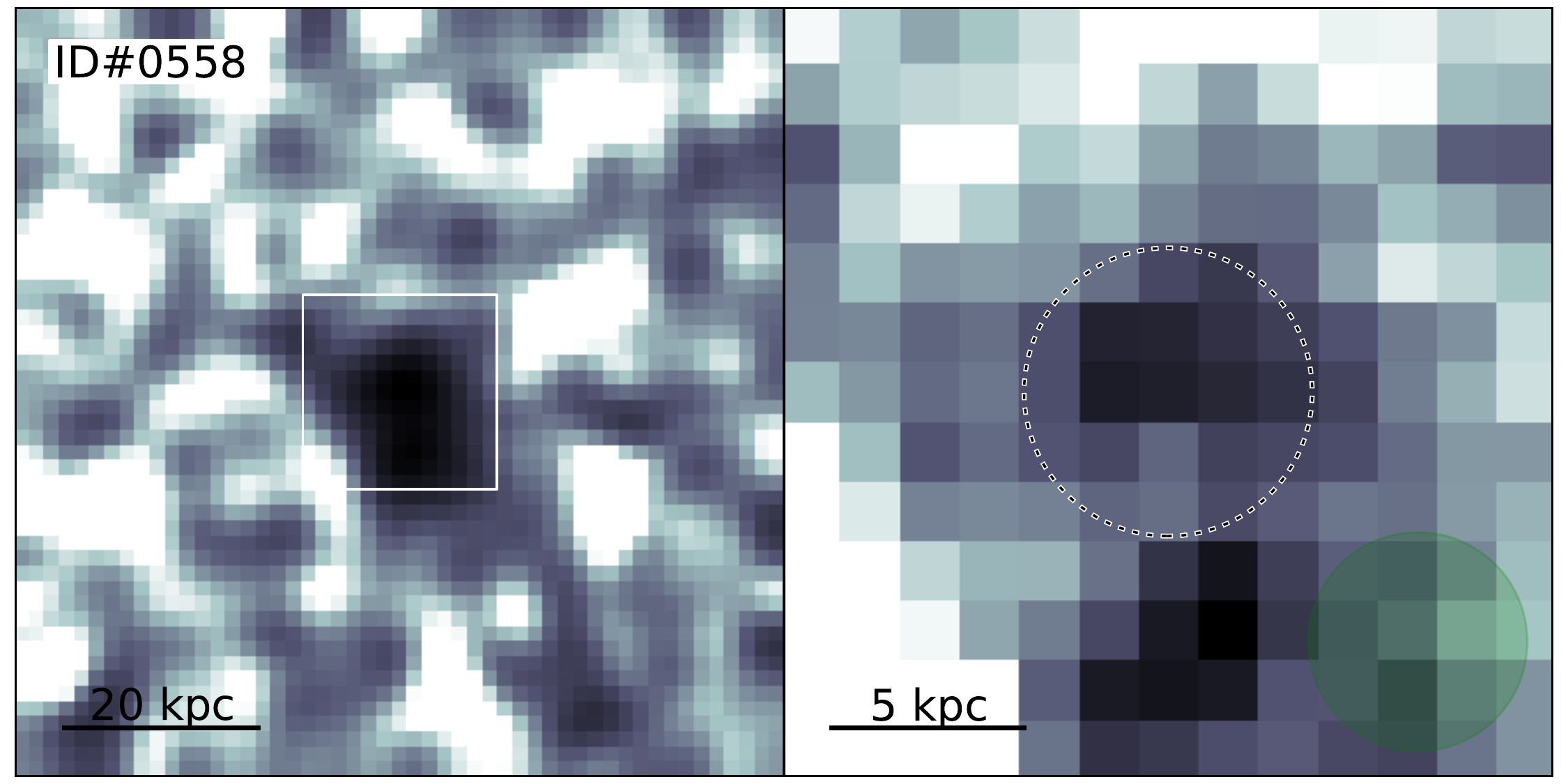}}
\put(0,1){\includegraphics[width=91.0mm]{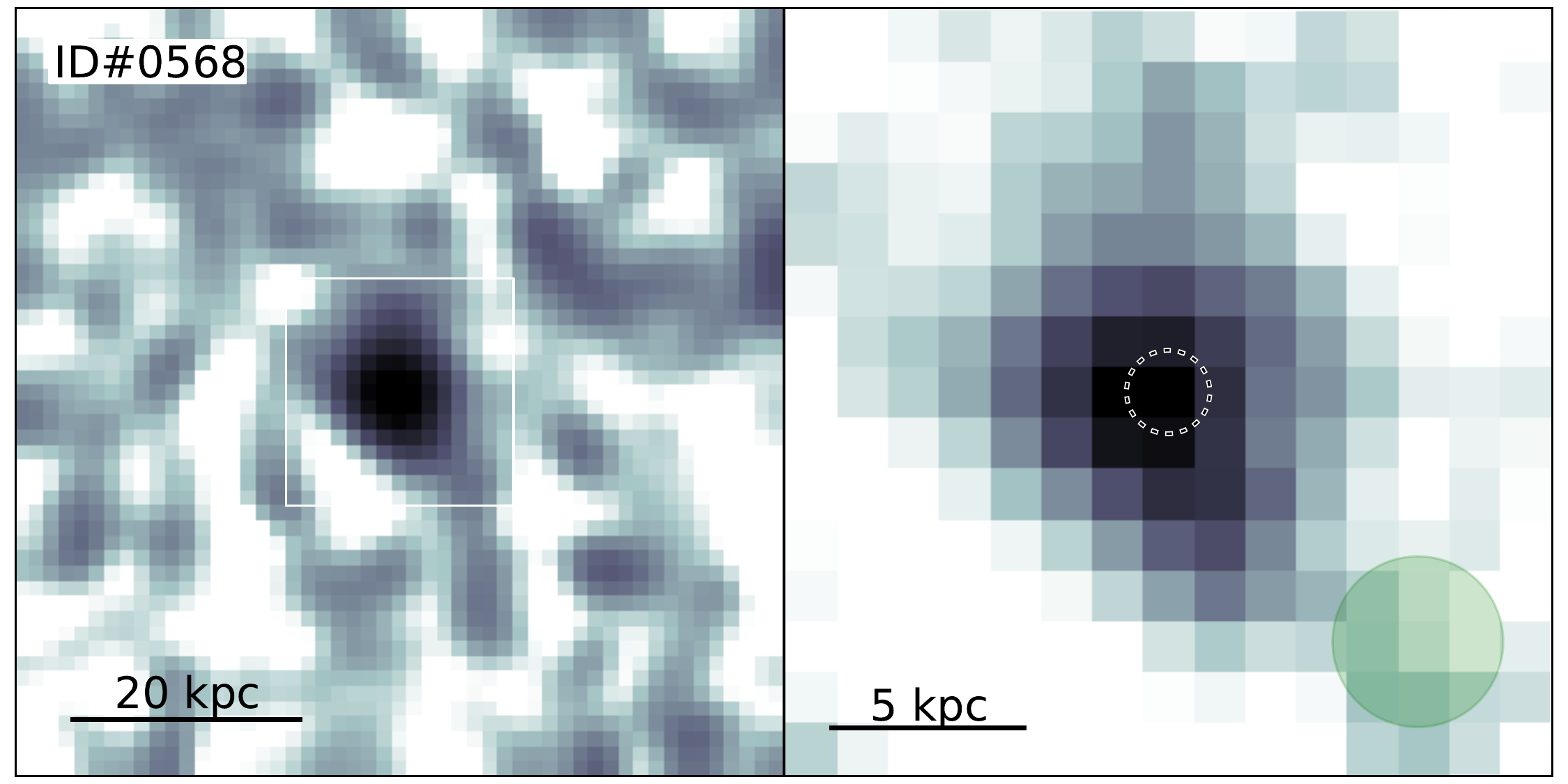}}
\end{picture}
\caption[]{\rev{(continued) Note that these five LAEs have no HST counterparts.}}
\end{figure*}

In this section we so far imposed strong constraints of smoothness and azimuthal symmetry. While this is reasonable for our main goal to obtain robust estimates of halo scale lengths and luminosities, it is nevertheless interesting and important to assert how far these imposed regularities agree with the observed \lya\ distribution in the individual images. We first subject the data to a visual inspection in this regard, then we report on the results from a few simple quantitative test.

Figure~\ref{fig:lya-grey+red} displays greyscale images of all significantly detected \lya\ haloes in our sample. Note that these images differ from those in Fig.~\ref{fig:ima+prof} in that here only the extended haloes are shown. We achieved this by subtracting from each \lya\ NB image a model of the PSF-convolved compact continuum-like \lya\ component. Each object is presented twice, to the left in a large field view with fixed cuts emphasizing the outer low-level features, and to the right in a zoom view of the high-SB inner regions with dynamically adjusted cuts. To suppress small-scale noise without blurring the gradients in the \lya\ distribution, we also subtracted a model image of the extended exponential halo, smoothed the residual with a Gaussian kernel, and added back the exponential halo model. For the large field view we used a kernel width of 0\farcs7 yielding a target resolution of approx.\ 1\arcsec. For the zoom view a kernel of half that size (0\farcs35 or 1.75 MUSE pixels) was used, leading to almost no degradation of resolution. This smoothing procedure achieved a substantial gain in visual contrast between signal and noise, while preserving essentially all structures on the relevant spatial scales.

To help localize the centroid of each object, its HST counterpart (whenever available) is superposed in red in both panels, in the large field view as a parametrised model image, and in the zoom view by the actual HST F814W pixel data, with fore- and background sources masked out. 

A visual survey of Fig.~\ref{fig:lya-grey+red} confirms that the detected \lya\ haloes for the most part are not strongly violating our symmetry assumption at the seeing-limited resolution of the MUSE data. There are no indications that the scale lengths (indicated by the circles in the zoom view panels) and fluxes derived from the SB model fitting might be heavily biased because of gross asymmetries. We quantify this statement below.

Unsurprisingly, it becomes also clear from these images that our observed \lya\ haloes do not exactly conform to ideal circular exponentials, especially for some of the brighter, high S/N haloes. While a detailed study of the morphology of individual \lya\ haloes is beyond the scope of this paper, Fig.~\ref{fig:lya-grey+red} demonstrates that such investigations are now becoming possible. Just as an example, note the `tripolar' shape of the inner halo of object ID\#112, at SB levels very significantly above the effective noise.

In order to put the discussion of the degree of asymmetries in the \lya\ distribution on more quantitative grounds we have run some simple tests to measure first-order deviations from axial symmetry. Recall that our baseline model enforces the exponential halo component to be circular as well as concentric with the continuum-like component. We now relaxed these constraints and considered, in turn: (i) Models where the halo centroids were allowed to deviate from the compact component, (ii) where the haloes could be elongated, and (iii) where both conditions were removed. The results from these tests are reassuring. The fluxes of both halo and continuum-like components were essentially unaffected by any of these additional degrees of freedom and changed on average by less than 3\% in $F_\mathrm{h}$ and by $\sim$13\% in $F_\mathrm{cl}$, respectively. The halo scale lengths came out systematically larger (as expected), but only by 5\% for test (i) and 30\% for test (iii), averaged over the sample; in most objects these changes were still within the error bars. 75\% of the objects came out with axis ratios $> 0.5$ when left free, and a similar fraction showed halo to continuum-like centroid displacements of $<0\farcs2$ (= 1 MUSE pixel). The highest deviations were always found among the faintest and lowest S/N objects and in a few cases resulted from clear overfitting of the data. These results gave us additional confidence in the axially symmetric baseline models. Since these are clearly more robust against noise (especially given the hidden low-level systematics mentioned in Sect.~\ref{sec:prof-err-noise}), we decided to restrict the subsequent analysis to the results of the baseline model fits.

As a final comment to this subsection we note that Fig.~\ref{fig:lya-grey+red} also illustrates very impressively the relative sizes between \lya\ haloes and their parent galaxies. This is the topic of the next section.

\subsection{Summary of modelling results}
\label{sec:sbmod-summary}

Our two-component model has some useful conceptual features. Firstly, it permits us to incorporate the results from the HST image analysis as a meaningful prior. Of course this prior should not be overinterpreted -- the \lya\ emission on small scales is almost certainly more complex than what we assume here. But since substructures on scales of less than $\sim$1~kpc are erased completely by the MUSE PSF, we can with some justification link the inferred presence of a compact \lya\ component to the known morphological properties of the UV continuum sources. 

We demonstrated that the resulting exponential halo scale lengths are quite robust, also against the level of asymmetries found in in the \lya\ surface brightness distribution. These sizes can be directly compared to published  scale lengths estimated by stacking analyses of \lya\ haloes. Note that such comparisons are meaningful to the extent that those other measurements are unaffected by or corrected for the blurring by atmospheric seeing. 

And finally, our models equip us with a simple but statistically adequate parametrisation to describe the \lya\ surface brightness distributions over the full observed range. Below we make use of this parametrisation to derive alternative estimates of halo `size' for comparison with the literature.

%%%%%%%%%%%%%%%%%%%%%%%%%%%%%%%%%%%%%%%%%%%%%%%%%%%%%%%%%%%%%%%%%%%%%%%%%%%%%%%%

%\input{LAE-ext_5-sizes}

\begin{figure*}
\includegraphics[width=\hsize]{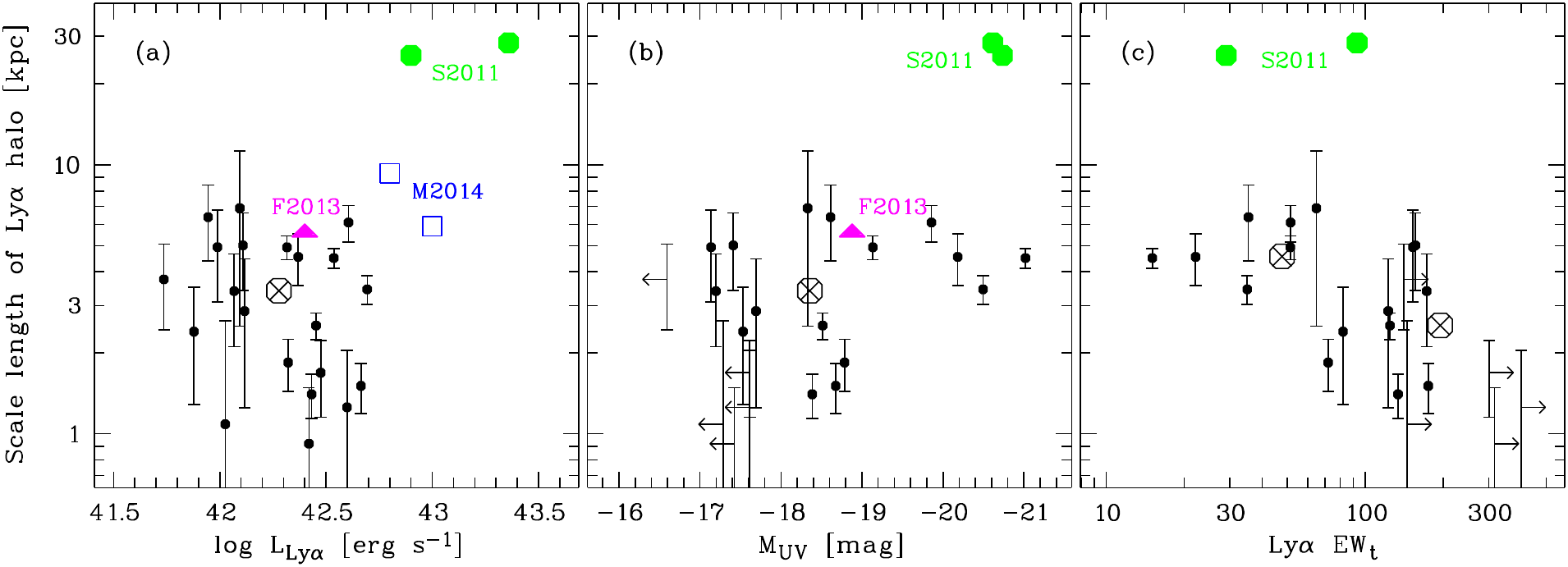}
\caption[]{Exponential scale lengths of the \lya\ haloes plotted against (a) total Ly$\alpha$ luminosities, (b) absolute UV continuum magnitudes, (c) total \lya\ rest frame equivalent widths. Lower limits on $M_\mathrm{UV}$ are indicated by arrows. The crossed circles give the mean values of each quantity in our sample (separately for EW$_\mathrm{t}<100$~\AA\ and EW$_\mathrm{t}>100$~\AA\ in panel c). Other coloured symbols show values from the literature based on the stacking of narrowband images -- green filled circles: \citet[][$\left<z\right> = 2.65$, LAE and non-LAE subsamples shown separately]{Steidel:2011jk}, magenta filled triangles: \citet[][C-O3 subsample, $z = 3.1$]{Feldmeier:2013fx}, blue open squares: \citet[][$z=3.1$ and 5.7 subsamples]{Momose:2014fe}.
}
\label{fig:r0halo1}
\end{figure*}

\section{Sizes of high-redshift Ly$\alpha$ haloes}
\label{sec:sizes}

In this section we investigate the spatial extents of the detected \lya\ haloes and relate them to other quantities, always restricting the sample to objects with formal halo detections ($p_0 < 0.05$). Besides the scale lengths from the SB fitting we also explore other measures of `size', especially with the goal of comparing our results with the few existing previous measurements.

\subsection{Exponential scale lengths}
\label{sec:sizes-scalelengths}

In Fig.~\ref{fig:r0halo1}(a) we show the exponential scale lengths \rlyah\ of the \lya\ haloes plotted against total \lya\ luminosities. Within our sample  \rlyah\ is essentially uncorrelated with $\log L_{\sublya}$, but the dynamic range in luminosities is also quite small. In the same diagram we also include the scale lengths determined by previous high-$z$ studies from the stacking of narrowband images of LAEs (with the cautionary note that the methods and assumptions used to estimate these scale lengths differ widely). In cases when the mean luminosities were not provided by the authors, we estimated them from the corresponding sample description papers. Since our HDFS dataset probes lower \lya\ luminosities than most comparison samples from the literature, the dynamic range is significantly increased by this compilation. Considering the entire range of luminosities, there appears to be a trend for more luminous LAEs to show, in the mean, more extended \lya\ haloes. The scatter is however considerable. As another cautionary note we remind the reader that aperture effects make a clean definition of `total' luminosities in extended systems a nontrivial task.

In panel (b) we plot \rlyah\ vs.\ UV absolute magnitudes. The dynamic range of our emission-line selected sample is much larger in $M_\mathrm{UV}$ than in $\log L_{\sublya}$, and we can compare the \lya\ halo sizes of continuum-bright and continuum-faint objects. The main difference is that UV-luminous galaxies with $M_\mathrm{UV}<-19$ appear to have \lya\ haloes with $\rlyah \ga 3$~kpc, whereas the \lya\ haloes around UV-faint galaxies cover a wider range of sizes; our sample is too small for a stronger statement. The comparison to the literature does also not lead to a conclusive picture. A simple compilation of sample averages suggests again that \rlyah\ might be correlated with $M_\mathrm{UV}$. Yet our UV-brightest objects are as luminous in $M_\mathrm{UV}$ as the LBG sample mean in S2011, while the \lya\ haloes of our objects are an order of magnitude smaller than those around the S2011 galaxies. Considering the high significances of the measured scale lengths in both S2011 and our data, this is unlikely to be a result of methodical differences in the analysis. 

Panel (c) of Fig.~\ref{fig:r0halo1} shows the distribution of \rlyah\ against total \lya\ rest frame equivalent widths. This plot seems to be almost a mirror image of panel (b), with the low-EW objects having a narrower range of halo sizes and a slightly larger mean \rlyah\ than the high-EW galaxies. In fact EW$_\mathrm{t}$ is tightly correlated with $M_\mathrm{UV}$ in our sample (cf.\ Table~\ref{tab:table1}), which explains the similarity of panels (b) and (c). However, if we also include the results by S2011 it appears that the \lya\ halo scale lengths do not depend much on the total EWs. 

\begin{figure}
\includegraphics[width=\hsize]{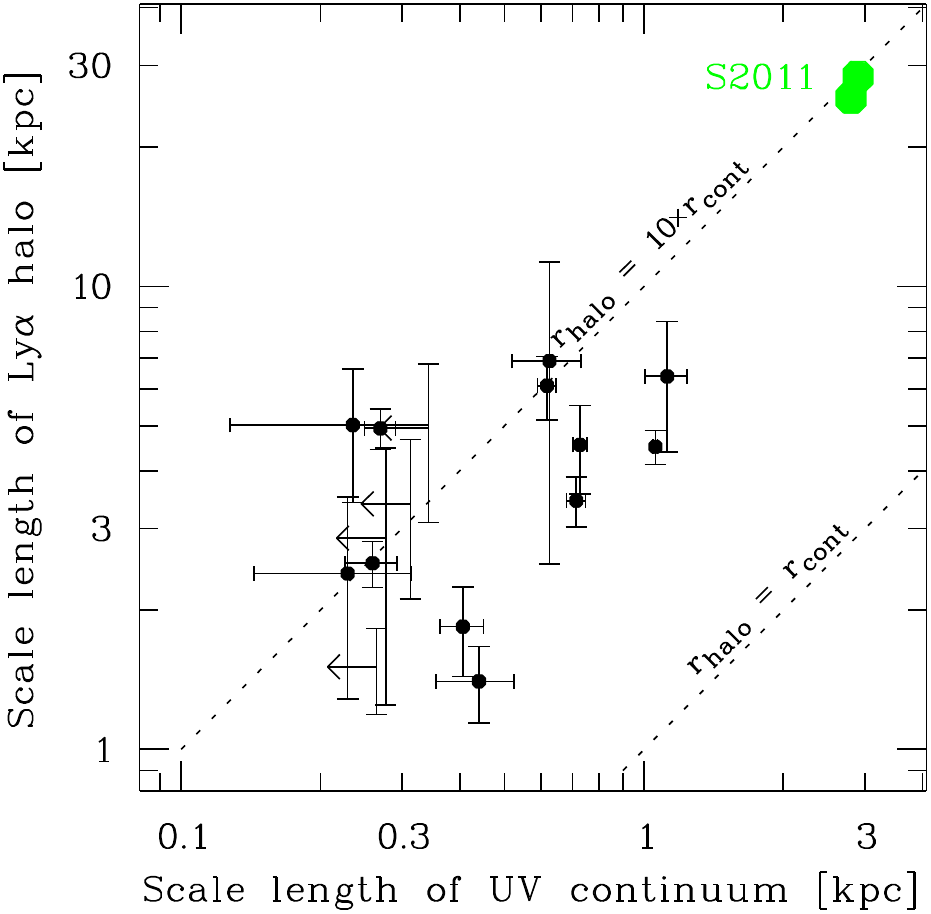}
\caption[]{\lya\ halo scale lengths vs.\ scale lengths in the UV continuum. Only objects detected by HST/WFPC2 in F814W are shown, with conservative upper limits for HST-unresolved sources. The green filled circles show the ratio obtained by \citet{Steidel:2011jk} from stacking of \lya-bright LBGs, for their LAE and non-LAE subsamples. The dotted lines indicate halo/continuum size ratios of 10 and 1, respectively.
}
\label{fig:r0halo2}
\end{figure}

Figure~\ref{fig:r0halo2} presents the relation between UV continuum and \lya\ halo scale lengths. This plot is restricted to the 16 objects with $p_0 < 0.05$ that have HST counterparts. 12 of these are resolved by HST, while 4 are point sources and hence have only upper limits to their continuum sizes. While the scatter is again substantial (as are the error bars!), the haloes are always \rev{between $\sim$ 5 and $>$15} times larger than the continuum regions, with a mean size ratio of $\ga$10 (this is a lower limit because of the upper limits on the continuum sizes; the median ratio is 9.8). Interestingly, this ratio is very close to the value found by S2011 from their stacking of much larger and much more luminous LBGs. 

\rev{Figure~\ref{fig:r0halo2} suggests that the \lya\ halo scale lengths are not independent of the apparent galaxy sizes in the UV continuum, albeit with a lot of scatter between individual objects. On the other hand, \citet{Matsuda:2012fp} presented evidence that  the \lya\ halo sizes of galaxies depend, on average, on the large-scale environment in which they reside. Since the objects in the S2011 sample are known to be located in overdense regions, their haloes may well be more extended compared to galaxies of otherwise similar properties but in poorer environments. Much better statistics on \lya\ haloes around individual galaxies will be required to disentangle the dependencies due to intrinsic properties and due to the environment.}

\subsection{Other size measures}
\label{sec:sizes-others}

Given the small number of publications so far that contain size estimates of \lya\ haloes around galaxies, the variety of approaches and conventions adopted by different authors is quite impressive. We now compare our data with those studies that avoided exponential scale lengths. 

Perhaps closest to our sample in detection methodology, luminosity range, and surface brightness sensitivity are the faint LAEs of \citet{Rauch:2008jy}, detected blindly in a single ultra-deep longslit exposure. Rauch et al.\ characterised the size of the \lya\ emission along the slit as the `radius' (= distance from the centroid) where the fitted surface brightness profile reaches a value of $1\times 10^{-19}$~erg~s$^{-1}$ cm$^{-2}$ arcsec$^{-2}$. Unfortunately, such isophotal radii are subject to cosmological SB dimming and thus not practical for comparisons between samples with different redshift ranges. Their radii were also not corrected for blurring by the seeing, which can be important if the PSF has extended wings. We nevertheless applied the same criterion to estimate isophotal radii $r_{-19}$ of our objects; these values are listed in Table~\ref{tab:table2} for readers wishing to make a detailed comparison. Here we only observe that the isophotal size distributions of the two samples differ markedly: While the median \emph{apparent} radius $r_{-19}$ of our LAEs is $1\farcs8$, the Rauch et al.\ sample has a median of $1\farcs2$, despite their poorer seeing and also despite the fact that the median redshift of their sample is considerably smaller. Correcting for any of these effects would only increase the discrepancy. Most likely the explanation for these differences is a mix of reasons: Small sample sizes (for both samples); the low S/N of the Rauch et al.\ data; and possibly also biases in their procedure to measure the radial surface brightness in a longslit for objects that are not centred in the slit. We therefore caution that the \lya\ size distribution function derived by \citet{Rauch:2008jy} probably underestimates the true sizes and incidence rates of the haloes.

\citet{Hayes:2013jc} adopted Petrosian radii $r_\mathrm{P20}$ with \mbox{$\eta = 20$\%} as size measure for their low-redshift LARS galaxies. This quantity is defined as the radius at which the azimuthally averaged SB reaches a fraction $\eta$ of the mean SB inside the aperture enclosed by that annulus \citep{Petrosian:1976ea}. Petrosian radii have some attractive features: No knowledge of the asymptotic behaviour of the SB profile is required, since only quantities measured at $r < r_\mathrm{P20}$ are evaluated, and the radii are insensitive to cosmological SB dimming because the SB normalisation cancels out. Note that for an ideal circular exponential profile, $r_\mathrm{P20} = 3.62$~$r_\mathrm{s}$. Applied to our MUSE data, Petrosian radii have the additional advantage that they reach sufficiently far out as to not being heavily affected by the seeing (although some influence remains, see below). On the other hand, the low S/N ratios at these distances make the measured SB ratios too noisy for determining $r_\mathrm{P20}$ directly from the data. We therefore applied the same recipe to the analytic SB model fits instead of to the measured SB profiles, which produced robust estimates of $\eta = 20$\% Petrosian radii for our LAEs.

\begin{figure}
\includegraphics[width=\hsize]{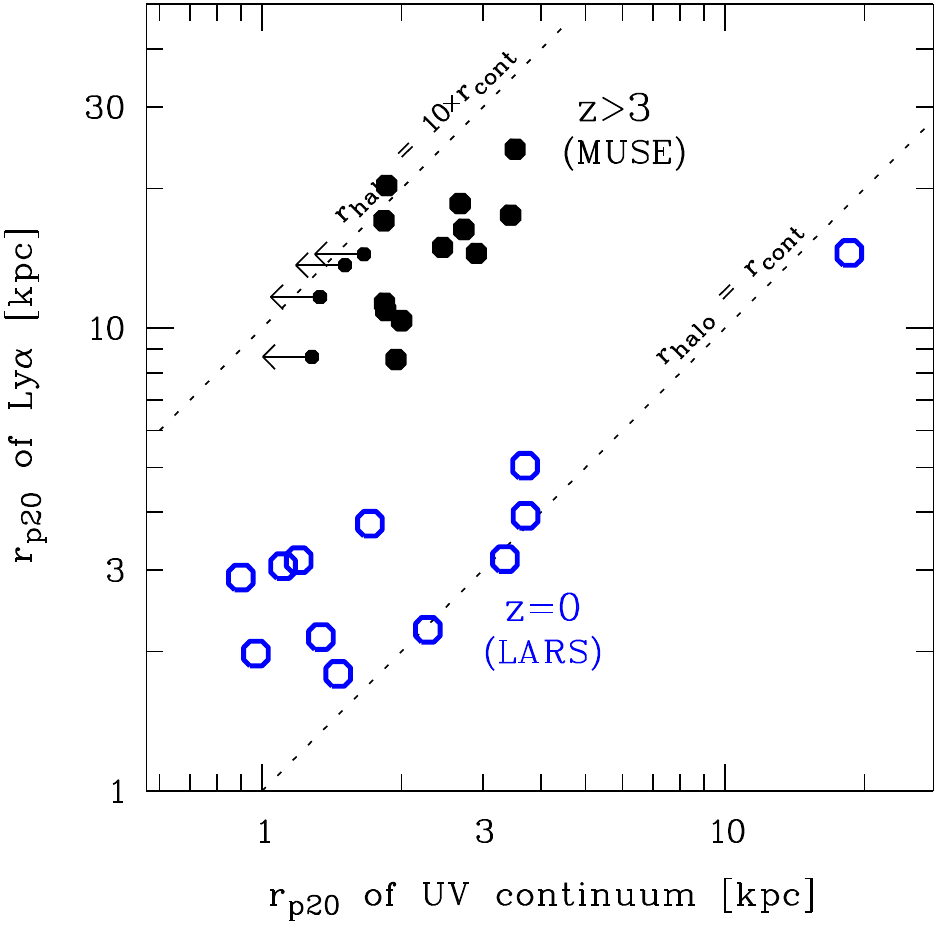}
\caption[]{\lya\ vs.\ continuum size relation for the same objects as in Fig.~\ref{fig:r0halo2}, expressed using Petrosian radii $r_\mathrm{P20}$ (filled black symbols). The blue open circles show the corresponding relation for the $z\approx 0$ galaxies in the LARS sample \citep{Hayes:2013jc,Guaita:2015kr}. The dotted lines again denote size ratios of 10 and 1, respectively.
}
\label{fig:rp20}
\end{figure}

In Fig.~\ref{fig:rp20} we present the relation between the sizes of the \lya\ light distributions and the UV continuum regions, measured by their Petrosian radii. The values of $r_\mathrm{P20}$ are listed in Table~\ref{tab:table2}. The relation looks very similar to Fig.~\ref{fig:r0halo2}, except that it appears somewhat compressed in range, which can be explained by the fact that the smaller values on both axes are probably biased high because of PSF blurring. The size ratios between haloes and continuum regions are again between 3 and $>$10, with a mean of 7.0 (which is again a lower limit because of the upper limits on the continuum radii; the median is 6.5).

The advantage of Fig.~\ref{fig:rp20} over Fig.~\ref{fig:r0halo2} is, of course, that we can now compare our individually estimated Petrosian radii of galaxies at $z > 3$ with the corresponding measurements of $z\approx 0$ galaxies in LARS \citep{Hayes:2013jc,Guaita:2015kr}, shown by the blue circles. It is striking that the two sets of points look extremely similar, but with a vertical offset of about a factor 5 in halo size at given continuum size. Note that the \emph{angular} extents of the LARS galaxies are much larger than of our objects, and that their Petrosian radii are therefore not hampered by PSF effects. In particular, those high-$z$ LAEs for which we can only set upper limits on the continuum radii $r_\mathrm{P20,c}$ likely have true sizes similar to the smallest LARS galaxies, which would then again imply \lya\ halo/continuum size ratios of well above 10.

Both low- and high-redshift samples suggest that the \lya\ halo-continuum size relation may not follow a simple proportionality. While for the continuum-brighter and larger galaxies the size ratio is consistent with more or less a constant plus some scatter, the relation appears to flatten for more `dwarfish' galaxies, resulting in \lya\ haloes that are very modest in terms of their absolute dimensions, but huge given the tiny sizes of the central objects. We caution, however, that so far these are just weak trends obtained from small samples.

\subsection{Size evolution}
\label{sec:sizes-evol}

\begin{figure}
\includegraphics[width=\hsize]{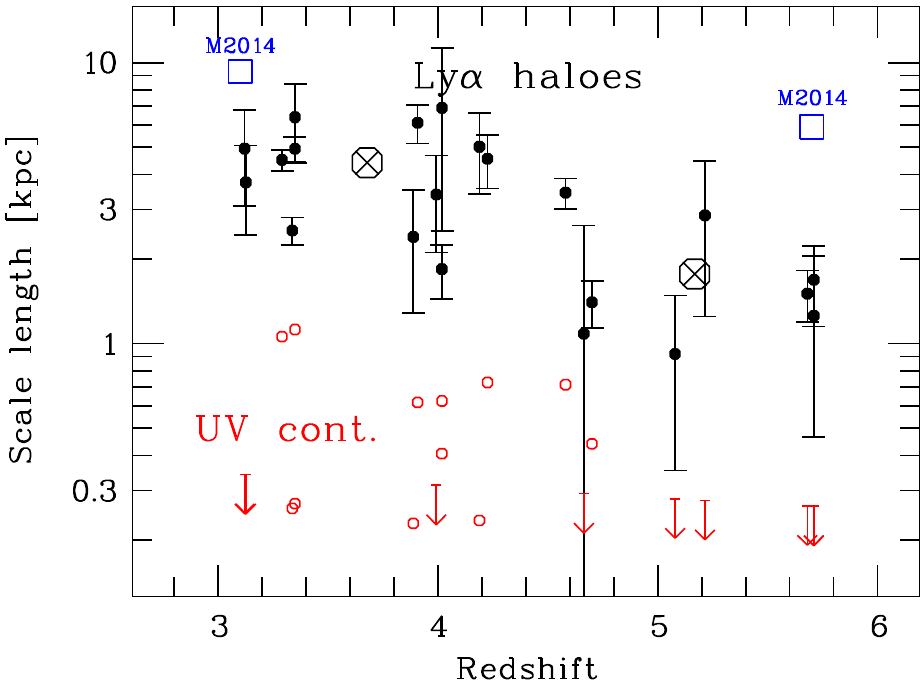}
\caption[]{Redshift dependence of scale lengths for \lya\ haloes (filled black symbols) compared to the UV continuum (red open circles). The large crossed circles give the mean \rlyah\ values $z<4.5$ and $z>4.5$, respectively. The blue open squares show the scale lengths from \citet[][$z=3.1$ and 5.7 subsamples]{Momose:2014fe}.}
\label{fig:rhalo-z}
\end{figure}

Since \lya\ haloes seem to be rather ubiquitous for high-$z$ galaxies, it is of interest to see whether the halo properties change with redshift. So far very little is known in this respect, except for the NB stacking study by \citet{Momose:2014fe} who estimated mean halo scale lengths at 4 different redshifts. They did not discern a clear evolutionary trend in $\left<r_\mathrm{s,h}\right>$, with values of (7.9, 9.3, 5.9, 12.6)~kpc at $z$ $=$ (2.2, 3.1, 5.7, 6.6).

In Fig.~\ref{fig:rhalo-z} we plot our estimated halo scale lengths against redshifts. The data show a significant reduction of scale lengths towards higher $z$. If we divide the sample at mid-range ($z=4.5$) into two sets, the corresponding subsample averages differ by a factor of $\sim$2 between $z\simeq 3.7$ and 5.1 (large crossed circles). A similar trend is seen in the $z=3.1$ and 5.7 stacks of \citet{Momose:2014fe}; but recall that their data behave differently outside the redshift range $3<z<6$.

Is the trend observed in our data real? Our scale lengths are mainly determined from the inner high-S/N datapoints of the \lya\ profiles which should be quite robust against cosmological SB dimming. We also looked at the redshift evolution of Petrosian radii $r_{\mathrm{P20,Ly}\alpha}$ -- which are formally independent of cosmological SB dimming -- and found a very similar behaviour. We therefore presume that the differences are real. The significance of the trend is supported by a Spearman rank-order test, giving a probability of only 0.004 for the null hypothesis that $z$ and $r_\mathrm{s,h}$ are uncorrelated. The trend becomes even more substantial when considering the fact that the luminosities of our flux-limited sample actually \emph{increase} slightly with redshift. 
 
However, the observed trend could be mainly a by-product of the general size evolution of galaxies \citep[e.g.][]{Shibuya:2015bj}. The red open circles in Fig.~\ref{fig:rhalo-z} show the continuum scale lengths and upper limits for our HST-detected galaxies, which appear to change by a similar factor as the \lya\ haloes. Indeed, the \emph{ratios} of halo to continuum scale lengths in our sample come out to be uncorrelated with $z$. The observed redshift evolution of \lya\ halo scale lengths therefore suggests that continuum and halo sizes of $z\gtrsim 3$ galaxies are physically linked -- or that both relate to a more fundamental quantity such as stellar or dark matter halo mass. 

This evolutionary pattern changes drastically when we broaden the redshift baseline to include the local universe. As demonstrated in the previous subsection, $z\approx 0$ \lya\ haloes are $\sim$5 times smaller at given continuum size when compared to galaxies at $z\approx 3$--6. We thus have to distinguish carefully between the evolution processes \emph{at} high redshifts and those connecting the young universe with present-day galaxies.

%%%%%%%%%%%%%%%%%%%%%%%%%%%%%%%%%%%%%%%%%%%%%%%%%%%%%%%%%%%%%%%%%%%%%%%%%%%%%%%%

%\input{LAE-ext_6-luminosities}

\begin{figure*}
\includegraphics[width=\hsize]{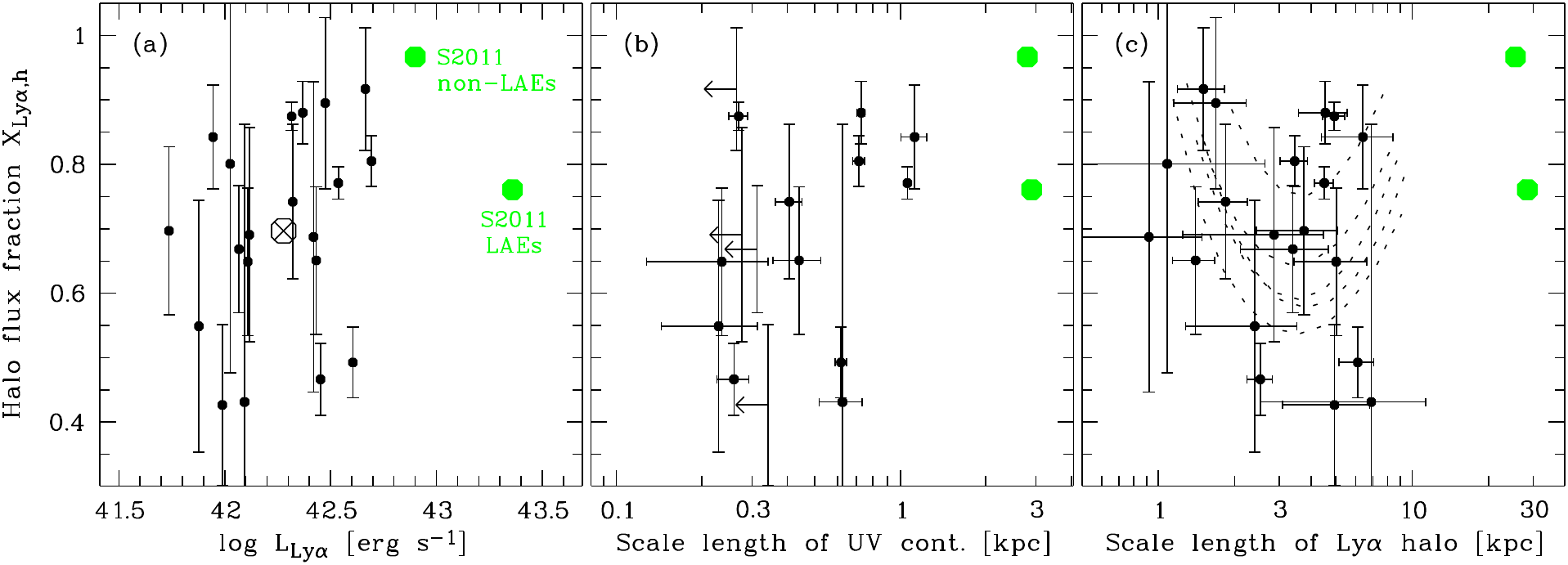}
\caption[]{Halo flux fractions $X_{\sublya,\mathrm{h}}$ in relation to (a) total \lya\ luminosities, (b) scale lengths of the UV continuum, and (c) \lya\ halo scale lengths. The green filled circles indicate the estimated corresponding \lya\ halo fractions from \citet{Steidel:2011jk}, for \lya-strong and \lya-weak galaxies. The dotted lines in panel (c) additionally show the upper limits on $X_{\sublya,\mathrm{h}}$ as a function of $r_\mathrm{s,h}$ estimated in Sect.~\ref{sec:sbmod-ul} for the 5 objects with $p_0 > 0.05$.
}
\label{fig:xfhstat}
\end{figure*}

\section{Ly$\alpha$ luminosities and equivalent widths}
\label{sec:lum}

In this section we investigate the fluxes and luminosities inferred from our surface brightness model fits in relation to other observables. Of particular interest is the flux ratio between the halo and the continuum-like component, or alternatively the fraction of the total \lya\ luminosity in the extended halo.

\subsection{Halo flux fractions}
\label{sec:lum-frac}

The halo fraction of the total \lya\ flux, expressed in terms of our model parameters, is $X_{\sublya,\mathrm{h}} = F_\mathrm{h}/(F_\mathrm{h}+F_\mathrm{cl}) = 1/(1+F_\mathrm{cl}/F_\mathrm{h})$. Our fits resulted in $F_\mathrm{h}/F_\mathrm{cl}$ ratios between $\sim$0.7 and 11 (see Table~\ref{tab:table2} and Fig.~\ref{fig:expfits}; leaving out object ID\#547 with a fitted value of $F_\mathrm{cl} = 0$ and no counterpart in HST). This implies that between 40\% and $\ga$90\% of the observed \lya\ emission comes from an extended halo (70\% in the mean). 

In their stacking analysis of LBGs, S2011 found a mean \lya\ halo fraction of 76\% for their subset of strong \lya\ emitters, very similar to our sample mean. Since S2011 did not directly specify halo fractions, we estimated $X_{\sublya,\mathrm{h}}$ for the S2011 galaxies from Table~2 in their paper, replacing $F_\mathrm{cl}/F_\mathrm{h}$ by the ratios of the \lya\ equivalent widths between small spectroscopic aperture and total EW, i.e.\ integrated over the stacked halo images. Using this prescription, the `non-LAE' subset of S2011 has a mean halo fraction of 97\%.

Figure~\ref{fig:xfhstat} presents the distribution of $X_{\sublya,\mathrm{h}}$ for our objects, plotted against total \lya\ luminosities, UV continuum scale lengths, and \lya\ halo scale lengths, respectively. We also considered UV absolute magnitudes (not shown here). In each panel we also show the means of the two S2011 subsamples. It is certainly noteworthy that while the bright LBGs of S2011 and our faint LAEs differ by an order of magnitude in luminosities and sizes, their \lya\ halo luminosity fractions are nearly the same. The somewhat broader range of $X_{\sublya,\mathrm{h}}$ occupied by our individually measured galaxies, extending down to halo fractions of 40\%, is probably just averaged out in the S2011 data. 

On the other hand, the non-LAE galaxies in S2011 with $X_{\sublya,\mathrm{h}}$ of nearly unity are actually dominated by absorption in the central arcsec$^2$, becoming net \lya\ emitters only when integrated over very large apertures. Such objects are missing in our current sample, which by construction contains only objects classified as LAEs already through small spectroscopic apertures. We are therefore also lacking galaxies with inferred halo fractions very close to one. Note however that because of the much smaller physical scales of our objects compared to the LBGs of S2011, already an aperture of the size of the seeing disc will always contain a non-negligible fraction of the halo flux. Even if the \lya\ flux from such a tiny galaxy should experience net absorption over scales of the continuum-emitting region, the blurring by the PSF would merge this region with the inner halo, and one possibly always sees a \lya\ emitter. A case of marginally visible central absorption may be present in object ID\#95, which shows an inflection in the two innermost points of its \lya\ SB profile, which at a radius of 0.3\arcsec\ already encompass $\sim$2 continuum scale lengths (see Fig.~\ref{fig:ima+prof}).

In panel (c) of Fig.~\ref{fig:xfhstat} we also show the upper limits derived in Sect.~\ref{sec:sbmod-ul} for the five objects with no significantly detected \lya\ halo. This plot demonstrates conclusively that the non-detections are not highlighting objects with unusual properties; the halo fractions these objects are perfectly consistent with several cases of galaxies with significantly detected haloes.

\subsection{Central \lya\ equivalent widths}
\label{sec:lum-ew}

If we could achieve HST angular resolution also in \lya, we might find cases among our objects where the \lya\ SB goes to zero in the very centre, similar to the centrally absorbed LBGs in the S2011 sample. Given the seeing-limited data that we have, we take the parameter $F_\mathrm{cl}$ in our two-component model as a proxy for the \lya\ flux from the central region (where `central' here means, on scales of the UV continuum source). The rest-frame equivalent width EW$_\mathrm{cl}$ of the continuum-like component then provides a measure of central \lya\ strength. These values are also listed in Table~\ref{tab:table2}.

\begin{figure}
\includegraphics[width=\hsize]{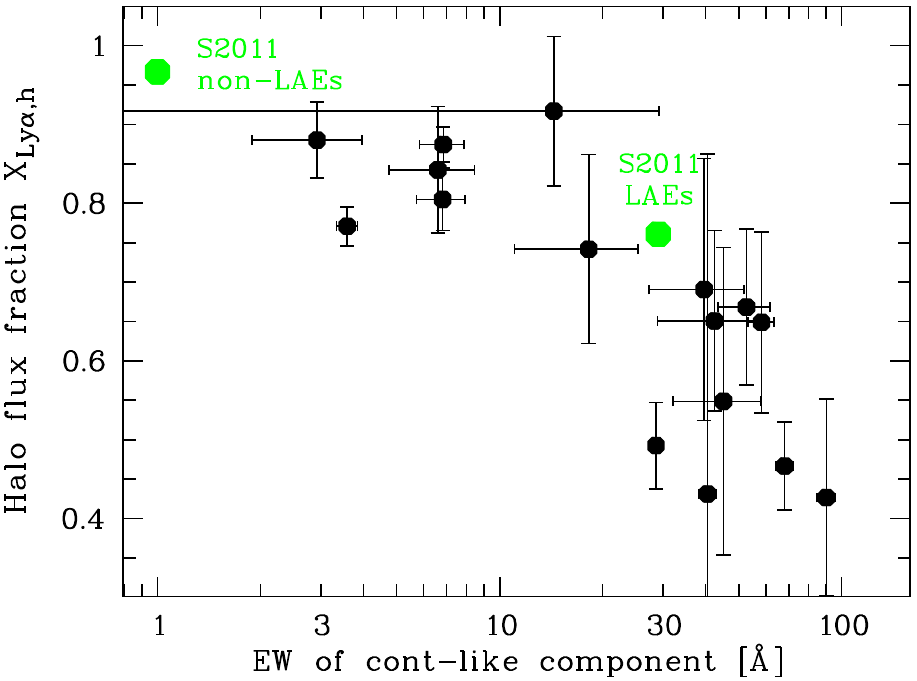}
\caption[]{\lya\ halo fractions plotted against rest frame equivalent widths of the continuum-like components. Only galaxies with detected HST counterparts are shown. The green filled circles show again the stacking results by \citet{Steidel:2011jk}.}
\label{fig:ewc-xfh}
\end{figure}

In Fig.~\ref{fig:ewc-xfh} we plot the \lya\ halo flux fraction $X_{\sublya,\mathrm{h}}$ against EW$_\mathrm{cl}$, revealing a clear anticorrelation: A high halo fraction of $X_{\sublya,\mathrm{h}} \ga 80$\% occurs \emph{only} when EW$_\mathrm{cl} < 20$~\AA, and vice versa. In other words, galaxies with weak central \lya\ emission have most of their total \lya\ luminosity in the halo, whereas strong central \lya\ implies that only $\sim 40$\%--50\% of the overall \lya\ flux comes from an extended region.

Essentially the same trend was found by S2011 by comparing small-aperture central spectra of LBGs to the large-scale \lya\ emission visible in their stacked NB images. Their LAE (non-LAE) subsample has a mean spectroscopic EW of 29~\AA\ (1\AA), while the corresponding halo fractions are 76\% (97\%), with $X_{\sublya,\mathrm{h}}$ estimated as described in the previous subsection. Including these points in Fig.~\ref{fig:ewc-xfh} shows that they line up very well with our individually measured values. This excellent agreement also supports our approach to use the simple 2-component fits to quantify the central \lya\ strength.

We conclude that there is a significant anticorrelation between the equivalent width of the central \lya\ emission and the \lya\ halo luminosity fraction in a galaxy, which  persists over a factor of at least 10 in galaxy sizes. The fact that faint LAEs and luminous LBGs appear to show such a scale independence in this behaviour will provide strong constraints on possible physical models of the extended \lya\ emission. We return to this point in the next section.

%%%%%%%%%%%%%%%%%%%%%%%%%%%%%%%%%%%%%%%%%%%%%%%%%%%%%%%%%%%%%%%%%%%%%%%%%%%%%%%%

%\input{LAE-ext_7-discussion}

\section{Discussion}
\label{sec:disc}

\subsection{Star formation rates and galaxy masses}
\label{sec:disc-galprop}

Of all the correlations investigated in the previous sections, the most convincing one is probably the fact that the \lya\ halo scale lengths of our high-$z$ galaxies are roughly proportional to the sizes of the stellar UV emission, with a proportionality factor of $\sim$10. This relation appears to be valid over the full luminosity and redshift range covered by our sample, and it furthermore extends to include the much larger and (on average) more luminous galaxies studied by S2011. 

In order to search for more fundamental relations it would be desirable to extend this analysis to stellar masses of our galaxies. This is however not possible with our present dataset. The HDFS is unfortunately rather poor in deep multiwavelength photometric coverage. In particular, only ground-based near-infrared data are available which do not nearly reach deep enough for our extremely continuum-faint sample. We therefore do not have a baseline to estimate individual  stellar masses or extinctions. While there are indeed better suited fields in the sky, we remind the reader that the MUSE observations of the HDFS were obtained while the instrument was still under commissioning, and the choice of this particular field was driven entirely by the accesible right ascension range at the time of observation.

In the few brightest galaxies where we can measure the shape of the UV continuum directly in the MUSE data, the spectral slopes come out close to the canonical dust-free value of $\beta = -2$ ($f_\lambda \propto \lambda^{\beta})$. Since the fainter galaxies in our sample are unlikely to be dust-richer, the extinction corrections are probably small in most objects.

Neglecting extinction, we can use both the observed \lya\ and the UV continuum to estimate star formation rates using the calibrations by \citet{Kennicutt:1998vu}. The resulting star formation rates are between 0.3 and 16~$M_\odot$~yr$^{-1}$ from the UV continuum (median 0.7~$M_\odot$~yr$^{-1}$, including the upper limits in the cases of the HST-undetected objects) and between 0.4 and 4.5~$M_\odot$~yr$^{-1}$ from the total \lya\ luminosities (median 1.2~$M_\odot$~yr$^{-1}$). 

If we assume that these galaxies form stars with \emph{specific} star formation rates in broad agreement with the cosmic average at those redshifts \citep[e.g.][]{Ilbert:2013dq,Salmon:2015iz}, we predict stellar masses in the range of $\sim$$10^8$--$10^9\:M_\odot$. We are thus dealing with galaxies much less massive than the $L^\star$ LBGs in the survey by S2011; in the local universe, most our objects would qualify as genuine dwarf galaxies. Using recent prescriptions for the relation between stellar and dark matter halo masses \citep{Moster:2010ep,Behroozi:2013fg} and assuming these to be valid at $z\simeq 3$--6, we estimate that our galaxies reside in dark matter haloes with masses $M_\mathrm{DMH}\la 10^{11}\:M_\odot$.

\subsection{Origin of the extended emission}
\label{sec:disc-orig}

In view of the evidence that essentially all star-forming galaxies at $z\ga 3$ appear to have \lya\ haloes much more extended than their stellar bodies, the question arises how this extended halo emission is generated. Since \lya\ photons are certainly produced in large numbers in the \ion{H}{ii} regions of each of these galaxies, and since these photons are prone to resonant scattering off neutral hydrogen atoms, the most natural explanation might be to assume that at least a good fraction of the total \lya\ luminosity has its origin inside the galaxy as recombination radiation from \ion{H}{ii} regions, which is then scattered outwards. As discussed already in the introduction, such scattering processes are very complex and difficult to model. Here we restrict ourselves to consider the viability of the basic hypothesis that the observed \lya\ emission may be powered by young stars.

We thus have to compare the measured \lya\ fluxes with the production rates of Lyman continuum (LyC) photons. Under the idealised conditions of `case B' recombination \citep[a medium optically thick to all Lyman series photons;][]{Baker:1938gh}, the intrinsic production rates should have a nearly constant ratio of $\mathcal{N}_{\sublya} = 0.68\,\mathcal{N}_\mathrm{LyC}$ \citep[for a derivation of this factor and its validity range see][]{Dijkstra:2014iq}. Since \lya\ photons may be destroyed by dust, the condition to be tested is an inequality: As long as $\mathcal{N}_{\sublya} \le 0.68\,\mathcal{N}_\mathrm{LyC}$, recombination radiation following stellar photoionization can do the job. However, predicting the total LyC photon production rate from observables requires several assumptions: The local radiation field must be specified, which in turn depends on the spectra of the hot stars (thus on metallicity), the adopted IMF, and on the star formation history; furthermore, the amounts of internal and external dust extinction go into the calculation. Once given the spectral shape of the UV continuum, one can convert the above inequality into a similar condition for the \lya\ equivalent width, EW$_{\sublya}$ $\le$ EW$_\mathrm{max}$, where EW$_\mathrm{max}$ depends on the various adopted ingredients. Since these are all quite uncertain, it is perhaps not surprising that different authors use different numbers for EW$_\mathrm{max}$. \citet{Charlot:1993cb} showed that for a range of typical conditions, the emerging maximum (dust-free) values for EW$_{\sublya}$ are between $\sim$50 and 200~\AA; this range is confirmed in the recent semianalytic models by \citet{Garel:2015fi}.

All our galaxies have measured values of (or lower limits to) the total \lya\ equivalent width EW$_\mathrm{t}$ that obey these bounds, apart from three objects with undetected HST counterparts and EW$_\mathrm{t} \ga 300$~\AA. Recall that EW$_\mathrm{t}$ includes the contribution of the extended halo to the integrated \lya\ flux. Energetically, we have thus no strong need to invoke other \lya\ generation mechanisms than recombination radiation from UV-bright star formation, at least for most of our galaxies, unless a significant fraction of this internally produced \lya\ radiation is destroyed by dust. As argued in the previous subsection, these low-mass galaxies are unlikely to be very dust-rich, although the lack of multiwavelength data for our sample currently inhibits a more definite statement. The few extreme EW objects are certainly interesting, and one possibility is very metal-poor underlying stellar populations \citep[e.g.,][]{Raiter:2010hs}. Another option might be that AGN radiation contributes to the \lya\ production, which in that case could easily reach much higher EWs. However, none of the usual AGN features (line broadening, high-ionisation lines) are visible in the spectra of these galaxies. There are also no sufficiently deep X-ray data for the HDFS to perform an independent check. We leave this issue open and conclude that in the vast majority of our galaxies, the stellar UV luminosities appear to be \emph{sufficient} to power the observed \lya\ emission, including the haloes.

In addition to star formation, other mechanisms could contribute to the \lya\ emission from the haloes, in particular gravitional cooling radiation or \lya\ fluorescence by the metagalactic UV background. While the latter is generally predicted to be far below our sensitivity threshold unless boosted by a nearby strong UV source such as a luminous quasar \citep{Cantalupo:2005hq,Kollmeier:2010cy}, \lya\ emitted by accreted intergalactic gas is actually an expected signature of galaxy growth \citep{Haiman:2000di,Furlanetto:2005jq}. Cooling radiation has been proposed as the main power source for the so-called \lya\ `blobs' \citep[e.g.][]{Fardal:2001ds,Goerdt:2010fp} and may also produce spatially extended \lya\ emission around normal high-redshift galaxies \citep[e.g.][]{FaucherGiguere:2010dd,Rosdahl:2012bt}. However, current predictions of the relevance of \lya\ cooling radiation in comparison to \lya\ powered by young stars depend on many assumptions and are still very uncertain.

\subsection{Implications for the demographics of \lya\ emitters}
\label{sec:disc-demo}

Most demographic studies of high-redshift galaxies are based on photometric or spectrophotometric measurements. It is worth spending a few thoughts on the consequences of the fact that much of the \lya\ emission of even a low-mass galaxy at $z>3$ comes from an area $\sim$25--100 times larger than the continuum-emitting region. 

A minor but still relevant effect is that the required larger flux-integration apertures make total \lya\ measurements more uncertain. Increasing an aperture from $r=1\arcsec$ to $r=3\arcsec$ will nearly double the enclosed flux for an average LAE (see Fig.~\ref{fig:gc}), but inflate the error bar by a factor 4--5 (mainly driven by uncertainties in the local background level); this error increase severely compromises the sensitivity for faint sources. Statistical aperture corrections are not much of an option, given the diversity of scale lengths and halo flux fractions in our sample. Each survey for \lya-emitting galaxies will have to negociate its own compromise between S/N and aperture losses.

The typical flux limit in wide-field narrowband imaging surveys for LAEs is around 1--$2 \times 10^{-17}$~\flcgs. LAEs from such surveys would already be among the brightest sources in our sample, and the expected aperture effects are therefore at least as pronounced as for our objects. While the photometric methods employed in past surveys varied a lot, systematic flux losses of the order of 50\% are easily conceivable, and actually unavoidable for the frequent choice of an integration aperture with $r = 1\arcsec$. This would lead to an offset of the observed luminosity function in horizontal direction by $\sim$0.3~dex from its `true' value, corresponding to \rev{a factor of several} in space density around L$^\star$, much more than the statistical error bars attained by modern surveys. 

Perhaps more worrying than such global offsets is the possibility for differential effects. The \lya\ haloes found in this study are not huge in absolute terms; their angular sizes are of the order of the seeing disc, which implies that aperture losses may depend strongly on the halo properties of an object. Since we have shown that the halo scale length is roughly proportional to the size of the continuum source, an intrinsically smaller galaxy suffers less aperture losses and therefore will automatically achieve a higher fraction of the total \lya\ emission inside the aperture -- an effect which will be especially strong for small apertures such as spectrograph masks. This can have several undesired consequences: 

(i) For any sample of continuum-selected galaxies, the fraction of objects above a certain \lya\ EW threshold will depend on the sizes of the galaxies investigated, which in turn are expected to scale with luminosities and stellar masses. This may lead to an apparent luminosity dependence in the fraction of \lya\ emitters (in the sense that fainter, hence smaller galaxies show a higher \lya\ fraction) even if that fraction was intrinsically constant when measured in \emph{total} luminosity. We do not claim that this bias alone can explain the observed luminosity dependence \citep[e.g.][]{Ando:2006jg,Pentericci:2009jo,Stark:2010hv}, but it goes in the same direction as the observed trend.

(ii) A decrease in size with increasing redshift of both haloes and UV continuum regions can lead to the same effect, namely boosting the apparent \lya\ fraction within a fixed aperture towards higher values at higher redshifts. Again, we do not wish to dispute the observed evidence for evolution in the \lya\ fraction \citep[e.g.][]{Stark:2010hv,Mallery:2012em,Cassata:2015gz}, but if the factor 2 for the size evolution between $z\sim 5$ and $z\sim 3$ suggested by Fig.~\ref{fig:rhalo-z} should be roughly correct, non-negligible biases for the apparent \lya\ fraction in spectroscopic surveys are conceivable.

(iii) Likewise, the determination of the luminosity and redshift dependence of \lya\ equivalent widths and of the \lya\ escape fraction in emission-selected LAE samples may be prone to such aperture biases, in the sense of artificially increasing the escape fractions for smaller (i.e., lower luminosity and higher redshift) galaxies. 

The magnitude of \lya\ aperture effects and possible biases in any particular demographic study of high-$z$ galaxies depend strongly on the details of the sample and on the setup of the observations, in particular the sizes of the spectroscopic apertures. A quantitative treatment of these effects is clearly beyond the scope of the present paper, but we alert the reader that some of the observed trends in the \lya\ properties of high-$z$ galaxies may require significant corrections for previously unsuspected aperture effects.

%%%%%%%%%%%%%%%%%%%%%%%%%%%%%%%%%%%%%%%%%%%%%%%%%%%%%%%%%%%%%%%%%%%%%%%%%%%%%%%%

%\input{LAE-ext_8-conclusions}

\section{Summary and conclusions}
\label{sec:concl}

In this paper we present MUSE observations of a sample of 26 \lya-emitting galaxies at $z = 3$--6 in the Hubble Deep Field South, revealing their extended diffuse \lya\ haloes on an individual basis, and allowing us for the first time to map these haloes and perform a quantitative study of their properties. The galaxies are mostly very faint in the continuum ($M_\mathrm{UV} \gtrsim -19$), corresponding to stellar masses in the range of $\sim$$10^8$--$10^9\:M_\odot$. The results of our analysis can be summarised as follows:

\begin{enumerate}\setlength{\itemsep}{0.5ex}

\item We detect significantly extended \lya\ emission in most objects of the sample. The extended nature of \lya\ is suggested already by the photometric growth curves in the extracted \lya\ pseudo-narrowband images (Fig.~\ref{fig:gc}), and confirmed by an analysis of the radial surface brightness profiles of the objects (Fig.~\ref{fig:ima+prof}). We tested the null hypothesis that there is no halo, i.e.\ that the spatial distribution of the \lya\ emission follows that of the stellar UV continuum. For 21 of the 26 objects, the null hypothesis is rejected with a probability $p_0 < 0.05$. For the galaxies with formally undetected haloes we derived upper limits to the halo fluxes (Fig.~\ref{fig:upperlim}), which show that most likely these objects are not lacking a halo, but that their total \lya\ emission is too faint for a significant detection.

\item The measured \lya\ surface brightness distribution of all galaxies can be approximated quite well by the superposition of two components, one that traces the shape of the compact UV continuum, and one that represents the extended halo by an exponential function, each convolved with the point spread function due to the atmospheric seeing (Fig.~\ref{fig:expfits}). We also explored one-component Sersic models as an alternative, but discarded this approach as it led to unphysical results in several objects. From the model fits we obtained exponential scale lengths and other size measures for the haloes, as well as the integrated \lya\ fluxes of the two components (Table~\ref{tab:table2}).

\item The detected \lya\ haloes have scale lengths between $\sim$1 and 7~kpc. The UV continuum scale lengths are mostly $< 1$~kpc, and several of the galaxies are even unresolved by HST, implying sizes $\la 300$~pc. The \lya\ haloes are therefore consistently much larger, by a factor $\sim$5--15 (median = 9.8), than the corresponding UV continuum regions. While our sample alone does not show any significant correlation between halo scale lengths and either \lya\ or UV luminosity, the combination with previous stacking results suggests that more luminous galaxies have larger haloes, albeit with a very large scatter (Figs.~\ref{fig:r0halo1} and \ref{fig:r0halo2}).

\item We find evidence for significant redshift evolution in the sizes of \lya\ haloes, with however very different trends for different redshifts domains. \emph{Within} our sample, haloes tend to be smaller at higher $z$, by a factor $\sim$2 between $z=3.7$ and $z=5.1$. A similar trend is observed for the UV continuum sizes of these galaxies, and thus the size \emph{ratios} are consistent with no evolution (Fig.~\ref{fig:rhalo-z}). On the other hand, a comparison with \lya\ haloes at $z\approx 0$ reveals the \lya\ emission around our high-redshift galaxies to be $\sim$5 times more extended, at fixed size in the UV continuum (Fig.~\ref{fig:rp20}).

\item Between 40\% and $\gtrsim$90\% of the total \lya\ emission is attributed to the extended halo. This halo luminosity fraction appears to be uncorrelated with almost any other observable currently available for our sample, also when considering previous stacking results (Fig.~\ref{fig:xfhstat}). The only exception is an anticorrelation between the halo luminosity fraction and the equivalent width of the continuum-like \lya\ component (Fig.~\ref{fig:ewc-xfh}).

\end{enumerate}

We show in this paper that galaxies at $z>3$ are essentially always surrounded by large amounts of circumgalactic gas. This is not completely surprising; absorption line studies with background sources close to foreground galaxies have already provided strong evidence for the existence of a significant circumgalactic medium around high-redshift galaxies \citep[e.g.][]{Lanzetta:1995hk,Chen:2001hj,Adelberger:2003fi,Steidel:2010go,Turner:2014gf}, although such studies have mostly not yet reached the redshift and galaxy mass regime of the present investigation. Furthermore, as summarised in the introduction, many previous observations reported indications that the \lya\ emission of galaxies is spatially extended \citep[e.g.][]{Moller:1998eo,Rauch:2008jy,Nilsson:2009ib}. Most notably, the stacking method \citep{Steidel:2011jk,Matsuda:2012fp,Feldmeier:2013fx,Momose:2014fe} has revealed \emph{average} properties of \lya\ haloes that are broadly consistent with our results for individual sources. 

The order-of-magnitude sensitivity gain in our data over narrowband imaging observations is remarkable. Authors using the stacking approach have recurringly expressed the view that mapping \emph{individual} \lya\ haloes around normal galaxies is currently not possible and, furthermore, that substantial improvements are out of reach for the present generation of telescopes. We demonstrated here that a state-of-the-art integral field spectrograph such as MUSE on the ESO-VLT already provides the sensitivity to detect diffuse \lya\ emission down to a 1$\sigma$ SB limit of $1\times 10^{-19}$~\sbl\ -- about the same depth as previously obtained in stacks of $\gtrsim$100 images. Perhaps even more important than the lowest detectable SB level is the achieved S/N of $\sim$10 at a SB of $10^{-18}$~\sbl\ in a radial distance of $1''$, enabling us to measure the properties of $z>3$ \lya\ haloes on an individual basis.

However, this study benefitted also from a number of other key improvements over previous endeavours: (i) The large field of view and spectral range of the MUSE instrument, essential for capturing many galaxies simultaneously in a single ultradeep pointing. (ii) Deep HST imaging observations covering the full field of view. (iii) Homogeneous sub-arcsec spatial resolution for all objects. (iv) A high-quality PSF calibrator star inside the field of view. Only the combination of all these factors equipped us with the means to go significantly beyond a mere detection experiment.

Our \lya\ halo images (Fig.~\ref{fig:lya-grey+red}) provide the first maps of the circumgalactic medium around individual normal (non-AGN) galaxies at $z>3$. While the origin of the extended \lya\ emission still needs to be established, it is clear that the circumgalactic gas must be at least partly neutral. Our images also show that the extended \lya\ emission is typically quite symmetric (albeit with interesting deviations in some objects) -- information which is necessarily erased in any stacking process. It will be intriguing to compare the \lya\ halo maps obtained by MUSE with simulated views from (possibly competing) model predictions.

In the context of our simple phenomenological model, two main parameters govern the overall appearance of a \lya\ halo at $z>3$: The halo scale length, typically $\sim$10 times larger than the scale length of the central continuum source, and the fraction of the total \lya\ radiation emitted by the halo, typically between 40\% and $>$90\%. These two quantities are essentially uncorrelated in our sample, which may indicate that they are physically determined at different scales. The size of a halo would then mainly depend on the spatial distribution and kinematic structure of the circumgalactic gas, independently of how much \lya\ it emits. On the other hand, the halo luminosity fraction will depend on the origin of the \lya\ radiation, but also on the importance of resonant scattering of \lya\ photons from inside the galaxy into the halo. 

Energetically, our measurements are consistent with the idea that most of the observed \lya\ radiation is produced within the UV-bright star-forming regions of the galaxies and then scattered outwards. However, additional processes for powering the \lya\ emission in the extended haloes are not excluded, and more theo\-retical and modelling work will be required to explore the relative importance of different mechanisms. Additional clues will come from spatially resolved spectroscopy of \lya\ haloes, which MUSE will enable us to perform at least for the brighter sources (or in strongly lensed objects where the magnification helps to reach even higher spatial resolution; see \citealt{Patricio:2015} for a first case study using MUSE).

By a remarkable coincidence, individual \lya\ haloes of nearby ($z\approx 0$) galaxies have just now become a subject of detailed scrutiny through the LARS project \citep{Hayes:2013jc,Ostlin:2014bs,Hayes:2014jv,Guaita:2015kr}. While the LARS galaxies can be resolved in much greater detail than our faint HDFS sources, the overall comparison of high- and low-redshift \lya\ halo properties shows some similarities, but also intriguing differences. In particular, the huge discrepancy between the sizes of \lya-emitting regions around $z>3$ and $z\approx 0$ galaxies (Fig.~\ref{fig:rp20}) suggests that the nature of the circum\-galactic medium has changed drastically from the epoch of galaxy formation to the present-day universe.

For the first time, MUSE has allowed us to assess the level of diversity in the \lya\ halo properties of high-redshift galaxies. While admittedly our sample is too small to go very far in this direction, much more can be expected in the future, as MUSE has just commenced operation. Furthermore, the MUSE data need to be complemented by observations with other instruments and in other wavelengths. In this respect our HDFS material is certainly not optimal, and we look forward to extend our studies towards other and richer regions in the sky.

%%%%%%%%%%%%%%%%%%%%%%%%%%%%%%%%%%%%%%%%%%%%%%%%%%%%%%%%%%%%%%%%%%%%%%%%%%%%%%%%
%%%%%%%%%%%%%%%%%%%%%%%%%%%%%%%%%%%%%%%%%%%%%%%%%%%%%%%%%%%%%%%%%%%%%%%%%%%%%%%%

\begin{acknowledgements}
The authors express their appreciation of the enthusiastic support provided by ESO staff during the MUSE commissioning activities. L.W., J.K. and T.U. acknowledge support by the Competitive Fund of the Leibniz Association through grants SAW-2013-AIP-4 and SAW-2015-AIP-2. R.B., J.B.C. and F.L. acknowledge funding from the ERC advanced grant 339659-MUSICOS. J.S. acknowledges support by the ERC starting grant 278594-GasAroundGalaxies. V.P. is supported by the ERC starting grant 336736-CALENDS. S.K. and P.M.W. acknowledge funding through BMBF Verbundforschung, grants 05A14BAC and 05A14MGA.
\end{acknowledgements}

\bibliographystyle{aa}
\bibliography{LAE-ext,husser,vera}

\end{document}